\documentclass{cernyrep} 
\pdfoutput=1

\usepackage{texnames}
\usepackage{wrapfig}
\usepackage[T1]{fontenc}
\usepackage{amsmath}
\usepackage{mathptmx}

\DeclareMathAlphabet{\mathitbf}{OML}{cmm}{b}{it}
\DeclareMathSymbol{\ell}{\mathalpha}{letters}{"60}
\DeclareMathSymbol{\gamma}{\mathalpha}{letters}{"0D}

\newcommand{\qbar}  {\ensuremath{\overline q}\xspace}
\newcommand{\fbar}  {\ensuremath{\bar f}\xspace}

\newcommand{\ubar}  {\ensuremath{\overline u}\xspace}

\newcommand{\dbar}  {\ensuremath{\overline d}\xspace}

\newcommand{\cbar}  {\ensuremath{\overline c}\xspace}

\newcommand{\bbar}  {\ensuremath{\overline b}\xspace}

\newcommand{\tbar}  {\ensuremath{\overline t}\xspace}
\newcommand{\ttbar} {\ensuremath{t\overline t}\xspace}
\newcommand{\ststbar}{\ensuremath{\tilde{t}\overline{\tilde{t}}}\xspace}
\mathchardef\Upsilon="7107
\def\Y#1S{\ensuremath{\Upsilon{(#1S)}}\xspace}

\newcommand{\CP} {\ensuremath{C\!P}\xspace}
\newcommand{\LambdaCutoff}{\ensuremath{\Lambda_{\mbox{\footnotesize cut-off}}}\xspace}

\newcommand\invpb{\ensuremath{\mathrm{\,pb}^{-1}}\xspace}
\newcommand\invfb{\ensuremath{\mathrm{\,fb}^{-1}}\xspace}
\def\Lumi#1#2{$10^{#1#2}$\,cm$^{-2}$s$^{-1}$\xspace}

\newcommand{\Kbar    }{\kern 0.2em\overline{\kern -0.2em K}{}\xspace}

\newcommand{\Kz      }{\ensuremath{K^0}\xspace}
\newcommand{\Kzb     }{\ensuremath{\Kbar^0}\xspace}
\newcommand{\KzKzb   }{\ensuremath{\Kz \kern -0.16em \Kzb}\xspace}
\newcommand{\Kp      }{\ensuremath{K^+}\xspace}
\newcommand{\Km      }{\ensuremath{K^-}\xspace}

\newcommand{\KpKm    }{\ensuremath{\Kp \kern -0.16em \Km}\xspace}

\newcommand\eps{\ensuremath{\varepsilon}\xspace}

\newcommand{\piz}{\ensuremath{\pi^0}\xspace}

\newcommand{\ee}{\ensuremath{e^+e^-}\xspace}

\newcommand{\ppb}{\ensuremath{p\overline{p}}\xspace}
\newcommand{\ppbar}{\ppb}

\newcommand{\cm}{\,cm\xspace}
\newcommand{\m}{\,m\xspace}
\newcommand{\mum}{\,$\mu$m\xspace}
\newcommand{\km}{\,km\xspace}
\newcommand{\mb}{\ensuremath{\mathrm{\,mb}}\xspace}
\newcommand{\pb}{\ensuremath{\mathrm{\,pb}}\xspace}
\newcommand{\nb}{\ensuremath{\mathrm{\,nb}}\xspace}
\newcommand{\ns}{\ensuremath{\mathrm{\,ns}}\xspace}
\newcommand{\tev}{\ensuremath{\mathrm{\,Te\kern -0.1em V}}\xspace}
\newcommand{\gev}{\ensuremath{\,{\rm Ge\kern -0.1em V}}\xspace}
\newcommand{\met}{\ensuremath{E_T^{\rm miss}}\xspace}
\newcommand{\Emiss}{\ensuremath{\not\!\!E}\xspace}
\newcommand{\ETmiss}{\ensuremath{\not\!\!E_T}\xspace}
\newcommand{\MeV}{Me\kern -0.1em V\xspace}
\newcommand{\GeV}{Ge\kern -0.1em V\xspace}
\newcommand{\TeV}{Te\kern -0.1em V\xspace}
\newcommand{\mev}{\ensuremath{\mathrm{\,Me\kern -0.1em V}}\xspace}
\newcommand{\kev}{\ensuremath{\mathrm{\,ke\kern -0.1em V}}\xspace}
\newcommand{\ev}{\ensuremath{\mathrm{\,e\kern -0.1em V}}\xspace}
\newcommand{\gevc}{\ensuremath{{\mathrm{\,Ge\kern -0.1em V\!/}c}}\xspace}
\newcommand{\mevc}{\ensuremath{{\mathrm{\,Me\kern -0.1em V\!/}c}}\xspace}
\newcommand{\gevcc}{\ensuremath{{\mathrm{\,Ge\kern -0.1em V\!/}c^2}}\xspace}
\newcommand{\mevcc}{\ensuremath{{\mathrm{\,Me\kern -0.1em V\!/}c^2}}\xspace}

\newcommand{\bei}{\begin{itemize}}
\newcommand{\eei}{\end{itemize}}
\newcommand{\ben}{\begin{enumerate}}
\newcommand{\een}{\end{enumerate}}
\newcommand{\beq}{\begin{equation}}
\newcommand{\eeq}{\end{equation}}
\newcommand{\beqn}{\begin{eqnarray}}
\newcommand{\eeqn}{\end{eqnarray}}
\newcommand{\beqns}{\begin{eqnarray*}}
\newcommand{\eeqns}{\end{eqnarray*}}

\newcommand{\ea}{\etal}
\renewcommand{\ie}{i.e.\xspace}
\renewcommand{\eg}{e.g.\xspace}
\def\@citex[#1]#2{\if@filesw\immediate\write\@auxout{\string\citation{#2}}\fi
  \@tempcnta\z@\@tempcntb\m@ne\def\@citea{}\@cite{\@for\@citeb:=#2\do
    {\@ifundefined
       {b@\@citeb}{\@citeo\@tempcntb\m@ne\@citea
        \def\@citea{,\penalty\@m\ }{\bf ?}\@warning
       {Citation `\@citeb' on page \thepage \space undefined}}%
    {\setbox\z@\hbox{\global\@tempcntc0\csname b@\@citeb\endcsname\relax}%
     \ifnum\@tempcntc=\z@ \@citeo\@tempcntb\m@ne
       \@citea\def\@citea{,\penalty\@m}
       \hbox{\csname b@\@citeb\endcsname}%
     \else
      \advance\@tempcntb\@ne
      \ifnum\@tempcntb=\@tempcntc
      \else\advance\@tempcntb\m@ne\@citeo
      \@tempcnta\@tempcntc\@tempcntb\@tempcntc\fi\fi}}\@citeo}{#1}}

\def\@citeo{\ifnum\@tempcnta>\@tempcntb\else\@citea
  \def\@citea{,\penalty\@m}%
  \ifnum\@tempcnta=\@tempcntb\the\@tempcnta\else
   {\advance\@tempcnta\@ne\ifnum\@tempcnta=\@tempcntb \else
\def\@citea{--}\fi
    \advance\@tempcnta\m@ne\the\@tempcnta\@citea\the\@tempcntb}\fi\fi}
\newenvironment{myquote}
               {\list{}{\leftmargin0cm\indent}%
                \item\relax}
               {\endlist}
\newcommand\allFontSize{\footnotesize}
\newcommand\detailsSize{\allFontSize}
\newenvironment{details}%
{\begin{myquote}\detailsSize}{\end{myquote}}

\begin{document}

\title{\LARGE Commissioning and early physics analysis with the ATLAS and CMS experiments  \\[0.3cm]
       {\em\normalsize Lecture notes, 2009}\thanks{Lecture notes from the 5$^{\rm th}$ Latin American School
of High-Energy Physics, Recinto Quirama, Antioquia Region, Colombia, March 15--28, 2009.}}
\date{\today} 
\author{Andreas Hoecker}
\institute{CERN, Geneva, Switzerland}

\maketitle 

\vspace{10cm}

{\small
\begin{abstract}
These lecture notes for graduate students and young postdocs 
introduce the commissioning and early physics programme of the 
high-transverse-momentum experiments ATLAS and CMS, operating 
at the Large Hadron Collider (LHC) at CERN. 

\end{abstract}
} 

\vspace{2cm}

{\footnotesize
\tableofcontents
} 
\newpage


\subsection*{\em Preface}

{\em 
This writeup of lectures given in March 2009 at the 5$^{\rm th}$ Latin American School
of High-Energy Physics, Recinto Quirama, Colombia,  provides an
overview of the various commissioning phases pursued by the ATLAS and CMS 
experiments to thoroughly prepare the detectors and data acquisition systems 
for physics. As an ATLAS member, 
the access to the relevant information from my own experiment was so invitingly 
easy that the document features an intolerable emphasis on ATLAS.
I can only sincerely apologise to my CMS colleagues, and state that changing 
all figures shown into the corresponding ones from CMS would not alter
the message the lectures seek to convey. In spite of their very different design,
ATLAS and CMS have similar physics potential. Wherever significant 
performance differences exist, they are pointed out throughout these notes. 
Most of the analyses discussed here are taken from the vast ATLAS 
and CMS detector, performance, and physics 
reports~\cite{atlasdetpaper,cmsdetpaper,atlascscbook,cmsphysicstdr}. No explicit
reference is given when using results from these papers. 
While finalising these notes, the LHC restarted the commissioning
programme in November 2009, after a year of repair and consolidation,
achieving for the first time proton--proton collisions at 900\,\GeV 
centre-of-mass injection energy, and --- for short periods ---
even the new world record energy of 2.36\,\TeV. Results from the analyses
of collision data, which were not available at the time of the lectures,  
are not included in these notes.
}

\newpage
\section{Motivation for a huge machine}
\label{sec:motivation}

\begin{wrapfigure}{R}{0.385\textwidth}
  \vspace{-15pt}
  \begin{center}
	  \includegraphics[width=0.385\textwidth]{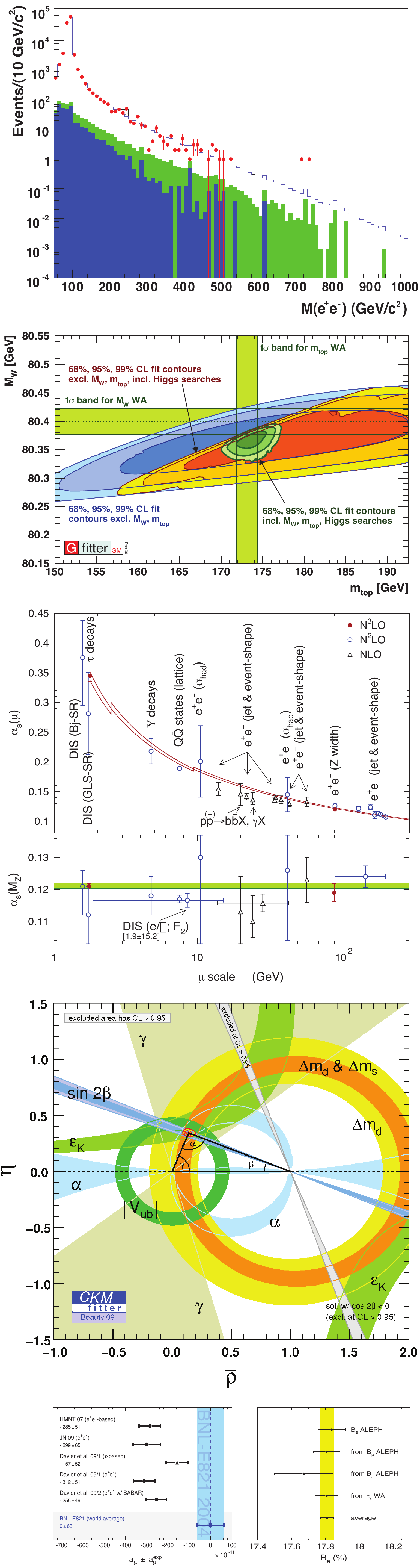}
  \end{center}
  \vspace{-15pt}
  \caption{Tests of the Standard Model.}
  \label{fig:smtests}
  \vspace{-20pt}
\end{wrapfigure}
The Large Hadron Collider (LHC) at the European Laboratory for Particle Physics 
Research (CERN) is the most powerful proton accelerator ever built. It 
collides two beams of protons accelerated to 7\,\TeV each and bent by dipole magnets 
with 8.3\,T magnetic field strength  within the 26.7\,km circular collider,
immersed in a ca. 100\,m deep tunnel between the Lake Geneva and the French 
Jura mountains. 
If the proton were an elementary particle, that is, if it were 
point-like, the 14\,\TeV centre-of-mass energy released 
by the collision could be fully transformed into mass. Dependent on quantum numbers 
and conservation laws (symmetries), for example one heavy particle of 14\,\TeV or a 
particle--antiparticle pair of 7\,\TeV each (masses of particles and antiparticles are 
identical) could be produced. Since the heaviest known particle is the top quark with 
mass of 173\,\GeV, any heavier particle found would be a discovery. These new heavy particles 
might decay to other new particles, still heavier than the top quark, and henceforth 
a full cascade of new particles could be discovered. 
The proton is, however, 
not an elementary particle, but is made out of a cloud of quarks and gluons (partons).  
The collision of two protons can thus be regarded as collisions between partons with 
momentum fractions that follow a density distribution with long tails towards one. 
Unlike for instance at \ee colliders, increasing the number of recorded 
collisions increases the probability for the occurrence of very hard parton scattering 
involving large fractions of the proton--proton centre-of-mass energy. A high-luminosity 
14\,\TeV proton--proton collider therefore allows the experiments to deeply explore the 
\TeV scale.

What does \TeV scale signify? Let us recall the relevant atomic, nuclear, and particle 
physics scales. The only known massless elementary particles are photons and gluons 
(bosons), which propagate the electromagnetic and strong forces, respectively. The 
lightest fermions are the neutrinos with masses probably lower than a few eV. 
This is below the atomic binding energy, which reaches tens of eV. The next orders 
of magnitudes are represented by the electron mass (1\,\MeV),  nuclear binding energy 
(up to 10\,\MeV),  pion and muon masses (100\,\MeV), the heaviest known lepton as well as 
proton, neutron, and vector-meson masses (1\,\GeV), the $c\cbar$ and $b\bbar$ resonances 
and heavy-quark mesons (10\,\GeV), and finally the electroweak unification scale, 
represented by the masses of the {\em Z} and {\em W} weak-interaction bosons, the top quark, 
and (presumably) the Higgs boson (100\,\GeV) and the Higgs vacuum expectation value 
(246\,\GeV). No particles beyond that scale are known to date. 

However, as we 
will see later, the requirement of a stable Higgs sector suggests the
existence of new phenomena at the \TeV scale, which is precisely the area of sensitivity 
of the LHC. Little is known beyond that scale. Will new symmetries arise, the breaking 
of which generates new particles? The seesaw mechanism accommodating massive neutrinos 
predicts heavy right-handed Majorana neutrinos of mass up to $10^{14}$\,\GeV. Unification 
of the electroweak and strong interactions may occur at $10^{16}$\,\GeV. Finally, 
gravitation becomes strong at the particle level at the Planck scale of order
$10^{18}$\,\GeV, requiring a quantum field theory that includes gravitation.
The minimal Standard Model (assuming massless neutrinos) of unified electroweak 
and strong interactions includes 19 free parameters, among which are 3 coupling 
constants, 1 spin-1 and 1 spin-0 boson mass, 9 fermion masses, 3 weak quark mixing 
angles, 1 \CP-violating weak phase, and 1 \CP-violating strong phase, which is 
either tiny or zero. Including a massive neutrino sector increases the number 
of free parameters by at least 9, depending on the nature of the neutrinos. 

The dynamical predictions of the Standard Model have been verified to extreme 
precision in the past thirty-five years at a large number of very different experiments. 
Let us recall a few eminent examples.
 The cross section of lepton pair production has been measured to order 1\,\TeV and 
      found in agreement with the {\em Z} resonance being the highest particle decaying 
      into two leptons, and Drell--Yan production being the dominant process beyond the {\em Z}
      (\cf topmost plot in Fig.~\ref{fig:smtests}~\cite{cdfd0dileptons}). 
 Electroweak unification has been tested by globally fitting the Standard Model 
      prediction to precision measurements 
      obtained at the high-energy \ee colliders LEP (CERN) and SLC (SLAC), and at the 
      $p\overline p$ collider Tevatron (FNAL). The second plot from the top in 
      Fig.~\ref{fig:smtests} shows the relation between measured and predicted 
      {\em W}-boson mass versus the top-quark mass~\cite{gfitter}.

 The universality of weak interactions has been verified at the 0.3\% level by 
      comparing the tau branching fractions to electron and muon plus neutrinos and 
      to the tau-lepton lifetime (\cf bottom left plot in Fig.~\ref{fig:smtests}~\cite{alephtaubrs}). 
 The asymptotic freedom property of QCD has been verified at the 1\% level by 
      measuring the evolution (`running') of the strong coupling at various energy 
      scales, the most precise of which being the ones at the $\tau$ and the {\em Z} mass 
      scales (third plot in Fig.~\ref{fig:smtests}~\cite{davieralphas}).
 The Standard Model predicts that all \CP-violating phenomena involving weak charged currents 
      originate from a single phase in the quark mixing matrix. This has been verified 
      by relating different measurements of \CP violation in the {\em B}-meson and 
      kaon sectors to each other, all showing compatibility 
      (fourth plot in Fig.~\ref{fig:smtests}~\cite{ckmfitter}).
 The \CP-violating electric dipole moment of the electron has been found to 
      be smaller than $10^{-27}$\,ecm as predicted by the Standard Model.
 The anomalous magnetic moment of the muon has been measured to the parts-per-million 
      level, verifying the predicted contributions from electromagnetic, weak, and 
      hadronic loop corrections. A small deviation from the expectation is currently
      not at a sufficiently significant level to draw conclusions 
      (\cf bottom right plot in Fig.~\ref{fig:smtests}~\cite{pdgg-2review}). 
Many more examples all confirm the Standard Model. So, what's the problem?

As explained in much detail by John Ellis~\cite{johnellis} and others at this school, the 
Standard Model --- though describing so gloriously the experimental data --- is, at best, 
incomplete. Firstly, the Higgs boson, the last elusive Standard Model ingredient, has not 
yet been discovered. Even if it were discovered, it would be the only elementary scalar 
particle in the Standard Model, which ---  for many physicists --- is conceptually 
unsatisfactory. A popular question is the origin of the large mass hierarchy 
between fermions of different generations, amounting to more than 4 orders of 
magnitude between top and up quarks. Many astrophysical observations have 
established the presence of cold dark matter in the galaxies and galactic halos. 
Moreover, spurious repulsive `dark energy' appears to accelerate the expansion of 
the universe. In particle physics, we can use the standard quantum field theory 
renormalisation groups to predict the energy-scale dependence of the electroweak and 
strong coupling constants. Evolving the three couplings to $10^{16}$\,\GeV, they 
{\em almost} converge towards a single unified coupling --- almost, but not quite. 
While unification might be considered an aesthetic requirement, stability of the 
Higgs sector is not. Indeed, the virtual loop corrections, in particular from 
top-pair vacuum polarisation, diverge quadratically with their high-energy cut-off. 
Also, perturbativity of the Higgs quartic coupling and stability of the Higgs 
potential require the Higgs mass to lie within a small allowed window, if the 
Standard Model is to survive up to the (reduced) Planck scale 
$M_P \simeq 2\cdot10^{18}$\,\GeV. Moreover, how would the unification of the 
Standard Model and gravitation be established at that scale? A subtle, but no 
less intriguing problem is the apparent smallness of the strong-\CP parameter, 
tightly bound from measurements of the neutron electric dipole moment, although 
no mechanism such as a symmetry in the Standard Model suggests such a small or 
even vanishing value. While the Standard Model features \CP violation in the 
charged weak current, theoretical calculations show that the amount of
\CP violation is insufficient by many orders of magnitude to be at the origin 
of the matter--antimatter asymmetry currently observed in the visible part 
of the universe.\footnote
{
   We could thus ask ourselves what the role of the weak phase is in the  
   evolution of the universe. Does it carry a hidden purpose? Or is weak \CP violation
   a meaningless `accident of nature': because there are three generations
   and because all quark flavours have mass there is quark mixing with four 
   parameters of which three are three Euler angles and one is a \CP-violating
   phase. The phase is not constrained by a symmetry and thus of order one 
   ($68^\circ$~\cite{ckmfitter}). Perhaps without major implications for nature. 
}

The instability of the mass of the scalar Higgs boson against radiative corrections
is denoted by the term `gauge hierarchy problem', which also sets the scale at 
which new physics can be expected. It is --- beyond the Higgs discovery and the 
strong Standard Model research programme --- a primary motivation for the 
construction of the LHC. 
Indeed, if a Higgs boson with mass $<$1\,\TeV is discovered, the Standard Model 
is complete. However, when computing radiative corrections to the Higgs propagator, 
modifying the bare Higgs mass, such as $t\tbar$ vacuum polarisation diagrams, 
or boson self-energies including the Higgs self coupling, the corresponding loop 
integrals diverge. To solve them, a cut-off parameter \LambdaCutoff is 
introduced to which the integrals are quadratically proportional. The cut-off 
parameter sets the scale where new particles and physical laws must come in, 
regularising the diverging integral.\footnote
{
   In a renormalisable quantum field theory, divergences in single loop 
   integrals frequently occur, but they are always cancelled to all perturbative
   orders by other diagrams contributing to the full matrix element of the 
   scattering process under study.    
} 
However, above the electroweak scale we know only of two scales exhibiting 
new physics: grand unification of the electroweak 
and strong forces ($\approx$$10^{16}$\,\GeV) and the Planck scale. 
A cut-off at such large energies would require an enormous amount of 
fine-tuning to keep the physical Higgs mass small and stable.  What could be a 
`natural' value for the scale \LambdaCutoff? The following three diagrams 
give the largest contributions to the Higgs radiative corrections and hence to 
the physical Higgs mass: $t\tbar$ loop: 
$-(3/8\pi^2)\lambda_t^2\LambdaCutoff^2\approx(2\,{\rm \TeV})^2$; gauge-boson loop: 
$(9/64\pi^2)g^2\LambdaCutoff^2\approx(0.7\,{\rm \TeV})^2$; and Higgs loop: 
$-(1/16\pi^2)\lambda^2\LambdaCutoff^2\approx(0.5\,{\rm \TeV})^2 $, where we have 
used $\LambdaCutoff=10$\,\TeV everywhere, and where $\lambda_t$, $g$, $\lambda$
are respectively CKM, weak, and quartic Higgs couplings. The total mass-squared of the 
Higgs is the sum of these contributions and the tree-level term. What would be 
the cut-off (= new physics) scales if only small ($\sim$10\%) fine-tuning were allowed? 
We would find $\Lambda_{\rm top} < 2$\,\TeV, 	 $\Lambda_{\rm gauge}< 5$\,\TeV, and 
$\Lambda_{\rm Higgs} < 10$\,\TeV. To naturally cancel these divergences, new 
physics at the \TeV scale should couple to the Higgs and should be related to 
the particles in the loop (top, gauge, Higgs) by some symmetry.

The gauge hierarchy problem denotes this fine-tuning of parameters, and the strong 
dependence of physics at the weak scale on the physics at (presumably) much higher 
scale: if the Higgs radiative corrections are cut off at the scale of gravity, 
why is the scale of electroweak symmetry breaking so different from the scale of 
gravity? Why is $m_W \ll M_P$? Equivalently, why is gravity so weak? Possible 
solutions to the hierarchy problem include: $(i)$ new physics appears not much above 
the electroweak scale and regularises the quadratic divergences, $(ii)$ new physics modifies 
the running of the couplings, approaching grand unification to the electroweak scale, 
$(iii)$ gravity is not as weak as we think, it is only diluted in our four-dimensional 
world but it is as strong as electroweak interactions in, \eg, five or more 
dimensions with Planck scale $M_P^{(5D)} ~ {\cal O}($\TeV$)$, or $(iv)$ the theory 
is fine-tuned and the explanation for the parameter values is statistical rather 
than dynamic (anthropic principle).

From the above discussion we retain that the Standard Model is in crisis. Most 
Standard Model extensions, developed with the goal to solve the hierarchy problem 
and/or to provide a dark matter candidate,
introduce new particles at the \TeV scale. To find these, we need a new, huge 
collider providing hard particle collisions with centre-of-mass energy well above 
1\,\TeV.


\section{The Large Hadron Collider}
\label{sec:lhc}

In principle, one could accelerate protons circulating in a magnetic ring
almost illimitably to higher and higher energy by continuously passing them 
through a radio-frequency field. The energy loss through synchrotron radiation of a 
proton in the Large Hadron Collider (LHC) amounts to a few keV per turn (compared 
to a few \GeV per turn for electrons in the $\ee$ collider LEP2), which is about 
one hundred times smaller than the acceleration the proton receives per turn. In 
practise however, the proton energy 
in the collider ring is limited by the superconducting dipole magnets that guide 
the circular beams: $E_{\rm proton} \simeq 0.3\cdot B\cdot r$. Because the 
radius of the LHC is fixed ($r=4.3$\km), one must use as strong fields as possible 
(8.3\,T, compared to approximately 4\,T at the HERA and Tevatron colliders), 
and fill all free LHC sections with dipole magnets ($\approx$2/3).\footnote
{
   More precisely, the total number of dipole magnets in the LHC is 1232, each 
   of which has a magnetic length of 14.3\,m, giving a total length of 17\,618\m. The 
   effective `bending radius' amounts thus to: $17\,618/(2\pi) = 2804$\m, and 
   hence $E_{\rm proton} \simeq 0.3\cdot B\cdot r\approx 7$\,\TeV.
}
Because the effective centre-of-mass energy of the hard parton collision depends
on the parton energy density distributions in the proton, with long tails towards 
a large energy fraction, accumulating larger statistics due to a high instantaneous 
luminosity effectively increases the available kinematic reach of the proton--proton 
collider. High luminosity (beyond \Lumi33), and good machine and data-taking 
efficiency (of the order of $10^7$ seconds good-quality data taking per year), are
also required to search for rare events, such as processes involving the Higgs boson, 
especially if the Higgs is light (Higgs production is an electroweak process with large
momentum transfer, which has a cross section roughly a billion times smaller than 
inelastic QCD (so-called `minimum bias') processes), and also for studies of the 
nature of new physics phenomena if discovered. To achieve high 
luminosity ($L$), strong currents are necessary, requiring dense proton bunches 
containing up to $N=110$ billion protons each (for comparison: $1$\,cm$^3$ of hydrogen contains 
$\approx$$10^{19}$ protons), and as many LHC bunches ($k$) as possible filled with protons
(maximum of $k=2808$ bunches out of a total of 3564 bunches). The bunches are spaced by 
25\ns from each other, corresponding to a distance of 7.5\,m. High luminosity also 
requires that the protons be transversely squeezed by magnetic lenses to a small spot to 
increase the probability that two protons collide. The typical transverse beam size, 
determined by the square-root of the product of an amplitude function characterising 
the beam optics (varying throughout the ring), and the constant phase space volume 
(emittance), amounts to $\sigma_x^\star=\sigma_y^\star=16$\mum at 7\,\TeV 
beam energy (for smaller beam energies, the beam emittance increases with 
$\eps\propto 1/\gamma_p$, as does the beam spot size as 
$\propto \sqrt{\eps}$).\footnote
{
   The free `volume' occupied by each proton in the interaction point is of the 
   order of $10^{-4}\,\mu{\rm m}^3$, which is huge compared to the size of 
   an atom, so that strong-interaction collisions between protons are still rare. 
   The probability of two protons colliding can be estimated to be approximately 
   $4\cdot 10^{-21}$, so that with $1.1\cdot 10^{10}$ protons per bunch one 
   finds $\approx50$ interactions per bunch crossing, of which, however, only 
   one-half are inelastic. 
} 
The luminosity value is obtained from the formula 
\beq
     L = \frac{k N^2 f}{4 \pi \sigma_x^\star\sigma_y^\star}\,,
\eeq
where $f=11.25\,{\rm kHz}$ is the revolution frequency determined by the LHC 
circumference and the speed of light of the protons. We thus obtain $L=3.5\cdot\,$\Lumi30
per bunch, reaching \Lumi34 when all bunches are filled.  

\begin{figure}[t]
  \begin{center}
	  \includegraphics[width=0.55\textwidth]{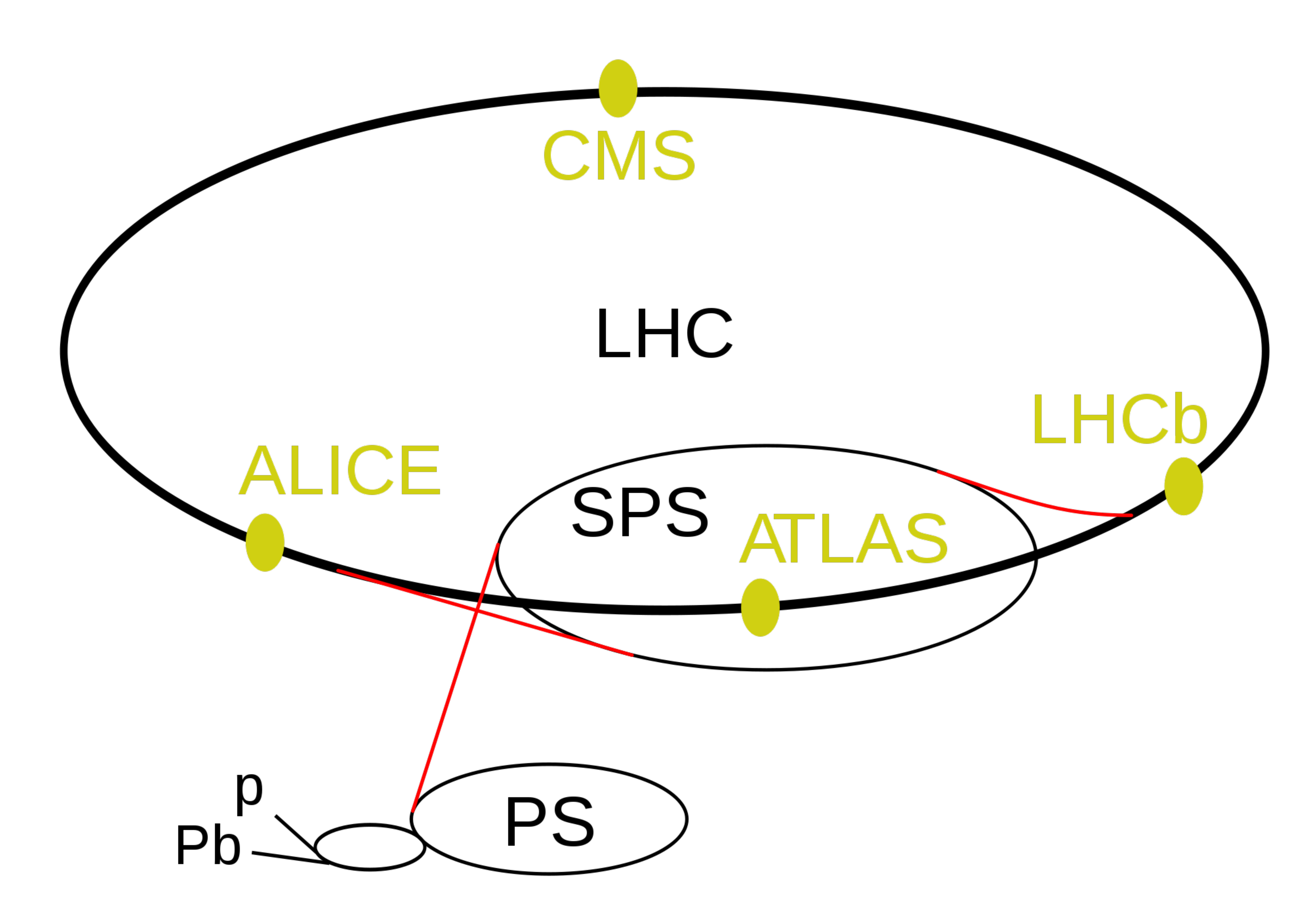}
  \end{center}
  \vspace{-0.5cm}
  \caption[.]{Schematic view of the main elements of the LHC accelerator complex 
                   (see text) and the location of the four largest LHC experiments 
                   ALICE, ATLAS, CMS and LHCb.} 
\label{fig:lhccomplex}
\end{figure}

The LHC acceleration chain involves several steps (see Fig.~\ref{fig:lhccomplex} for 
a schematic view). The injector complex
consists of the LINAC-2, preaccelerating the protons to 50\,\MeV, followed by 
the Proton Synchrotron Booster (PSB) consisting of four superimposed rings accelerating 
the protons to 1.4\,\GeV. Two large circular rings further accelerate the protons
to 26\,\GeV (Proton Synchrotron -- PS) and 450\,\GeV (Super Proton Synchrotron -- SPS),
which is the LHC injection energy. The beams are transferred from the SPS to 
the LHC via two newly built 3\km transfer lines. The PSB--PS--SPS complex 
required significant upgrades to be able to provide beams with the appropriate 
intensity, size, and bunch distance. The injection chain is particularly delicate
because any increase of beam emittance during injection will be `remembered' by 
the protons in the LHC and lead to a reduction of the available peak luminosity 
and/or beam lifetime (thus increasing beam-related backgrounds and reducing the 
integrated luminosity the LHC can deliver during a proton fill). We note that 
in each acceleration step, the energy increase lies between a factor of 10 and 20, 
which are reasonable ranges for the dipole magnets. The injector also 
has the task of creating the proton bunches and (fixed) bunch pattern for the 
LHC. The chain is as follows: 6 booster bunches are injected into the PS; 
each of these is split into 12 smaller bunches giving a total of 72 bunches 
at extraction; between 2 and 4 batches of 72 bunches are injected into 
the SPS giving from 144 up to 288 bunches; finally, a sequence of 12 
extractions of (up to) 288 SPS bunches is injected into the LHC, giving a 
maximum of 2808 bunches (39 groups of 72 bunches). The filling scheme (difference
between the 3564 possible and 2808 actually filled bunches) foresees
a number of short gaps for, \eg, kicker magnet rise times in the 
injection chain, and one long gap of 119 empty bunches ($3\,\mu{\rm s}$) for 
the rise time of the LHC beam dump kicker magnet. Once injected into the LHC, 
the protons are accelerated from 450\,\GeV to 7\,\TeV in a 20-minute
acceleration process, during which the protons receive an average 
energy gain of 0.5\,\MeV per turn when passing the electrical radio-frequency (RF)
fields created in 8 superconducting cavities per beam with a peak accelerating
voltage of 16\,MV.

\begin{figure}[t]
  \begin{center}
	  \includegraphics[width=0.90\textwidth]{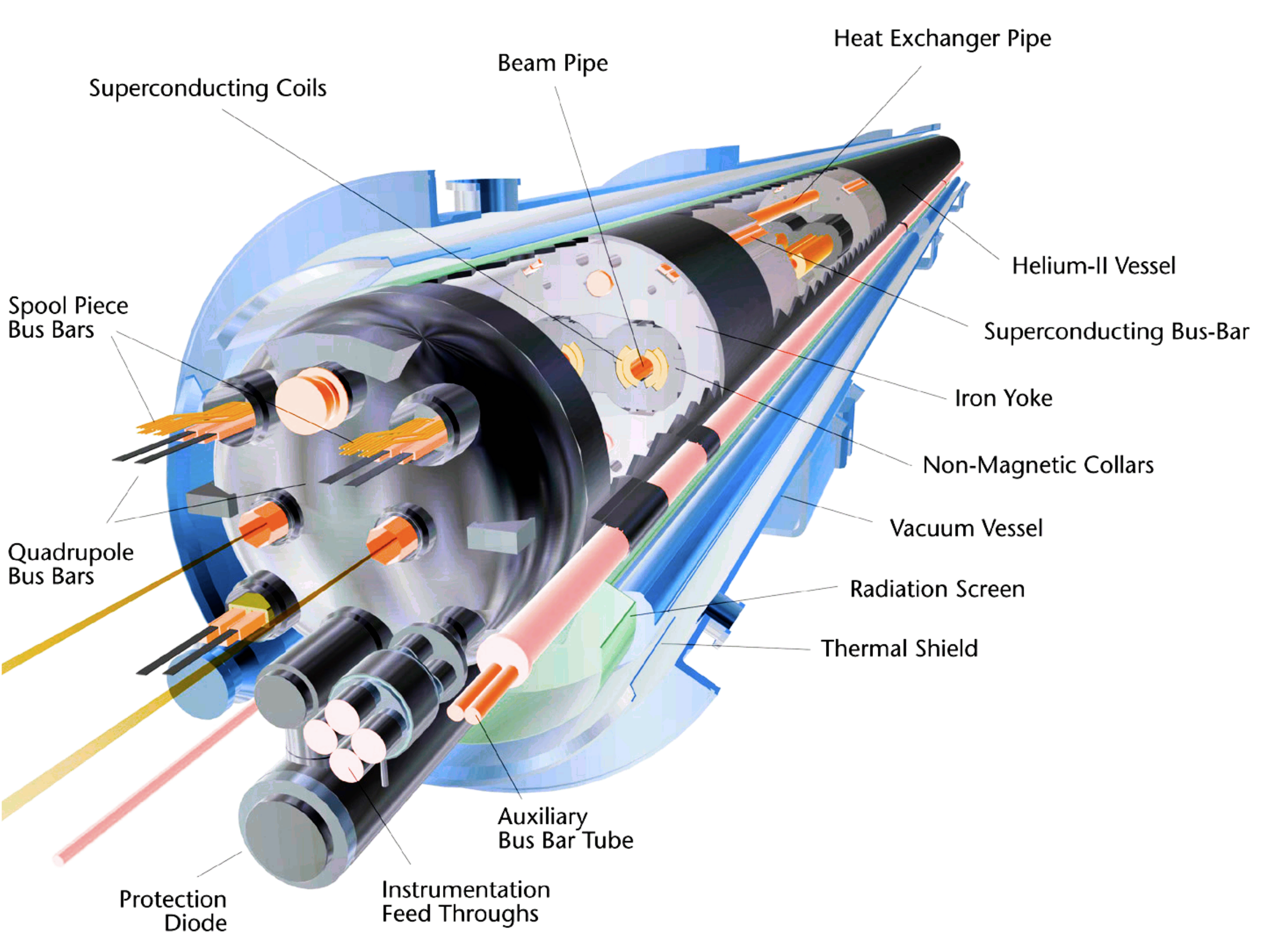}
  \end{center}
  \vspace{-0.3cm}
  \caption[.]{Section view of a superconducting LHC dipole magnetic. The two beam 
                  pipes are wrapped into two oppositely poled superconducting coils. }
\label{fig:lhcdipoles}
\end{figure}
The LHC consists of eight 2.45-km-long arcs with bending dipole magnets (see 
Fig.~\ref{fig:lhcdipoles} for a schematic drawing of a dipole section),\footnote
{
   The LHC magnet systems consists of a total of 1232 superconducting dipoles
   (cooled with 120 tons of superfluid Helium down to 1.9\,K), in 
   which currents of 12\,kA create the required 8.33\,T magnetic field; 
   392 focusing quadrupoles; and 3700 multipole corrector magnets.
} 
and eight 545-m-long straight sections. Four particle detectors have been constructed 
and are housed in huge underground caverns located at four of the straight 
sections. They record the objects left by collision debris by interacting
with them. The detectors are: 
ATLAS (A Toroidal LHC ApparatuS), 
CMS (the Compact Muon Solenoid), 
ALICE (A Large Ion Collider Experiment), and 
LHCb (study of physics in {\em B}-meson decays at the LHC).\footnote
{
   In addition, there are two smaller experiments: 
   TOTEM (Total Cross Section, Elastic Scattering and Diffraction Dissociation at the LHC) 
   and LHCf (Large Hadron Collider forward) for very low-$p_T$ physics. 
}
The remaining four straight sections are used by the RF cavities, the beam dump,
and by two beam-cleaning systems using chains of collimators to absorb off-beam 
protons that would provoke magnet quenches and create so-called beam-halo backgrounds
in the experiments. Although the energy of a single 7\,\TeV proton corresponds to 
only that of a flying mosquito (1\,$\mu$J), the total stored energy of 2808 bunches 
each filled with $10^{11}$ 7\,\TeV protons amounts to 360\,MJ.\footnote
{
   The stored energy is sufficient to heat up and melt 12 tons of copper. It is 
   equivalent to an Airbus A380 flying at 700\,km/h speed, to 90\,kg of TNT, 
   8 litres of gasoline, or 15\,kg of chocolate.   
} 
It is a huge challenge to control this energy and avoid damage to accelerator 
and experiments.


\section{The high-$p_T$ general-purpose detectors ATLAS and CMS}
\label{sec:detectors}

The broad range of physics opportunities and the demanding experimental environment 
at high-luminosity 14\,TeV proton--proton collisions impose unprecedented performance 
requirements and technological constraints upon the LHC particle detectors. 
ATLAS and CMS are general-purpose detectors, capable of adequately covering the 
entire physics programme reachable with high-luminosity 14\,\TeV proton--proton 
collisions: from charm and beauty physics at lowest transverse momenta ($\sim$3\,\GeV),
to new physics searches up to the highest reachable scales ($\sim$4\,\TeV). 
The cross sections of the dominant QCD processes and those representing the primary 
physics channels for research differ by many orders of magnitude. For example, while
at 14\,\TeV centre-of-mass energy, the total inelastic $pp$ cross section amounts
to approximately 70\,mb (giving a 1\,GHz event rate at $L=\,$\Lumi34),\footnote
{
   Recall that $1\,\mb^{-1}=10^{27}$\,cm$^{-2}$.
} 
hard quark and gluon scattering into pairs of jets (or more) occurs roughly a 
thousand times less frequently; inclusive $b$-hadron production has a
cross section of approximately 0.5\,mb; 
inclusive $W\to\ell\nu$ and $Z\to\ell\ell$ boson production and decay have cross 
sections times branching fractions of approximately 20\,nb and 2\,\nb,\footnote
{
   Because of the proton quark structure, producing more $u\dbar$ than $\ubar d$ 
   quarks in scattering reactions, roughly a quarter more $W^+$ than $W^-$ are 
   produced at the LHC~\cite{hep-ph/9907231} (while equal amounts of both charges 
   are produced at the \CP symmetric Tevatron collider).

} 
respectively (compared to roughly a factor of 8 smaller at the Tevatron);
top and antitop production has a cross section of almost 1\,nb (rate of 10\,Hz),
two orders of magnitude higher than at the Tevatron; inclusive Higgs-boson production, 
dominated by gluon-gluon-to-Higgs fusion via a triangular top-quark loop,
has a Higgs-mass dependent cross section between 45\,\pb ($m_H=120$\,\GeV) and 
20\,pb (180\,\GeV); and the production via gluon--gluon scattering of 1\,\TeV 
supersymmetric squarks and gluinos has a cross section of a few pb. These vast 
disparities, rendering physics analysis at the LHC like searching for needles in 
a giant haystack, drive the detector design. 

Let us list some of the most outstanding LHC conditions and derive from these 
the corresponding design challenges.
\begin{itemize}

\item  The 40\,MHz bunch crossing rate\footnote
       {
         For comparison, the bunch crossing rates at LEP and the Tevatron are 45\,kHz 
         and 2.5\,MHz, respectively, while the {\em B} factory PEP-II, an \ee 
         collider, has achieved 240\,MHz, and the CLIC design foresees 2\,GHz.
         }
       requires a fast trigger decision, precise timing 
       and `pipeline' electronics, locally storing readout data until the Level-1 
       (hardware) trigger response signal has been derived. For a pipeline memory depth
       of 100 bunch crossings, the Level-1 trigger latency must not exceed 2.5\,$\mu$s.

\item  The interaction rate of up to 1\,GHz at maximum peak luminosity of \Lumi34
       (LEP and Tevatron: $L_{\rm max}=$\,\Lumi32 and $3.5\cdot\,$\Lumi32, 
       respectively), corresponding 
       to approximately 25 inelastic interactions piling up in a single collision event,
       requires efficient pattern recognition to reduce the event rate from  
       1\,GHz to 75\,kHz (Level-1 output, high-level trigger input) to approximately 
       200\,Hz (HLT output rate, events written to disk). 

\item  The approximate data size of 1.5\,MB per event together with the 200\,Hz accepted
       trigger rate provides an average raw data throughput of 300\,MB per second. 
       Storage, worldwide distribution, prompt reconstruction and reprocessing of these 
       data require adequate storage media, and powerful network and computing resources. 
       The paradigm of distributed computing chosen by the LHC experiments requires the 
       availability of several (order 10) large-scale computing centres (Tier-1s,
       demarcating `computing clouds'), with 
       resources similar to those at CERN, and located representatively for the 
       collaborations' geographical extensions. These clouds embrace smaller 
       computing centres for user analysis and simulation production. 

\item  The irradiation rate after 10 years of successful LHC operation is expected
       to reach $5\cdot10^{14}$ neutron equivalents per cm$^2$ (300\,kGy), requiring
       radiation-hard inner tracker (pixel detector with large signal-to-noise ratio
       and small silicon volume close to the interaction point) and forward 
       calorimeter technology.

\item  The high charged multiplicity of up to 1000 tracks per event ($4\cdot10^{10}$ tracks 
       per second) requires the use of high-granular pixel/silicon or fine-grained straw 
       tracker technologies. Three-dimensional pixel technology, replacing traditional 
       silicon strip detectors close to the beam pipe, is mandatory to provide sufficient
       pattern recognition capability. 

\item  Large background rates from beam-gas interactions, beam-halo muons, 
       thermal neutrons and photons (`cavern background', bathing the detector 
       during event pileup and afterwards due to activation of materials in the 
       detector, its support structure, and the cavern), require 
       precise muon timing, redundant pattern recognition, and radiation hardness.

\end{itemize}

Similarly, the detector design reflects the challenges posed by the physics programme. 
\begin{itemize}

\item  The search for rare {\em B}$_{s(d)}\to\mu\mu$ decays, which have Standard Model branching
       fractions of $3.3\cdot10^{-9}$ and $1.1\cdot10^{-12}$, respectively, and the 
       measurement of time-dependent \CP violation and the unitarity triangle angle 
       $\beta_s$ using (among others) flavour-tagged {\em B}$_s\to J/\psi\phi$ decays, 
       require good trigger efficiency and purity for muon tracks with transverse 
       momenta as low as 3\,\GeV. To achieve sufficient purity, the HLT tracking algorithm
       must reconstruct charges as well as the {\em B} vertex and mass.

\item  Measuring the $W$ mass to a precision better than the current world 
       average~\cite{wmassave} of 
       $(80.399\pm0.023)$\,\GeV, requires excellent alignment of the tracking
       detectors, good track reconstruction efficiency, calorimeter uniformity, and 
       missing transverse energy resolution.

\item  A precision measurement of the top mass needs --- apart from a better 
       theoretical understanding of the nature of the measured top mass --- excellent 
       jet energy calibration, resolution and uniformity, as well as excellent 
       $b$-tagging purity and efficiency. 

\item  A sensitive search for the Higgs boson in the most promising final states
       $2e(\mu)2\nu$, $4e(\mu)$, $2e2\mu$, $\gamma\gamma$, $\tau\tau$ (via weak 
       boson fusion accompanied by forward jets) requires very pure and efficient particle 
       identification, excellent electromagnetic and hadronic calorimeter resolution 
       and uniformity, efficient high-luminosity tracking, and efficient 
       reconstruction of forward jets.

\item  Searching for the multifaceted signatures from supersymmetry requires excellent 
       jet and missing transverse energy resolution, low calorimeter noise, excellent 
       $\tau$ identification and reconstruction, as well as maximum detector acceptance.

\item  The search for heavy resonances of masses beyond 1\,\TeV, as they are predicted
       in models with excited weak bosons or extra spatial dimensions, requires
       good tracking (including charge reconstruction) and calorimeter resolution, 
       and a large dynamic range (small calorimeter saturation) up to the highest 
       reachable energies.

\end{itemize}

\subsection{Detector design}

\begin{figure}[t]
  \begin{center}
	  \includegraphics[width=0.99\textwidth]{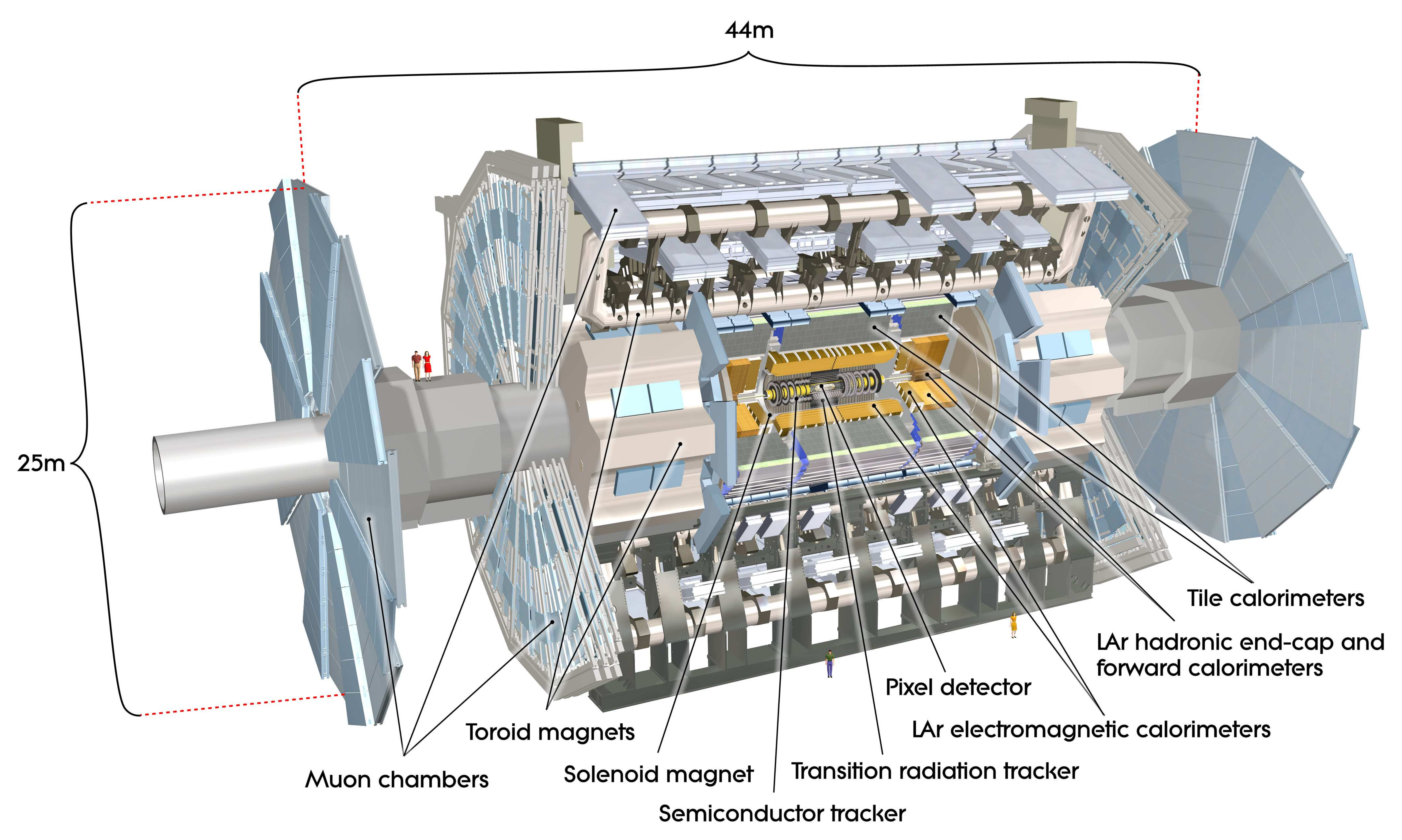}
     \vspace{0.2cm}

	  \includegraphics[width=0.99\textwidth]{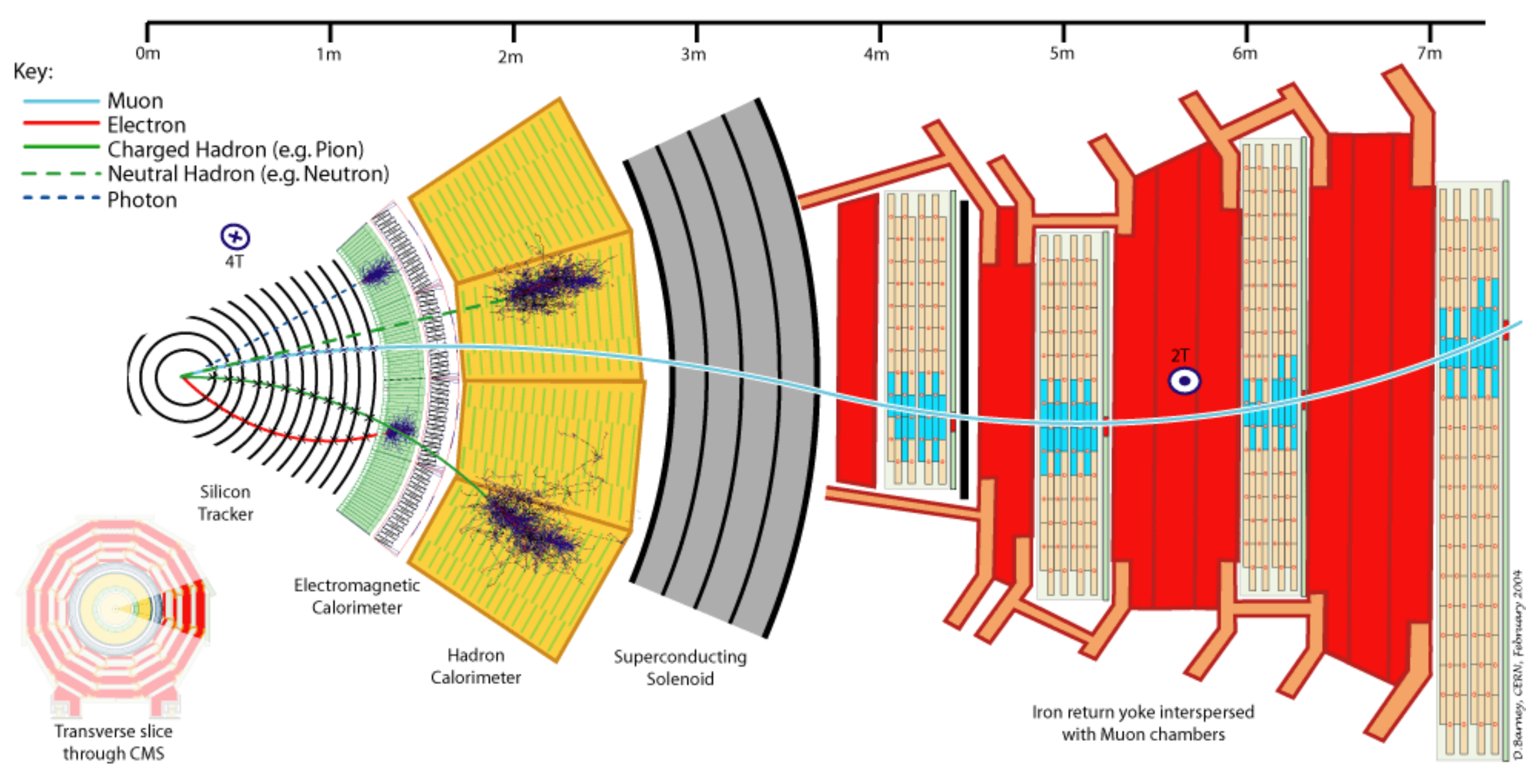}
  \end{center}
  \vspace{-0.3cm}
  \caption[.]{Schematic drawings of the ATLAS detector (upper) and a slice of the 
                   CMS detector (lower), showing the trajectories of charged and neutral 
                   particles interacting with the various detector layers. } 
\label{fig:atlascmsdet}
\end{figure}

The high-$p_T$ detectors, ATLAS and CMS, are designed as a result of careful
optimisation processes to respond as well as possible to these unprecedented
and sometimes conflicting requirements, while respecting budget limitations 
(approximately 550 million Swiss francs per detector). Both detectors have fast, 
multi-level trigger systems allowing one to select complex signatures, fast data 
acquisition based on broadband network switches, 
excellent inner tracking devices allowing efficient high-$p_T$ tracking 
and secondary ($b$) vertex reconstruction in a high-luminosity environment; 
fine-grained, high-resolution electromagnetic calorimeters for excellent electron
and photon reconstruction, complemented by full coverage hadronic calorimetry
for accurate jet and missing transverse energy measurements, and an efficient
identification of semileptonic $\tau$ lepton decays; as well as high-precision 
muon systems with standalone tracking
capability~\cite{atlasdetpaper,cmsdetpaper,jordannash,sphicasfroid}. 
Schematic drawings of the ATLAS and CMS detectors are shown in Fig.~\ref{fig:atlascmsdet}.

The most striking difference between ATLAS and CMS, strongly determining the 
entire detector design, is the {\bf magnet structures}. CMS has a single, albeit 
huge solenoid (inner diameter 5.9\m, thickness 60\cm, axial length 12.9\m), 
fully immersing the inner tracking systems and electromagnetic and hadronic 
calorimeters in a 3.8\,T axial magnetic field\footnote
{
     The solenoid is designed to deliver a 4\,T field. Longevity considerations have 
     however led to the decision to decrease the current from 19.5\,kA to 18.2\,kA,
     reducing the field to 3.8\,T.
}
(18.2\,kA current), and providing muon momentum measurement via the $\sim$2\,T field 
in the flux return yoke made out of 10\,000 tonnes of steel. 
ATLAS has three different magnet systems: a thin solenoid
(inner diameter 2.46\m, thickness 5\cm, axial length 5.8\m, axial magnetic field 
2\,T at the centre of the tracking volume, 7.7\,kA
current) around the inner tracking system, and 8 barrel and 2$\times$8 endcap air-core 
toroid magnets (magnetic fields between 0.5\,T and 4\,T, strongly varying with the radial 
distance from the toroids, 20.5\,kA currents), arranged radially around the hadron 
calorimeters such that the Lorentz force bends charged tracks along their $z$ coordinates. 
The toroid magnets do not affect the central solenoid field. All magnet systems 
are superconducting. 

The {\bf inner tracking systems} are made out of semiconducting silicon pixel and 
silicon strip detectors for the inner and outer layers (disks in the endcaps), 
respectively, comprising approximately 80 million channels. Pixel systems close 
to the collision impact point are mandatory to cope with the large track density. 
The innermost barrel pixel layer, of a total of 3 layers, is as close as 5.0\cm 
(ATLAS) and 4.4\cm (CMS) to the beam line. The design $R\phi$ position resolution 
of the pixel system is 10\mum. In CMS silicon strip technology is used to cover the entire
inner detector between pixel and electromagnetic calorimeter (radius of the outermost
layer: 107--110\cm), providing a total of 14 measurement points. The ATLAS silicon 
strip detector, being shorter in radius, provides 8 measurement points. A
transition radiation tracker made of 350\,000 Kapton straw tubes of 4\,mm diameter,
providing on average 35 measurement points for pseudorapidity\footnote
{
   The pseudorapidity is defined by 
   \beq
      \eta\equiv-\ln\left(\!\tan\!\frac{\theta}{2}\right)
          = \frac{1}{2}\ln\!\left(\frac{|{\bf p}|+p_L}{|{\bf p}|-p_L}\right)\,,
   \eeq
   where $\theta$ is the polar angle between the particle momentum ${\bf p}$ and the 
   beam axis ($z$), and $p_L$ is the longitudinal component of ${\bf p}$.
   In hadron collider physics, the pseudorapidity is preferred over the use of 
   the polar angle because particle production is constant as a function of the 
   pseudorapidity.
} 
lower than 1.8 (resolution of 
130\mum per straw), and between 18 and 35 $R\phi$ measurement points (no 
$\eta$ measurement) between $1.8\le|\eta|\le2.5$, is inserted
between the silicon strip tracker and solenoid. Transition radiation
with 8\,keV photons on average, emitted when charged ultrarelativistic particles 
traverse the boundary of two different dielectric media (foil and air), increases
the signal size so that dual readout with low and high thresholds allows the 
identification of $\beta=1$ particles (electrons). 

Owing mainly to the stronger solenoid magnetic field, CMS has better momentum 
resolution with $\sigma(p_T)\simeq1.5\%$ ompared to $3.8\%$ (ATLAS) for 
100\,\GeV tracks at $\eta=0$. At low momentum, multiple scattering that occurs due 
to the significant material in the tracking systems of both detectors (varying 
between $0.3X_0$ at $\eta\simeq0$ and $1.4X_0$ at $\eta\simeq1.5$) reduces this 
difference. 

The {\bf electromagnetic calorimeters} consist of a lead and liquid-argon sampling 
technique, radially shaped as an accordion to minimise inhomogeneities and cracks, 
chosen by ATLAS, versus high-granular lead tungstate (PbWO$_4$) scintillating crystals in 
CMS (61,200 crystals in the barrel and 7,324 in each endcap). 
Both calorimeters have a geometry pointing towards the collision point, which 
simplifies the energy reconstruction of the incident particles. The lead absorber
in the ATLAS calorimeter reduces the available light yield for energy measurement,
thus limiting the stochastic resolution to $\sigma(E)\simeq10$--$12/\sqrt{E}$ with a constant
term of 0.2--0.35\%, compared to $\sigma(E)\simeq3$--$5.5/\sqrt{E}$ and a constant 
term of 0.5\% for the CMS crystals. The influence of the constant term, originating 
from non-uniformities in the calorimeter response due to inhomogeneities and non-linearities,
is small for ATLAS, while it becomes a limiting factor at energies beyond 40\,\GeV for CMS
(hence, for example affecting the measurement of $H\to\gamma\gamma$). While CMS has only
a single electromagnetic layer, the ATLAS calorimeter is longitudinally segmented in 
four layers (including the presampler, which corrects the measured energy for early 
electromagnetic showers in solenoid and cryostat), permitting one to measure the shower 
development and so distinguish electromagnetic from hadronic showers. It 
also allows one to reconstruct the direction of the incoming particle. The cell 
granularity for the ATLAS main sampling layer is 
$\Delta\eta$$\times$$\Delta\phi=0.025^2\,{\rm rad}$, improved 
in $\eta$ by fine strips with $\Delta\eta=0.003$ (barrel number) in front of the main 
sampling layer to help identifying $\pi^0$. CMS has a crystal granularity of 
$\Delta\eta$$\times$$\Delta\phi=0.017^2\,{\rm rad}$ in the barrel, and 
0.018$\times$0.003 to 0.088$\times$0.015 in the endcaps.
Saturation of the energy reconstruction occurs for
energy depositions beyond 3\,\TeV (ATLAS) and 1.7\,\TeV (CMS). Biases due to saturation 
are corrected but lead to a decrease in the energy resolution. 

The {\bf hadronic calorimeters} use similar sampling techniques, based on iron 
(ATLAS) and brass absorbers (CMS) and scintillating tiles read out via wavelength shifting
optical fibres guiding the light to photomultiplier tubes. The main difference in 
performance originates from the strong constraint imposed by the maximum achievable 
size of the CMS solenoid, resulting in a barrel hadronic calorimeter with insufficient 
absorption (radiation length of $7.2\lambda$ at $\eta=0$ for all calorimeter layers
including the crystals, compared to $9.7\lambda$ for ATLAS) before the coil. A 
tail catcher had to be added around the CMS coil to complete the hadronic shower
reconstruction and provide better protection against punch-through to the muon 
system, faking muons. The reduced sampling fraction of CMS versus ATLAS leads
to an approximately twice worse jet resolution of $100\%/\sqrt{E}$ for CMS, and a 
worse constant term of up to 8\% in the barrel. It similarly affects the missing 
transverse energy resolution. Energy flow algorithms, attempting
to replace charged hadrons in the shower by the corresponding measurement in the 
inner tracker, improve the energy resolution for hadrons and jets, in particular
at low energies. 

Hermeticity of the detectors for an excellent missing transverse energy measurement,
but also to tag forward jets occurring, for example, in weak boson fusion processes, requires 
calorimeter coverage up to the very forward direction. The {\bf forward calorimeters} 
of ATLAS and CMS extend the energy measurement to pseudorapidities of 5 (polar angle 
of 0.77 degrees). They are located in different parts of the detector. The ATLAS forward 
calorimeter, made of copper and tungsten absorbers with gaps filled with liquid argon,
is fully integrated into the cryostat that houses the end-cap calorimeters,
which reduces the neutron fluence in the muon system and, with careful design, has 
minimal impact on the neutron fluence in the inner tracker. The CMS forward calorimeter, 
made out of steel and quartz fibres and operating with Cherenkov light, is situated 
11\m from the interaction point, thereby minimising the amount of radiation and 
charge density during operation.

Driven by the design of the magnets, the {\bf muon systems} strongly differ between
ATLAS and CMS. While CMS measures muons within the instrumented flux return, 
requiring the extrapolation of the track into the inner tracker, ATLAS has standalone 
muon tracking inside the large area spanned by the air-core toroids.  Both experiments
use drift tubes and cathode strip chambers (forward direction) for the precision muon 
measurements, and fast resistive plate chambers (thin gap chambers in the ATLAS endcaps)
for fast muon Level-1 trigger signals. The pseudorapidity coverage amounts to 
$|\eta|<2.7$ ($2.4$) for ATLAS (CMS) for muon measurements, lowering by 0.3
units for triggering. The combined momentum resolution for a 100\,\GeV 
(1\,\TeV) track at $\eta=0$, reconstructed in the inner tracker and muon systems, is 
$\sigma(p_T)\simeq2.6\%$ (10.4\%) (ATLAS) and $\sigma(p_T)\simeq1.2\%$ (4.5\%) (CMS).
The resolution significantly deteriorates in CMS for forward muons due to the 
reduced bending power of the solenoid (6\,T.m at $|\eta|=2.5$ compared to 16\,T.m at
$\eta=0$).

Apart from these main detector systems, both ATLAS and CMS have dedicated 
luminosity detectors in their forward regions. 

In summary, we may recall that ATLAS has put emphasis on excellent jet and missing
transverse energy resolution, particle identification, and standalone muon measurement,
while CMS has prioritised excellent electron, photon and tracking (muon) resolution. 
Both detectors have good hermeticity (very few `cracks'). 

References~\cite{atlasdetpaper,cmsdetpaper} present the essential performance parameters 
of the ATLAS and CMS experiments, sub-divided into track reconstruction, muon, electron and 
photon identification and reconstruction, jet and hadronic tau reconstruction, $b$-flavour 
tagging and the trigger selection (see below). Many of the results given are supported by 
existing test beam and cosmic ray measurements (also discussed in these lecture notes), 
in particular for the single-particle response of the detector elements to electrons, 
photons, pions and muons at various benchmark energies. Others rely on the simulation of 
the detector response and the underlying physics processes. They are affected by numerous 
uncertainties also due to hard-to-quantify soft-QCD and machine background effects. 

\subsection{Trigger and data acquisition}

In former times, when particle physics experiments used bubble and cloud chamber
techniques, data acquisition (DAQ) was made by means of stereo photographs. There
was effectively no trigger. Instead, each bubble expansion was photographed based on the 
constant (and known) accelerator cycle. The high-level trigger was {\em human}, 
realised by scanning teams operating worldwide with varying trigger efficiencies
(rumours claim that physicists had the worst scanning efficiency). The slow operation 
rate of this setup allowed one to measure only the most common processes. Later, 
electronic signals were used to trigger the camera to photograph an event
(a single trigger level). The dead-time occurred while the film advanced after 
a trigger. 

The trigger~\cite{nickellis} is a function of the fast detector response to a collision event 
providing a binary accept or reject signal. Its task is to look at (almost) all 
bunch crossings and select the most interesting ones. Data acquisition (DAQ) collects 
all detector information and stores it for offline analysis. Requirements for 
a DAQ system are the provision of online services, such as a state machine 
(`Run Control'), governing the run sequences, and data quality monitoring.
It must keep records of the detector configuration and run conditions, 
avoid corruption or loss of data (and hence verify the data sanity),
be robust against imperfections in the detector and associated electronics
and readout systems, and minimise dead time.\footnote
{
   Dead time is the fraction of time where valid interactions could not be 
   recorded for various reasons. Typical system-imminent dead time is of 
   the order of up to 10\%. It originates from the readout and trigger system,
   from operational dead time (\eg, the time to start and stop a run or
   to configure the detector systems), trigger or DAQ down-time (\eg, following
   computer failure), or detector down-time (\eg, following a high-voltage trip).
   For a multi-level trigger, the total dead time is the sum of the dead times
   of all levels. The trigger dead time for a given level is computed from the 
   product of the trigger rate of the previous level and the latency for this level. 
   The readout dead time is given by the product of the final (highest-level)
   trigger rate and the local readout time. 
   Note that trigger dead-time logic is {\em required} to prevent triggering
   another event before the detector has been fully read out. Given the 
   investment in the accelerators and the detectors for a modern HEP experiment, 
   it is clearly important to keep dead time to a minimum.
} 
Because the trigger latency even for the fastest level is longer than the 25\ns 
bunch crossing period, the electronics signals need to be saved locally in 
so-called pipelines until the trigger signal arrives.

A problem for any trigger at the LHC is that one cannot (and does not want to) save 
all events. `Old' (known) physics occurs more often than `new' physics, \ie, the 
new physics is buried under huge amounts of old physics. We have seen 
that the interesting physics occurs at rates of 10\,Hz (for top antitop production) 
and below at highest peak luminosity. The remit is thus to keep {\em all} of those 
events, while rejecting most of the others. One exception to this is low-mass flavour 
physics, which --- although being `old' --- has still important potential for discoveries. 
We hence must aim at fitting the best possible physics cocktail into the available 
bandwidth. Efficient selection and background rejection requires one to include the 
response of the entire detector in the trigger decision. This can only be achieved 
by splitting the trigger decision into several levels with increasing complexity. 
The first level has short latency and high efficiency and must only aim 
at the rejection of the `obviously' uninteresting events (once rejected, events
are rejected forever!). Later levels, which can be slower thanks to the rejection 
in the previous level, perform fine-grained selection and rejection. 

The trigger systems of ATLAS and CMS are separated into a first-level
ultra-fast hardware trigger, based on information from the calorimeters and dedicated
muon systems only. The detector data are transferred to large buffer memories 
after a Level-1 accept. The data rates to DAQ and the next level triggers are 
massive: with approximately 1\,MB event size at 100\,kHz event rate one 
has a rate of 100\,GB/s (\ie, 800 Gbit/s). The subsequent high-level trigger (HLT)
uses partial event data read out or powerful network switches to feed
reconstruction and software selection algorithms running on farms
with several thousand central processing units. In ATLAS the HLT is separated 
into two independent steps. A fast Level-2 trigger using only detector information
from so-called `regions of interest', which are sections along azimuthal and 
pseudorapidity cuts around triggered Level-1 objects, and including only the detector
systems required by the Level-2 algorithm. The Level-2 decision must come
within a few milliseconds and reduce the outgoing Level-1 rate from 75\,kHz to 
2\,kHz, which is the input rate to the event builder requiring to read out the 
full detector. A subsequent Level-3 trigger (`Event Filter') then further 
reduces the event rate to approximately 200\,Hz, which is written to disk.
These events are promptly reconstructed at CERN and, in parallel, distributed 
to 10 worldwide computing centres. In CMS, the large HLT input rate is tamed by 
factorising the event building into a number of slices each of which sees only 
a fraction of the rate. This requires a large and expensive total network 
bandwidth, but avoids the need for a very large single network switch.

An important requirement for the event building is a proper timing-in of the 
various detectors. Indeed, within the 40\,MHz bunch crossing rate, particles 
can only travel 7.5\m through the detectors, which are significantly larger than 
that (ATLAS has a height of 2$\times$11\,m and a length of 2$\times$23\,m). 
In addition, the collection of the detector signals, notably in the large muon
drift tubes, can take up to 40
bunch crossings (1\,$\mu$s). To properly collect the signals belonging to the 
same bunch crossing (\ie, `event') and to keep the exposure time per event
as small as possible, trigger-decision and detector response 
collection delays must be aligned to a few nanoseconds. Timing-in is one 
of the first commissioning tasks for all detector systems.


\section{Detector commissioning --- Overview}
\label{sec:detectorCommissioningOverview}

All detector systems, as well as the performance and physics groups developed
detailed commissioning strategies for initial running with colliding beams. 
Even before beams collide in the LHC, as more and more systems are being 
installed, extensive stand-alone and combined studies with comic ray events and 
detector calibrations are performed. These studies as well as dress rehearsals using simulated 
data exercise the full data acquisition chains, including the online and offline data quality 
assessment tools, and the streaming of the events into several physics and calibration 
streams based on the trigger decision. 

The cosmic ray data provide important information to align the detectors relative to 
each other (but not relative to the beam axes). They set an initial reference geometry 
for most of the barrel muon detector, and will be used to correct the alignment based 
on precise survey data and optical sensors. Muons from beam halo data taken during 
single-beam LHC commissioning runs will be used as an initial validation of the end-cap 
muon detector alignment. For example, in ATLAS the magnetic field strengths of the toroids, 
determining the muon energy scale, are known to better than 0.5\% versus $\phi$ from 
survey data of the measured coil positions. Later the precision can be improved to 0.1--0.2\%,
using a system of Hall probes. The field of the solenoid immersing the inner detector 
has been mapped to a precision of a few Gauss, which approaches the design goal.

Charge injection or pulsed calibrations of the electronic boards and pedestal 
runs provide initial settings for channel thresholds, ramp and delay values, pedestals, 
etc. for the various systems, and are used to map noisy and to some extent dead channels. 
Hadronic calorimeters also perform calibration with laser-light and radioactive caesium 
sources. These tasks together with test beam measurements contribute to achieving a 
sufficient quality of the first collision data. 

\begin{table}[t]
\setlength{\tabcolsep}{0.0pc}
\begin{tabular*}{\textwidth}{@{\extracolsep{\fill}}lrr} 
\hline\noalign{\smallskip}
{\bf ATLAS subdetector}	& {\bf Number of channels}	& {\bf Operational fraction (\%)} \\
\noalign{\smallskip}\hline\noalign{\smallskip}
Pixel Tracker	& 80 million	& 97.9 \\
Silicon Strip Tracker&	 6.3 million	& 99.3 \\
Transition Radiation Tracker&	 350\,000	& 98.2 \\
Liquid-Argon Electromagnetic Calorimeter	& 170\,000	& 98.8 \\
Tile Hadronic (Extended) Barrel Calorimeter	& 9800&	 99.2 \\
Hadronic Endcap Liquid-Argon Calorimeter	& 5600	& 99.9 \\
Forward Liquid-Argon Calorimeter	& 3500	& 100 \\
Muon Drift Tubes	 & 350\,000	& 99.7 \\
Muon Cathode Strip Chambers &	 31\,000&	 98.4 \\
Barrel Muon Trigger	& 370\,000&	 98.5 \\
Endcap Muon Trigger&	 320\,000	& 99.4 \\
Level-1 Calorimeter trigger&	 7160 &	 99.8 \\
\noalign{\smallskip}\hline
\end{tabular*}
\caption{Number of channels and operational status as of autumn 2009 of the ATLAS subdetectors.  }
\label{tab:ATLASOperationalStatus}
\end{table}

As an example, the ATLAS operational status as of autumn 2009 is given in 
Table~\ref{tab:ATLASOperationalStatus}. The experiment's start-up and ultimate design 
goals in terms of the tracking and calorimeter performance are summarised in 
Table~\ref{tab:calib_perf}.

With the start-up of the LHC,\footnote
{
   All event numbers given in this overview section refer to 14\,\TeV LHC
   centre-of-mass energy. The impact from lower centre-of-mass energies 
   (10\,\TeV and 7\,\TeV) is briefly discussed in Section~\ref{sec:outlook}.
} 
and after timing-in the detector systems with the 
colliding LHC bunches and the trigger signal, minimum bias triggers
from scintillator counters will provide Level-1 accepts for initial physics studies at 
a luminosity less than or equal to \Lumi31. These events can be used to provide 
first occupancy tests of the inner tracking systems, and to refine the dead 
channel maps. Copious isolated tracks from minimum bias events will allow the 
experiments to refine 
the inner detector alignment using the distributions of residuals between measured 
hits and fitted tracks, and the comparison of $E/p$ for pions of opposite charge. 
Alignment monitoring information will also be derived from $K_S^0$ and $\Lambda$ 
invariant mass and azimuthal decay vertex distributions. The $K_S^0$ invariant mass  
together with the known, ideally uniform decay-angle distribution can be used for a 
data-driven determination of the tracking efficiency. In ATLAS, high and low 
threshold transition radiation hits from isolated pion tracks will be compared to 
the expectation from simulation. Minimum bias events will help both experiments to 
monitor the uniformity of the calorimeter response, which can be done azimuthally 
and by comparing positive and negative pseudo-rapidity regions. In this initial phase 
it will also be possible to some extent to validate the calorimeter simulation 
by comparing shower shapes for isolated hadronic tracks and low energetic jets.
The statistics corresponding to a few days of low-luminosity data taking 
without toroid fields will allow the collection of enough straight muon tracks to 
calibrate the ATLAS muon optical alignment system to better than 100\mum. It will be
improved to up to 30\mum at higher luminosity, which is required to take full 
advantage of the spatial resolution of 40\mum per muon chamber, providing a 10\% 
measurement of 1\,\TeV muon tracks.
\begin{table}[t]
\setlength{\tabcolsep}{0.0pc}
\begin{tabular*}{\textwidth}{@{\extracolsep{\fill}}lccc} 
\hline\noalign{\smallskip}
                                 & {\bf Start-up of LHC }    & {\bf Ultimate goal}     & {\bf Physics goals}   \\  
\noalign{\smallskip}\hline\noalign{\smallskip}
EM energy uniformity             &  1--2\%      & 0.7\%        & $H\to\gamma\gamma$  \\
Electron energy scale            &  $\sim$2\%   & 0.02\%       & $W$ mass            \\
Hadronic energy uniformity       &  2--3\%      & $<1$\%       & Missing $E_T$       \\
Jet energy scale                 &  $<10$\%     & 1\%          & Top-quark mass            \\
Inner detector alignment         & 50--100\mum & $<$10\mum     & $b$ tagging           \\
Muon spectrometer alignment      & $<$$200_{\rm barrel}$\mum
                                                & 30\mum     & $Z^\prime\to\mu\mu$ \\
Muon momentum scale              &  $\sim$1\%   & 0.02\%       & $W$ mass            \\
\noalign{\smallskip}\hline
\end{tabular*}
\caption{Expected calibration and alignment accuracies at the LHC start-up
	       and the ultimate design goals for the ATLAS experiment. Examples for physics 
          channels or measurements driving the requirements are 
	       given in the last column. }
\label{tab:calib_perf}
\end{table}

While the trigger system is being commissioned, simple inclusive Level-1 calorimeter and muon 
triggers will be included first, followed by more complex Level-1 triggers, involving, for example, 
isolation and missing transverse energy. At the same time, the HLT systems will begin to operate, 
initially in pass-through mode, allowing the experiments to test the algorithms,
and later using the full power of the HLT, while continuing to run pre-scaled triggers 
in pass-through mode. Combinations of pre-scaled multi-threshold triggers will be 
used to determine efficiency curves for the three trigger levels (so-called `bootstrapping 
method'). The data collected with the complete low-luminosity trigger menu will 
contain copious quantities of low-energy leptons from heavy quark decays and 
also from direct $J/\psi$ and $\Upsilon$ production. Approximately 5000 
$W\to \mu\nu$ and 500 $Z\to \mu\mu$ decays should be reconstructed per 1\,\invpb 
of integrated luminosity (the expected rates are somewhat lower for electrons).
The low-luminosity trigger menu will also provide abundant samples of high-$p_T$
jets, prompt photons mainly from $\gamma$-jet events, and semileptonic $\tau$ decays. 

All these events will be crucial for the initial validation of the detector performance. 
More specifically, the inner detector material can be mapped with photon-to-\ee 
conversions to order 1\% with the statistics available after a few months of data taking. 
This procedure can be validated by studying the momentum dependence of 
the reconstructed invariant masses of low-mass resonances. Inclusive 
electrons can be used to test bremsstrahlung recovery in the inner detector.
The inner detector alignment is expected to converge to the relative design accuracy of 
approximately 10\mum soon after the full detector commissioning has started (the alignment
with cosmic ray events will be insufficient in the endcaps), allowing 
the constant term in the tracking resolution to be below 20\% of the full resolution. 
Local inner detector misalignment can be studied with the use of resonances with known 
masses and lifetimes decaying to lepton pairs, and with high-$p_T$ muons in combined track fits
with the muon spectrometer. 

Preliminary electromagnetic inter-calibration can be obtained at low luminosity using 
the azimuthal and $\pm\eta$ symmetry of inclusive isolated electrons from various sources. 
It is, however, not clear whether this procedure improves the intrinsic electromagnetic 
calorimeter inter-calibration determined in test beams at the higher energy scales of 
interest for most of the physics analyses (it will be useful for CMS where only 9 out 
of 36 supermodules of the electromagnetic calorimeter could be calibrated in the H4 
test beam, see Section~\ref{sec:testbeams}). The ultimate high-energy electromagnetic 
inter-calibration will use $Z\to ee$ events, requiring about 100\,\invpb recorded 
integrated luminosity to significantly improve the expected initial uniformity of 1--2\% 
to a statistical precision of $\sim$0.7\% (ATLAS) with high granularity, provided the inner
detector material is well enough understood. These events will also serve to calibrate 
the global electromagnetic energy scale.  

Hadronic track and jet inter-calibration will employ $E/p$ measurements (assuming an aligned 
inner tracker) and $E_T$ balancing in di-jet,  $\gamma$-jet and also $Z$-jet events, versus 
$\phi$. The latter two channels also determine the global jet energy scale with an expected 
precision better than 5\% after a few months of data taking. Di-jet events will also be 
used to validate the forward $E_T$ scale and resolution. The expected number of $\sim$500
fully reconstructed $t\overline t$ events for 100\,\invpb with one $W$ decaying 
hadronically and the other one leptonically (electron or muon) allows a first calibration 
of narrow jets with invariant mass fits to $W\to q q^\prime$ decays. It will 
also be important to study the stability of the electromagnetic and hadronic cluster 
reconstruction with respect to varying calorimeter noise (significant event pileup is 
expected to occur only above peak luminosities of ${\cal O}$(\Lumi33) for the nominal
LHC bunch pattern scheme). 

The performance of heavy-flavour jet tagging crucially relies on locally aligned
silicon detectors. Flavour tagging will be calibrated using 
$t\overline t$ events, but initially also using orthogonal information from tagging 
algorithms based on track fits and soft-muon reconstruction in di-jet events.

One of the most difficult detector observable to measure accurately is missing transverse
energy (\ETmiss).
Because it is sensitive to many new physics signatures, the tails of its distribution, 
dominated by resolution and instrumental effects, must be precisely calibrated with 
data before they can be used for discrimination and reconstruction purposes. The 
computation of \ETmiss requires the cleaning of the event from beam halo muons, beam gas 
collisions, cavern background, and cosmic rays. Moreover, the calorimeter cells must be 
calibrated (for both electromagnetic and hadronic showers), and deficient calorimeter 
cells (including noise) must be mapped and corrected. Initial data-driven \ETmiss
studies will use minimum bias events, analysing the \ETmiss resolution as a 
function of the $E_T$ sum and comparing it with the expectation from simulated data, 
the transverse $W$ mass in $W\to e(\mu)\nu$ events, $Z\to ee(\mu\mu)$ events, and, 
with rising statistics, mass-constrained $t\overline t$ and $Z\to\tau\tau$ events decaying 
to charged leptons and hadrons (approximately 7000 of the latter events with 
$p_T(\mu)>15$\,\GeV are expected in 100\,\invpb, allowing one to calibrate the absolute
\ETmiss scale to about 5\%).

\begin{figure}[t]
  \begin{center}
	  \includegraphics[width=0.97\textwidth]{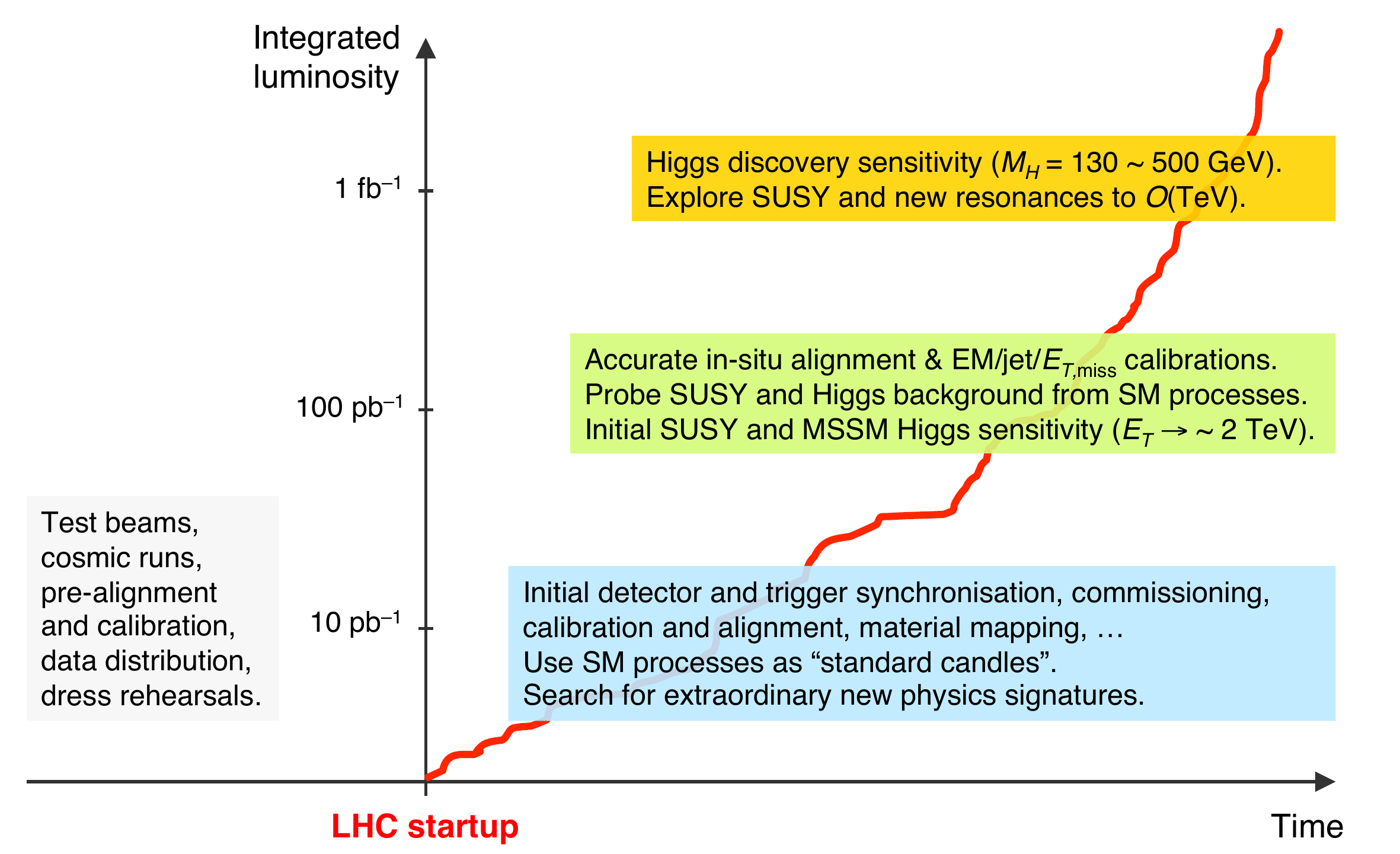}
  \end{center}
  \vspace{-0.5cm}
  \caption[.]{Sketch for a commissioning and early physics roadmap at the LHC. }
\label{fig:commissioning_roadmap}
\end{figure}
For muon tracks, the correlation of muon spectrometer and inner detector provides 
powerful reconstruction cross-checks for both systems. The muon reconstruction 
efficiency for stand-alone (muon spectrometer or inner detector only) and combined tracks
can be determined with $Z\to\mu\mu$ events by reconstructing one muon and probing the 
reconstruction of the other one (`tag-and-probe method'). The muon fake rate, 
expected to be negligible at low luminosity, will become significant above \Lumi33, 
due to the neutron and photon background in the cavern. The fake rate concentrates, 
however, at very low $p_T$, and remains small enough so that the impact on most 
physics analyses should be negligible. The overall muon energy scale will be calibrated 
with $Z\to\mu\mu$ events, where a statistical precision of 0.8\% and reasonable
geometrical granularity can be reached with 100\,\invpb integrated luminosity. 
With more data available, local misalignment problems in towers of chamber 
triplets could also be resolved with $Z$-mass constraints. A sketch for the
commissioning and early physics roadmap at the LHC is displayed in 
Fig.~\ref{fig:commissioning_roadmap}.

Initial physics measurements will primarily focus on Standard Model processes with 
high cross-sections. Among these are the multiplicity and 
pseudo-rapidity distribution of minimum bias events and cross sections of events with jets.
Low-$p_T$ physics mainly dedicated to the study of {\em B}$_s$ decays will begin by
measuring $J/\psi$ to $\Upsilon$ cross section ratios, which involves the
validation of vertexing tools, and cross sections and lifetimes of {\em B}, {\em B}$_s$ and 
$\Lambda_b$ mesons using decays to $J/\psi$. Statistically competitive lifetime 
measurements for these mesons can be expected with $\sim$100\,\invpb integrated luminosity.
The cross section of $t\overline t$ production using semileptonic decays can be 
measured to a precision better than 20\% with 100\,\invpb integrated luminosity,
without requiring $b$ tagging. Moreover, a significant 
single-top signal is expected to be seen in this data sample. Analyses aiming at 
searches for new phenomena will initially concentrate on the understanding of 
the detector performance and Standard Model processes, using calibration channels
and studying phase space areas where new physics contamination is expected to be 
small. 

The subsequent sections describe in some detail several of the commissioning and 
early physics studies mentioned above. 


\newpage
\section{Commissioning with test beams}
\label{sec:testbeams}

Both ATLAS and CMS have performed series of measurements with test beams of 
known energies and particle types. Electrons, photons, muons, pions, protons with 
energies between 1 and 350 GeV and varying magnetic field configurations were 
collected to test the tracking efficiency, alignment and particle identification, 
(inter-)calibrate the electromagnetic and hadronic calorimeters, test the muon trigger 
efficiency, tune Monte Carlo simulation, etc. 

\begin{figure}[t]
  \begin{center}
	  \includegraphics[width=0.99\textwidth]{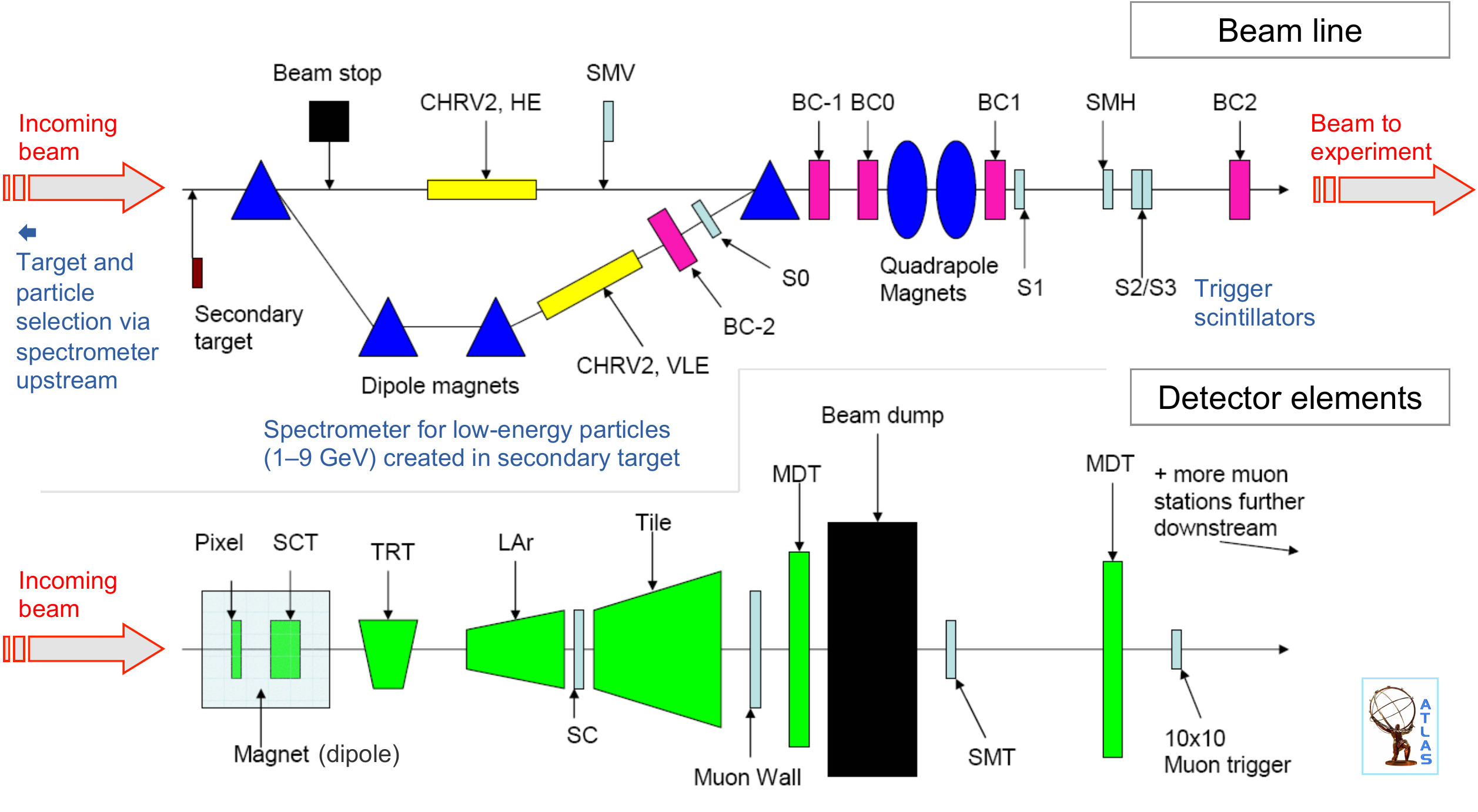}
  \end{center}
  \vspace{-0.5cm}
  \caption[.]{H8 beam line of the ATLAS combined test beam 2004.
                   Protons from the SPS, accelerated to 450\,\GeV energy, hit a target producing
                   hadrons, electrons and muons with energies in the range of 1 to 350\,\GeV,
                   which are selected upstream by a mass spectrometer. The composition 
                   of the incoming monochromatic particle beam is measured with Cherenkov 
                   counters (upper picture). The beam is focused and passes trigger scintillators 
                   before entering a complete ATLAS barrel slice (lower picture) with realistic 
                   geometry composed of Pixel and silicon strip detector (SCT) layers, 
                   immersed in a 1.4\,T magnetic dipole field parallel to the beam, a 
                   transition radiation tracker (TRT) module outside the magnetic field, 
                   liquid-argon electromagnetic and tile hadronic calorimeter layers, 
                   interleaved with a scintillator to measure the energy lost in the 
                   liquid argon cryostat, and a series of muon drift tube and resistive 
                   plate chambers before and after a beam dump block. }

\label{fig:h8beamline}
\end{figure}

Figure~\ref{fig:h8beamline} shows
a sketch of the `H8' beam line used for the ATLAS 2004 combined test beam.
A full barrel slice, from the innermost tracking detectors and magnetic field 
to the outermost muon spectrometer, was exposed to the particle beams. 
The experimental setup was kept as close as technically possible 
to the ATLAS geometry. The distance between subdetectors, the pointing geometry, 
and the magnetic field orientation were preserved where permitted. The most important 
goals of this test beam campaign were: $(i)$ test the detector performance with 
final or close to final electronics equipment, data acquisition and trigger infrastructure 
and reconstruction software, $(ii)$ validate the description of the data by Monte Carlo 
simulations down to energies of 1\,\GeV to prepare the simulation of the ATLAS data, 
and $(iii)$ perform combined studies in a setup very close to that of ATLAS (\eg, 
combined electromagnetic and hadronic calorimetry, and combined tracking and calorimetry).

\subsection{Energy reconstruction in the ATLAS liquid-argon electromagnetic calorimeter}

\begin{figure}[t]
  \begin{center}
	  \includegraphics[width=0.7\textwidth]{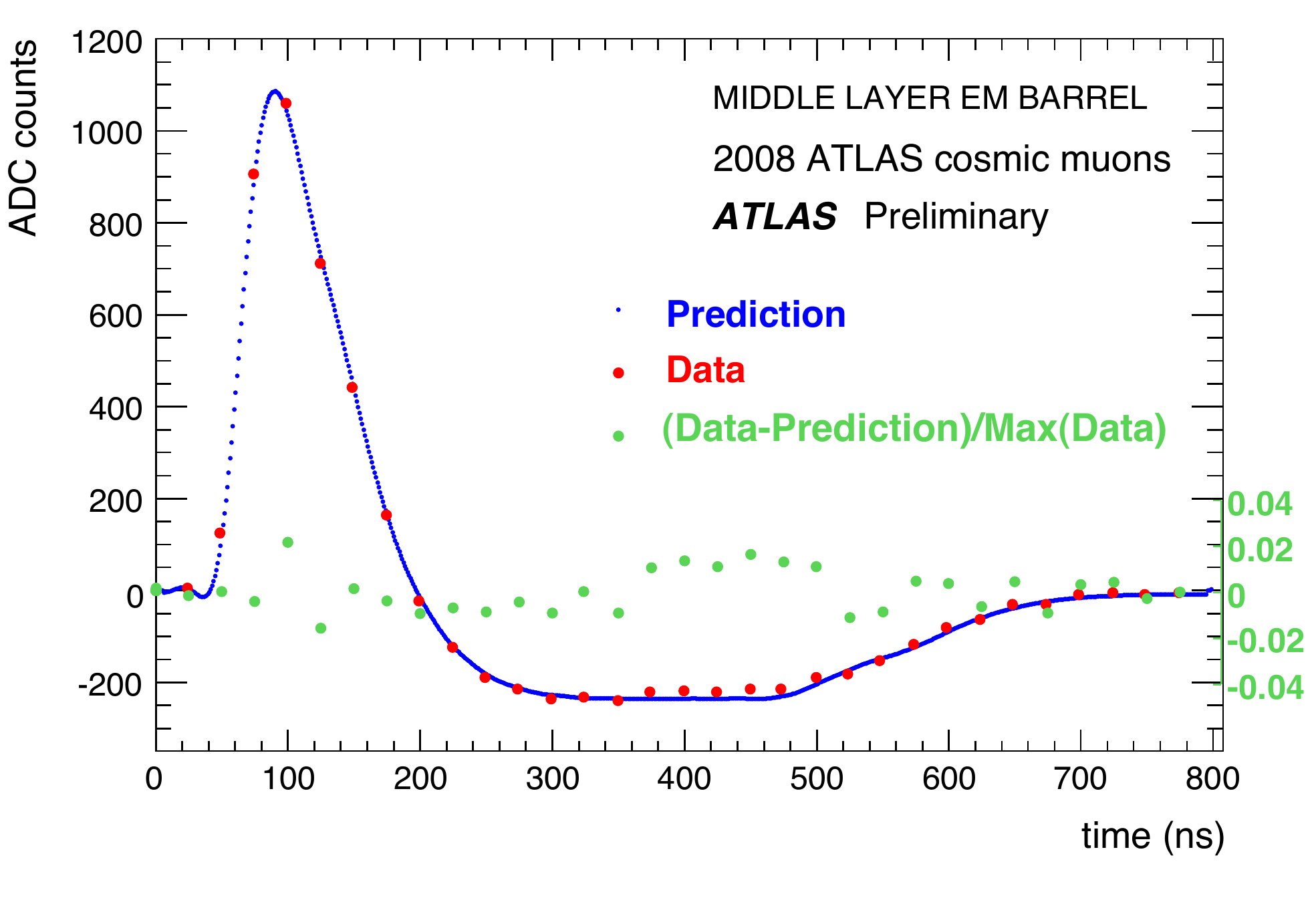}
  \end{center}
  \vspace{-0.8cm}
  \caption[.]{Digitised bipolar ionisation pulse shape of a 15\,\GeV cosmic-ray signal measured 
                   in the middle layer of the ATLAS electromagnetic calorimeter. The signal is 
                   shaped and sampled with 40\,MHz frequency, corresponding to a sample 
                   period of 25\,ns, and a total sampling window of 800\,ns (during normal 
                   data-taking only 5 samples (125\,\ns) are read out). The study of the pulses 
                   measured with 32-sample readout allows one to determine the drift time in the 
                   liquid argon gaps related to the undershoot of the pulse, and the electrode
                   position related to the rise at the end of the pulse.
                   The curve shows the expectation agreeing to better than 2\% with the measurement. }
\label{fig:larpulseshape}
\end{figure}

The ionisation signal generated in the ATLAS electromagnetic calorimeter is collected from the 
readout electrodes and brought via cables to the front-end electronics where it is amplified, 
shaped and sampled at a 40\,MHz frequency. The samples (usually five) are stored 
in an analog pipeline until the arrival of a trigger accept decision. The samples 
belonging to the accepted event are 
then digitised and transmitted by the calorimeter backend electronics to readout driver
modules, where the signal amplitude is reconstructed and converted to \MeV. 
Figure~\ref{fig:larpulseshape} shows a fully digitised pulse shape with 32 samples from 
a cosmic-ray event with an unusually large energy deposit. The full cell-energy reconstruction 
from the digitised pulse samples is encoded in the following conversion formula
\beq
    E_{\rm cell} = F_{\mu{\rm A}\to{\rm \MeV}}\cdot 
                        F_{{\rm DAC}\to\mu{\rm A}}\cdot
                        \left(\frac{M_{\rm phys}}{M_{\rm calib}}\right)^{\!\!-1}\cdot
                        R\left(
                        \sum_{i=1}^{N_{\rm samples}}\!\!\!a_i\left(s_i-p\right)\right)\,,
\eeq
where the subscripts specify the conversion type. The sum over the digitised samples 
on the right-hand side is computed from the measured ADC counts, 
corrected for an overall pedestal ($p$), obtained in regular calibration runs --- together 
with noise and autocorrelation terms --- from random triggers in physics events, and 
multiplied by the sample-specific `optimal filtering coefficients' $a_i$, obtained 
from so-called `delay' runs where calibration signals are injected to measure
the pulse shape. The sum is taken as an argument to the ADC-to-DAC ramp function $R$,
obtained from dedicated electronics calibration runs, where known charges are
injected and the corresponding ADC output is measured and fit to a linear function.
Differences between the calibration and physics pulse shapes are 
corrected via the $M$ ratio. The DAC values are then converted to $\mu$A, which
is related to the calibration injection resistance and computed taking into account 
cable and other attenuation effects. Finally, the $\mu$A signal is converted to \MeV by 
applying the corresponding current-to-energy conversion factor, and by correcting
the energy lost in the absorber material (sampling fraction). 
Once the cell energies are reconstructed, cells are summed to form 
a cluster over all three longitudinal compartments and the presampler of the 
electromagnetic calorimeter. 

This procedure provides the electromagnetic energy scale. Physics events such as
$Z\to ee$ will be used to achieve absolute energy calibration. For hadrons and jets, 
one needs to account for hadronic shower corrections, that is, one must pass from 
the electromagnetic to the hadronic energy scale.

\subsection{Electromagnetic energy resolution}

\begin{figure}[t]
  \begin{center}
	  \includegraphics[width=1\textwidth]{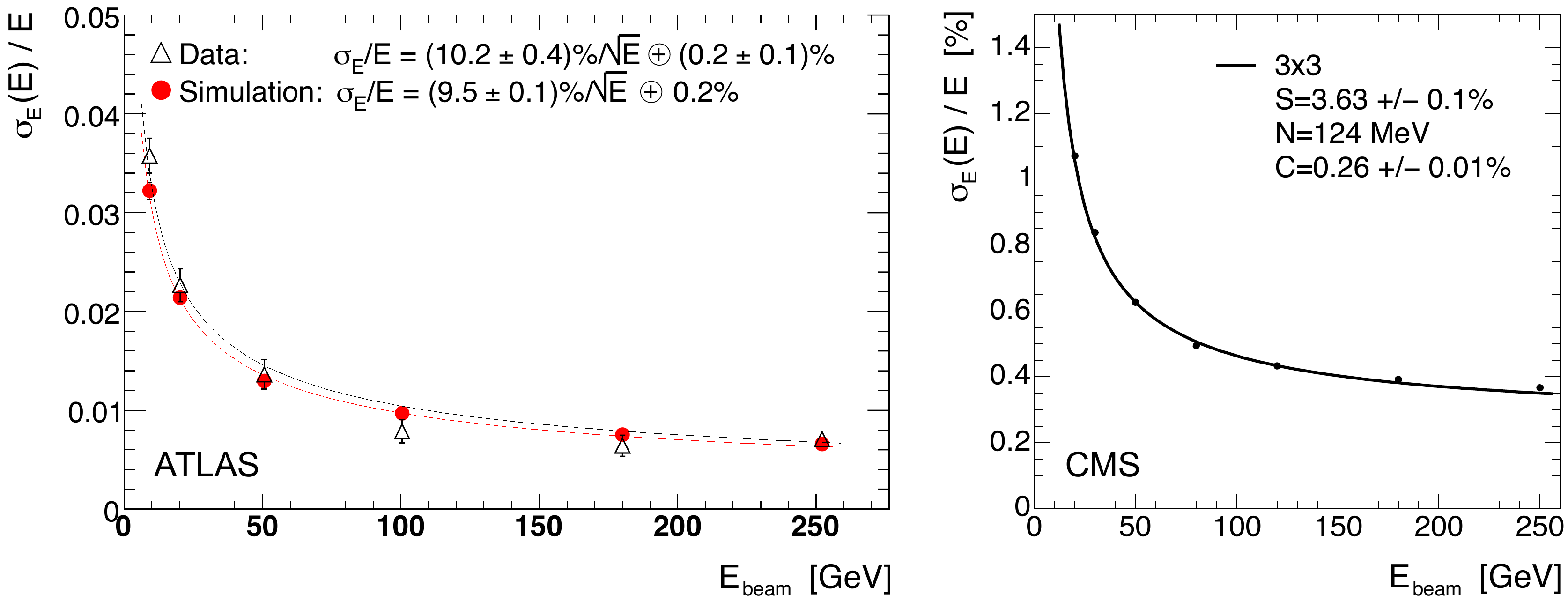}
  \end{center}
  \vspace{-0.35cm}
  \caption[.]{Fractional electromagnetic energy resolution versus the incident energy 
              obtained from electron test beams. 
              \underline{Left}: result for the ATLAS liquid-argon calorimeter obtained,
              behind $1.6X_0$ material thickness, from the H8 combined test beam in 2004. 
              \underline{Right}: result for the CMS crystal calorimeter obtained 
              without upstream material for 9 out of 36 tested supermodules at the H4 
              test beam in 2006. The energy was measured in an array of 3$\times$3
              crystals with electrons impacting the central crystal. }
\label{fig:emresolution}
\end{figure}

The resolution of an electromagnetic calorimeter is driven by the amount of active material
in which the electromagnetic shower develops, and by the shower containment.
Containment requires a calorimeter thickness of many radiation lengths $X/X_0>20$, where 
the radiation length $X_0$ is a material characteristic related to the energy loss of 
high-energy particles interacting electromagnetically with the material.\footnote
{
   The radiation length is both the mean distance over which a high-energy electron loses 
   all but 1/$e$ of its energy by bremsstrahlung, and 7/9 of the mean free path for pair 
   production by a high-energy photon. 
}
Test beams with known particle content and energy allow the experiments to measure 
resolution, linearity and uniformity of the calorimeter energy response. The resolution 
results obtained by ATLAS and CMS for electron beams with different energies 
are shown in Fig.~\ref{fig:emresolution} (the measurements were obtained under different 
experimental conditions, see figure caption). Calorimeter resolution is conveniently 
expressed as a function of the incident electron/photon energy, $E$, by the expression
\beq
\label{eq:energyresolution}
       \frac{\sigma(E)}{E} = \frac{S}{\sqrt{E~({\rm \GeV})}} \oplus C \oplus 
                                        \frac{N}{E~({\rm \GeV})}\,,
\eeq
where the first term on the right-hand side determines the {\em stochastic} resolution
resulting from statistical fluctuations in the number of shower particles\footnote
{
    The number of particles produced in the shower is proportional to the energy of the 
    incident particle: $N_{\rm part}\propto E$. The error in the energy measurement 
    is due to statistical fluctuations in $N_{\rm part}$, \ie, $\sigma(E)\propto\sqrt{N_{\rm part}}$.
    One thus finds for the stochastic contribution to the energy resolution
    $\sigma(E)/E\propto 1/\sqrt{E}$. Because in sampling calorimeters the absorber material 
    does not contribute to the energy measurement, the electromagnetic energy resolution is 
    worse than for crystal calorimeters, provided that the crystals have sufficiently 
    large $X/X_0$ so that the full shower can be contained. This is the case for the PbWO$_4$
    scintillating crystals used by CMS, which have very high density so that the total calorimeter 
    has $28X_0$ (for comparison, the ATLAS calorimeter has $22 X_0$).
    The sampling fractions in the ATLAS electromagnetic calorimeter are $f_{\rm sampl}=0.17$
    (0.20) for $|\eta|\le0.8$ ($|\eta|>0.8$). The measured energy must thus be corrected
    for the dead material $E_{\rm true}=f_{\rm sampl}^{-1}E_{\rm meas}$, so that the 
    stochastic resolution becomes 
    $\sigma(E)/E\propto \sqrt{d_{\rm sampl}/f_{\rm sampl}}/\sqrt{E}\approx 3/\sqrt{E}$,
    where $d_{\rm sampl}$ is the thickness of the sampling layers (finer sampling 
    provides better resolution). Hence the approximately three times worse intrinsic
    electromagnetic energy resolution in ATLAS compared to CMS. 
} 
and in the shower containment, the second {\em constant} term is due to non-uniformities in 
the calorimeter response introduced by inhomogeneities and non-linearities, and the third 
{\em noise} term quantifies electronics noise and in-time physics pile-up. The `$\oplus$'
indicates that the different resolution terms are added in quadrature. Some numbers 
obtained for these terms from fits to electron test beam data are quoted on the plots 
in Fig.~\ref{fig:emresolution}. Taking into account the full detectors and materials, 
one expects for ATLAS (CMS) the following benchmark resolution parameters: 
$S=10$--12\% (3--5.5\%), $C=0.2$--0.35\% (0.5\%), $N=250$\,\MeV
(200--600\,\MeV), where the better (worse) numbers refer to the barrel (endcaps).\footnote
{
    \label{ftn:higgsconstantterm}
    With these parameters, a back-of-the-envelope calculation for $H\to\gamma\gamma$ 
    gives for the di-photon mass resolution as a function of the photon energy: 
    $\sigma(M_{\gamma\gamma})|_{E_\gamma} \propto M_H\sigma(E_\gamma)/(\sqrt{2} E_\gamma)\approx 1.2\,{\rm \GeV}~(0.7\,{\rm \GeV})$, for ATLAS (CMS) and where $M_H=120$\,\GeV has been assumed.  To obtain a realistic 
    estimate of the resolution one must also include the error on the opening angle 
    (photon directions), as well as $\gamma\to\ee$ conversions (20--60\% of all 
    photons from $H\to\gamma\gamma$ decays, strongly increasing for large $|\eta|$). 
    Both effects reduce the effective resolution difference between the experiments. 
} 
With the 9 out of 36 super-modules calibrated in the 2006 test beam, CMS also found 
excellent energy-response uniformity of 0.27\%.

\subsection{Hadronic energy resolution}

\begin{wrapfigure}{R}{0.43\textwidth}
  \vspace{-40pt}
  \begin{center}
	  \includegraphics[width=0.43\textwidth]{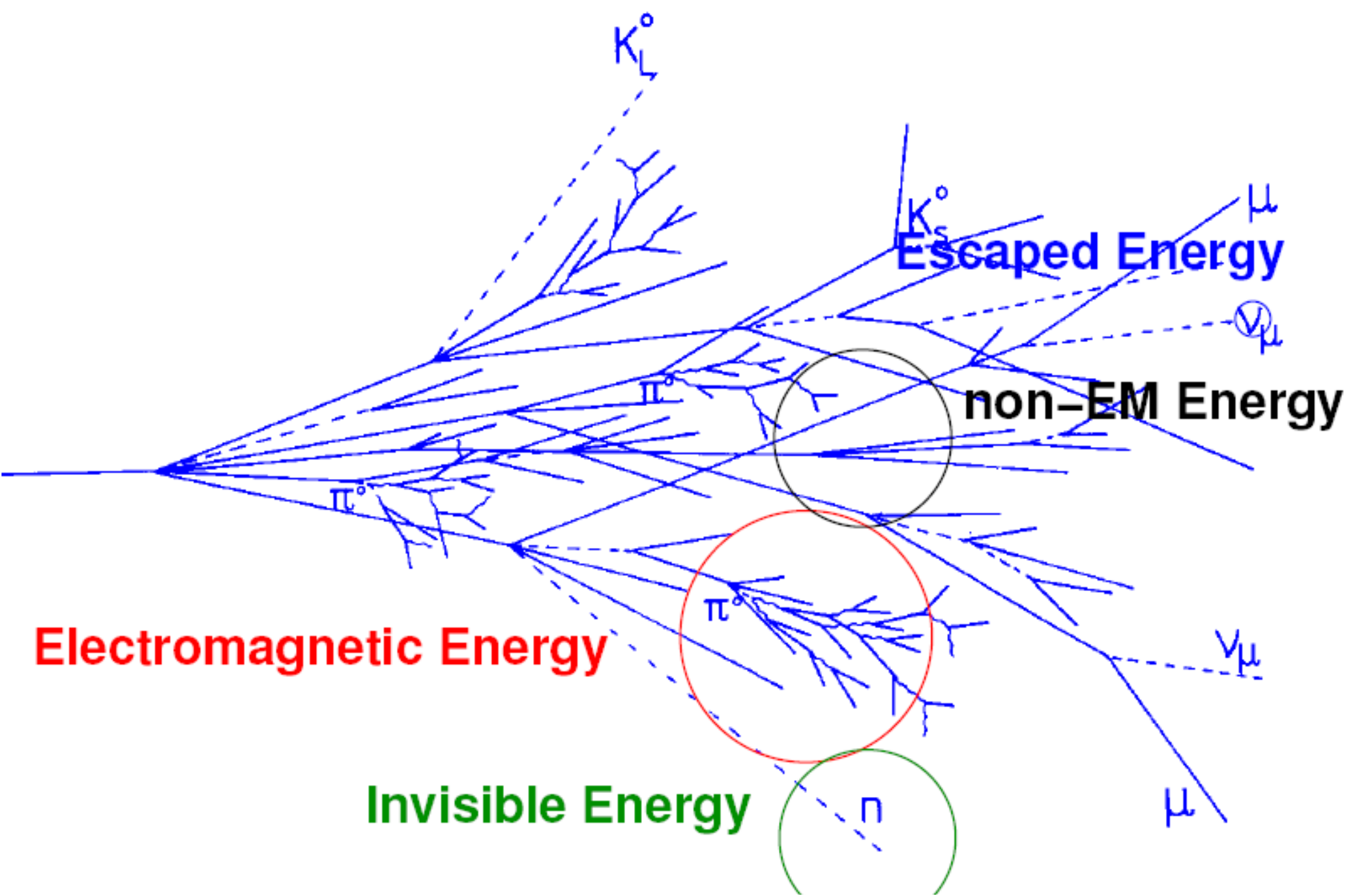}
  \end{center}
  \vspace{-15pt}
  \caption{Simulated hadron shower consisting of electromagnetic and non-electromagnetic,
           invisible and escaped energy. }
  \label{fig:hadronshower}
  \vspace{-10pt}
\end{wrapfigure}
\begin{figure}[t]
  \begin{center}
	  \includegraphics[width=0.73\textwidth]{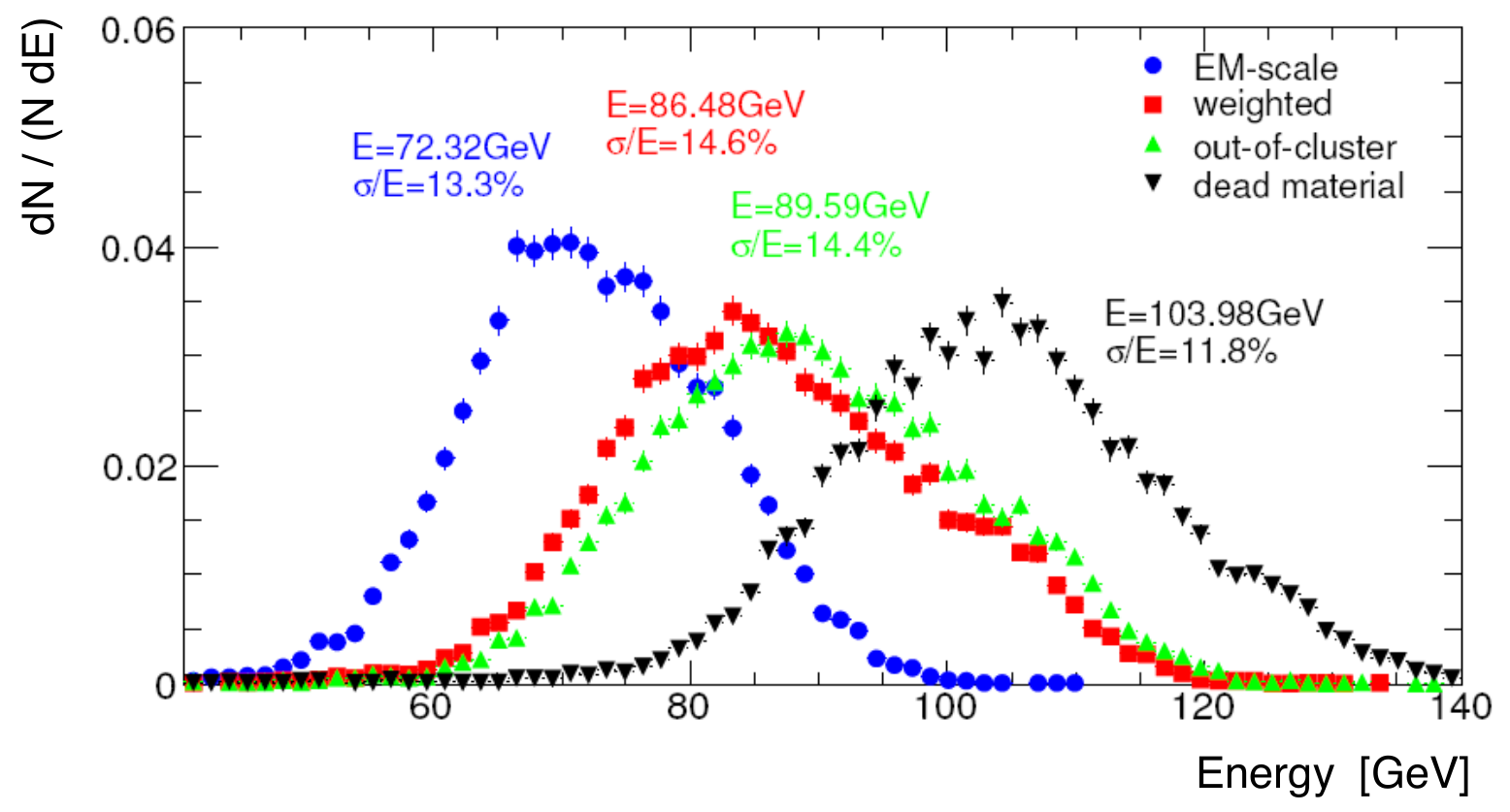}
  \end{center}
  \vspace{-0.4cm}
  \caption[.]{Reconstructed energy for 100\,\GeV test beam pions in a slice of the ATLAS 
              barrel electromagnetic and hadronic calorimeters. Shown are: raw measured 
              energy (circles), after reweighting from the electromagnetic to the hadronic
              scale (squares), after applying out-of-cluster corrections from shower 
              leakage (top-oriented triangles), and after dead-material corrections
              (bottom-oriented triangles). }
\label{fig:pionenergyreco}
\end{figure}
During the ATLAS H8 combined test beam campaign, pion beams with 6 discrete energies 
ranging from 10\,\GeV to 350\,\GeV were used to study the hadronic energy reconstruction 
in the calorimeters. Hadron showers originate from interactions of hadrons with 
nuclei. The density of hadron calorimeters is therefore appropriately expressed in terms of 
the nuclear interaction length $\lambda$, which quantifies the mean free path of hadrons in 
material between strong collisions. For example, silicon has $\lambda=45.5$\,cm, iron 
16.8\,cm, lead 17.1\,cm, and water 83.6\,\cm, to be compared to $X_0=0.56$\,cm for lead 
and 1.76\,\cm for iron. Hence $\lambda\gg X_0$ and one can separate electromagnetic showers,
which are short-ranged, from far-ranged hadronic showers, which also clarifies why 
calorimeters are called and arranged as they are: electromagnetic calorimeters fully absorb 
electromagnetic showers, but only parts of the showers initiated by hadrons; the following 
calorimeter layers (usually sampling calorimeters) entirely absorb the hadronic showers. 

\begin{figure}[t]
  \begin{center}
	  \includegraphics[width=0.55\textwidth]{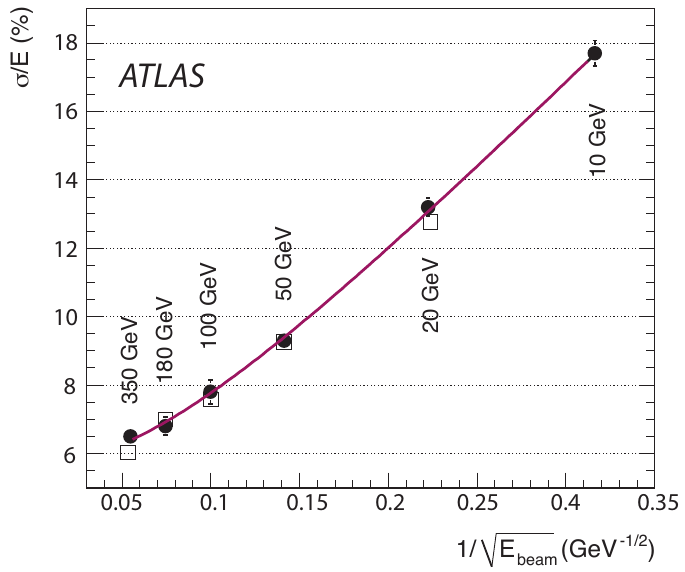}
  \end{center}
  \vspace{-0.3cm}
  \caption[.]{Fractional energy resolution for pions at 0.35 pseudorapidity 
              (equivalent calorimeter depth $7.9\lambda$), versus the 
              incident energy from test beam data in the ATLAS hadronic calorimeter
              (full circles), and compared to Monte Carlo simulation (open squares).}
\label{fig:hadronresolution}
\end{figure}
Hadronic showers (Fig.~\ref{fig:hadronshower}) consist of approximately 50\% 
electromagnetic energy (\eg, $\piz\to\gamma\gamma$), 25\% non-electromagnetic 
energy (such as $dE/dx$ from $\pi^\pm$, $\mu^\pm$, $K^\pm$), another 25\% 
invisible energy (nuclear fission and excitation, neutrons), and 2\% escaped energy 
(\eg neutrinos). Invisible and escaped energy causes worse resolution for hadronic 
showers than for electromagnetic ones. When uncorrected it also causes an underestimate
in the measured energy with respect to the true hadron energy. 
Figure~\ref{fig:pionenergyreco} shows the reconstructed energy in the ATLAS barrel 
calorimeter slice for 100\,\GeV pions from test beams. The raw measured energy 
at the electromagnetic scale undershoots by 28\% with the largest contributions
to the bias coming from invisible and escaped energy, and from dead material. 
While the various corrections recover the overall energy scale, they cannot improve
the resolution (unless event-by-event corrections as a function of the longitudinal
and transverse shower shapes are applied).

The final energy resolution obtained from pion test beam data for the ATLAS calorimeter
is shown in Fig.~\ref{fig:hadronresolution}, and compared to the expectation from 
Monte Carlo simulation (Geant-4). One finds benchmark values for single hadrons of 
53\%, 3\%, and 0.5\,\GeV, for the stochastic, constant and noise terms, respectively 
(\cf Eq.~\ref{eq:energyresolution}). For comparison, for central jets Monte Carlo 
simulation predicts 60\%, 3\%, and 0.5\,\GeV for the resolution parameters, and a 
missing transverse energy resolution of $\sigma(\met)/\sum E_T\approx55\%$. These 
values are somewhat worse in CMS due to the reasons mentioned in Section~\ref{sec:detectors}.

\newpage
\section{Commissioning with cosmic rays}
\label{sec:cosmics}

ATLAS and CMS have performed extensive campaigns of cosmic ray data-taking, initially 
with the individual systems, later including more and more detector systems with the 
completion of the installation in the pits. The goals of these studies are --- 
along with exercising the detector operation, and the full data taking, reconstruction and 
analysis chain --- tracking alignment (with and without magnetic field), deriving dead 
channel maps, measuring the muon trigger and tracking efficiencies, analysing calorimeter 
pulse shapes, improving the detector timing, tuning Monte Carlo simulation, etc. 

\begin{figure}[t]
  \begin{center}
	  \includegraphics[width=0.99\textwidth]{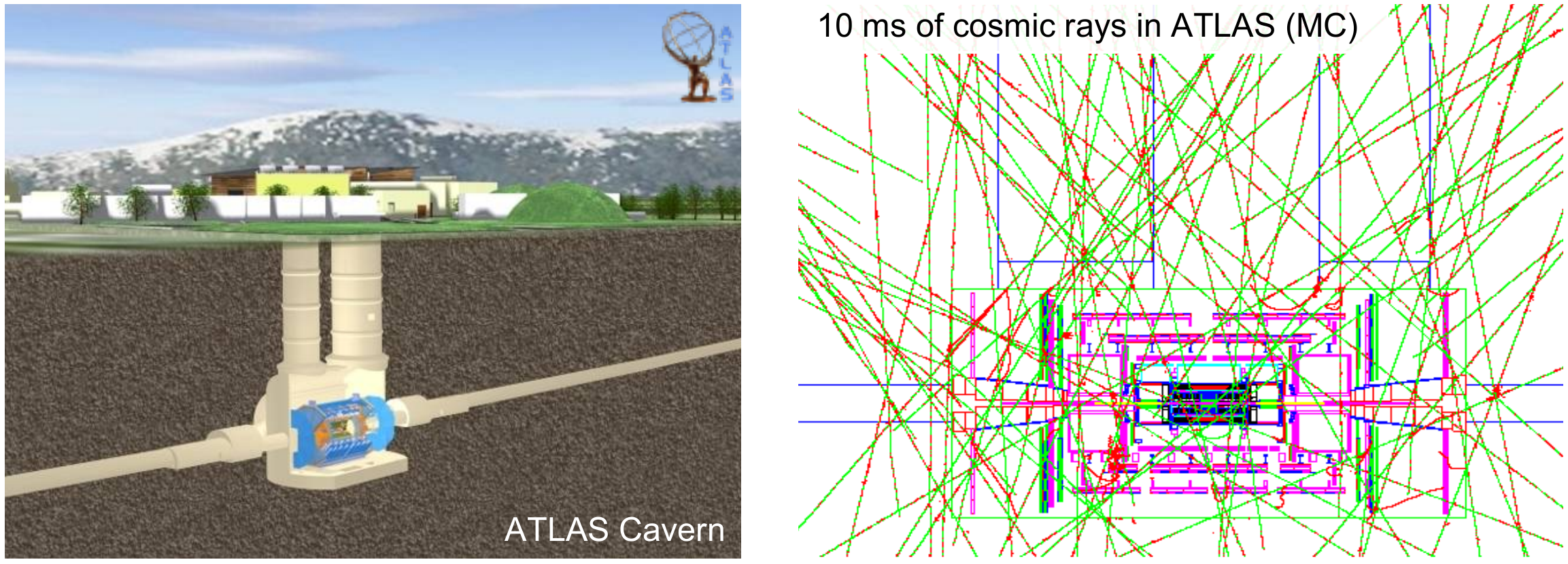}
  \end{center}
  \vspace{-0.3cm}
  \caption[.]{Schematic drawings of the ATLAS underground cavern with supply shafts
                  (left ---  two lateral elevator shafts are not drawn), and simulated cosmic
                  rays through ATLAS within 10\,ms exposure time (right). }
\label{fig:atlasschema10ms}
\end{figure}
\begin{wrapfigure}{R}{0.4\textwidth}
  \vspace{-24pt}
  \begin{center}
	  \includegraphics[width=0.4\textwidth]{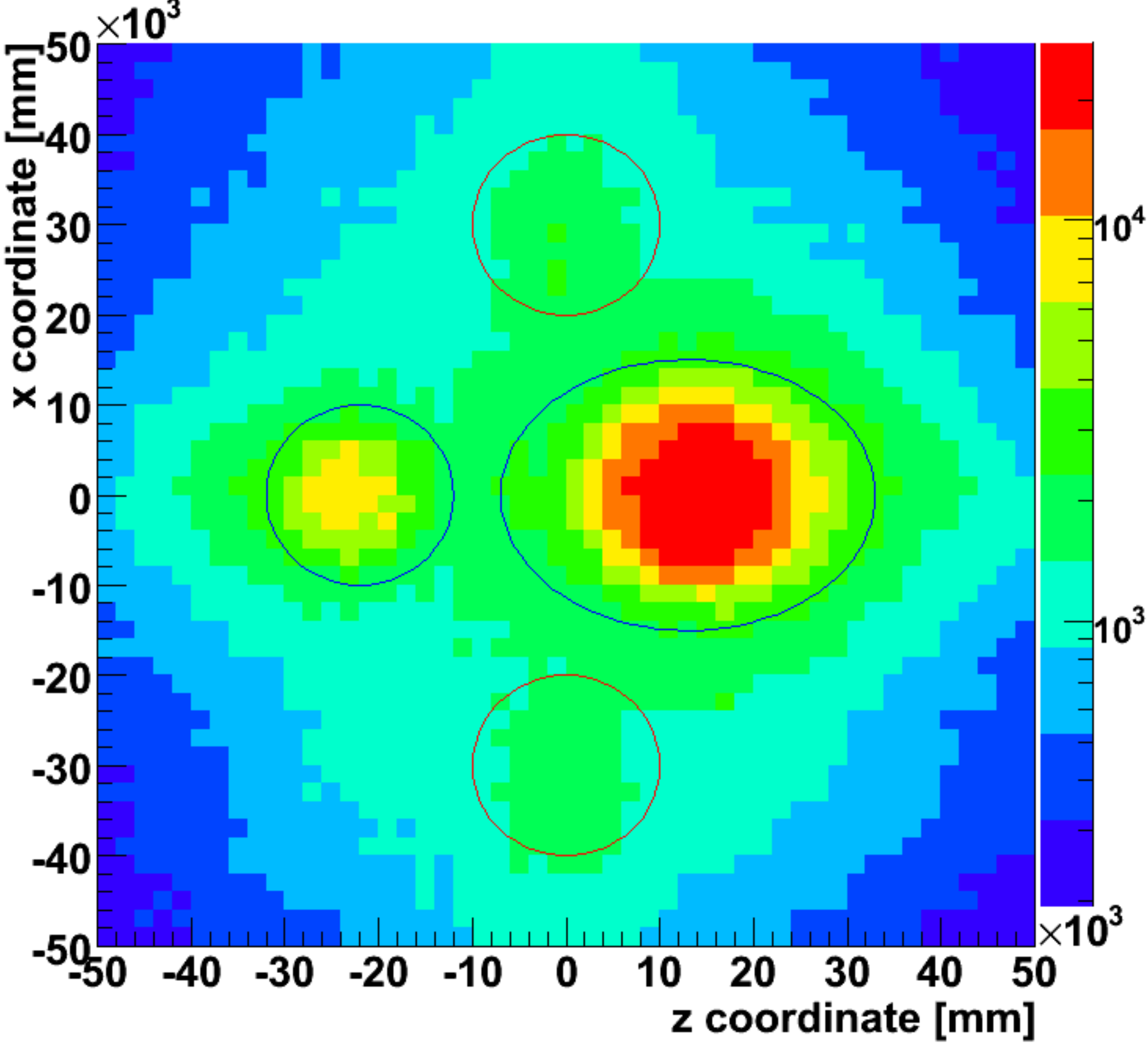}
  \end{center}
  \vspace{-15pt}
  \caption{Reconstructed cosmic tracks (6.6 million) in the ATLAS resistive plate chambers, 
           extrapolated to the surface. The ellipses indicate the supply and elevator 
           shafts.}
  \label{fig:cosmicsoccupancyatlas}
  \vspace{-6pt}
\end{wrapfigure}
Cosmic rays stem from cosmic nuclei (90\% protons, \ie, hydrogen nuclei) that interact 
strongly with the Earth's atmosphere, creating hadrons --- mainly pions and kaons with 
relative intensity 1:0.054~\cite{gaisser}, which decay to minimum ionising relativistic 
muons that reach sea level on Earth,\footnote
{
    Cosmic rays have been, and are still, sources of major discoveries in particle physics. 
    For example, in 1932, Anderson (Cal Tech, USA) discovered the antielectron
    (positron) in cosmic rays. Later in 1946, Rochester and Butler (Manchester, England)
    observed two tracks `out of nothing' in cosmic rays, which were
    pions from the decay of a neutral (`strange') kaon, thereby initiating 
    particle physics. Today, very high energy cosmic rays are extensively studied. 
} 
or which undergo nuclear interactions with nuclei in air. 
The muon flux at the surface is approximately 130\,Hz per m$^2$ for $E_\mu>1$\,\GeV,
and the average muon energy is about 4\,\GeV. The ATLAS detector, being separated
from the surface by 100\,m of earth and stone, receives a muon flux of approximately 
4\,kHz in the fiducial volume of the muon spectrometer, and 15\,Hz in the TRT barrel
(numbers from Monte Carlo simulation). The supply and elevator shafts (see left-hand 
plot of Fig.~\ref{fig:atlasschema10ms}) provide reduced shielding, which 
translates into an increased occupancy of the detector elements underneath the shafts
or close by (Fig.~\ref{fig:cosmicsoccupancyatlas}). 
The right-hand plot of Fig.~\ref{fig:atlasschema10ms} shows a simulated 10\,ms 
snapshot of the ATLAS detector bombarded by cosmic rays. High-energy cosmic rays 
sometimes also produce so-called `air showers' (and {\em extensive} air showers), 
where an avalanche of secondary scattering particles is created. Such air showers 
have been observed by the experiments, giving rise to events with large numbers of 
muons (order 10 to 100), jets, and large deposited energy (events with 6 jets, all
exceeding 20\,\GeV transverse energy, have been seen). 

\begin{figure}[t]
  \begin{center}
	  \includegraphics[width=0.99\textwidth]{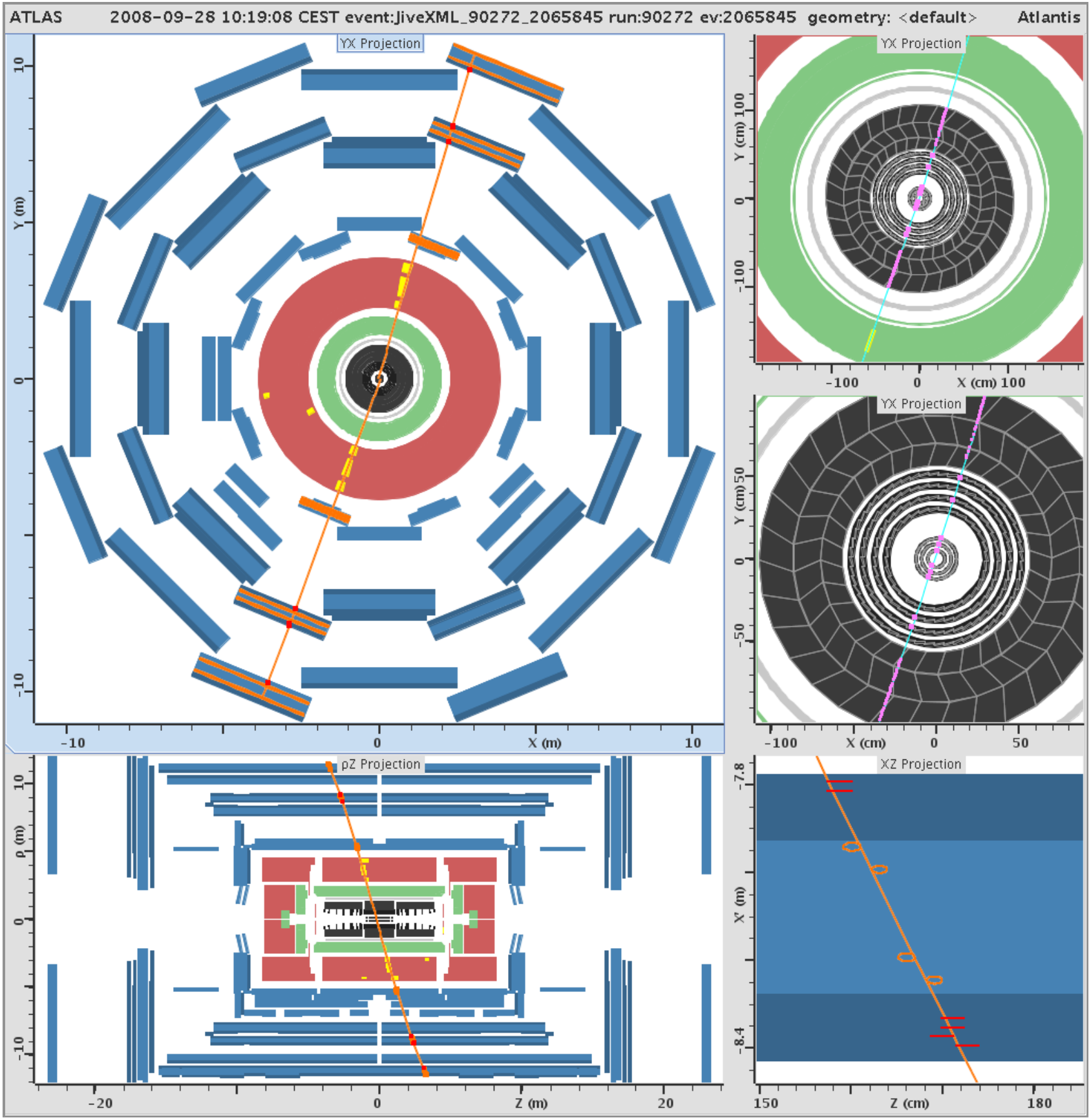}
  \end{center}
  \vspace{-0.3cm}
  \caption[.]{A cosmic ray muon measured by ATLAS. Seen are hits in the muon spectrometer and 
              the inner tracking systems, as well as energy deposits in the hadronic tile 
              calorimeter. All magnets were switched off in this run.  }
\label{fig:cosmicatlas1}
\end{figure}

\begin{figure}[p]
  \begin{center}
	  \includegraphics[width=0.7\textwidth]{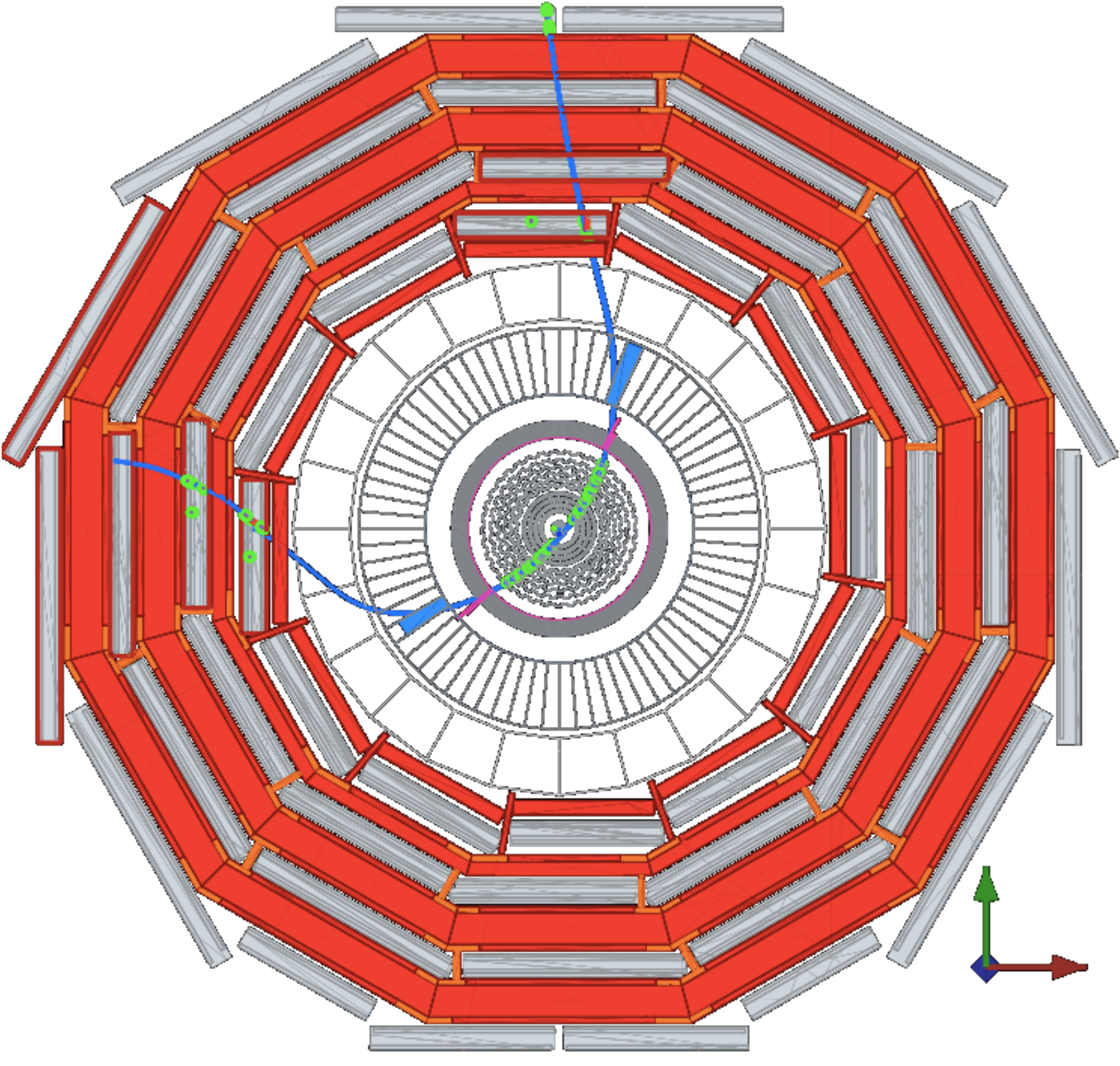}

     \vspace{0.5cm}
	  \includegraphics[width=0.8\textwidth]{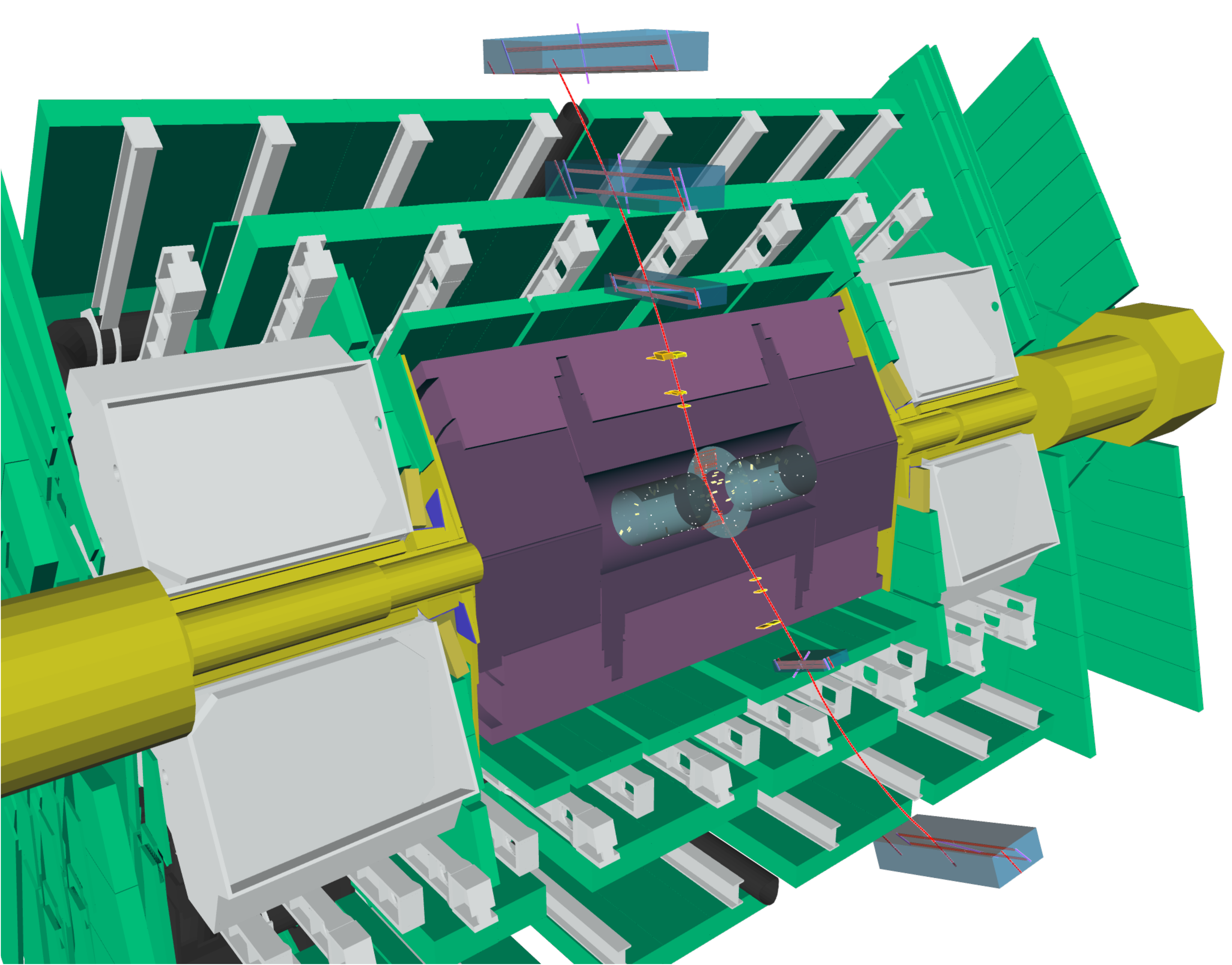}
  \end{center}
  \vspace{-0.0cm}
  \caption[.]{\underline{Top}: a cosmic ray muon measured by CMS, strongly 
              bent in the transverse plane by the 3.8\,T solenoid field.
              \underline{Bottom}:
              three-dimensional view of a cosmic ray muon in ATLAS.}
\label{fig:cosmicatlascms1}
\end{figure}

\begin{figure}[t]
  \begin{center}
	  \includegraphics[width=1.00\textwidth]{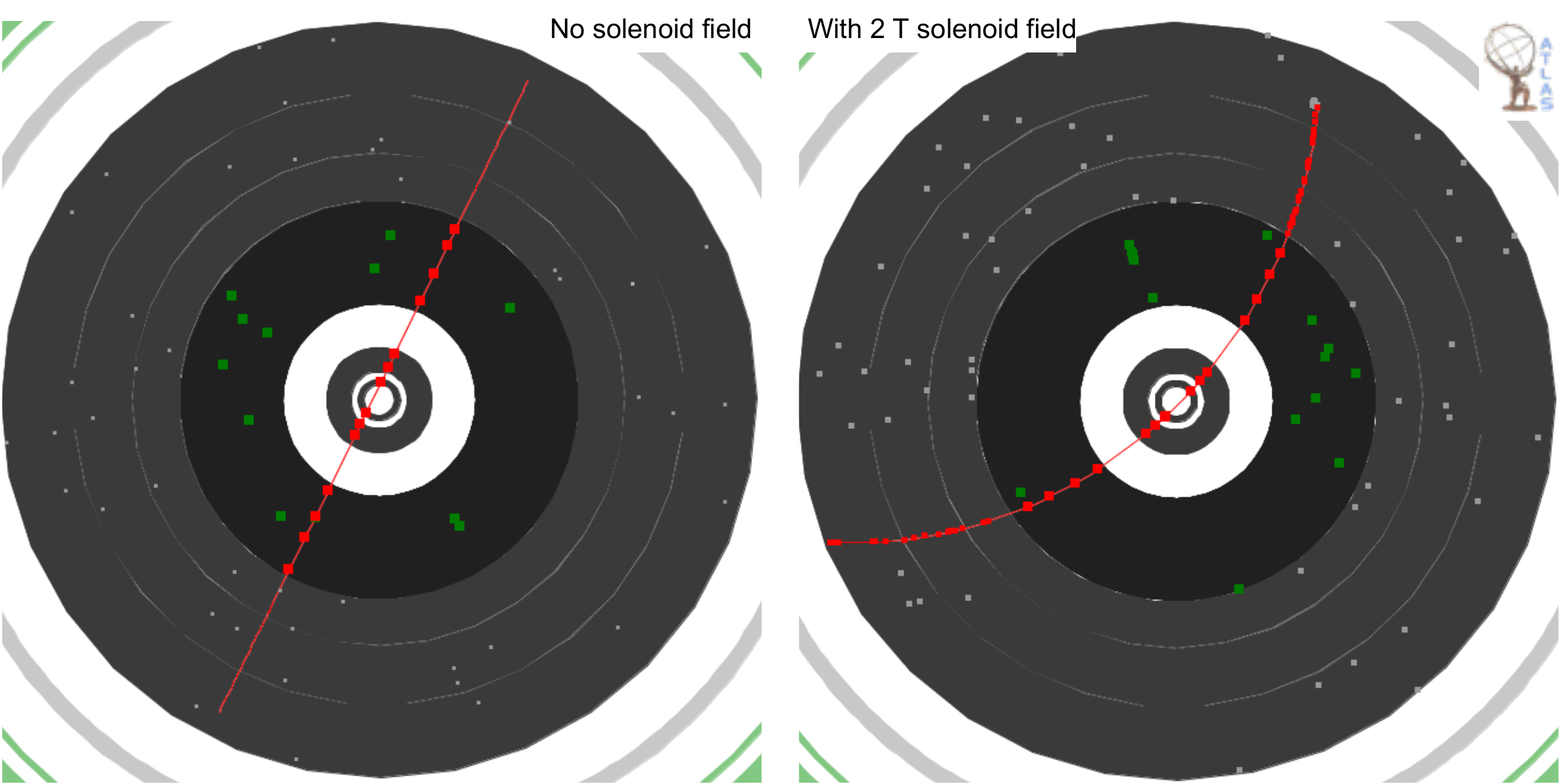}
  \end{center}
  \vspace{-0.3cm}
  \caption[.]{Transverse views of cosmic ray tracks measured in the ATLAS pixel 
              (the three innermost hits depicted by the dots) and silicon strip 
              detectors (four double hits at about half radius in the event displays). 
              The left (right) drawing shows a straight track measured with the 
              solenoid field off (on). The right plot shows also transition radiation 
              tracker hits. }
\label{fig:cosmicatlas3}
\end{figure}

\begin{figure}[t]
  \begin{center}
	  \includegraphics[width=1.00\textwidth]{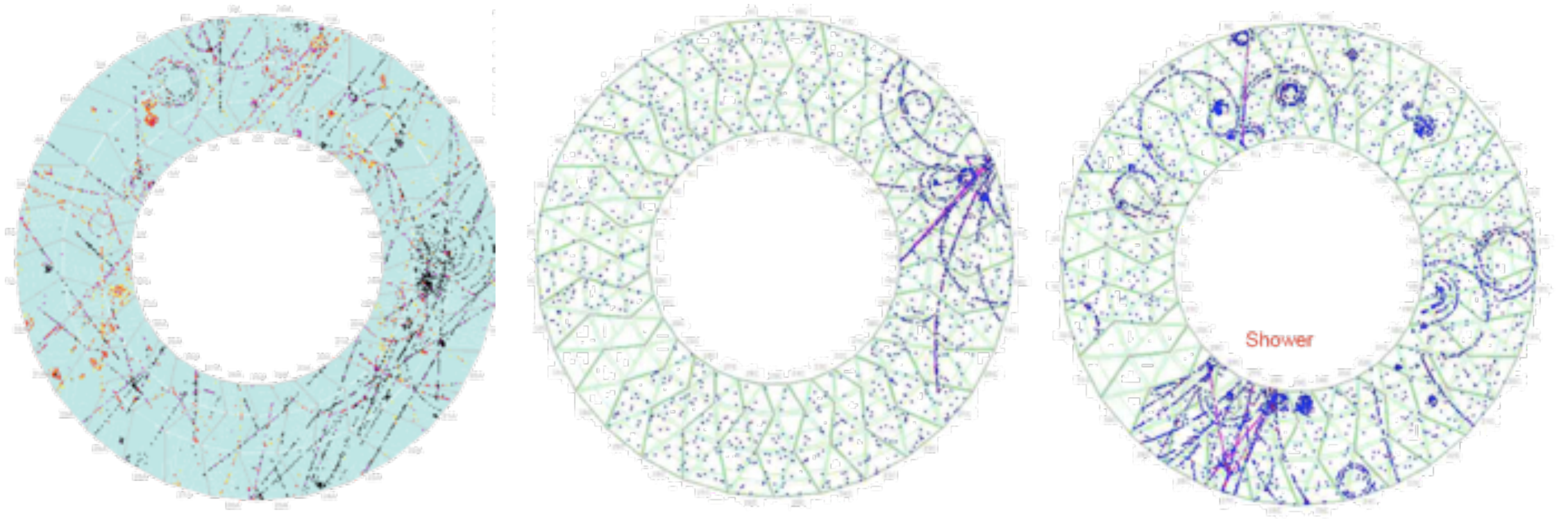}
  \end{center}
  \vspace{-0.3cm}
  \caption[.]{Cosmic ray shower tracks seen in the ATLAS transition radiation tracker. }
\label{fig:cosmicatlas4}
\end{figure}

Figures~\ref{fig:cosmicatlas1}--\ref{fig:cosmicatlas4} show 
event displays of cosmic rays in ATLAS and CMS, measured with the full detectors. ATLAS 
accumulated 580 million combined cosmic ray events between September 13 and October 
29, 2008, and in June/July and October/November 2009. CMS recorded 370 million combined 
events between October 13 and November 11, 2008 during the CRAFT exercise (many more cosmic
ray data have been recorded by CMS during other campaigns). All events have been promptly 
reconstructed at the CERN Tier-0 centre, reprocessed after software and conditions upgrades 
at the Tier-1 worldwide computing centres, and distributed for analysis on the LHC
Computing Grid. 

\begin{figure}[t]
  \begin{center}
	  \includegraphics[width=1\textwidth]{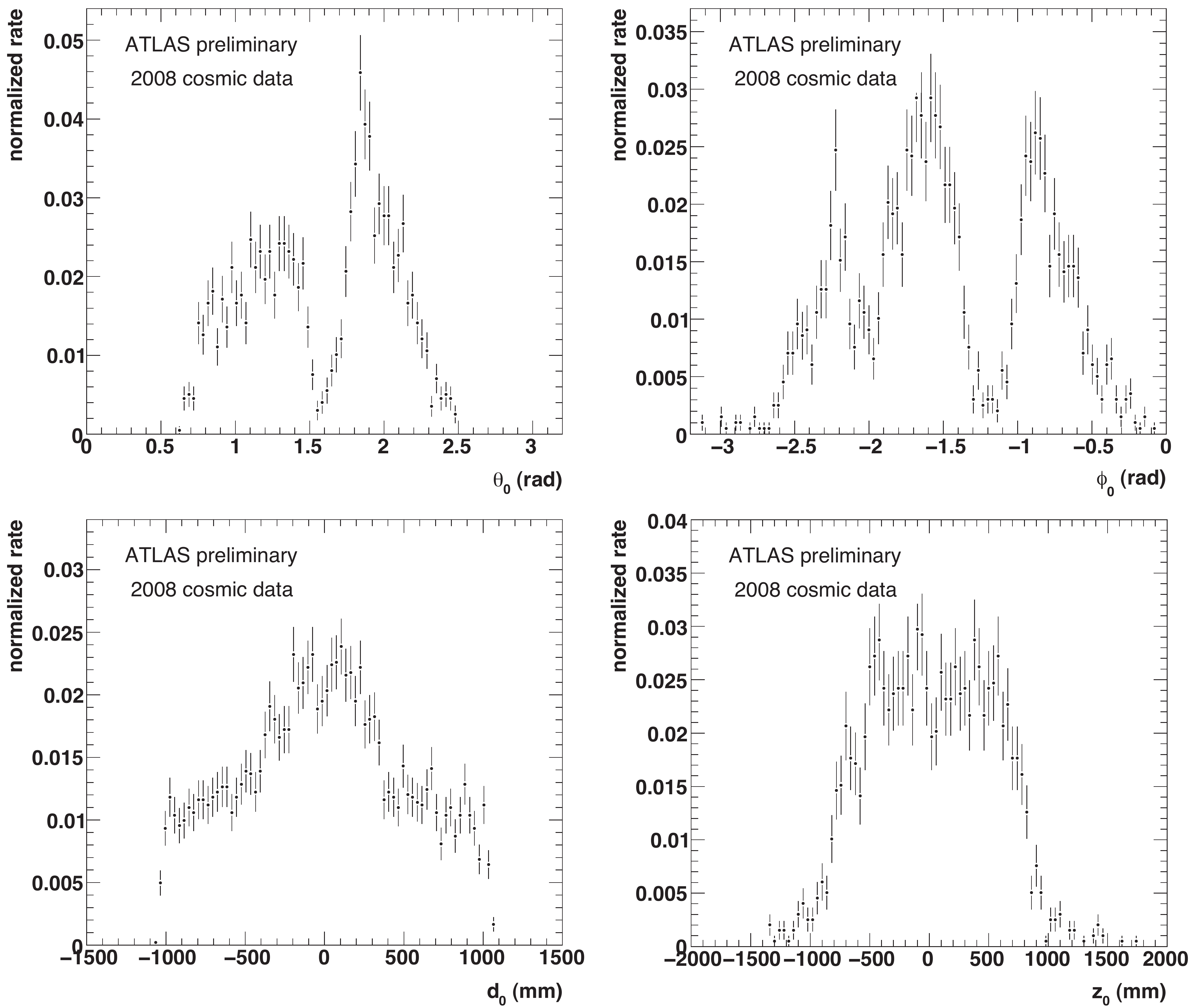}
  \end{center}
  \vspace{-0.5cm}
  \caption[.]{Track parameter distributions of cosmic muon tracks measured in the ATLAS inner
              tracker. Shown are the polar and azimuthal angles (upper plots) and transverse
              and longitudinal impact parameters (lower plots). The asymmetries reflect
              the top-down nature of cosmic tracks, and the shaft architecture of the 
              ATLAS cavern. }
  \label{fig:cosmicstrackdistatlas}
\end{figure}

\subsection{Cosmic ray spectra in the inner tracker}

Tracks bent in a magnetic field are characterised by five parameters. The parameters are 
defined with respect to a reference point, the perigee, which is the point of closest 
approach to the beam axis (along $z$). 
The impact parameters $d_0$ and $z_0$ are the signed distances to the $z$-axis and the 
$z$-coordinate of the perigee, respectively. Accordingly, the angles $\phi_0$ and $\theta_0$ 
are defined in the transverse plane and with respect to the $z$-axis at the perigee, 
respectively. The fifth parameter, $q/p$, is the charge of the cosmic muon divided by 
its momentum, defining curvature and orientation of the track helix. 

Figure~\ref{fig:cosmicstrackdistatlas} shows the angular and impact parameter distributions 
of cosmic muon tracks measured in the ATLAS inner tracker. The asymmetries reflect
the top-down nature of cosmic tracks, and the shaft architecture of the ATLAS cavern 
(Fig.~\ref{fig:cosmicsoccupancyatlas} on page~\pageref{fig:cosmicsoccupancyatlas}). 
For the $\theta_0$ and $z_0$ distributions, 
the tracks are required to have hits in the silicon detectors, because these parameters 
are not measured by the transition radiation tracker (barrel).

\subsection{Inner tracker alignment}
\label{sec:innertrackeralignment}

The high-precision tracking detectors of ATLAS and CMS, and the huge muons systems
(especially in ATLAS) challenge the accuracy with which the positions of the active detector
elements must be known. And although the detectors have been built and installed with the 
greatest care, it does not meet the requirements imposed by the detector performance and by
physics. Therefore the detectors have to be empirically {\em aligned}. 
Alignment signifies measuring the real detector positions and orientations 
from data, and correcting the reconstruction software accordingly. (It does not mean 
moving detector parts!). Several methods of varying complexity to solve alignment problems 
exist, and it is convenient to separate the alignment procedure into alignment levels, such 
as system, layer, and module, requiring increasing statistics due to an increasing number
of degrees of freedom. 

\subsubsection*{Alignment of the inner tracking systems}

The inner tracking systems of ATLAS and CMS (\cf Section~\ref{sec:detectors}) provide
excellent position resolution, with (ATLAS-barrel numbers) 10\mum ($r\phi$), 115\mum ($z$) 
for the Pixel device (total of 1744 modules), 17\mum ($r\phi$), 580\mum ($z$) for the silicon 
strip detector (4088 modules), and 130\mum ($r\phi$) per straw for the transition radiation 
tracker (2688 modules). A reasonable challenge is to align all parts of the detectors so that 
the track degradation due to misalignment not exceed 20\% of the intrinsic resolution. 
The sources of information used for alignment are fourfold:
($i$) assembly knowledge: construction precision and survey data, for the initial alignment 
precision, and for corrections and uncertainties; 
($ii$) online monitoring and alignment: lasers and optical cameras, before and during a run;
($iii$) offline track-based alignment: using physics and track residual information;
($iv$) offline monitoring: using physics observables, tracks and particle identification 
parameters.

Before coming to the alignment based on track residuals, let us briefly recall how a track 
momentum is measured. Charged particles are deflected in the homogeneous\footnote
{
    Not quite, as seen below.
}
axial field (\ie, the field is oriented parallel to the $z$ coordinate along the beam line) 
of the solenoid magnet. Since the Lorentz force 
is perpendicular to the magnetic ({\em B}) field and to the particle's flight vector, the particle 
trajectory projected onto the plane perpendicular to the {\em B} field describes a circle with 
radius $r\,[{\rm m}]=p_T\,[{\rm \GeV}]/(0.3\cdot B\,[{\rm T}])$. Thus, for transverse 
momenta between 10\,\GeV and 1000\,\GeV, one finds radii between 17\,m (9\,m) and 1700\,m (895\,m), 
for ATLAS (CMS), which are to be compared with the radius of $\sim$1\,m of the ATLAS and CMS 
inner tracking systems. Tracks with transverse momenta smaller than 0.3\,\GeV 
(ATLAS) or 0.6\,\GeV (CMS) become so-called `loopers', which travel a full circle in the 
inner tracker and do not reach the barrel electromagnetic calorimeter. The $r$ and $p_T$ 
values of a track are derived from the measurement of the  track's sagitta ($s$) by
$r\approx L/(8s)$ (if $s\ll L$), where $L$ is half the length of the transverse distance vector 
between the two extreme measurement points of the arc in the tracking system, and the sagitta 
determines the maximum distance between the intersection of the transverse distance vector
with the radius vector, and the arc (the sagitta measures the deviation of the arc from a 
straight line, $L$, \cf Fig.~\ref{fig:muonsagittasketch} on page~\pageref{fig:muonsagittasketch}). 
The smaller the sagitta $s$ the larger the 
radius and therefore the momentum of the track and, for constant precision on $s$,
the larger the relative error on the sagitta determination and hence on $p_T$: 
$p_T\propto s^{-1}$ and $\sigma(p_T)/p_T\propto p_T$. 

Track fitting in the LHC 
environment is very challenging. It must deal with ambiguities, hit overlaps, multiple 
scattering, bremsstrahlung, multiple vertices, etc. Track fitters take Gaussian noise 
(\eg, Kalman filter) and non-Gaussian noise (\eg, Gaussian sum filter) into account. 
Owing to the large number of tracks per event and because tracks are used for selection 
in the high-level trigger, the fits must be very fast. 

\begin{figure}[t]
  \begin{center}
	  \includegraphics[width=1.02\textwidth]{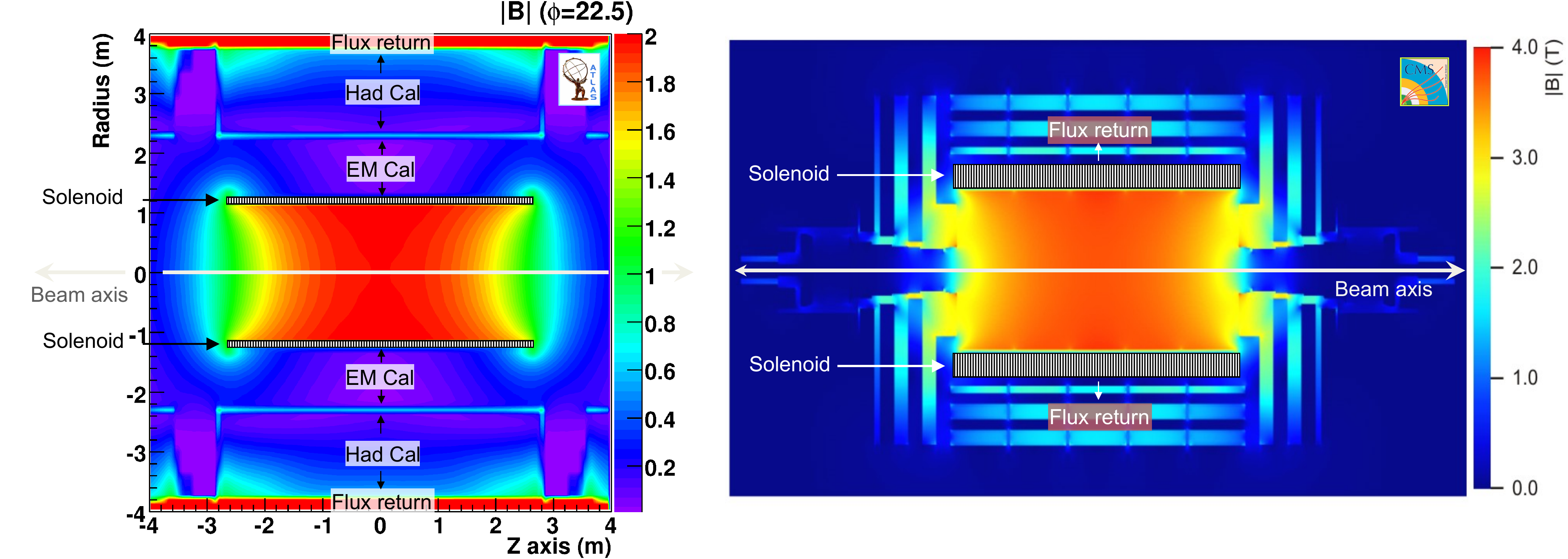}
  \end{center}
  \vspace{-0.3cm}
  \caption[.]{Solenoid fieldmaps for ATLAS (left) and CMS (right). The colour scales are
              indicated on the vertical axes. Because the CMS solenoid is much longer
              (axial length of 12.9\m compared to 5.3\m in ATLAS), the inner tracking detectors,
              with total active lengths of 5.6\m (ATLAS) and 5.4\m (CMS), see a more 
              homogeneous field in CMS than in ATLAS, where the inhomogeneities
              in the endcaps can reach up to 50\% (which are however accurately 
              mapped with magnetic field surveys and properly included in the 
              reconstruction).  }
\label{fig:solenoidfields}
\end{figure}
\begin{wrapfigure}{R}{0.35\textwidth}
  \vspace{-24pt}
  \begin{center}
	  \includegraphics[width=0.35\textwidth]{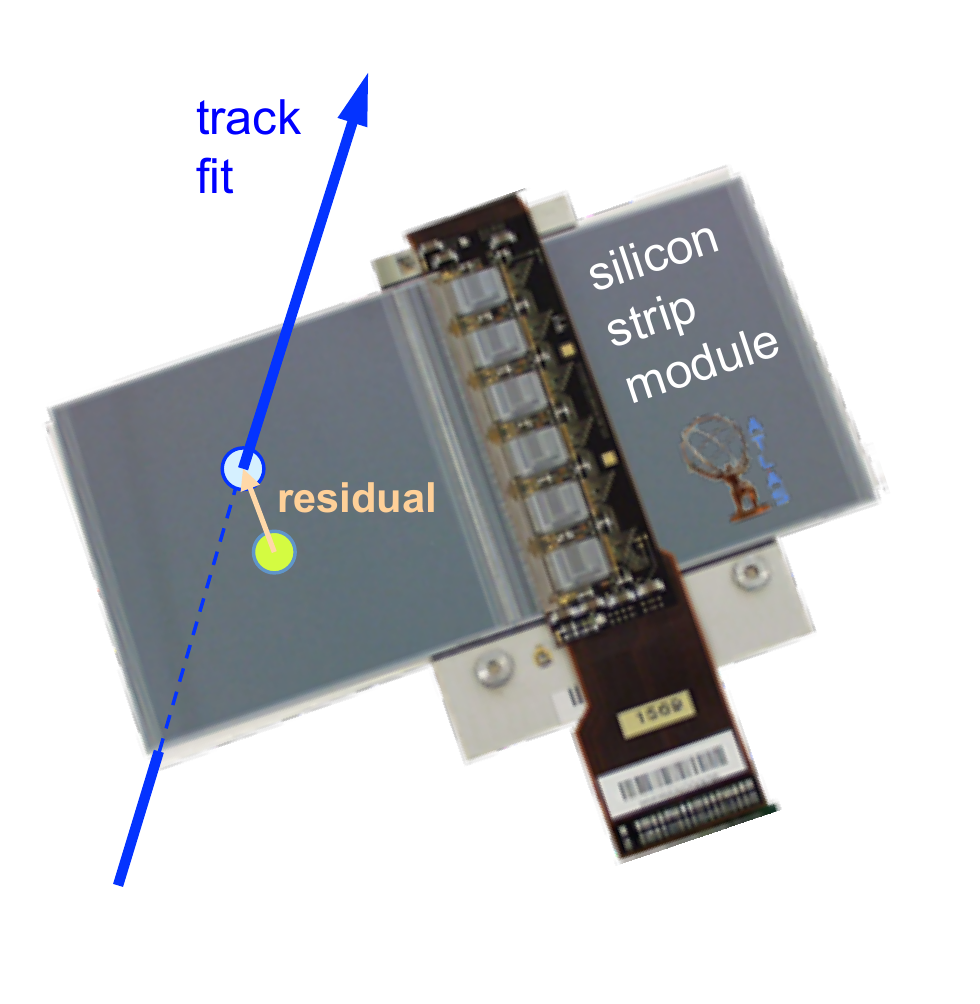}
  \end{center}
  \vspace{-30pt}
  \caption{Sketch of a track model through an ATLAS silicon strip tracker module, and 
           a measured close-by hit defining the hit residual. }
  \label{fig:moduleresidual}
  \vspace{-6pt}
\end{wrapfigure}
Figure~\ref{fig:solenoidfields} shows the superconducting solenoid field maps for ATLAS and 
CMS. Inhomogeneities in the magnetic field strengths occur towards the end of the solenoids,
which are strongly influenced by the magnetic structure of the nearby detector elements.
The $\sim$2\,T flux return yoke in CMS is used for muon momentum measurement.
(The ATLAS return yoke, integrated into the tile hadronic calorimeter and its support structure,
also produces a $\sim$2\,T.m azimuthal track deviation, which is, however, not measured 
precisely in the muon spectrometer and hence not used for momentum measurement.)

The alignment algorithm minimises the track residuals by fitting detector positions 
(layers and modules) to measured tracks (Fig.~\ref{fig:moduleresidual}). The fit 
minimises a global estimator, which could be written by
$\chi^2=\sum_{i\in {\rm hits}}(m(\vec{\alpha})-h_i)^2/\sigma_i^2$ ,
where the function $m$ corresponds to the model prediction (track) at module of hit $i$, 
$\vec{\alpha}$ is the vector of track parameters, and $h_i$ and $\sigma_i$ are the 
measured hits and their errors. The full global $\chi^2$ function must, however, also 
account for correlations so that it becomes:
$\chi^2=\sum_{\rm tracks}(r^T V^{-1}r)$, where the residuals $r$ are functions
of the track parameters, the alignment parameters and the hit measurements 
along a track. The $\chi^2$ function is simultaneously minimised with respect 
to the track and the alignment parameters. 

\begin{figure}[t]
  \begin{center}
    \includegraphics[width=0.32\textwidth]{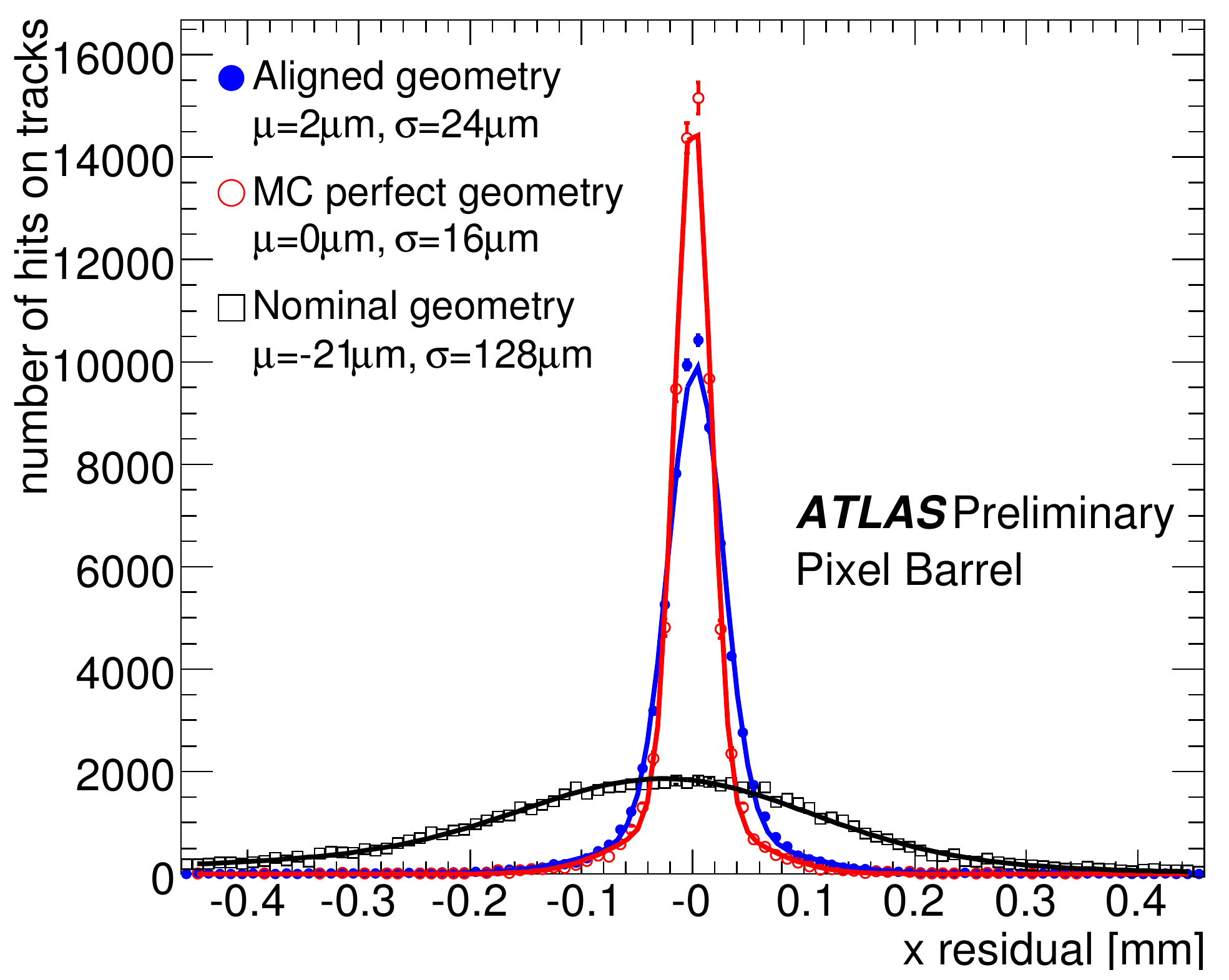}
    \includegraphics[width=0.32\textwidth]{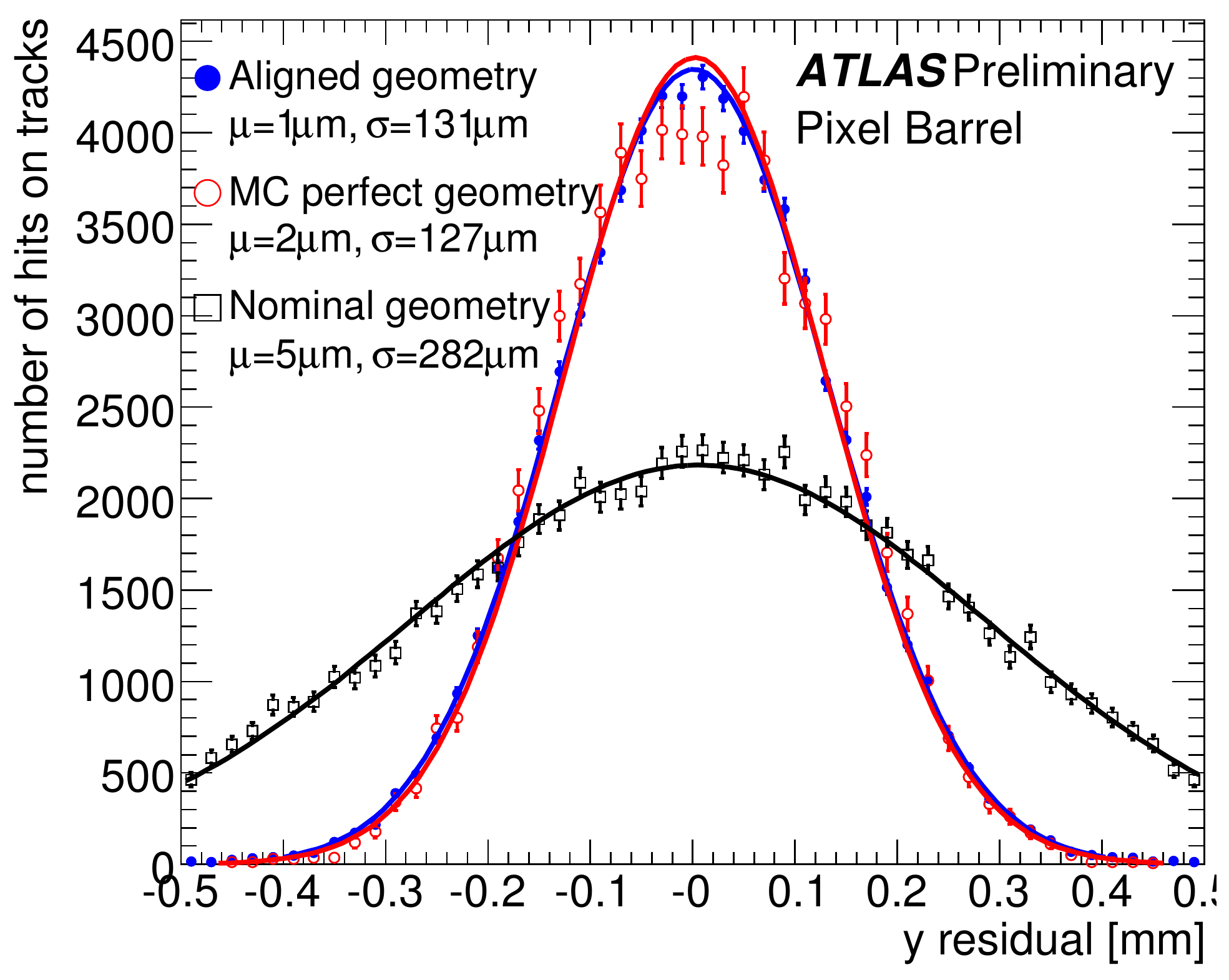}
    \includegraphics[width=0.32\textwidth]{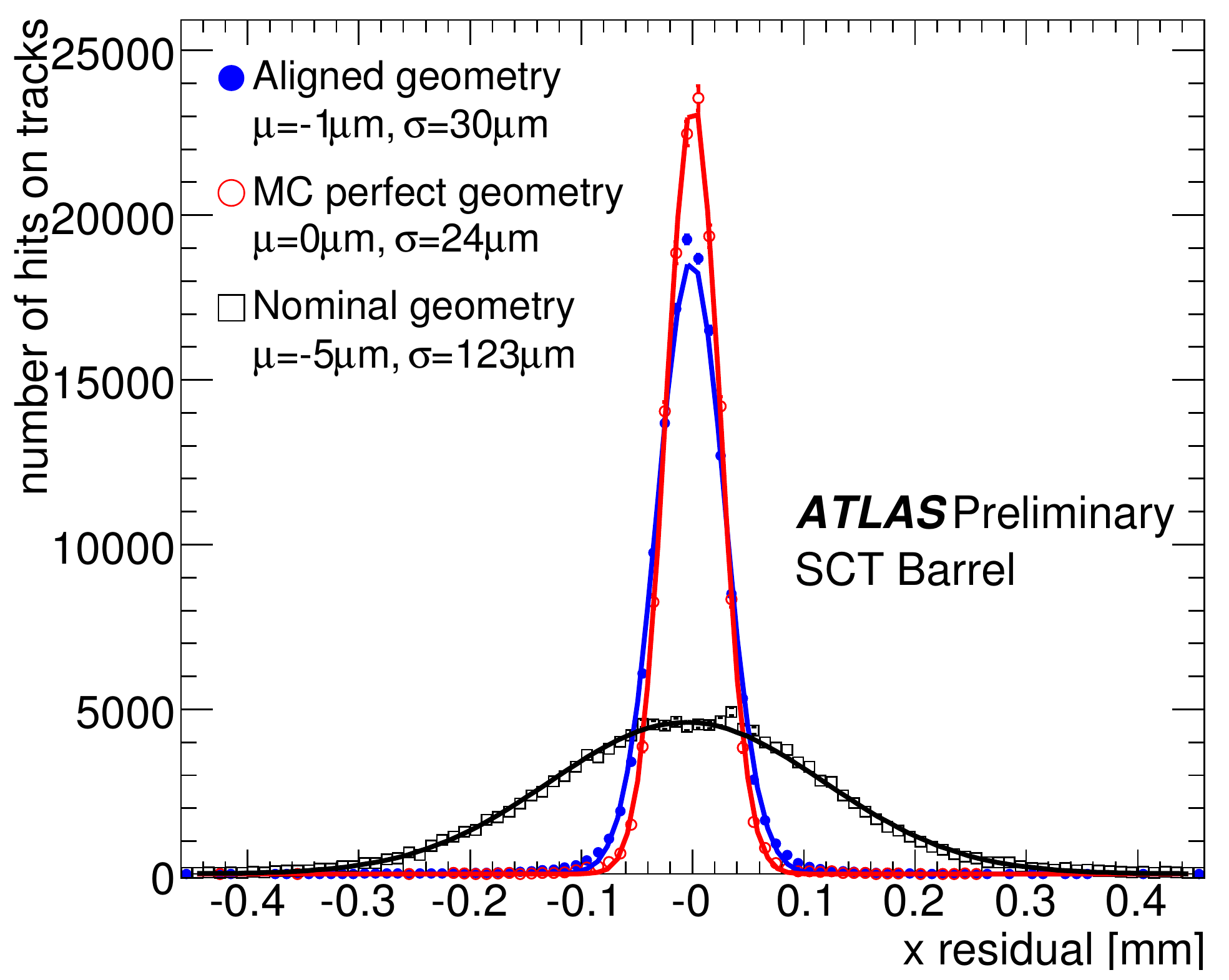} \\
    \includegraphics[width=0.321\textwidth]{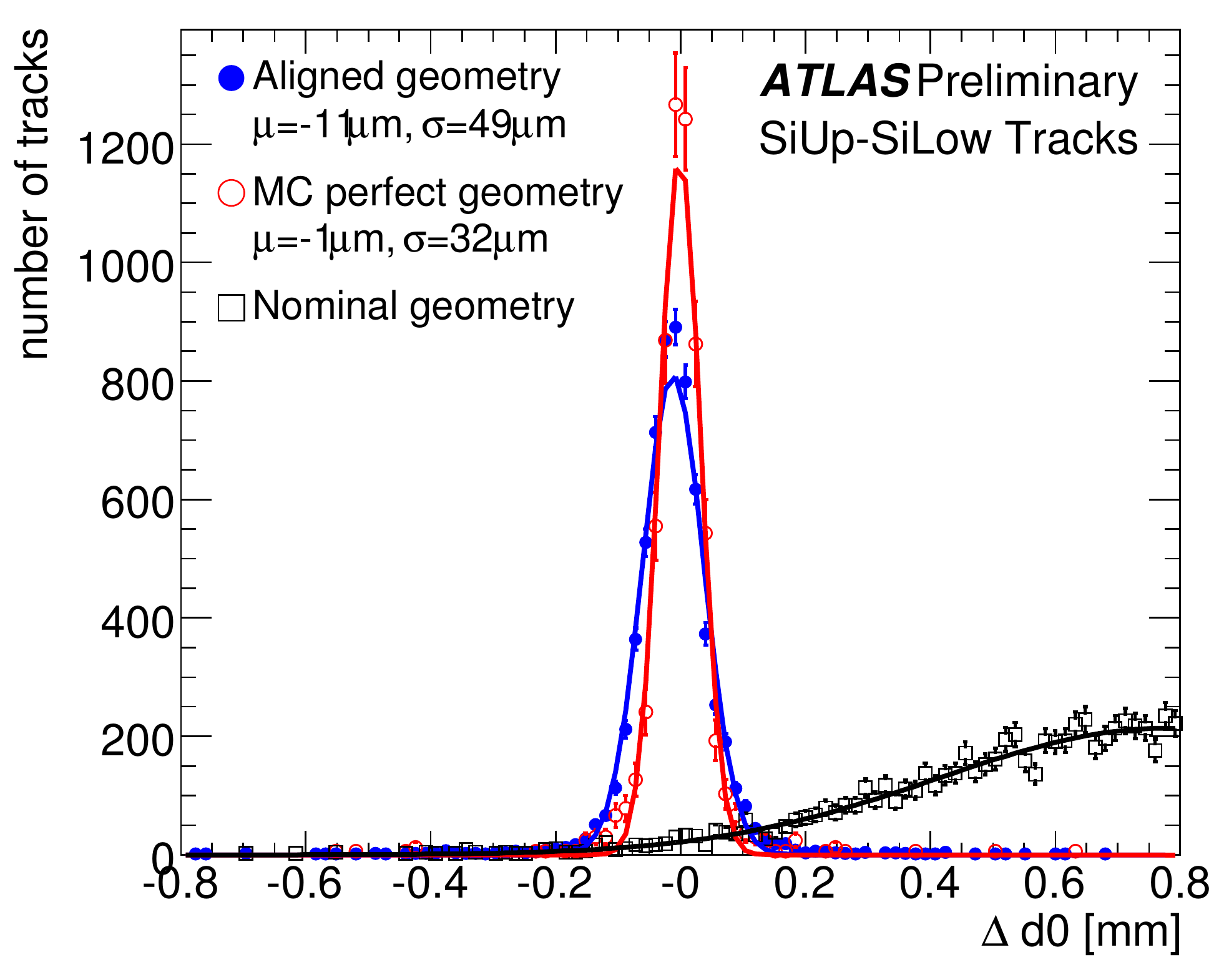} 
    \includegraphics[width=0.321\textwidth]{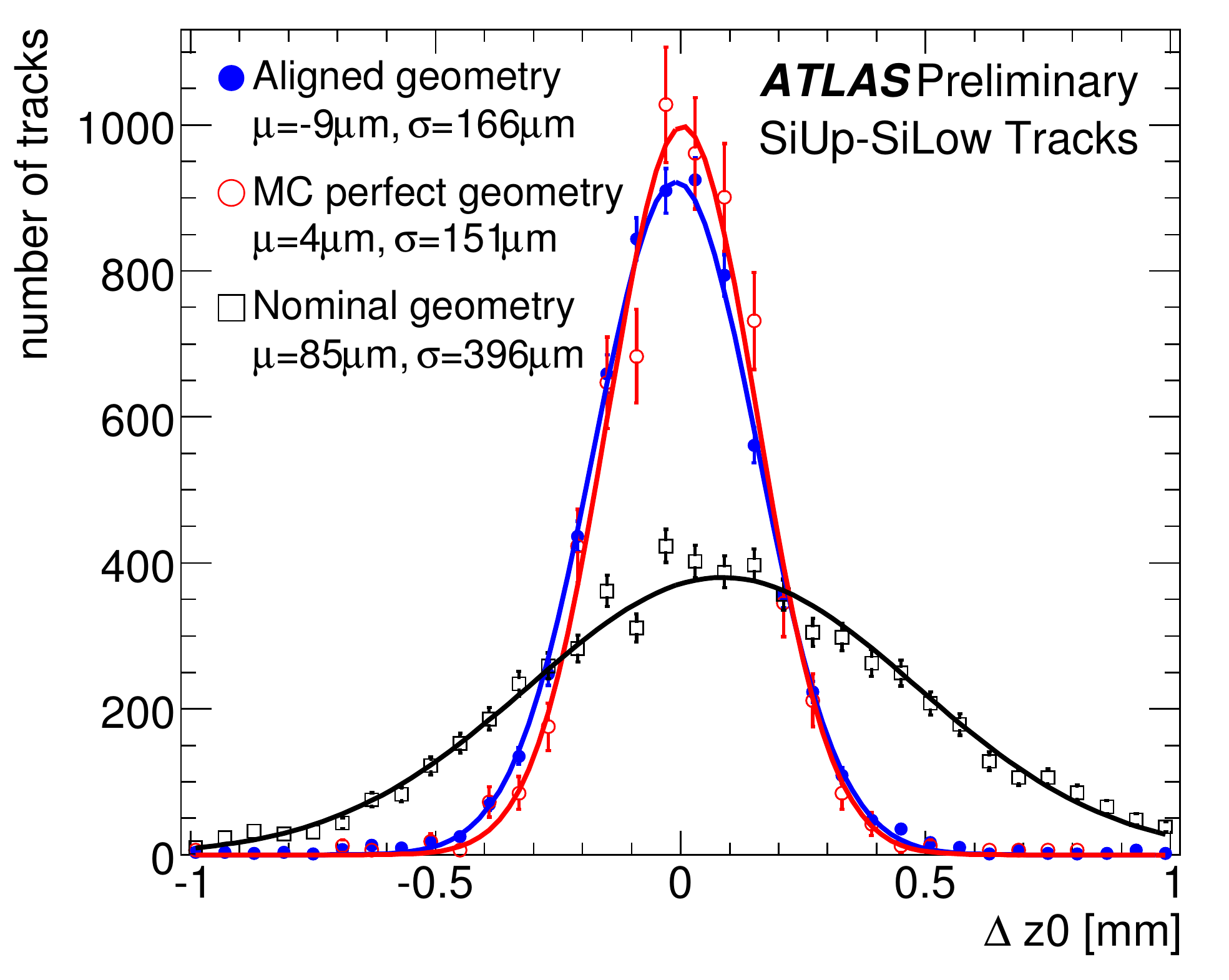}
    \includegraphics[width=0.321\textwidth]{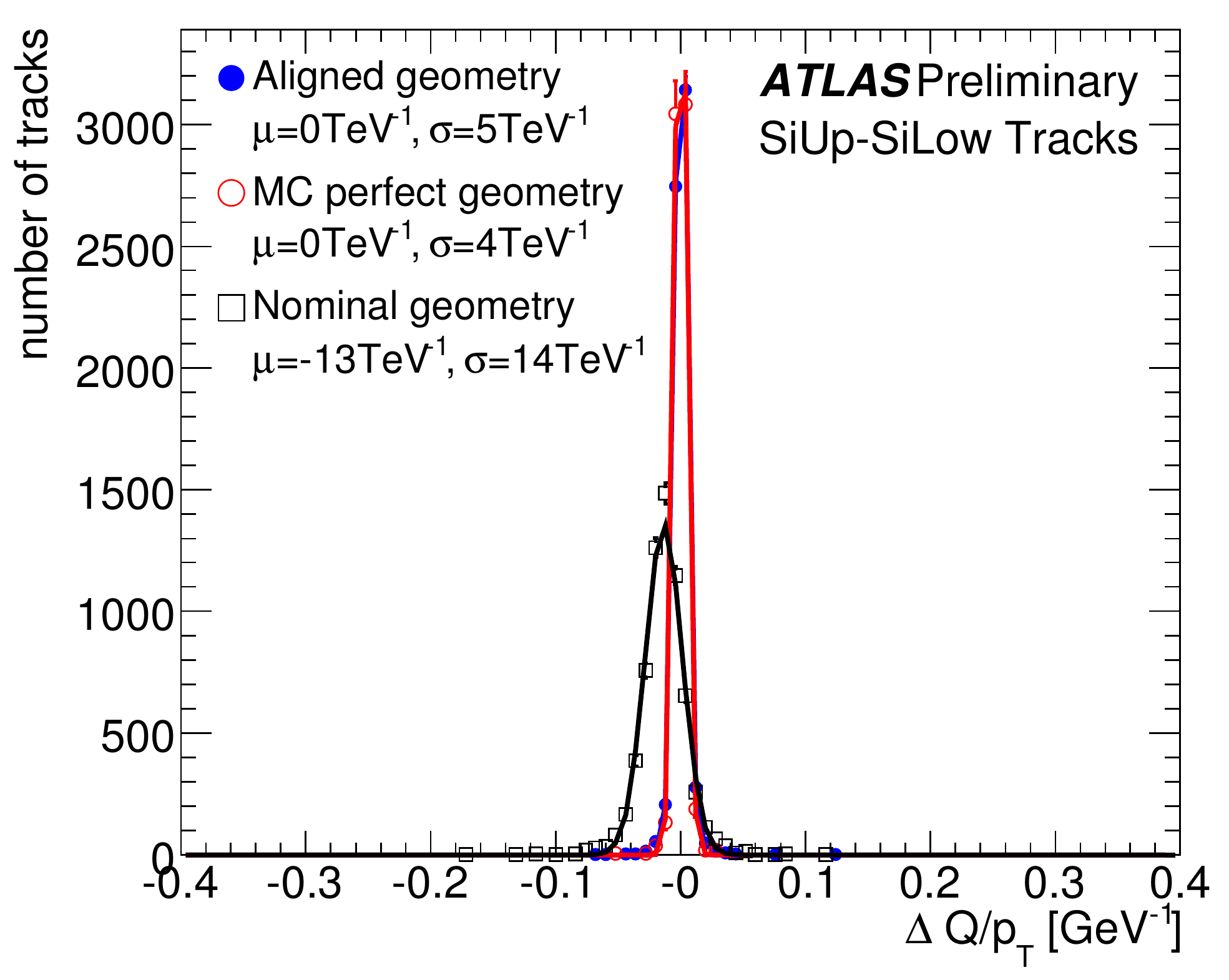}
  \end{center}
  \vspace{-0.4cm}
  \caption[.]{Hit residuals (upper plots) for the ATLAS pixel and silicon strip detectors 
              before alignment (open squares), after alignment with cosmic ray tracks 
              (full circles), and for ideal conditions from Monte Carlo simulation (open 
              circles). The lower plots give the impact parameter resolution for the same 
              three data samples. The resolution is obtained with the track-splitting 
              technique (see text). }
  \label{fig:trackalignmentatlas}
\end{figure}
The smallest movable object in the alignment procedure
is a module, which has 6 degrees of freedom: 3 translation coordinates and 3 rotation
angles. Taking into account the total number of modules of ca. 8500 (ATLAS number),
one obtains 51\,000 degrees of freedom that need to be
determined by the fit. Depending on the alignment level (whole barrel/endcap, 
layers/disks, modules) different techniques can be used, where for either 
of these the correlations between fit parameters are important ingredients to help 
the fit converge rapidly. Neglecting correlations may not lead to a wrong fit result, 
after full convergence, but it is less efficient. 

Figure~\ref{fig:trackalignmentatlas}
shows residual distributions for the ATLAS pixel and silicon strip detectors, as
well as impact parameter and $Q/p_T$ distributions, before and after alignment with 
cosmic ray tracks. The widths of these distributions are convolutions 
of the intrinsic hit and tracking resolution (seen under ideal conditions), and 
misalignment effects. The impact parameter and transverse momentum resolutions are
obtained by splitting a cosmic ray muon track traversing the full detector into 
two tracks that are re-fit independently and compared.\footnote
{
  The resolution is the RMS of the difference divided by $\sqrt{2}$.
} 
A total of 4.9 (2.7) million
tracks with solenoid field on (off) have been used by ATLAS (similar numbers of tracks
are used by CMS for alignment), of which 1.2 million (230 thousand) have silicon 
strip (pixel) track components so that they can be used to align these detectors. 
Alignment results close to ideal have been obtained. 

\subsubsection*{Weak modes}

\begin{figure}[t]
  \begin{center}
	  \includegraphics[width=0.6\textwidth]{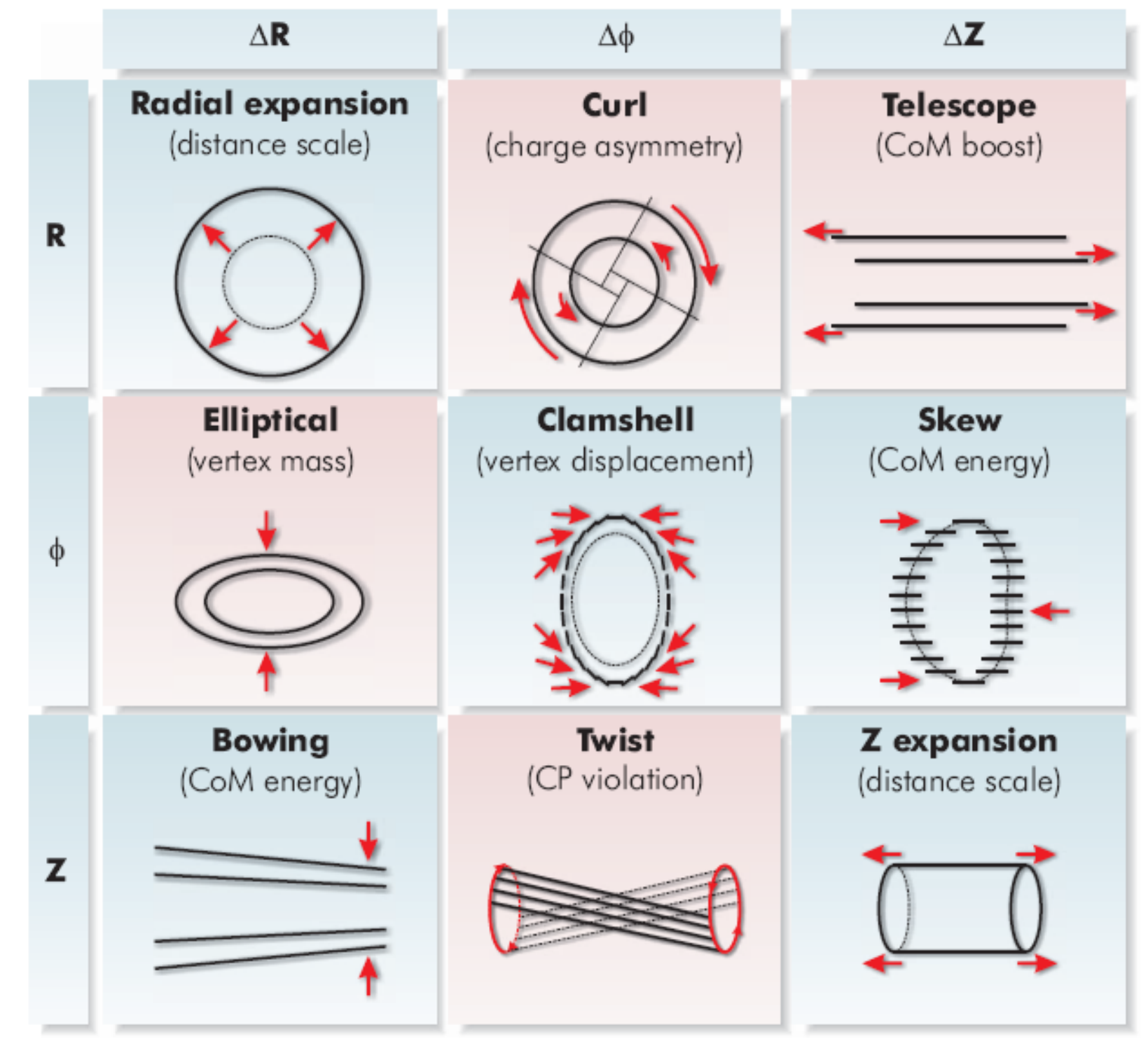}
  \end{center}
  \vspace{-0.3cm}
  \caption[.]{Different types of misalignment according to transverse distortions in 
              $R$, $\phi$, and deformations along the beam axis ($z$). 
              The pink types leave the $\chi^2$ estimator approximately 
              invariant (`weak modes').}
  \label{fig:alignweakmodes}
\end{figure}

Unfortunately, the minimisation of hit residuals does not guarantee that indeed the true
positions of the detector elements have been determined. This is because the residuals,
and hence the $\chi^2$ estimator, are insensitive against some types of misalignment, 
which may nevertheless impact the physics performance. Examples for such `weak 
modes' are elliptical skews, \ie, distortions of the type $\delta\phi=\lambda+\beta/R$
or $\delta z\approx R$. Figure~\ref{fig:alignweakmodes} summarises the various 
types of misalignment. The pink-coloured types represent weak modes in the global
residual-based $\chi^2$ estimator. Weak modes contribute to the lowest part of 
the eigenspectrum. Their deformations bias physics measurements and lead to 
systematic effects. The understanding of these effects is thus of utmost importance.
Weak modes can be constrained by adding more information to the fit, such as: 
($i$) cosmic ray and beam halo tracks (off-beam axis) in addition to beam collision data;
($ii$) vertex and beam-spot constraints;
($iii$) resonance masses ($Z$, $J/\psi$, $\Upsilon$, $K_S$, \dots);
($iv$) $E/p$ measurements for electrons; and 
($v$) survey data and mechanical constraints. 

\subsection{Inner tracker resolution}

\begin{figure}[t]
  \begin{center}
	  \includegraphics[width=1\textwidth]{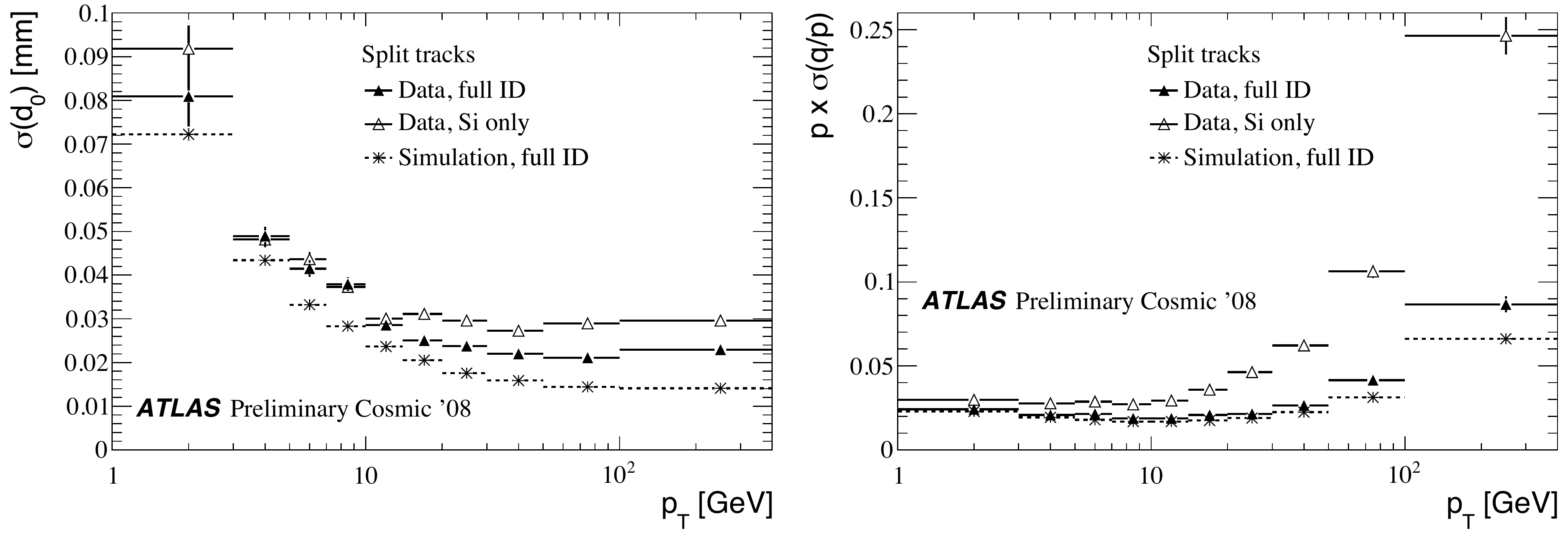}
  \end{center}
  \vspace{-0.5cm}
  \caption[.]{Transverse impact parameter resolution (left) and relative momentum 
              resolution (right) versus the transverse momentum for the ATLAS 
              inner tracker. The full (open) triangles give the results for all inner tracker
              detectors combined (only silicon pixel and strip detectors), and the asterisk 
              is the expectation from Monte Carlo simulation with ideal alignment conditions.}
  \label{fig:cosmicstrackresolutionatlas}
\end{figure}
The tracking resolution for cosmic ray muons in the inner tracker is studied by comparing 
track parameters at the perigees using the track-splitting technique. Because both tracks 
emerging from the splitting have errors, the quoted resolution is the RMS of the residual 
distribution of a track parameter divided by $\sqrt{2}$. Well reconstructed tracks are 
selected for these studies. ATLAS requires a minimum number
of hits in Pixel, silicon strip detector and transition radiation tracker of 2, 6 and 25, 
respectively, and $|d_0|<40$\,mm and $p_T>1$\,\GeV, and good timing properties. 
The left-hand plot of Fig.~\ref{fig:cosmicstrackresolutionatlas} shows the transverse impact 
parameter resolution versus the transverse momentum for the ATLAS inner tracker. In the low 
$p_T$ region, the resolution is dominated by multiple scattering. At higher momenta, the 
resolution becomes independent of the momentum as is expected for almost straight tracks. 
Including the transition radiation tracker information improves 
the resolution due to the extended lever arm. The difference between data and the Monte
Carlo prediction is a measure of the remaining misalignment. The right-hand plot of 
shows the relative momentum resolution versus $p_T$.  At intermediate momentum, reduced 
multiple scattering counterbalances the $p_T$-dependent rise of the error due to a 
decreasing relative accuracy of the sagitta measurement. This latter effect dominates 
at higher momentum. Again, the difference with respect to the Monte Carlo expectation 
stems from residual misalignment. 

\subsection{Muon spectrometer alignment}

\begin{wrapfigure}{R}{0.4\textwidth}
  \vspace{-24pt}
  \begin{center}
	  \includegraphics[width=0.4\textwidth]{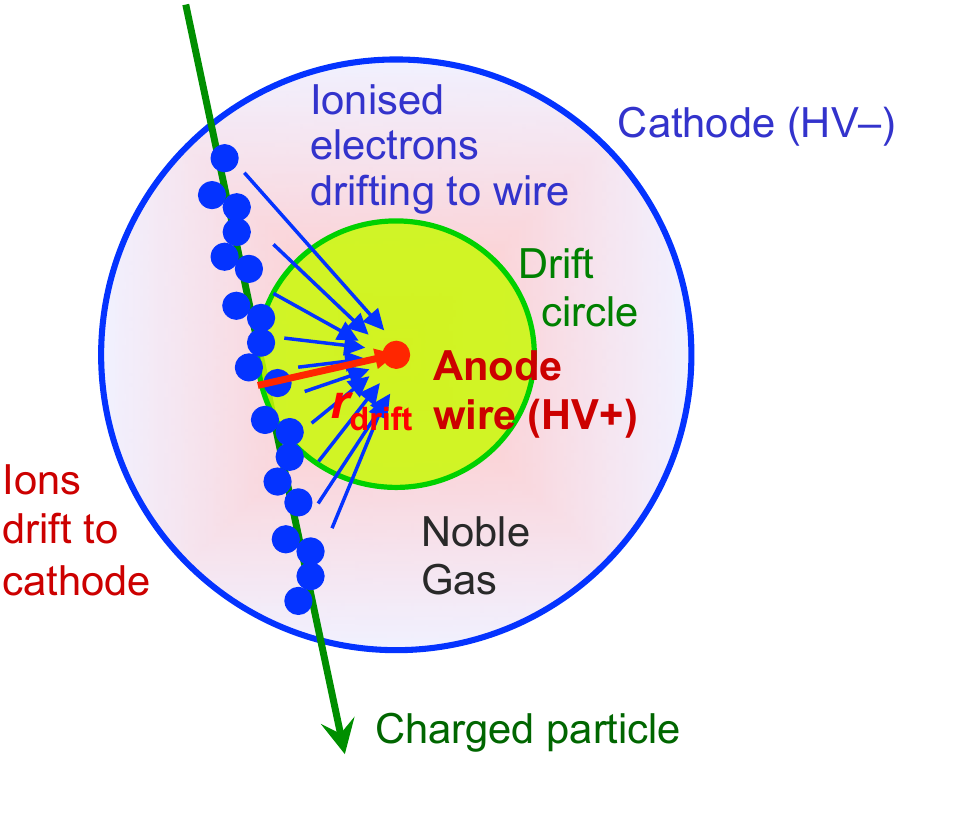}
  \end{center}
  \vspace{-20pt}
  \caption{Principle of a drift tube used for precision measurement in the ATLAS and CMS muon 
           systems, and also in the ATLAS transition radiation tracker.}
  \label{fig:drifttubesketch}
  \vspace{+70pt}
\end{wrapfigure}
The huge active volumes of the ATLAS and CMS muon spectrometers require a detailed 
understanding of the inhomogeneous magnetic fields (especially for ATLAS and the CMS endcaps) 
and the chamber positions to achieve design performance. To derive quantitative requirements,
let us briefly recall how the muon precision measurements are obtained. Both 
experiments use drift tubes, which are standalone coaxial cylindrical drift chambers
functioning similarly to proportional tubes, in the barrel (ATLAS also in the outer 
endcaps for $|\eta|<2.0$), and cathode strip chambers in the forward direction. 

The drift tubes 
in ATLAS (denoted `monitored drift tubes' --- MDT) are made of thin aluminium 
tubes with 3\,cm diameter (4\,\cm in CMS, 4\,mm for the ATLAS transition radiation tracker), 
filled with a 93\% argon and 7\% CO$_2$ gas mixture at 3\,bar pressure 
(Fig.~\ref{fig:drifttubesketch}). A 50\mum  gold-plated tungsten wire in the centre of 
each tube serves as anode with an applied potential of 3080\,V. A charged track traversing 
the tube ionises the gas and the ionised electrons drift in the electrical field to the wire, 
while the ions drift to the cathode (cylinder). From the measured hit time of the induced 
electrical pulse, and the known drift velocity (`space-drift time ($r$-$t$) relation'), it 
is possible to determine a {\em drift circle} around the anode wire, tangential to which 
the track has passed. 

The measurement of several adjacent layers of tubes provides the redundant information 
required for a full track fit. The measured drift time in a tube reaches up to 800\,ns 
corresponding to a drift velocity of approximately 18\,km/s. The average position resolution 
is 80\mum per tube (250\mum in CMS), but varies strongly along the drift radius: tracks very
far from the anode wire are measured with better precision than close tracks, due to the 
smaller dispersion in the drift time of the incoming electrons. 

\begin{figure}[t]
  \begin{center}
	  \includegraphics[width=0.99\textwidth]{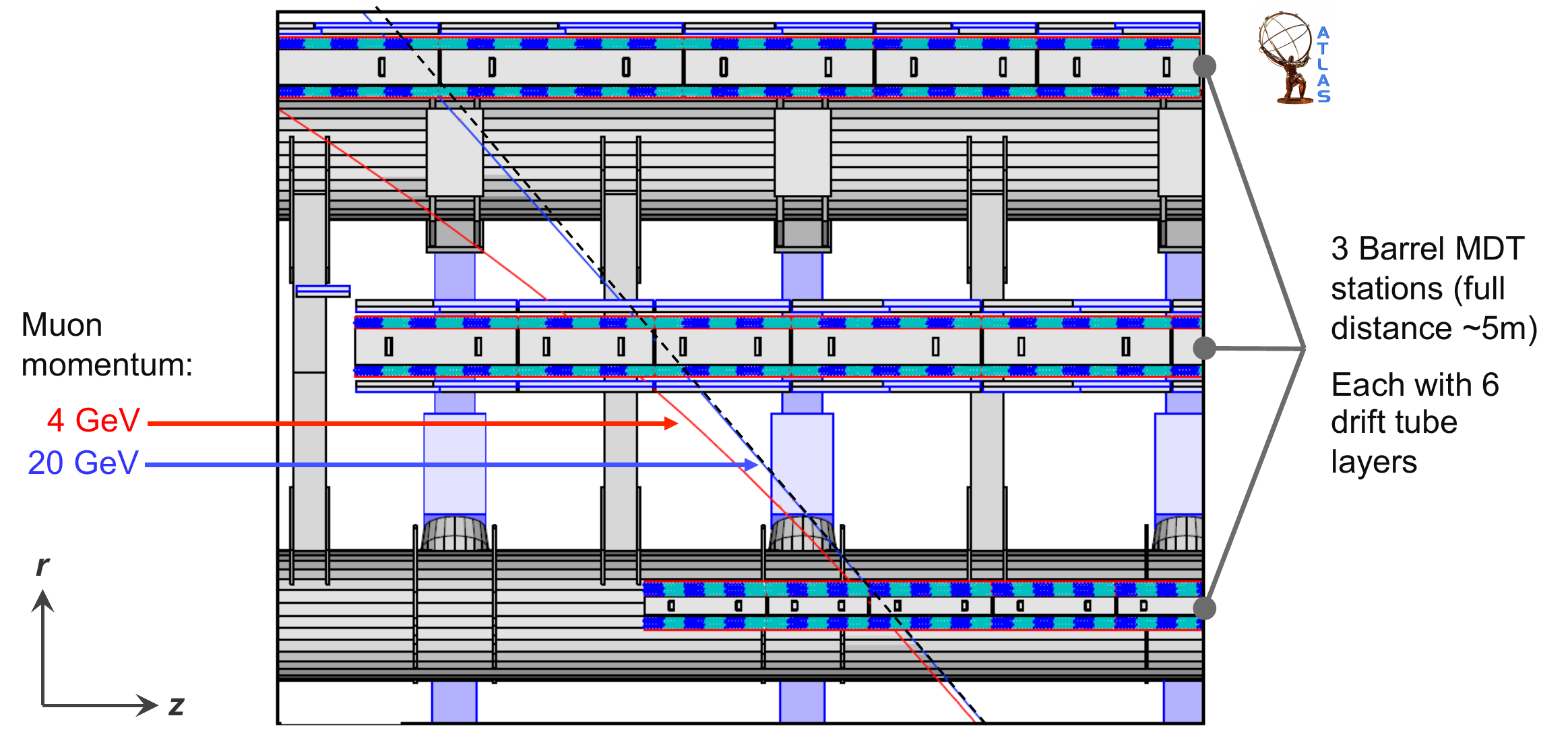}
  \end{center}
  \vspace{-0.3cm}
  \caption[.]{Drawing of ATLAS barrel monitored drift tube stations. There are three
              of these spanning a full radial distance of $\sim$5\,m. Shown by the curved
              lines are simulated muon tracks with 4\,\GeV (red) and 20\,\GeV (blue),
              bent in the $z$ direction by the toroidal magnetic fields. The curvature is 
              hardly visible for the latter track (see straight dashed line for comparison).
              The MDT system is designed to measure 1\,\TeV tracks with 10\% relative
              accuracy, requiring a position alignment of better than 40\mum.}
  \label{fig:mdtchambersatlas}
\end{figure}

\begin{wrapfigure}{R}{0.45\textwidth}
  \vspace{-24pt}
  \begin{center}
	  \includegraphics[width=0.45\textwidth]{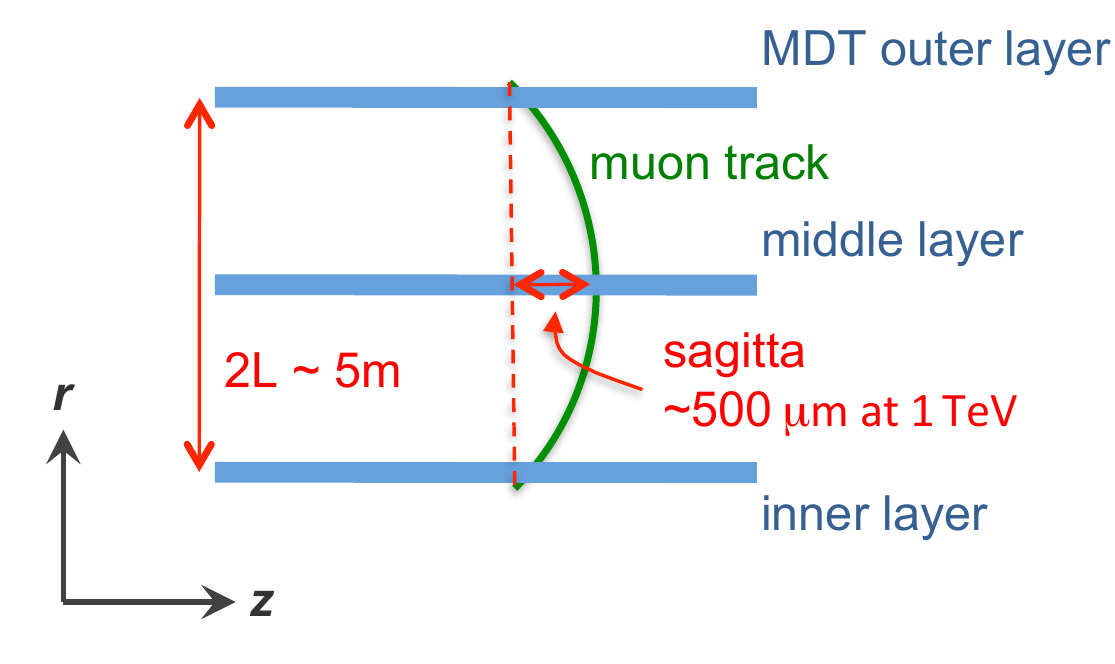}
  \end{center}
  \vspace{-15pt}
  \caption{Sketch for the muon sagitta measurement in ATLAS. For a 1\,\TeV track 
           the sagitta measures about 500\mum.}
  \label{fig:muonsagittasketch}
  \vspace{-6pt}
\end{wrapfigure}

The ATLAS drift tubes are arranged in large-sized MDT chambers
with six tube layers oriented along $\phi$ to allow for a precise measurement of the 
$z$ coordinate, the direction of which the charged particles are bent in the toroidal
magnetic fields. Three almost equally spaced stations of MDT chambers (inner, middle 
and outer) are installed in the barrel with about 2.5\,m radial distance from each 
other (Fig.~\ref{fig:mdtchambersatlas}). A 1\,\TeV track has a sagitta of about 
$s=500$\mum at $\eta=0$ (\cf sketch in Fig.~\ref{fig:muonsagittasketch}). 
A measurement of that sagitta with 10\% accuracy requires the error induced by 
misalignment to be significantly smaller than 50\mum. With 
$\sigma(s)\approx\sqrt{3/2}\cdot\sigma(z)$, one finds $\sigma_{\rm misallign}(z)\ll40$\mum, 
which represents a tremendous alignment challenge given the size of the system. 

\begin{figure}[t]
  \begin{center}
	  \includegraphics[width=0.99\textwidth]{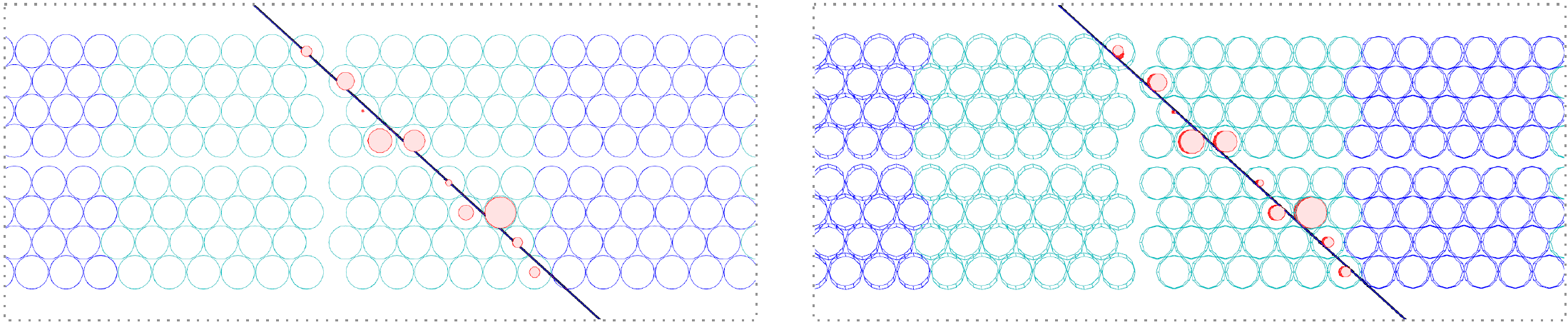}
  \end{center}
  \vspace{-0.3cm}
  \caption[.]{Example of a misaligned drift tube chamber in the ATLAS barrel muon spectrometer 
              (simulation). In the left-hand picture, without alignment corrections, 
              it is not possible to draw a straight line track through the drift circles. 
              After alignment (right-hand picture) the chambers 
              have been slightly tilted so that a good track fit can be obtained.}
  \label{fig:mdtalignmentsketchatlas}
\end{figure}

Figure~\ref{fig:mdtalignmentsketchatlas} shows an example of a misaligned MDT chamber 
in ATLAS (from simulation). In the left drawing, where no alignment corrections have 
been applied, the track is not tangential to all drift circles. The $\chi^2$ of the 
track fit is bad. In the right drawing the chambers have been aligned leading to a 
good track fit.

\newpage
\subsubsection*{Optical muon chamber alignment in ATLAS}

ATLAS implements a twofold alignment strategy for the muon system: fits to measured tracks 
from cosmic rays and collision events, in particular using straight tracks without 
the toroid fields, provide the absolute MDT chamber positions.\footnote
{
   Full alignment not only requires a proper positioning of the chambers and tubes in 
   the chambers, but one must also correct for the wire sag in the drift tubes, which has been 
   measured from survey data for a fraction of the tubes, and must be derived from track
   fits for the remaining ones. The wire-sag induced error in the position measurement 
   amounts to 20--30\mum, depending on the size of the MDT chamber. 
} 
Relative chamber movements due to temperature-dependent `breathing' and when 
switching on the toroid magnets, are monitored by means of an optical alignment system,
designed to detect slow chamber displacements, occurring at a timescale of hours or more. 
The system is based on optical and temperature sensors, and on alignment bars, which 
are up to 9.6\,m long instrumented aluminium tubes used as precision reference rulers. 
The information from the optical system together with the track-based alignment is 
used in the offline track reconstruction to correct for the MDT chamber misalignment.
Similar to ATLAS, CMS is instrumented with a precise and complex opto-mechanical 
alignment system that provides a common reference frame between tracker and muon 
detection systems by means of a net of laser beams. We discuss in the following 
the ATLAS system. 

To first order, only the relative alignment of triplets of chambers traversed by the same 
muon track is important for a precise sagitta measurement.  The barrel optical alignment 
system thus uses 3-point straightness monitors, which are installed on the inner, middle 
and outer chambers to form projective lines pointing to the interaction region.\footnote
{
   In the endcaps, projective lines cannot be installed because the cryostats of the endcap 
   toroid magnets block the way to the interaction region. The optical alignment system 
   thus relies on high-precision reference rulers and alignment bars forming an alignment 
   grid.
}
The straightness monitor creates a highly redundant image of a coded mask (for example a 
chess-like pattern) through a lens onto a charged-coupled device (CCD) acting as screen.
The mask is lit by infrared LEDs passed through a diffuser to minimise effects of 
imperfections in the light source. The relative position in transverse direction to the projective
lines is measured along the line mask, the optical centre of the lens, and the CCD camera. 
It is also possible to measure the (relative) rotation of the mask or the sensor, and 
the relative rotation around any axis of the mask with respect to the CCD camera.
Finally, by computing the actual image size and comparing it with the known mask 
size  (magnification), the position of the lens along the longitudinal axis can be obtained.
A total of 6000 (7000) optical lines have been installed in the ATLAS barrel (endcap). 
Not all of these are projective. In the barrel, praxial lines align adjacent chambers 
in each layer. In the endcaps there are bars, polar and proximity lines. 

\begin{figure}[t]
  \begin{center}
	  \includegraphics[width=0.7\textwidth]{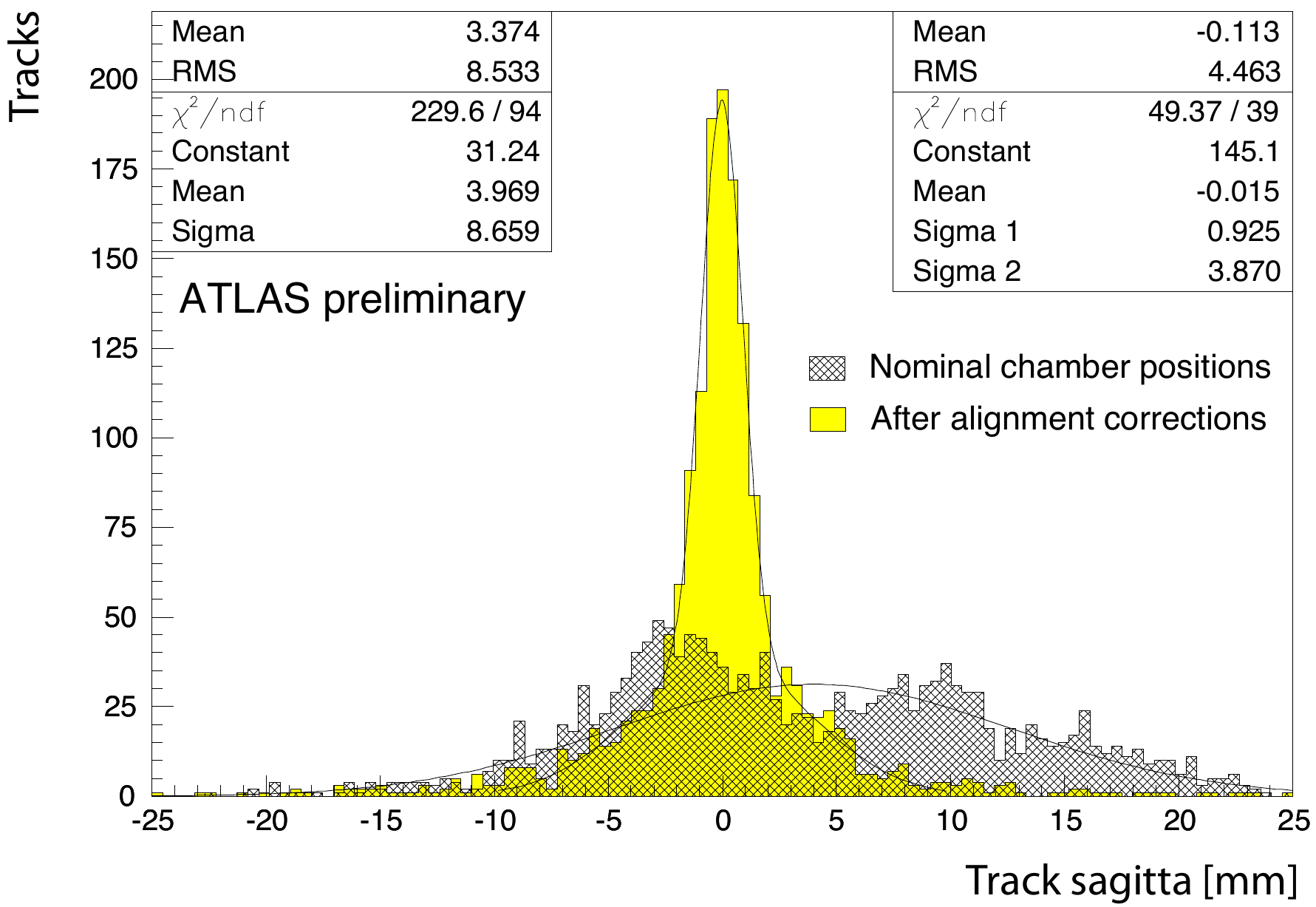}
  \end{center}
  \vspace{-0.3cm}
  \caption[.]{Track `sagitta' for straight cosmic ray muon tracks (toroid fields off) 
              in the ATLAS endcaps before 
              (dark shaded) and after (light shaded/yellow) applying the optical alignment.
              The sagitta is calculated from the distance in the precision coordinate of the 
              middle chamber segment from the line joining the inner and outer endcap 
              segments. After alignment, the resolution (width) is dominated by multiple 
              scattering effects. }
  \label{fig:muonalignmentresultatlas}
\end{figure}
\begin{figure}[t]
  \begin{center}
	  \includegraphics[width=0.9\textwidth]{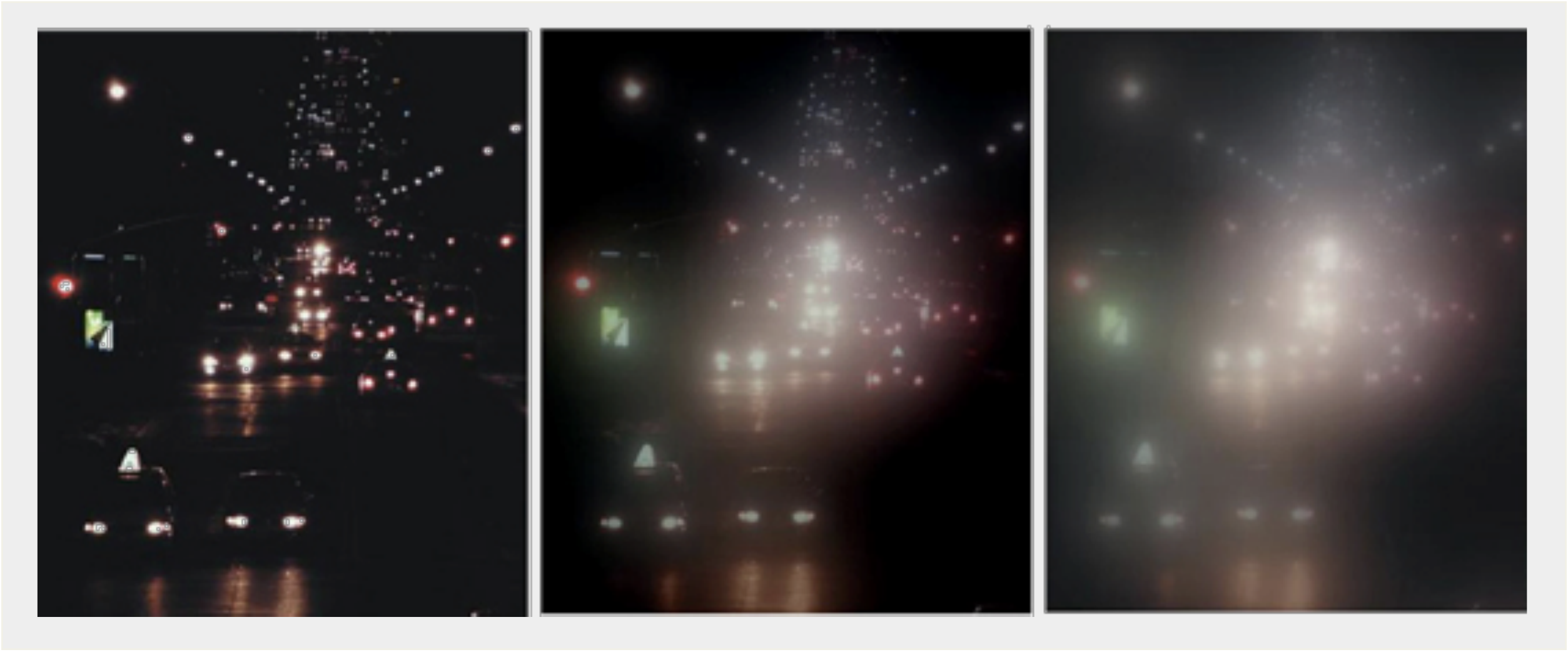}
  \end{center}
  \vspace{-0.3cm}
  \caption[.]{Multiple scattering (diffusion) of light passing through a wetter 
              and wetter windscreen (left to right).  }
  \label{fig:multiplescattering}
\end{figure}

The absolute resolution of the optical alignment system is of the order of 300--500\mum,
which is insufficient for precision measurements. Hence the necessity to rely on 
track measurements for absolute chamber positions. The relative optical alignment accuracy 
has been evaluated with simulated muon shifts of the H8 test beam arrangement 
and found to correct misalignment within 14\mum error (RMS) on the sagitta, which 
is well within the specified requirement~\cite{atlasmuonalign}. 
Figure~\ref{fig:muonalignmentresultatlas} shows the distribution of sagitta values
for straight cosmic muon tracks (the toroid magnets were turned off so the expected 
sagitta is zero) in the ATLAS
endcaps before and after applying the optical alignment. The sagitta is computed from 
the distance in the precision coordinate of the middle chamber segment from the line 
joining the inner and outer chamber segments.  The resolution found is compatible 
with the expectation. The tails in the sagitta distribution after alignment originate 
from {\em multiple scattering}. 
\begin{details}
{\bf\em Digression.} Multiple scattering denotes the deflection by (or convolution of) successive 
small-angle scatters of a charged particle traversing a medium. The multiple scattering
cross section, dominated by Coulomb scattering from nuclei, is proportional to 
$\sqrt{{\rm pathlength}/X_0}\cdot p^{-1}$, \ie, it is enhanced for soft particles
and dense matter. The angular distribution is 
approximately Gaussian at small angles (owing to the central limit theorem), but 
also large-angle Rutherford scattering occurs with a differential cross 
section $\propto \sin^{–4}(\theta/2)$.
Multiple scattering is analogous to diffusion. Figure~\ref{fig:multiplescattering} 
shows the effect of light diffusion on a wet windscreen. The more matter in terms of 
radiation lengths a particle traverses in the tracking volume, the more the detector
`sees' the particle as we see other cars at night in rainy weather with a broken 
wiper. Multiple scattering complicates the track fitting and limits the resolution
of the momentum measurement.
\end{details}

\begin{figure}[t]
  \begin{center}
 	  \includegraphics[width=1\textwidth]{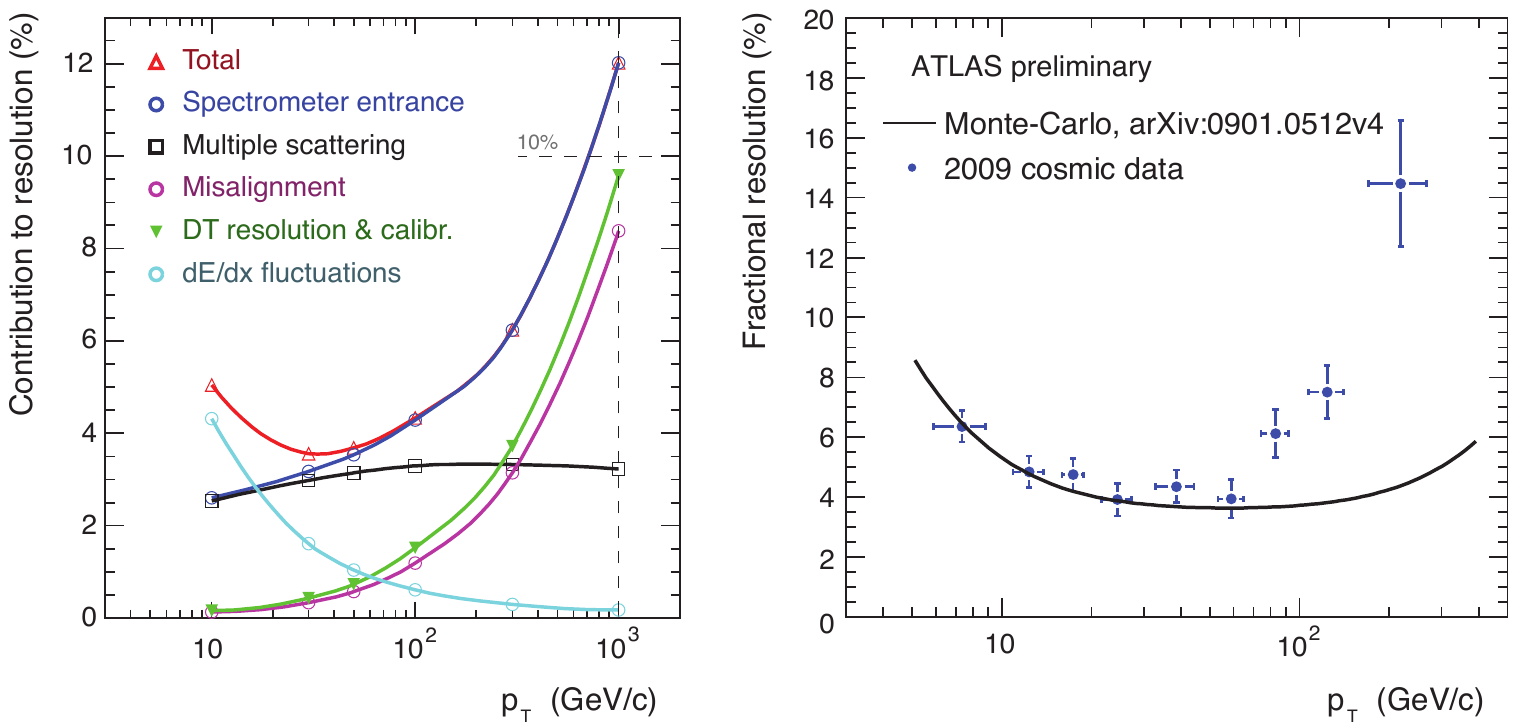}
  \end{center}
  \vspace{-0.5cm}
  \caption[.]{\underline{Left}: expected contributions to the standalone muon momentum resolution of 
                   the ATLAS barrel muon spectrometer (Monte Carlo simulation). See text for a discussion 
                   of the various terms. \underline{Right}: fractional standalone momentum resolution 
                   measured by comparing split top and bottom muon spectrometer tracks in cosmic ray 
                   data. The degradation of the measured resolution with respect to the expected one 
                   is mainly due to imperfect alignment, but also due to $r$-$t$ relation inaccuracies
                   due to the missing LHC clock reference. } 
  \label{fig:muonstandaloneresolutionatlas}
\end{figure}

Figure~\ref{fig:muonstandaloneresolutionatlas} (left) gives the contributions to 
the standalone muon momentum resolution versus the incident momentum of the ATLAS 
barrel muon spectrometer. Multiple scattering (black line) determines the resolution 
for momenta below $\sim$200\,\GeV. At very low momentum (below 20\,\GeV) the 
fluctuations in the energy loss of the muon traversing the calorimeters becomes 
the dominant effect (cyan coloured line --- the blue line indicates the 
resolution with respect to the entrance at the muon spectrometer). However, below 
100\,\GeV the momentum measurement is in any case dominated by the inner
tracking system. For high-momentum muons the contribution from the intrinsic drift 
tube resolution and $r$-$t$ calibration is of similar magnitude as the expected systematic 
error in the mechanical alignment, hence the challenge for the alignment system. 
The right-hand plot in Fig.~\ref{fig:muonstandaloneresolutionatlas} shows the
fractional standalone momentum resolution measured by comparing top and bottom 
muon spectrometer tracks in cosmic ray data (track splitting method). The measured resolution 
is compatible with the expected one from Monte Carlo simulation at transverse momenta 
below 100\,\GeV, and is degraded at higher momenta. The degradation is caused by 
imperfect alignment of the muon chambers and by limited timing accuracy because 
cosmic muons are not synchronous with the artificial LHC clock used in drift 
time measurements (no fixed time reference).

\subsection{Muon charge asymmetry in cosmic rays}

The charge ratio of positive to negative muons in cosmic rays, with momenta in the range 
10--300\,\GeV, has been measured to be 1.27 at sea level~\cite{muoncharge}, and is 
expected to increase somewhat with the muon momentum due to a growing influence 
from kaon decays (the charge ratio of pion decays is expected to be 
approximately 1.25, while it is 2 for kaons~\cite{minoschargeratio}). 

\begin{figure}[t]
  \begin{center}
 	  \includegraphics[width=1\textwidth]{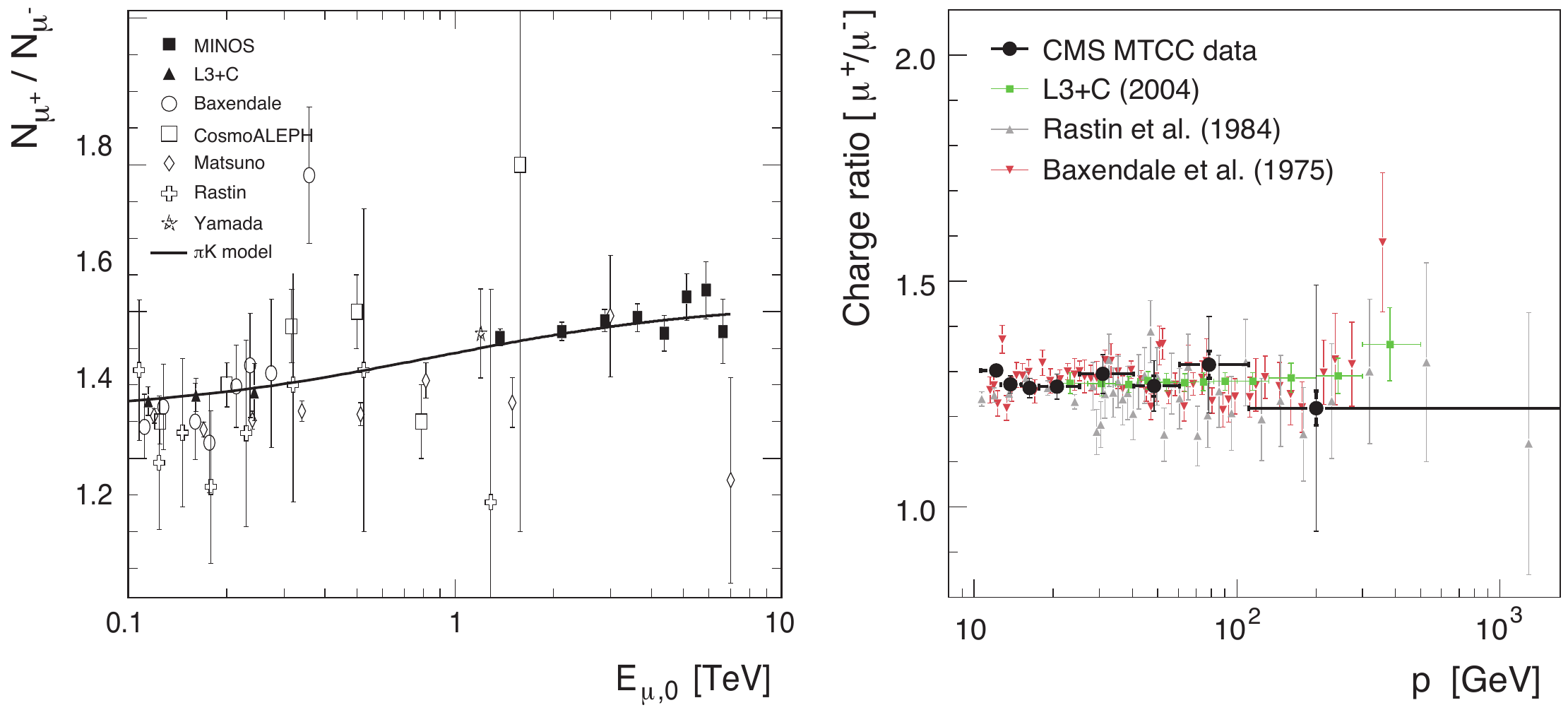}
  \end{center}
  \vspace{-0.4cm}
  \caption[.]{\underline{Left}: muon charge-ratio measurements compiled by the 
                    MINOS experiment~\cite{minoschargeratio}. \underline{Right}:
                    muon charge ratio measured by CMS (black dots) with statistical 
                    (bold bars) and systematic errors (thin bars), together with results 
                    from other experiments (see Ref.~\cite{cmschargerationote} for references).}

  \label{fig:muonchargeratio}
\end{figure}

In 2006, during the `Magnet Test and Cosmic Challenge (MTCC)', CMS performed a 
measurement of the muon charge asymmetry on the surface, using a 30$^\circ$ slice of the detector
including the muon drift tubes in presence of a 4\,T solenoid field~\cite{cmschargerationote}. 
Owing to the high muon rate at the surface, 337\,000 high quality tracks with hits in 
at least 3 (of 4) barrel stations and transverse momentum larger than 3\,\GeV could be selected.
The most important systematic effect on the charge measurement stems from the charge-dependent 
alignment uncertainty, in particular for high muon momenta. The resolution-induced charge 
misidentification probability is estimated from Monte Carlo simulation and also contributes 
significantly to the systematic error above 100\,\GeV (no inner tracking used). The total 
systematic error varies between 2\% below 10\,\GeV, $\sim$8\% at 100\,GeV, and up to and beyond 
20\% above 100\,\GeV. It exceeds the statistical errors at all muon momenta. To compare the raw 
charge ratio measurement with other measurements, the result is expressed in terms of the muon 
momentum before entering CMS using Monte Carlo simulation. The resulting momentum correction 
is about $+$7\,\GeV and almost independent of the muon momentum. Figure~\ref{fig:muonchargeratio}
(right plot) shows the charge-ratio measurements versus the corrected muon momentum,
together with results from other sources (see references in Ref.~\cite{cmschargerationote}).
Within their uncertainties, the CMS results can be regarded as independent of the muon 
momentum, giving the average $R_{\mu^+/\mu^-}=1.282\pm0.004\pm0.007$, where the 
first error is statistical and the second systematic. The left plot in Fig.~\ref{fig:muonchargeratio}
gives a compilation of previous muon charge-ratio data between 0.1 and 7\,\TeV taken 
from a MINOS publication~\cite{minoschargeratio}. Superimposed is the model expectation.

\subsection{Combining muon and inner tracker reconstruction}

\begin{figure}[p]
  \begin{center}
 	  \includegraphics[width=0.48\textwidth]{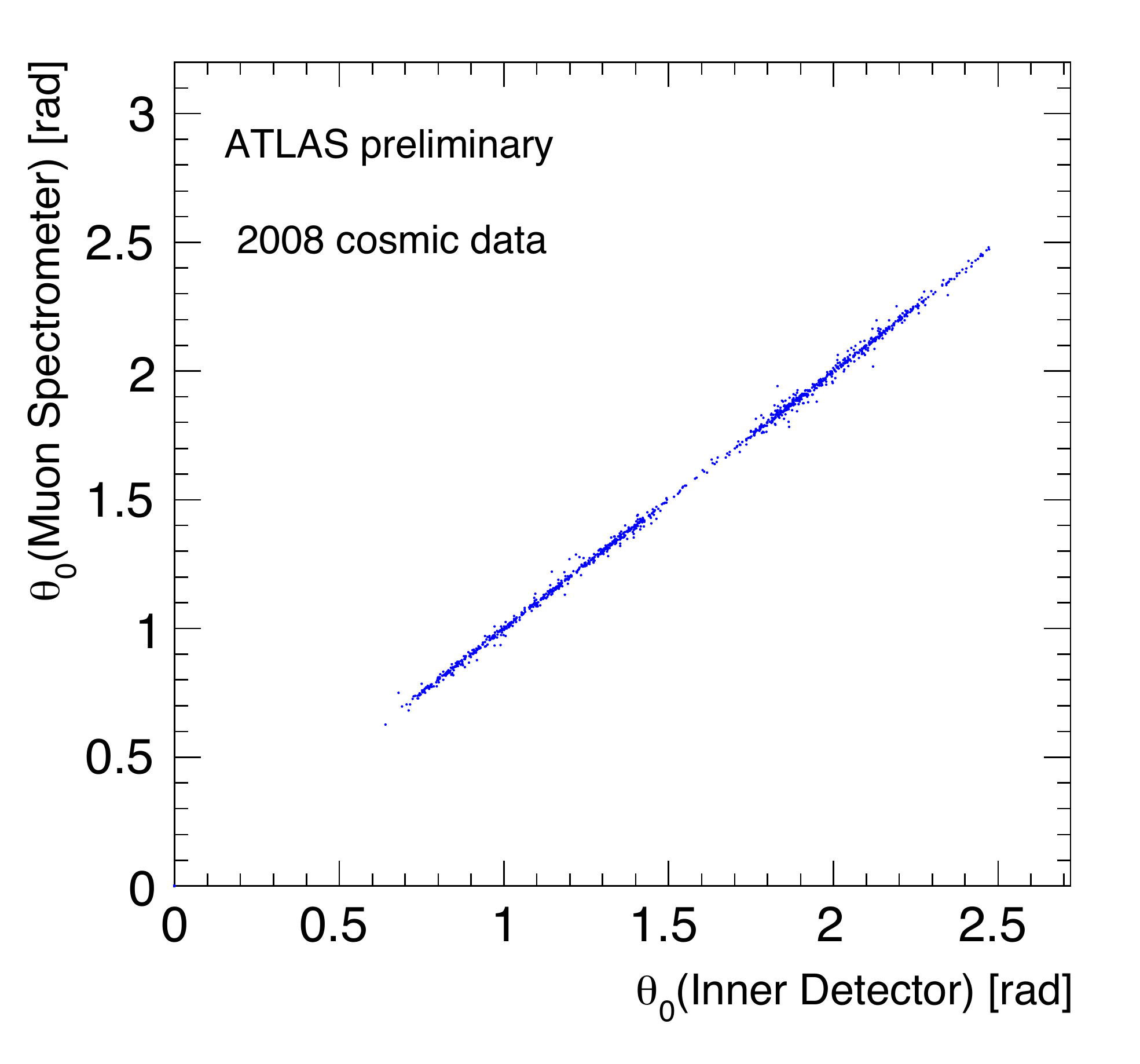} 
	  \includegraphics[width=0.48\textwidth]{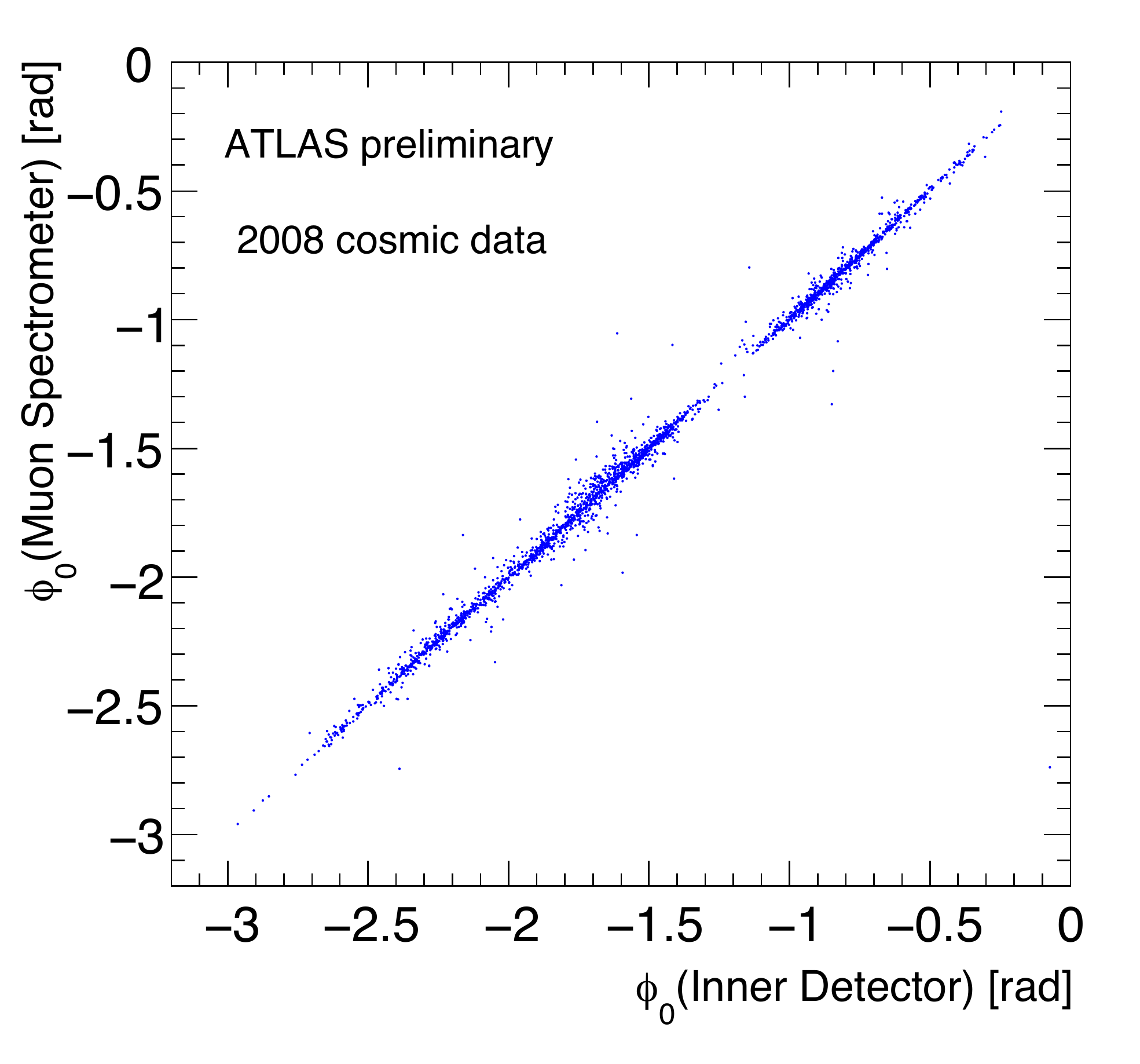}\\[-0.2cm]
	  \includegraphics[width=0.48\textwidth]{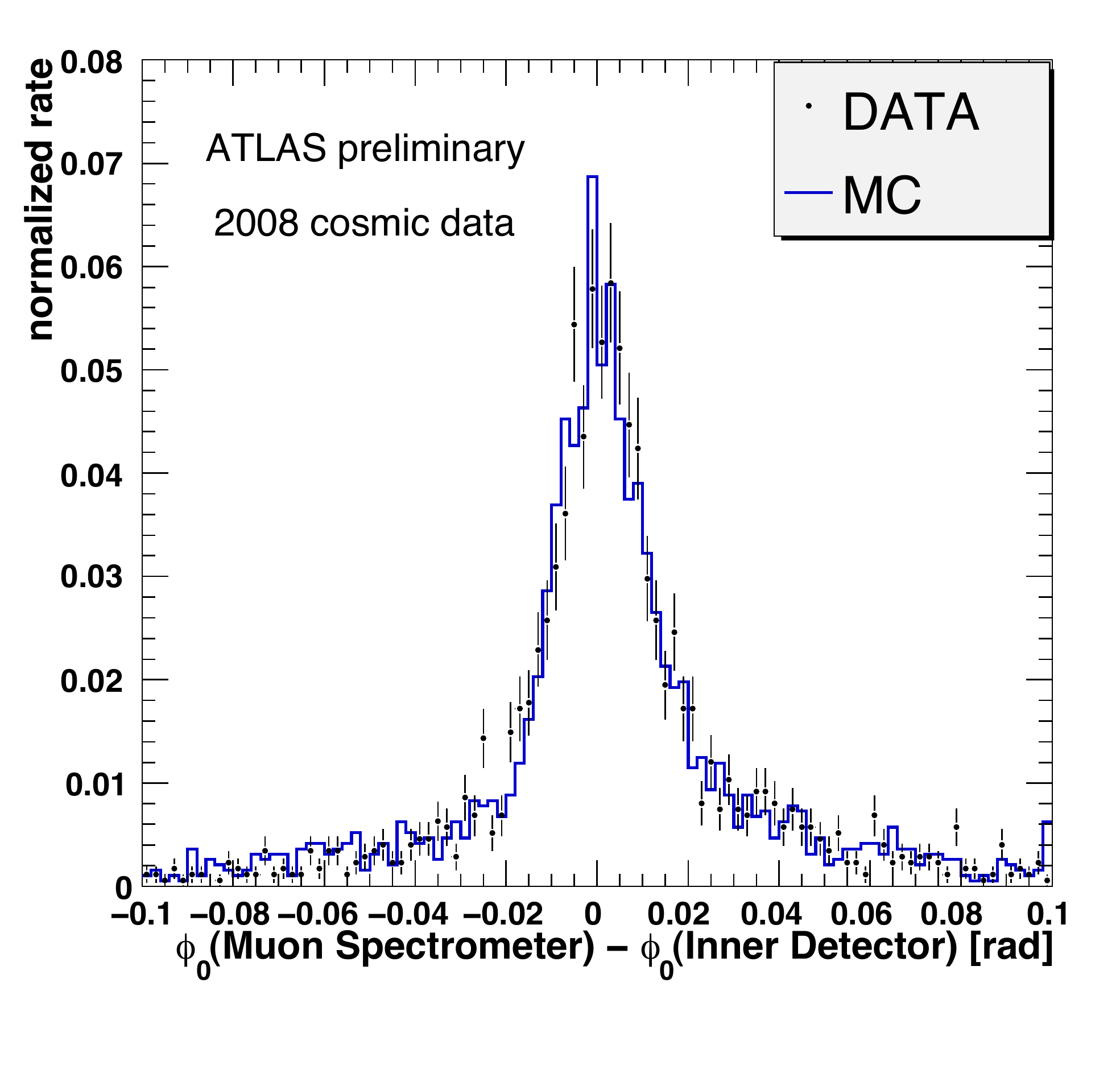}
	  \includegraphics[width=0.48\textwidth]{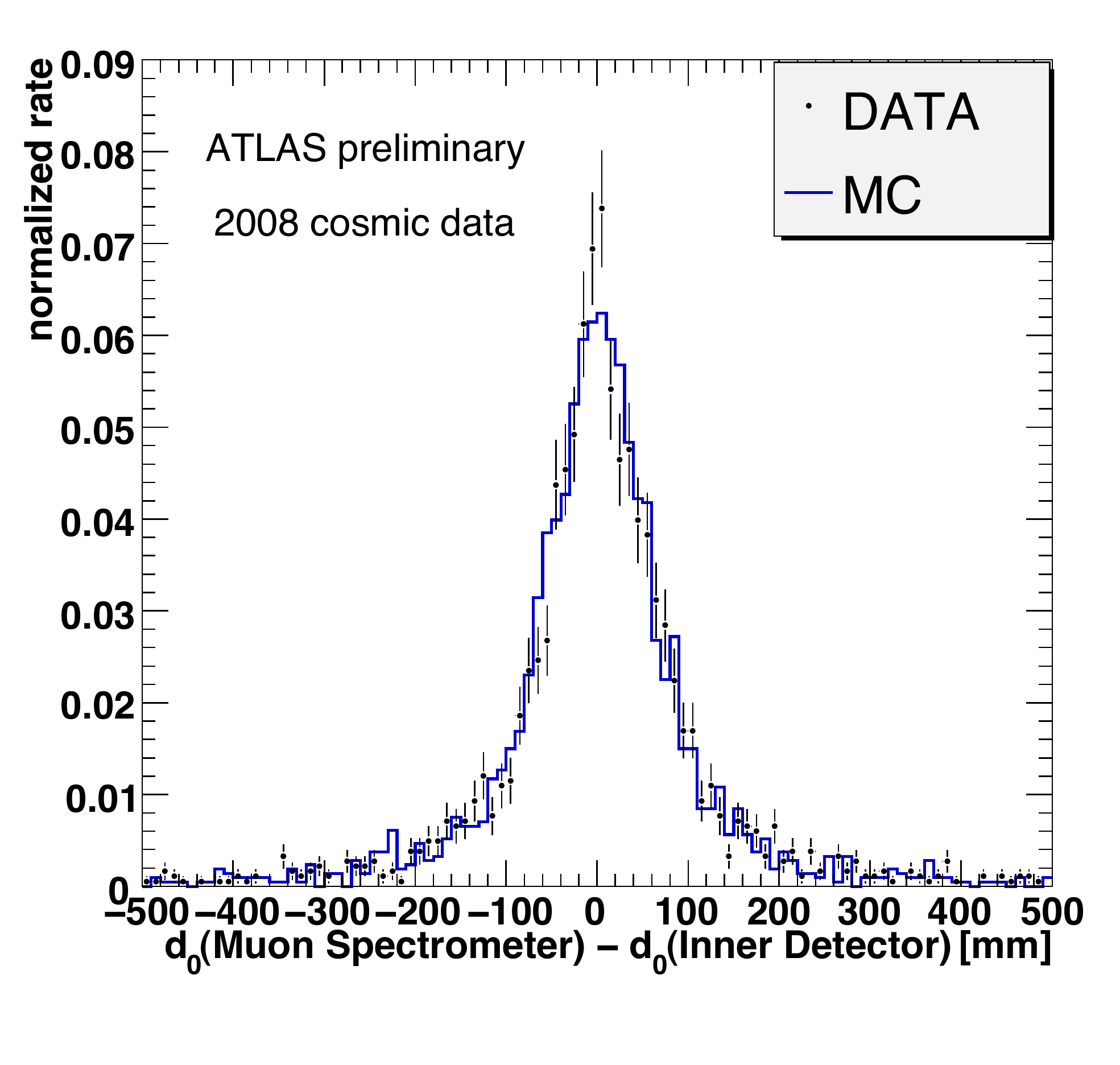}\\[-0.3cm]
	  \includegraphics[width=0.48\textwidth]{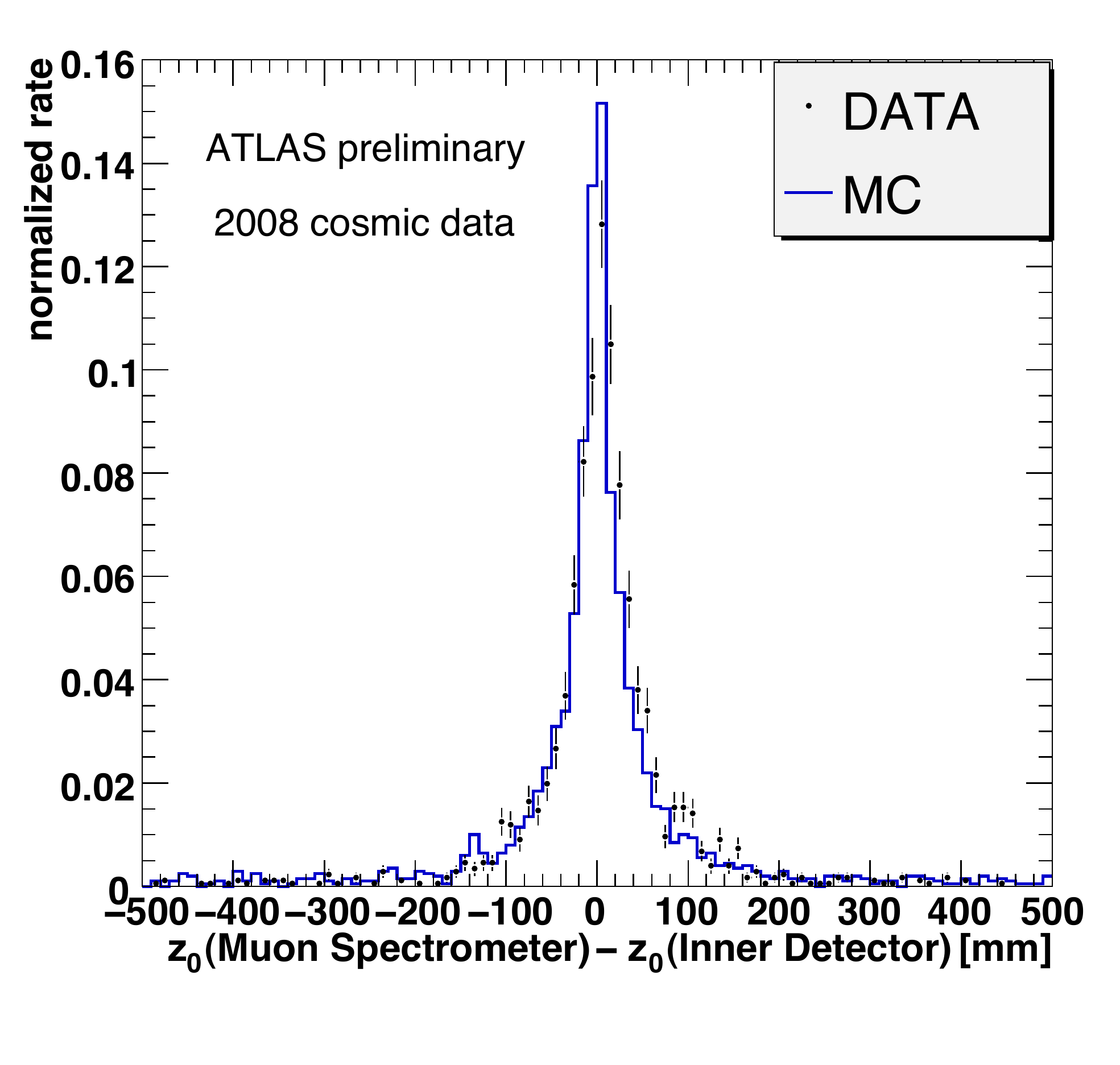}
	  \includegraphics[width=0.48\textwidth]{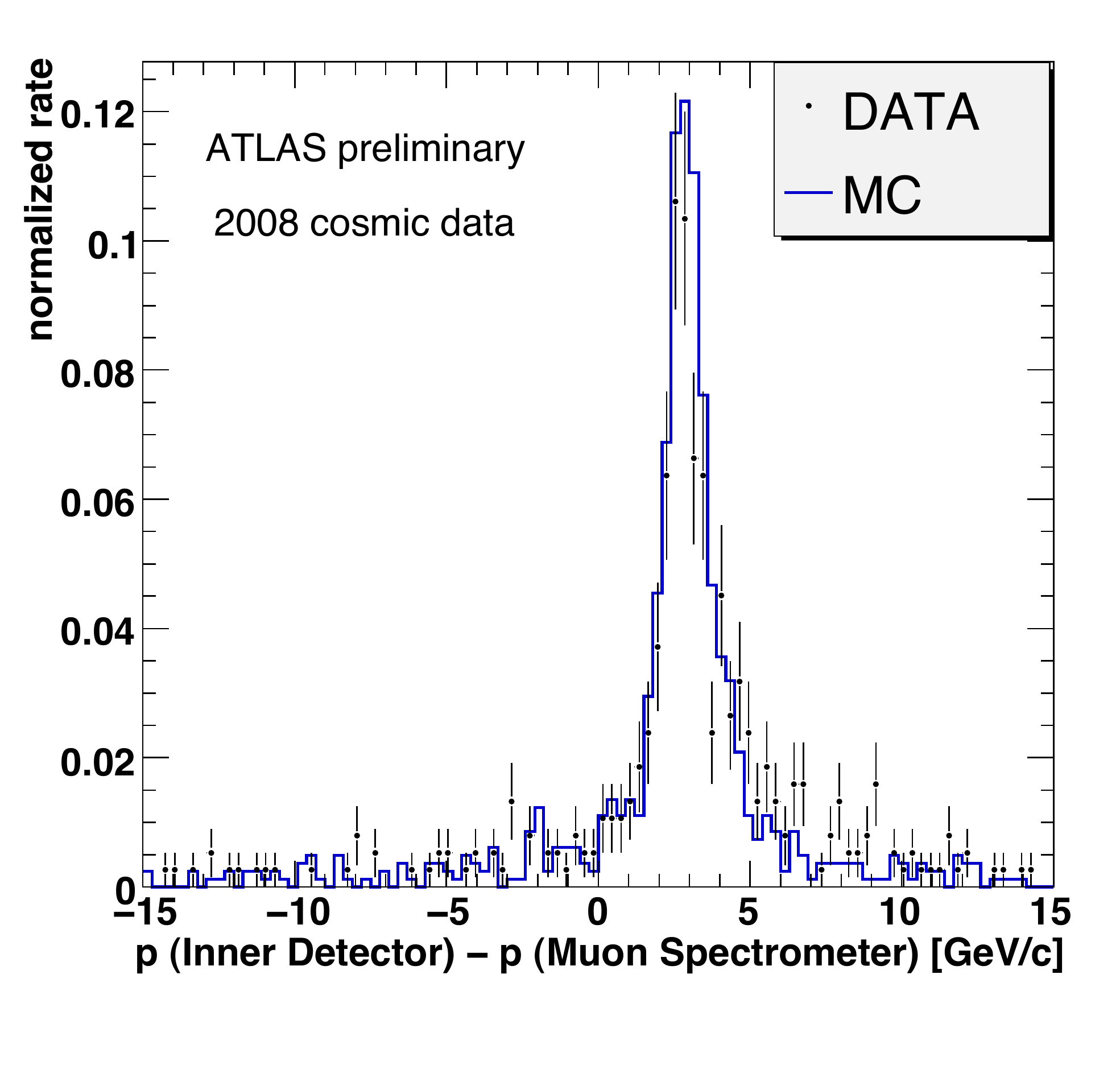}
  \end{center}
  \vspace{-0.8cm}
  \caption[.]{Comparison between standalone track fits to cosmic ray muons in the ATLAS 
                   muon spectrometer and the inner tracker.
                   Shown are the polar and azimuthal angle correlation (upper plots),
                   azimuthal angle and impact parameter differences (middle and lower left plot), and
                   momentum scale difference (lower right plot, sensitive to the energy loss of the 
                   muons when traversing the calorimeters). The dots are data and the 
                   histograms correspond to the Monte Carlo prediction. }

  \label{fig:muoninnertrackercomp}
\end{figure}
The comparison of cosmic muon track measurements in the muon system and in the inner tracker 
allows one to study the momentum scale and the energy loss in the calorimeters, and to tune
the Monte Carlo simulation. Figure~\ref{fig:muoninnertrackercomp} shows a comparison 
between standalone track fits to cosmic ray muons in the ATLAS spectrometer and the inner 
tracker. Shown are the polar and azimuthal angle correlation, the azimuthal angle and 
impact parameter differences, and the momentum scale difference. A satisfactory agreement 
is observed between the detectors, and between data (dots) and the Monte Carlo prediction 
(histograms), showing that the relative alignment and the momentum scales are understood 
within the available statistics (a single run was used for these plots).

The difference in the momentum scale of 3\,\GeV on average corresponds to the energy 
loss of the muons between spectrometer and inner tracker, mainly when traversing 
the calorimeters.\footnote
{
   One can attempt a back-of-the envelope calculation of the expected energy loss 
   to understand the magnitude of the effect. The barrel ATLAS hadronic calorimeter uses
   iron absorber and plastic scintillator tiles. Inserting the corresponding densities
   and $dE/dx$ expectations for cosmic ray muons one finds: 
   $\langle \Delta E({\rm Had~cal})\rangle\simeq
    200{\rm \cm}\cdot(0.4\cdot dE/dx|_{\rm Fe}\cdot 11.8\,{\rm g/cm}^3 + 
                                 0.6\cdot dE/dx|_{\rm C}\cdot   2\,{\rm g/cm}^3)\approx2.1$\,\GeV.
   Similarly one finds for the electromagnetic liquid-argon accordion calorimeter:
   $\langle \Delta E({\rm EM~cal})\rangle\simeq
    100{\rm \cm}\cdot(0.4\cdot dE/dx|_{\rm Pb}\cdot 16.9\,{\rm g/cm}^3)\approx1.0$\,\GeV,
   and for the contribution from the thin solenoid magnet: 
   $\langle \Delta E({\rm solenoid})\rangle\simeq
    5{\rm \cm}\cdot(0.4\cdot dE/dx|_{\rm Cu}\cdot 8.9\,{\rm g/cm}^3)\approx0.1$\,\GeV.
    The sum of all contributions gives roughly 3.2\,\GeV expected energy loss.
} 
It is well described by the simulation. 

\subsection{Cosmic ray muons in the inner tracker}

One of the first measurements performed with cosmic ray muons is the verification of the 
hit reconstruction efficiency in the silicon trackers, which is expected to be very high 
($>99\%$). The method is as follows.
\begin{enumerate}

\item Selection of good quality tracks by requiring a large number of silicon hits, 
      satisfying goodness-of-fit and a small incident angle.

\item To measure the efficiency of the $i$-th layer, the hits from this layer 
      (if there are any) are excluded, and the track is refitted without the $i$-th layer.

\item The hit efficiency is computed by searching for hits in the $i$-th layer within 
      a narrow road around the refitted track.

\end{enumerate}

\begin{figure}[t]
  \begin{center}
 	  \includegraphics[width=0.8\textwidth]{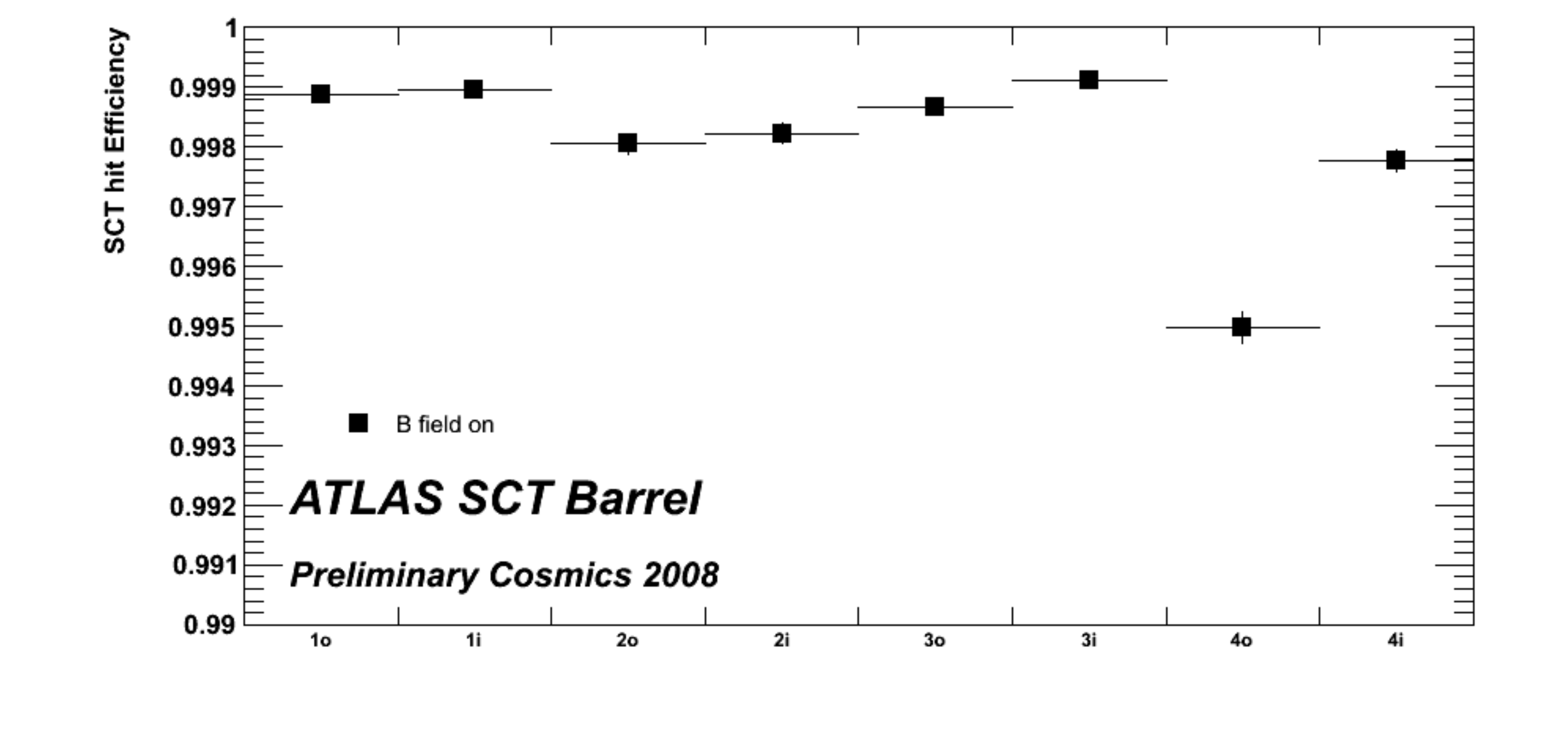} 
  \end{center}
  \vspace{-0.8cm}
  \caption[.]{Hit efficiencies for the ATLAS barrel silicon strip tracker as measured with cosmic 
                  muon tracks (see text for details of track requirements and procedure).}
  \label{fig:scthitefficiency}
\end{figure}

The hit reconstruction efficiencies obtained with this method for the ATLAS barrel silicon 
strip tracker are 
shown in Fig.~\ref{fig:scthitefficiency}. Here the  tracks were required to have at least 
10 hits in the silicon tracker, 30 hits in the transition radiation tracker, and 
an average $\chi^2$ per degree of freedom smaller than 2. Furthermore their intersection 
with the modules had to be within 40 degrees of normal incidence, and there had to be a 
hit of some kind on the track before and after the module being studied. Finally a 
guard region around the edge of the active silicon was excluded. The silicon efficiency 
was then found to be 99.75\% on average. Very similar results have been found for the
ATLAS pixel detector using the same measurement technique, and also for the CMS silicon 
pixel and strip detectors. 

The hit reconstruction 
efficiency per straw for the ATLAS transition radiation tracker depends on the distance
of the track to the anode wire (maximum distance 2\,mm). There is a plateau region 
below 1\,mm where the efficiency reaches 97.2\%, decreasing to $\sim$90\% (80\%) at 
1.5\,mm (1.8\,mm) and steeply dropping beyond that distance. 

\subsection{Measurement of the Lorentz angle}

\begin{wrapfigure}{R}{0.5\textwidth}
  \vspace{-24pt}
  \begin{center}
	  \includegraphics[width=0.5\textwidth]{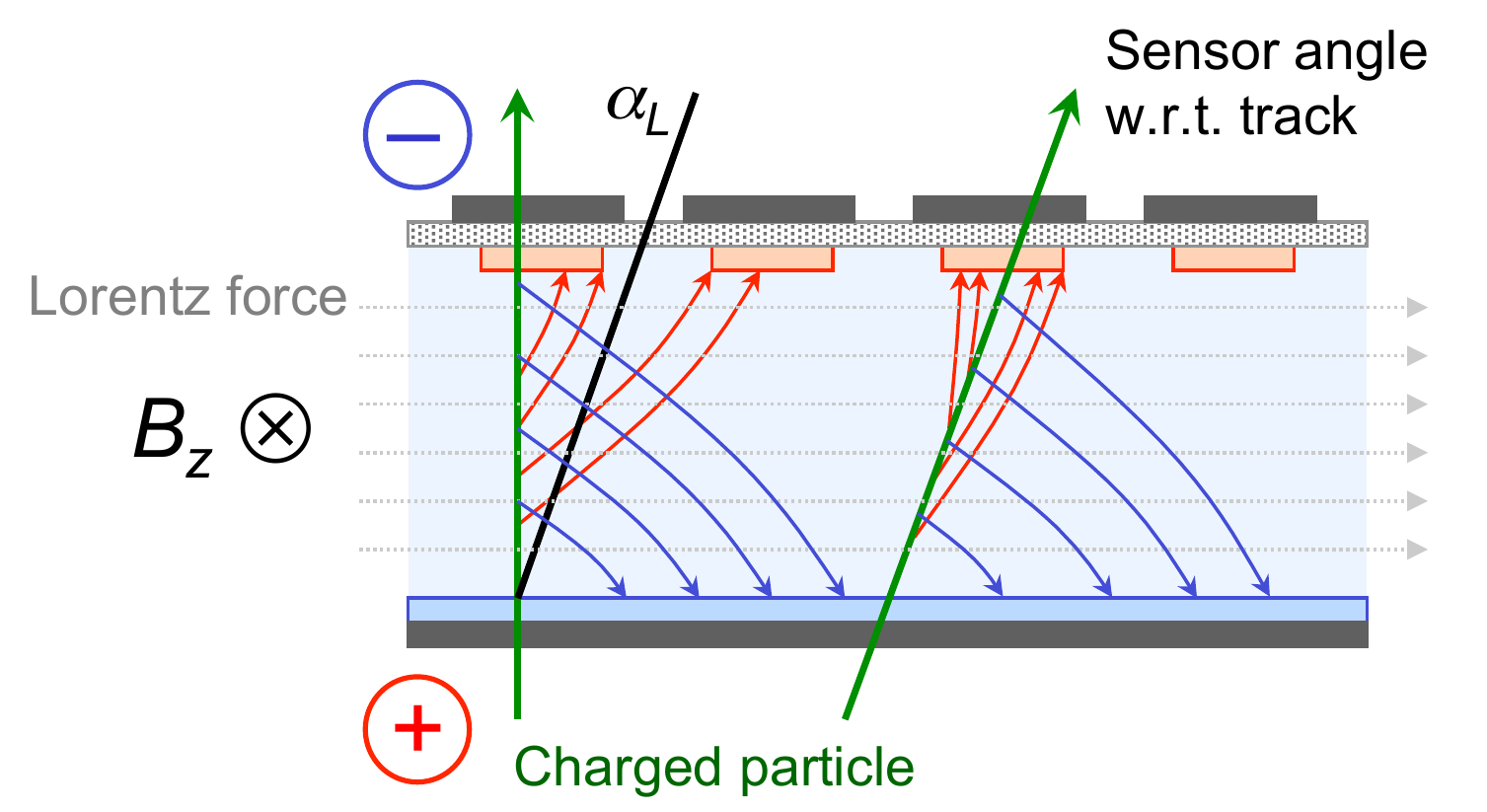}
  \end{center}
  \vspace{-10pt}
  \caption{Sketch illustrating the deflection of moving ionisation charges in the solenoid 
           field, leading to a bias in the position measurement. Tilting the modules 
           by the amount of the Lorentz angle $\alpha_L$ would correct for the bias. }
  \label{fig:sketchlorentzangle}
  \vspace{-6pt}
\end{wrapfigure}
The solenoid field applies a Lorentz force on moving charges that deflects the 
track-induced ionisation electrons and holes, travelling through the depleted substrate 
of the silicon junction along the high-voltage potential (Hall effect). 
The deflection angle is denoted {\em Lorentz angle}. The value of the Lorentz angle 
depends on the mobility of the charge carriers as well as the external magnetic field.
For silicon immersed in a magnetic field {\em B} the Lorentz angle $\alpha_L$ is given by 
$\tan\!\alpha_L=\mu_H B=\gamma\mu_d B$, where $\mu_H$ is the Hall mobility,
$\gamma$ represents the Hall factor which is of order unity, and $\mu_d$
is the drift mobility, which is a function of the ratio of drift velocity to the electric 
field induced by the bias voltage. The drift velocity for both electrons and holes 
saturates at high electric field. This leads to a drop in the mobility thus decreasing 
the Lorentz angle.\footnote
{
   The electron and hole mobility and hence the Lorentz angle also depend on the 
   temperature: increasing temperature reduces the mobility and thus
   the Lorentz angle. 
} 

Figure~\ref{fig:sketchlorentzangle} sketches the Lorentz deflection effect. 
Owing to the opposite charge of electrons and holes, both carriers
are deflected into the same transverse direction along the Lorentz force. The
deflection generates a bias in the position measurement (cluster barycentre) 
of the track incident in the silicon strip or pixel. The bias could be reduced by 
tilting the modules in the direction of the Lorentz angle, and indeed the modules 
in the ATLAS and CMS silicon detectors are tilted ({\em shingled}). The values 
for the tilts chosen are, however, due to technical reasons to allow overlaps 
between adjacent modules.\footnote
{
   In ATLAS the chosen tilts with respect to the pointing axis are 11 degrees 
   ($-20$ degrees) for the silicon strip tracker (pixel tracker), whereas the 
   Lorentz angle for non-irradiated modules is 4 degrees (13 degrees).
}
Instead of a mechanical solution, the position bias due to the Lorentz deflection 
is corrected by software. The correction must be recalibrated at regular intervals 
because the size of the depletion region in the semiconductor reduces with rising 
irradiation and constant bias voltage, thus reducing the position bias. 

\begin{figure}[t]
  \begin{center}
	  \includegraphics[width=0.6\textwidth]{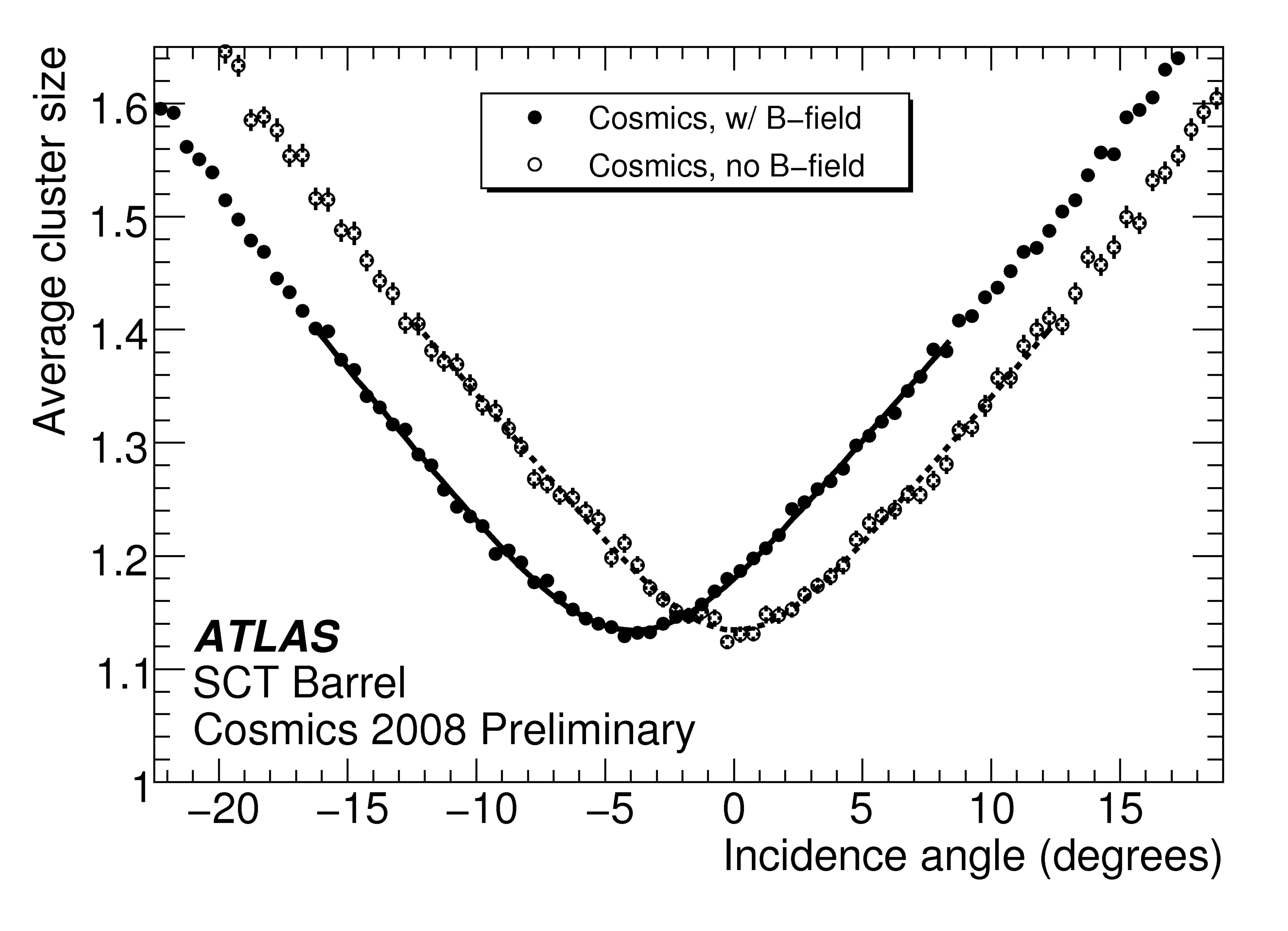} 
  \end{center}
  \vspace{-0.6cm}
  \caption[.]{Measurement of the mean cluster size versus the incidence angle with respect 
              to the module normal in 
              the ATLAS barrel silicon strip tracker, using cosmic ray muon tracks. 
              Measurements with and without magnetic field are shown (the Lorentz angle 
              vanishes without external field). The value of the Lorentz angle 
              is extracted from the position of the minimum cluster size.}
  \label{fig:sctlorentzangleatlas}
\end{figure}
The Lorentz angle is determined empirically by minimising the measured cluster width 
of hits on tracks. Figure~\ref{fig:sctlorentzangleatlas} shows the cluster width 
versus the cosmic muon track incident angle with respect to the module normal for 
the ATLAS barrel silicon strip tracker. Measurements with and without magnetic field 
are shown. The value of the Lorentz angle, extracted at the minimum cluster size, is 
found to be $\alpha_L=(3.93 \pm 0.03 \pm 0.10)^\circ$, where the first error is 
statistical and the second systematic (for comparison, the Lorentz angle for the 
ATLAS pixel device is $12.3^\circ$).

\subsection{Particle identification with transition radiation}

\begin{wrapfigure}{R}{0.45\textwidth}
  \vspace{-24pt}
  \begin{center}
	  \includegraphics[width=0.45\textwidth]{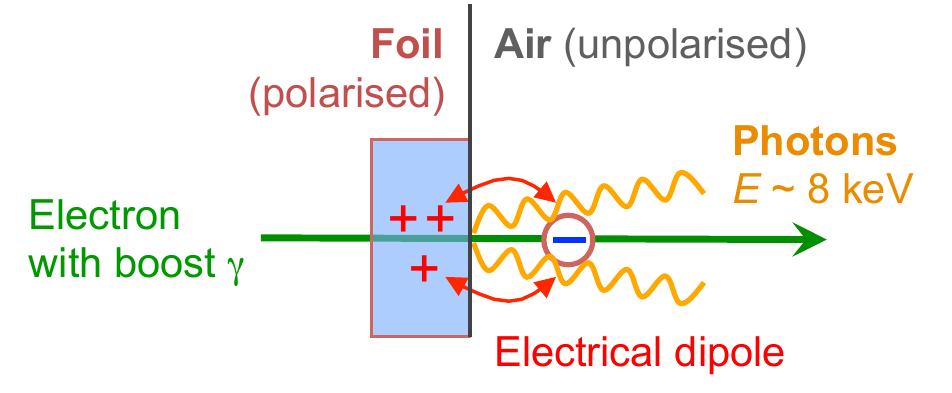}
  \end{center}
  \vspace{-15pt}
  \caption{Transition radiation is produced when charged ultrarelativistic particles 
           traverse the boundary of two different dielectric media (\eg, polymer fibres/foil 
           and air). The radiation is intense enough to be measured for $\gamma  > 1000$ 
           and more than 100 boundaries.}
  \label{fig:sketchtrt}
  \vspace{+120pt}
\end{wrapfigure}
Hits from ultrarelativistic particles, generating transition radiation photons in the keV 
range that contribute to the gas ionisation in the ATLAS transition radiation tracker 
(TRT), are identified via dedicated high-threshold readout. It turns on at a gamma factor 
above $\simeq$1000 (with $p=\beta\gamma m\simeq\gamma m$, the threshold momenta for 
$\gamma=1000$ are 0.5\,\GeV, 105\,\GeV and 139\,\GeV for electrons, muons and pions, 
respectively), and thus essentially only for electrons in the typical energy range, so 
that it can be used for electron identification. 

\begin{figure}[t]
  \begin{center}
	  \includegraphics[width=1\textwidth]{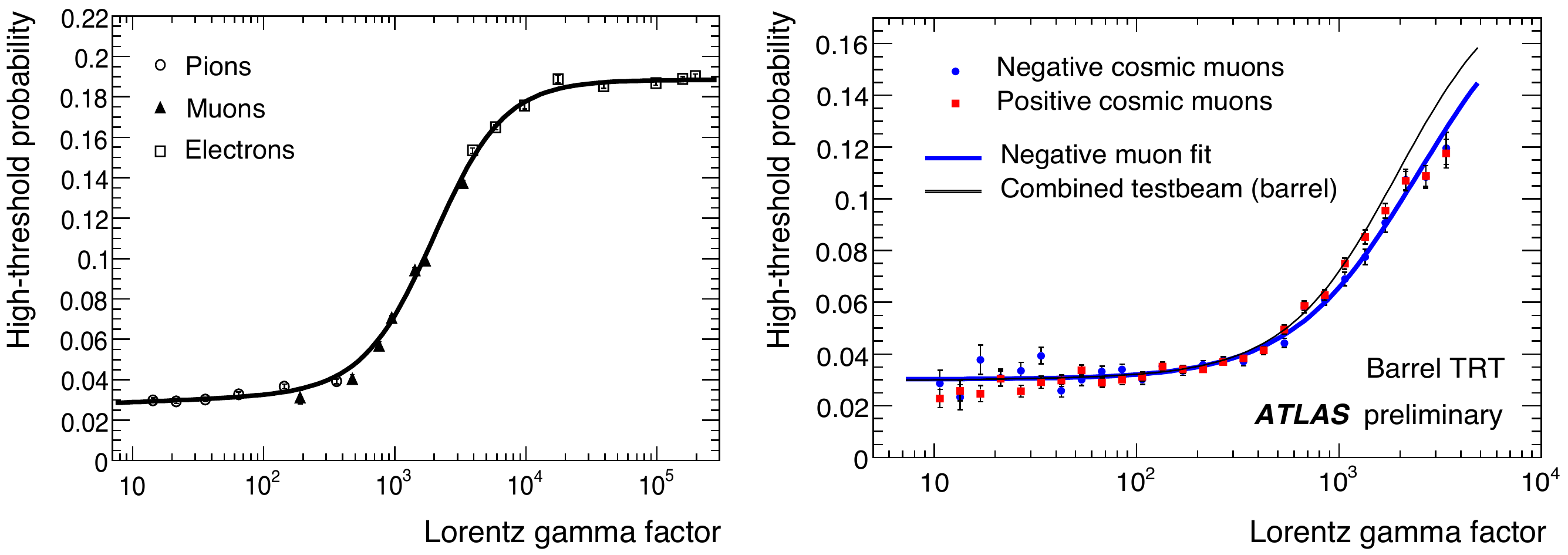}
  \end{center}
  \vspace{-0.3cm}
  \caption[.]{\underline{Left}: average probability of a high-threshold hit in the ATLAS barrel 
              transition radiation tracker (TRT) as a function of the Lorentz $\gamma$ factor for 
              electrons (open squares), muons (full triangles) and pions (open circles) in 
              the energy range 2--350\,\GeV, as measured in the 2004 combined test beam.
              \underline{Right}: transition radiation turn-on versus $\gamma$ in the 
              ATLAS barrel TRT for cosmic muon tracks. 
              The data points are shown for both muon charges (positive: red dots, negative: 
              blue dots) and are compared with test beam results (black line). The blue line 
              gives a fit to the results obtained with the cosmic data.}
  \label{fig:trtpidatlas}
\end{figure}

The principle of the creation of transition radiation via an electric dipole is sketched in 
Fig.~\ref{fig:sketchtrt}. Figure~\ref{fig:trtpidatlas} shows the high-threshold hit probability
obtained for the ATLAS barrel TRT from 2004 combined test beam data 
(\cf Section~\ref{sec:testbeams}) for different particle species (left plot), and for 
cosmic ray muons (right plot). The turn-on curves are found to be in good agreement. 

\subsection{Calorimeter performance with cosmic ray muons}

Cosmic ray muons have also been exploited by the calorimeter groups of ATLAS and 
CMS to study pulse shapes, and occupancy distributions, detect bad channels, 
understand the muon energy loss in the calorimeters and tails in energy distributions, 
and for energy inter-calibration purposes. 

\begin{figure}[p]
  \begin{center}
	  \includegraphics[width=1\textwidth]{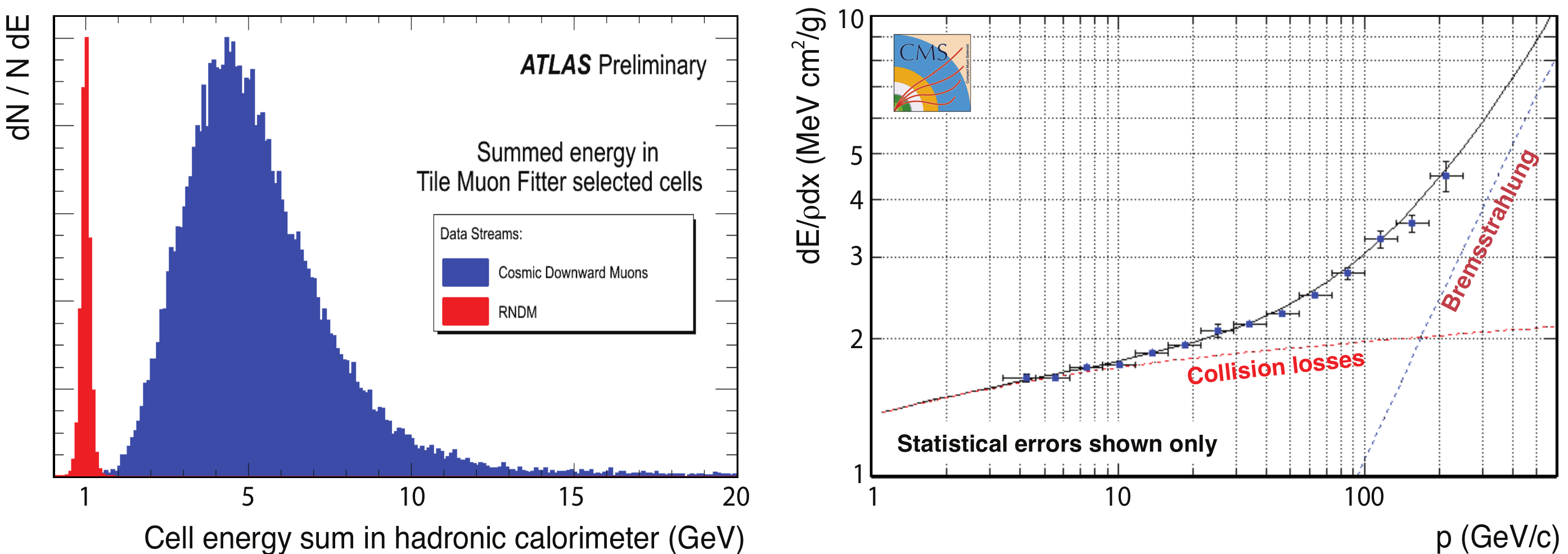}
  \end{center}
  \vspace{-0.4cm}
  \caption[.]{\underline{Left}: total energy sum of all cells along a muon track in the ATLAS hadronic 
                  calorimeter (blue) and the corresponding noise distribution obtained 
                  from randomly triggered events (red). The minimum ionising muon signal is 
                  well separated. 
                  \underline{Right}: average energy deposits in the CMS 
                  electromagnetic calorimeter versus the muon momentum measured in the 
                  tracking devices. Overlaid is the expected energy loss for the lead-tungsten 
                  calorimeter. Indicated by the dotted lines are the contributions to the energy 
                  loss from collisions (red) and bremsstrahlung (blue).  }
  \label{fig:etileatlasanddedxecalcosmicscms}
  \vspace{0.1cm}
  \begin{center}
	  \includegraphics[width=0.7\textwidth]{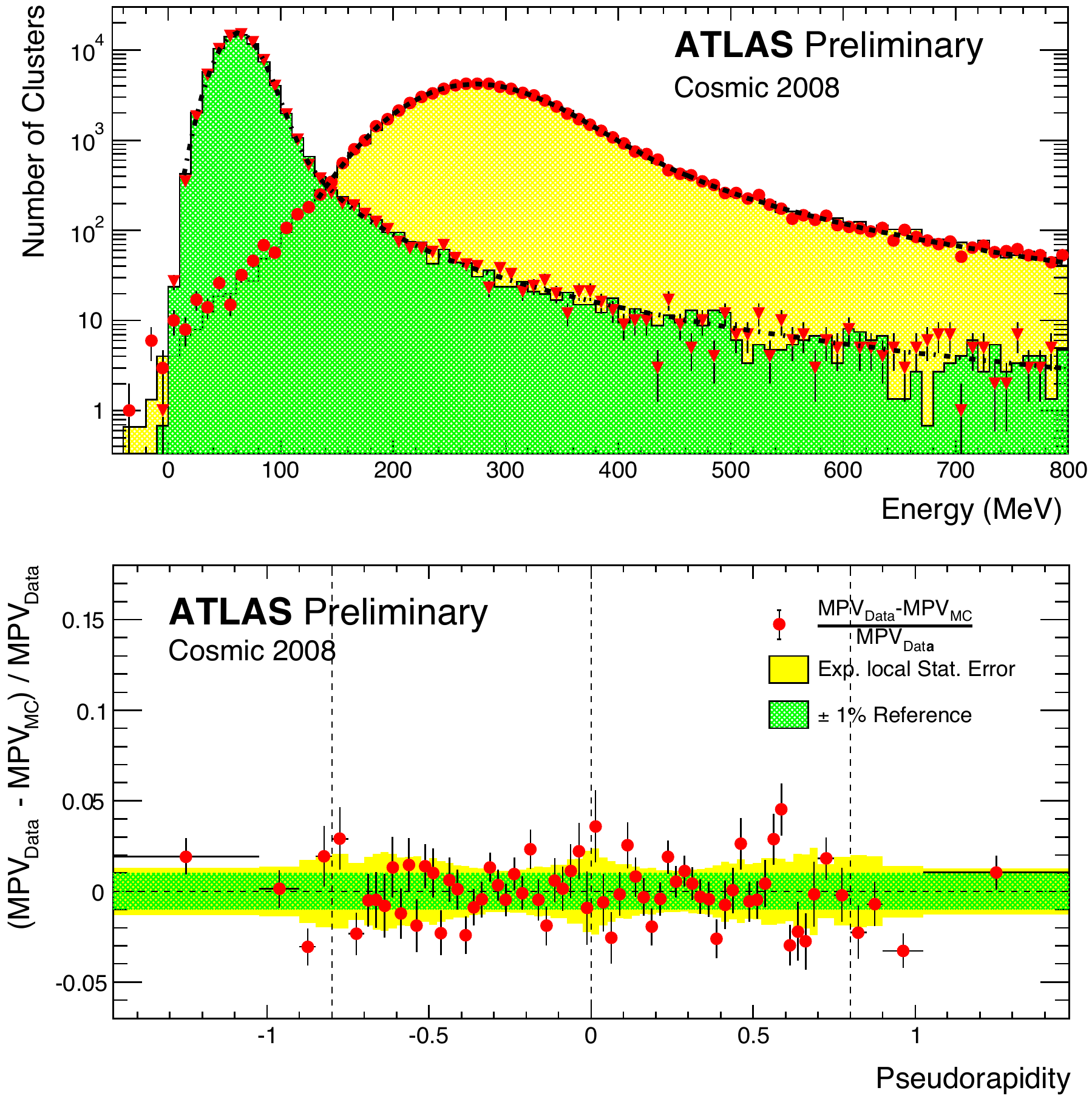}
  \end{center}
  \vspace{-0.4cm}
  \caption[.]{\underline{Top}: 
                  reconstructed cosmic muon energy in a 2$\times$1 cluster in the first 
                  layer (dark shaded/green histogram for Monte Carlo and 
                   triangles for data) and in a 1$\times$3 cluster in the second layer (light 
                   shaded/yellow histogram and dots for data) of the ATLAS electromagnetic
                   calorimeter. 
                   \underline{Bottom}: electromagnetic calorimeter energy response dispersion
                   between data and Monte Carlo simulation versus the pseudorapidity, as 
                   measured with cosmic muons for the second (main) layer of the ATLAS electromagnetic
                   calorimeter. The dark shaded (green) band indicates the $\pm1\%$ 
                   region for reference, and the light shaded (yellow) band indicates the expected 
                   statistical accuracy ($1\sigma$ error band) of the measurement.  } 
  \label{fig:eclusterlaruniformityatlas}
\end{figure}
The total energy sum of all cells along a muon track in the ATLAS hadronic calorimeter 
is shown in the left-hand plot of Fig.~\ref{fig:etileatlasanddedxecalcosmicscms}. The 
peak of the minimum-ionising particles (\ie a 
particle whose mean energy loss rate through matter is close to the minimum), is well
distinguished from the corresponding noise distribution obtained from randomly triggered
events. The energy of the cosmic ray muons deposited in the active parts of the hadronic 
calorimeters of ATLAS and CMS exceeds the one in the electromagnetic calorimeters by approximately 
a factor of 10. The ionisation energy loss of the muons when traversing the electromagnetic 
calorimeters is
measured by comparing the momenta between the muon system and the inner tracker
(\cf Fig.~\ref{fig:muoninnertrackercomp} on page~\pageref{fig:muoninnertrackercomp}).
It can be correlated on an event-by-event basis to the measured calorimeter energy 
deposits. This has been done by CMS in the right-hand plot of 
Fig.~\ref{fig:etileatlasanddedxecalcosmicscms}, where the 
average electromagnetic calorimeter deposits are drawn versus the muon momentum.
Overlaid is the expected energy loss, which is found to be in good agreement with the 
measurement. The results indicate the correctness of the tracker momentum scale and 
of the calorimeter energy scale calibrated with electrons at test beams. 

The energy deposition can also be directly compared to Monte Carlo simulation, as done 
by ATLAS in the upper plot of Fig.~\ref{fig:eclusterlaruniformityatlas}
(see Ref.~\cite{larpaper}), showing the energy reconstructed in the first and second 
layers for data and Monte Carlo cosmic ray events. 
Good agreement is observed up to the tails both for the shape and the absolute 
scale. This result can be used to measure the uniformity in the energy response
of the calorimeter versus the pseudorapidity by integrating over the response in the 
azimuth angle (the statistics is insufficient to make a full $\eta$$\times$$\phi$ uniformity 
map). The estimation of the muon energy is done with a fit of the cluster energy 
distribution using a Landau function, which accounts for fluctuations of the energy 
deposition in the ionisation process, and a Gaussian describing essentially electronic 
noise (and also cluster non-containment). The response uniformity is computed from
the RMS of the normalised difference between the data and Monte Carlo most probable
values (MPV) of the Landau distribution in each $\eta$ bin. The resulting distribution
for the second (and main) liquid-argon calorimeter layer in ATLAS is shown in the lower plot of 
Fig.~\ref{fig:eclusterlaruniformityatlas}. The observed dispersion is in agreement with 
statistical fluctuations, \ie, no significant non-uniformity is seen at the per cent level. 
Similar results have been obtained by CMS where an intercalibration with cosmic 
muons (aligned to the crystal axis and with a reference energy of 250\,\MeV (MPV))
achieved an intercalibration of better than 1.5\% in the barrel and better than 2.2\%
in the forward region. All 36 CMS crystal supermodules could thus be intercalibrated with 
cosmic muons, which was an important achievement because only 9 supermodules (25\%)
had been calibrated with electron test beam data prior to the calorimeter installation. 

\begin{figure}[t]
  \begin{center}
	  \includegraphics[width=1\textwidth]{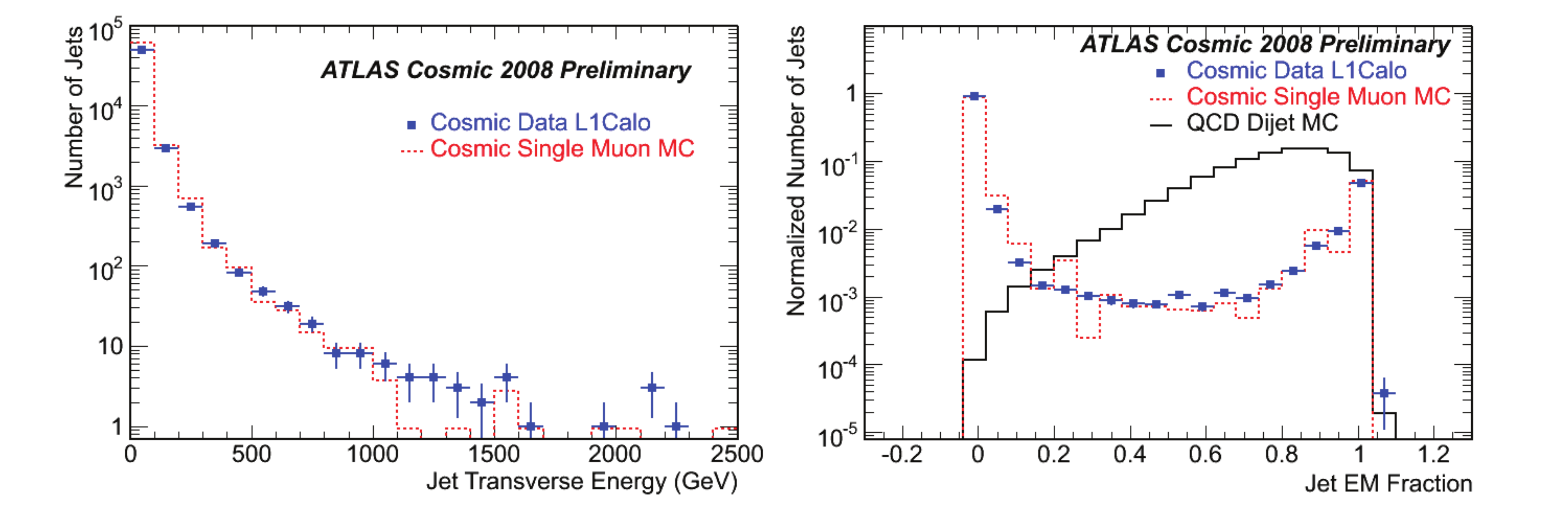}
  \end{center}
  \vspace{-0.4cm}
  \caption[.]{\underline{Left}: distribution of the jet energy for data (dots) and Monte 
                   Carlo simulation (dotted histogram). Only events with at least one jet that 
                   exceeds 20\,\GeV transverse energy are included. The Level-1 trigger 
                   inefficiency and cosmic air showers are not simulated. 
                   \underline{Right}: electromagnetic fraction of jets for data (dots) and 
                   Monte Carlo (dotted), where the fraction is defined by the ratio of energy 
                   deposited in the electromagnetic calorimeter divided by the total deposited 
                   energy.  The distributions are normalised to unity. Only jets with $E_T>20$\,\GeV 
                   are included. Shown by the solid histogram is the expected distribution for 
                   QCD di-jet events as they originate from proton--proton collisions.
 } 
  \label{fig:jetcosmicsatlas}
\end{figure}
The reconstruction of jets and missing transverse energy requires the electromagnetic 
and hadronic calorimeter responses to be combined. It can be studied with highly energetic 
cosmic muons releasing a Level-1 calorimeter trigger-accept signal. Jets from muon showers 
with energies exceeding the \TeV scale are found in the data. Figure~\ref{fig:jetcosmicsatlas} 
(left) shows the distribution of the jet energy for calorimeter triggered events for data and 
Monte Carlo simulation. Because the simulated data do not include the Level-1 trigger 
inefficiency, the Monte Carlo distribution is normalised to data in the 100--300\,\GeV range. 
Only events that have a jet with $E_T>20$\,\GeV are included in the figure. Good agreement 
between data and simulation is observed. A small excess at large transverse energy in data 
may be due to air-showers, not included in the simulation. The right-hand plot in 
Fig.~\ref{fig:jetcosmicsatlas} shows the electromagnetic (EM) fraction of jets for data and 
Monte Carlo, where the fraction is defined by the ratio of energy deposited in the 
electromagnetic calorimeter divided by the total deposited energy.  The distributions are
normalised to unity. As before, only jets with $E_T>20$\,\GeV are included. Also shown 
for comparison is the distribution expected for QCD di-jet events as they originate from 
proton--proton collisions. The most likely value for the EM fraction is 0 or 1 for fake jets 
from cosmics, because the high energy deposition from photons originating from highly energetic 
muons will localise either in the electromagnetic or the hadronic calorimeter. QCD jets 
have a broad distribution of the EM fraction with a maximum at around 0.8. Electromagnetic 
fractions less than 0 or larger than 1 are due to small negative energy contributions from 
noise. One concludes from the plot that good separation between QCD jets and fake jets from 
cosmic rays can be obtained by vetoing jets with EM fractions close to 0 and 1.

\newpage
\section{Commissioning with single proton beam data}

\begin{wrapfigure}{R}{0.35\textwidth}
  \vspace{-28pt}
  \begin{center}
	  \includegraphics[width=0.35\textwidth]{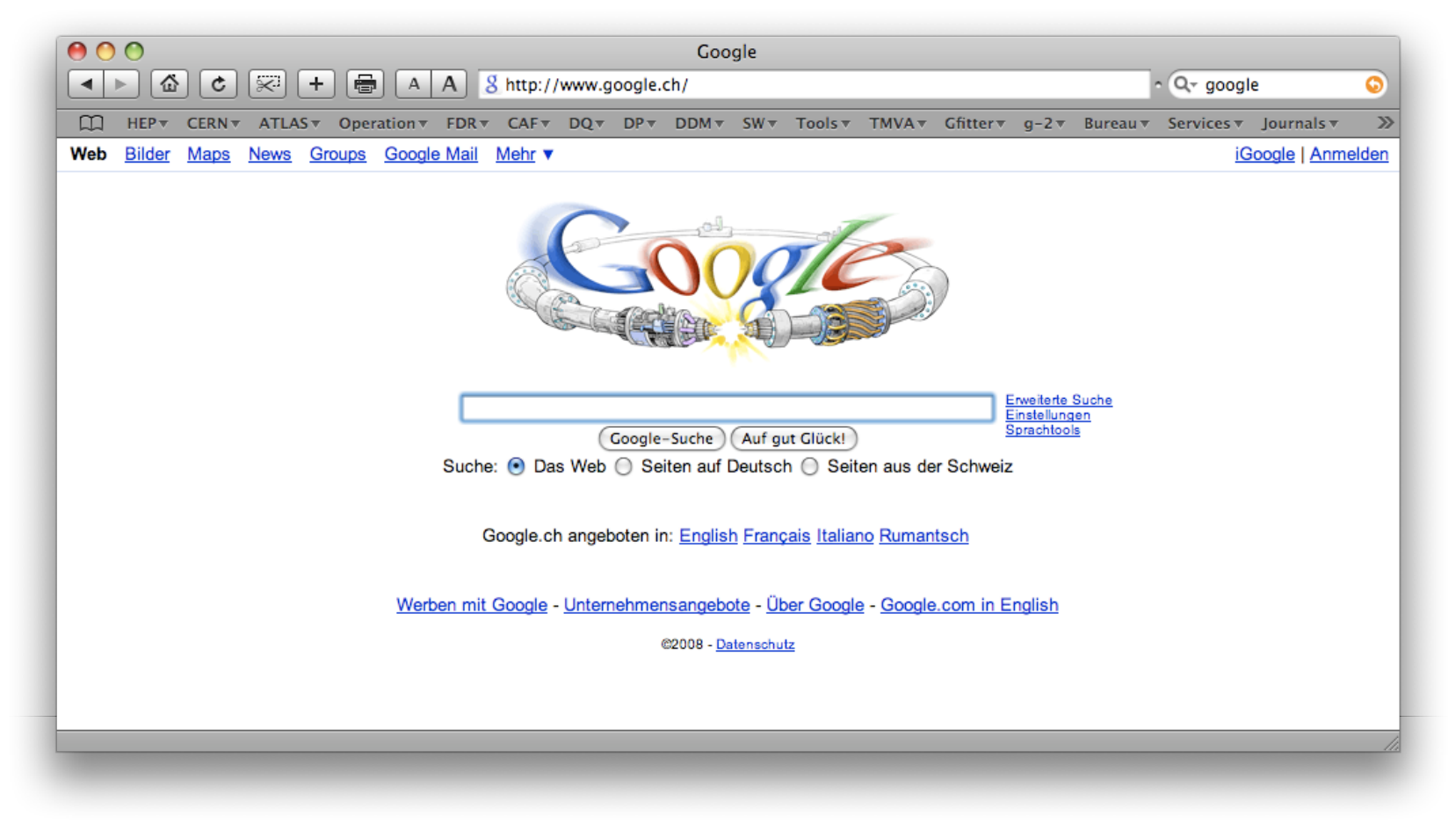}
  \end{center}
  \vspace{-15pt}
  \caption{The Google search page at `Jour J'  --- the LHC start-up, 10 September 2009.}
  \label{fig:googlejourj}
  \vspace{-6pt}
\end{wrapfigure}
A lucky period between September 10 and 13, 2008, with --- for the first time --- single 
beams of 450\,\GeV LHC injection energy circulating in both directions of the LHC, gave the 
experiments the opportunity to commission the detector and the 
data taking chain with proton-beam background 
in synchronisation with the LHC clock. A single `pilot' bunch containing approximately 3 billion 
protons --- radio-frequency captured and not, with closed and open collimators, stably circulating
or lost --- travelled through the injection chain, transfer lines and finally the LHC. The 
single-beam exercise at injection energy was briefly repeated in 2009, at the restart of the 
LHC after an accident that caused a one-year delay in the commissioning and physics schedule. 

\begin{figure}[t]
  \begin{center}
	  \includegraphics[width=1.01\textwidth]{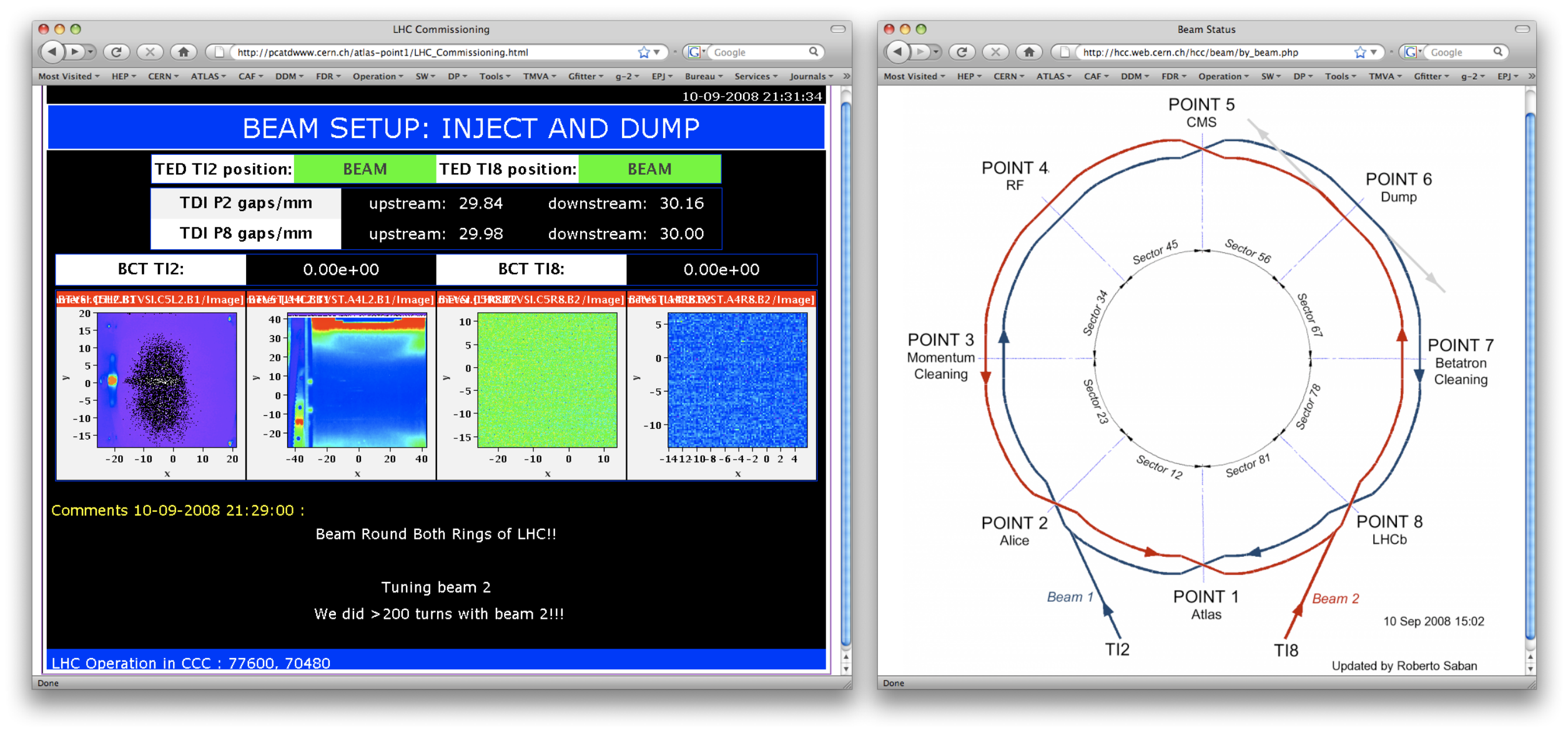}
  \end{center}
  \vspace{-0.5cm}
  \caption[.]{Main LHC information displays sent from the CERN Control Room (CCC, `Triple-C')
              to the experiment and the interested world. The left picture displays basic 
              quantities such as the currents (in number of protons per bunch) passing
              through the two transfer lines serving to inject the LHC beam lines. Apart 
              from displaying sometimes cryptic information displays and plots, it features
              useful operator comments on the bottom of the display: ``Beam 
              Round Both Rings of LHC !!'' (one notices the capital letters and the abundant
              use of exclamation marks, which appropriately reflect the mood of the day).
              The right panel is a sketch of the two LHC beams. The colour codes are important: 
              Beam Blue (1) must {\em always} be blue, and Beam Red (2) must {\em always}
              be red (source: Steve Myers, LHC coordinator). The detectors are located at 
              four out of eight straight sections: Point 1 (ATLAS), Point 2 (ALICE), 
              Point 5 (CMS) and Point 8 (LHCb). The remaining four straight sections serve 
              beam acceleration, beam cleaning and dump purposes (see Section~\ref{sec:lhc}).  }
\label{fig:lhcfirstpage}
\end{figure}

\begin{figure}[t]
  \begin{center}
	  \includegraphics[width=1\textwidth]{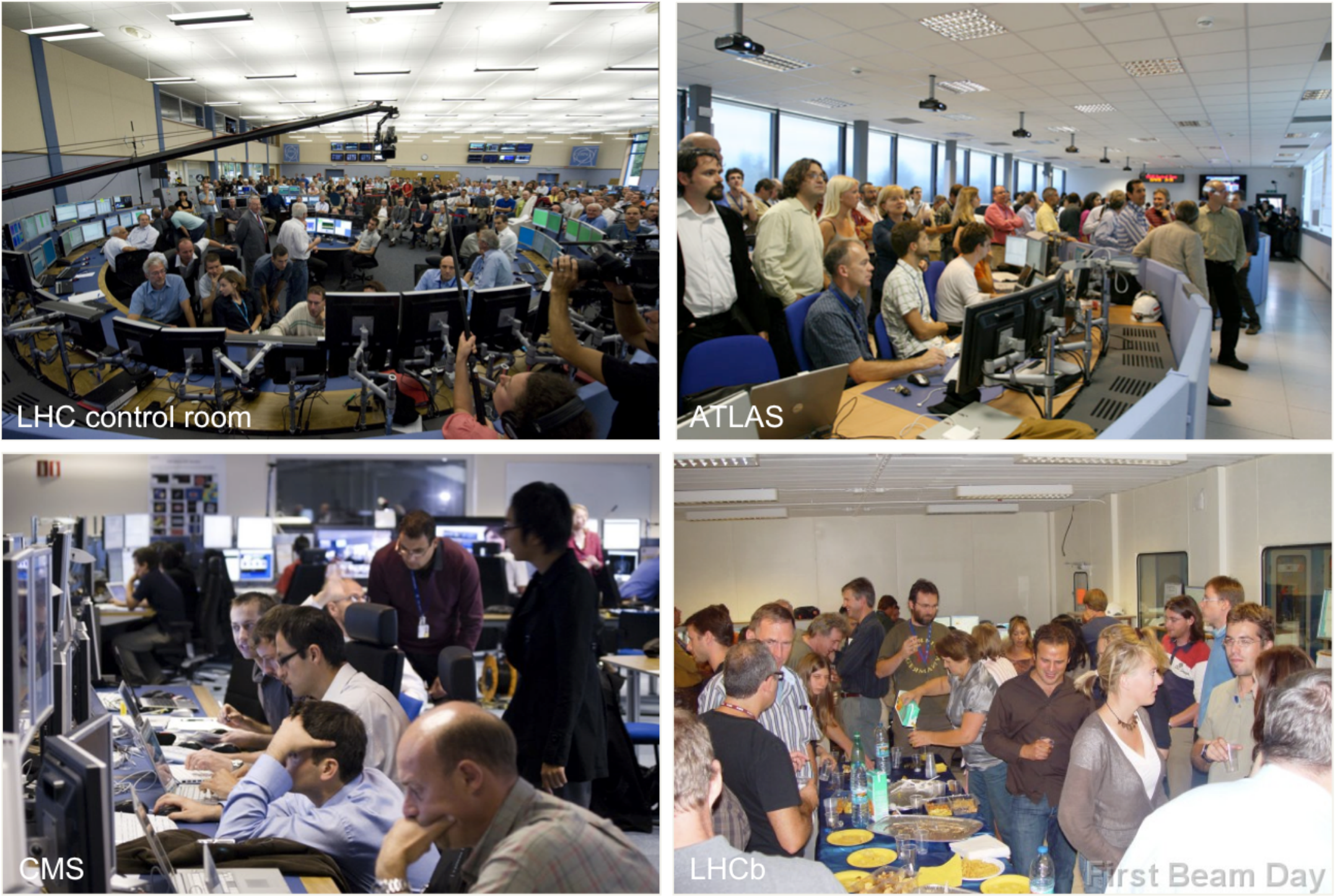}
  \end{center}
  \vspace{-0.3cm}
  \caption[.]{Snapshots taken on 10 September in the 
              LHC (upper left), ATLAS (upper right), CMS (lower left), and LHCb 
              (lower right) control rooms, exhibiting untypical occupancy. }
\label{fig:sep10photos}
\end{figure}

Figure~\ref{fig:lhcfirstpage} shows two of the most important information panels provided 
to the experiments (and the general public) by the LHC operators. One notices the particular
location of Point 1 (ATLAS cavern) on the right panel: both beams need to make a full turn 
before reaching ATLAS. It was hence the last experiment to see beam, and it is affected
by any problem along the beam line. A few photographs taken on 10 September in the 
LHC, ATLAS, CMS, and LHCb control rooms are shown in Fig.~\ref{fig:sep10photos}.

\begin{figure}[t]
  \begin{center}
	  \includegraphics[width=1\textwidth]{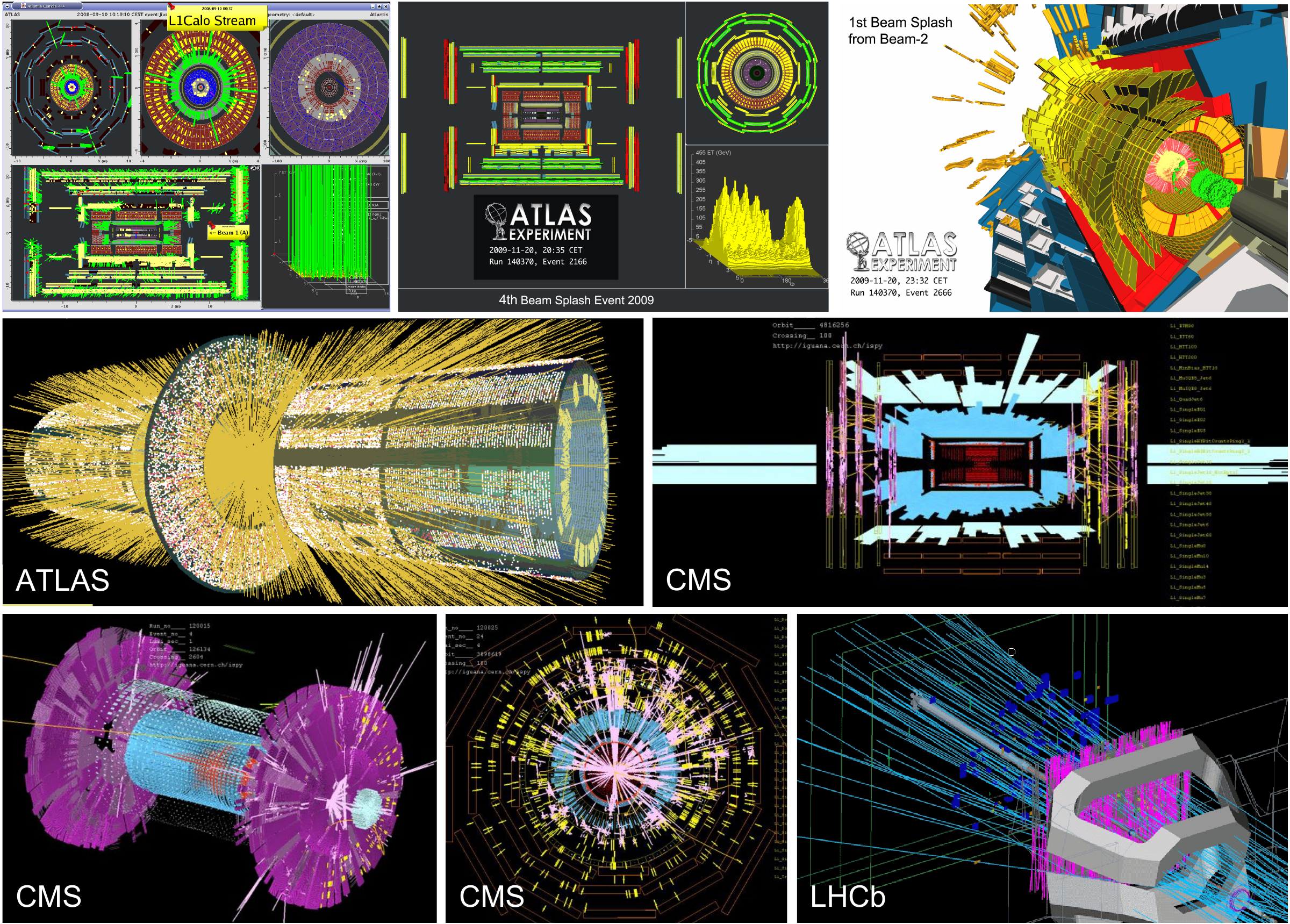}
  \end{center}
  \vspace{-0.3cm}
  \caption[.]{Event displays of beam--on--collimator `splash' events recorded by 
              ATLAS (upper plots and centre left), CMS (centre right, lower left and middle),
              and LHCb (lower right). }
\label{fig:splasheventdisplays}
\end{figure}

\subsection{Beam-on-collimator events}

\begin{wrapfigure}{R}{0.33\textwidth}
  \vspace{-24pt}
  \begin{center}
	  \includegraphics[width=0.33\textwidth]{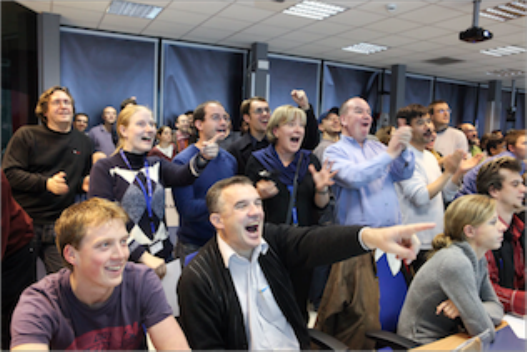}
  \end{center}
  \vspace{-15pt}
  \caption{A `beam splash' event being seen in the ATLAS control room.}
  \label{fig:splashphotoatlas}
  \vspace{-6pt}
\end{wrapfigure}

Somewhat unexpectedly and all of the sudden, events 
where the entire detector was lit appeared on the event displays. A few typical  
events are collected in Fig.~\ref{fig:splasheventdisplays}. The reaction in the ATLAS 
control room upon the arrival of the first event is witnessed by the photo in 
Fig.~\ref{fig:splashphotoatlas}. {\em What happened?}

The events seen belong to so-called `beam splash' type events, which originate from 
pilot-beam-on-collimator dumps. Collimators are placed at a distance of about 140\,m on 
both sides of the experiments. If they are closed, the beam dumps on them, producing 
an avalanche of scattered particles that reach the detector. For such an event occurring
every 42 seconds during a short period ATLAS
typically recorded 300\,000 silicon strip tracker hits (on lowered voltage for safety reason, 
reducing the hit efficiency; the pixel detector was switched off) and 350\,000 transition 
radiation tracker hits, approximately all passing high-threshold discrimination.
The sum of all calorimeter cells in these events exceeds 3000\, \TeV. Moreover 350\,000 
drift tube hits were recorded in the muon spectrometer and 320\,000 (65\,000) muon 
trigger hits in the barrel (endcaps). Apart from being 
spectacular, beam splash events are useful in many ways for the experiments.
Their main purpose is to serve timing-in the various detector parts and systems including 
the trigger with respect to each other. It is also interesting to correlate position and 
energy response in splash events, and to use them to identify dead channels.
In the November 2009 beam splash period, after the LHC restart, it was also possible 
to exercise, for the first time in realistic conditions, the ATLAS standalone beam abort system 
using the diamond Beam Condition Monitor (BCM) detectors. By lowering the abort 
thresholds, a deliberate BCM beam abort was triggered by a beam splash event 
reaching ATLAS. No fake abort was observed. Beam splash events have also been 
observed in the forward detectors of the experiments, designed to measure the 
relative luminosity. In total, ATLAS recorded about 70 beam splash events 
(of a total of approximately 100 delivered) in September 2008, and another 
106 events (all triggered) in November 2009. CMS received and recorded 
an order of magnitude more beam splash events.

\begin{figure}[t]
  \begin{center}
	  \includegraphics[width=1\textwidth]{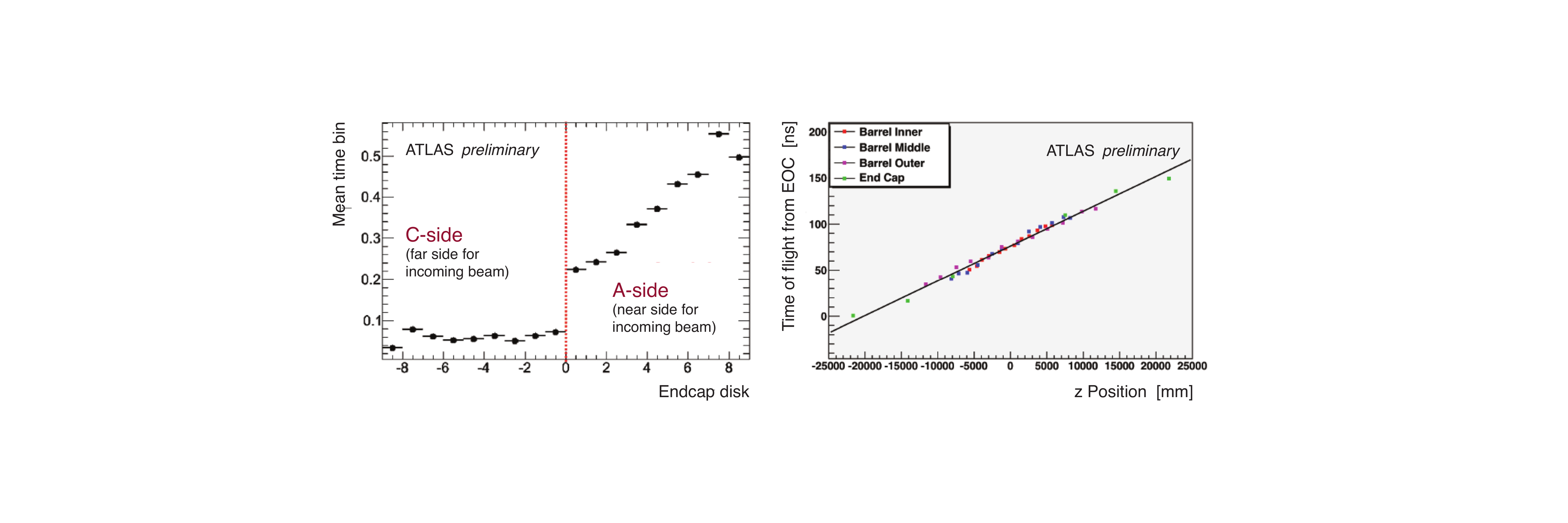}
  \end{center}
  \vspace{-0.4cm}
  \caption{\underline{Left}: timing properties of a single beam-splash event 
           originating from the A-side in the ATLAS silicon strip tracker 
           (see text). \underline{Right}: time residual 
           versus the $z$ coordinate along the ATLAS muon drift tube chambers
           for a beam splash event. The slope is determined by the speed of light.}
  \label{fig:splashtimingsctatlas}
  \begin{center}
    \includegraphics[width=1\textwidth]{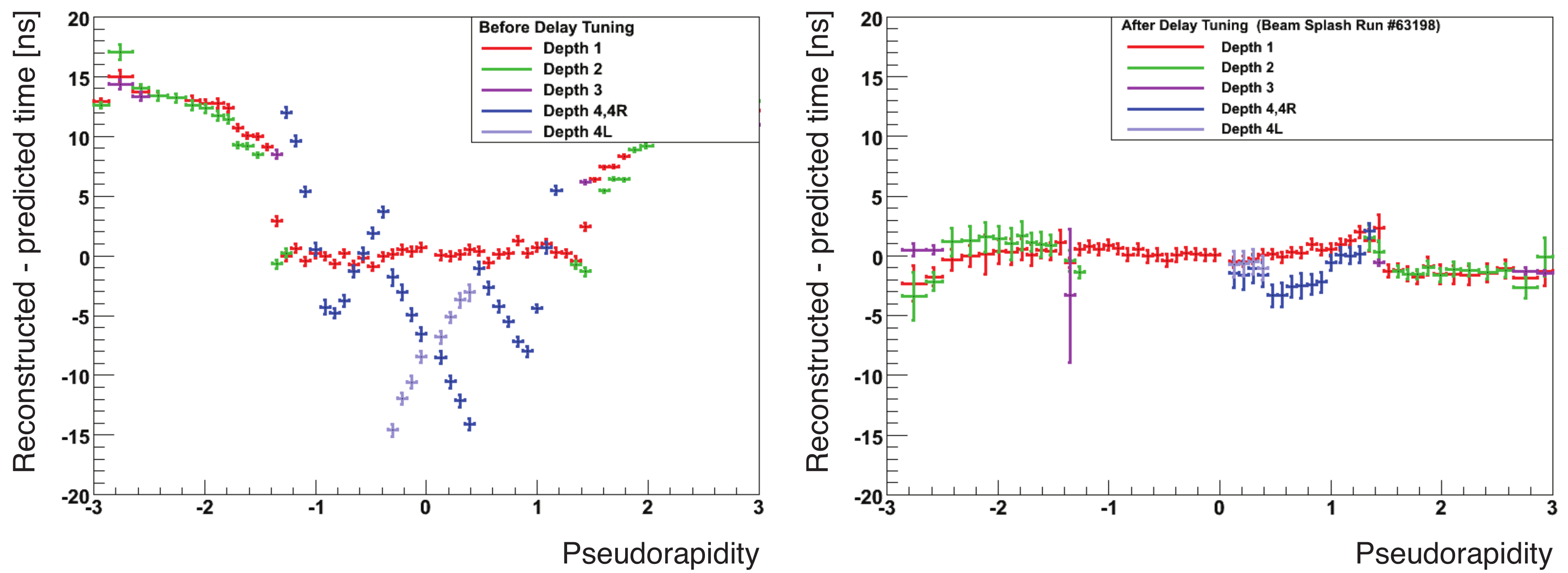}
  \end{center}
  \vspace{-0.4cm}
  \caption{Difference between reconstructed and expected cell time versus the 
               pseudorapidity for beam splash events in the various layers and 
               geometrical regions of the CMS hadronic calorimeter. Left is 
               uncorrected assuming collision timing, and right is after 
               correction with the use of previously observed events. }
  \label{fig:splashtiminghcalcms}
\end{figure}
An example for a timing study is given in Fig.~\ref{fig:splashtimingsctatlas}. Shown 
in the left plot is the mean hit time (expected minus measured) versus the endcap 
disk, where a larger absolute number corresponds to a larger absolute pseudorapidity. 
The measurement corresponding to a single beam-splash event is shown. The event
arrives from the A-side ($+z$ side) so that the hit time behaves as expected for a 
collision event for the far side (C-side), but wrongly for the A-side with respect 
to the expected collision timing used in the event reconstruction (the event
comes from behind and the hit time is thus anticipated). A similar behaviour 
is observed for all other detector systems. 

Beam splash events from both sides can be used to adjust the timing for both far sides. 
The right plot in Fig.~\ref{fig:splashtimingsctatlas} shows the mean time residual along 
the $z$ coordinate of all ATLAS muon drift tube chambers using the synchronous front 
of splash particles and the very large particle flux. A linear relation is found with 
a slope determined by the speed of light. A timing study with beam splash events in the 
CMS hadronic calorimeter is shown in Fig.~\ref{fig:splashtiminghcalcms}. Drawn are
the differences between reconstructed and expected cell times for beam splash events
before (left panel) and after timing adjustment (right) using previously measured beam 
splash events. The large deviations from zero 
in the left panel are due to collision time settings. 
CMS also correlated the energy deposits in the hadronic and 
electromagnetic calorimeters for beam splash events, reproducing nicely the 
expected linear dependence and a relative coefficient of $E_{\rm HCAL}\simeq6.5\cdot E_{\rm ECAL}$.

\subsection{Beam background events}
\label{sec:beambackgroundevents}

After the beam splash events, the collimators were all opened allowing the beam 
to circulate in the LHC and to pass by the experiments. Beam passages without 
interactions are measured primarily in the beam pickup detectors based on 
electrostatic current induction. These detectors are installed $\pm$175\,m 
away from the interaction points of the experiments (many more such beam pickups
are installed along the LHC for beam monitoring purposes). They provide input 
signals to the Level-1 triggers, indicating filled LHC bunches, and also a time 
reference for the detector systems. In case of beam collisions, the 
coincidence of signals in the two beam-pickup detectors can be used to identify colliding 
bunches and, more importantly, their timing difference (measured by an oscilloscope)
can be used as input to the beam `cogging', that is a relative radio-frequency 
phase adjustment of the bunches to ensure collisions in the interaction point
($z=0$) without longitudinal shift. In the Level-1 trigger the beam pickup signals are put 
in coincidence with the other triggers to reduce background from cosmic rays. 
This requires, however, a proper timing-in of the various trigger signals. 

\begin{figure}[t]
  \begin{center}
	  \includegraphics[width=0.55\textwidth]{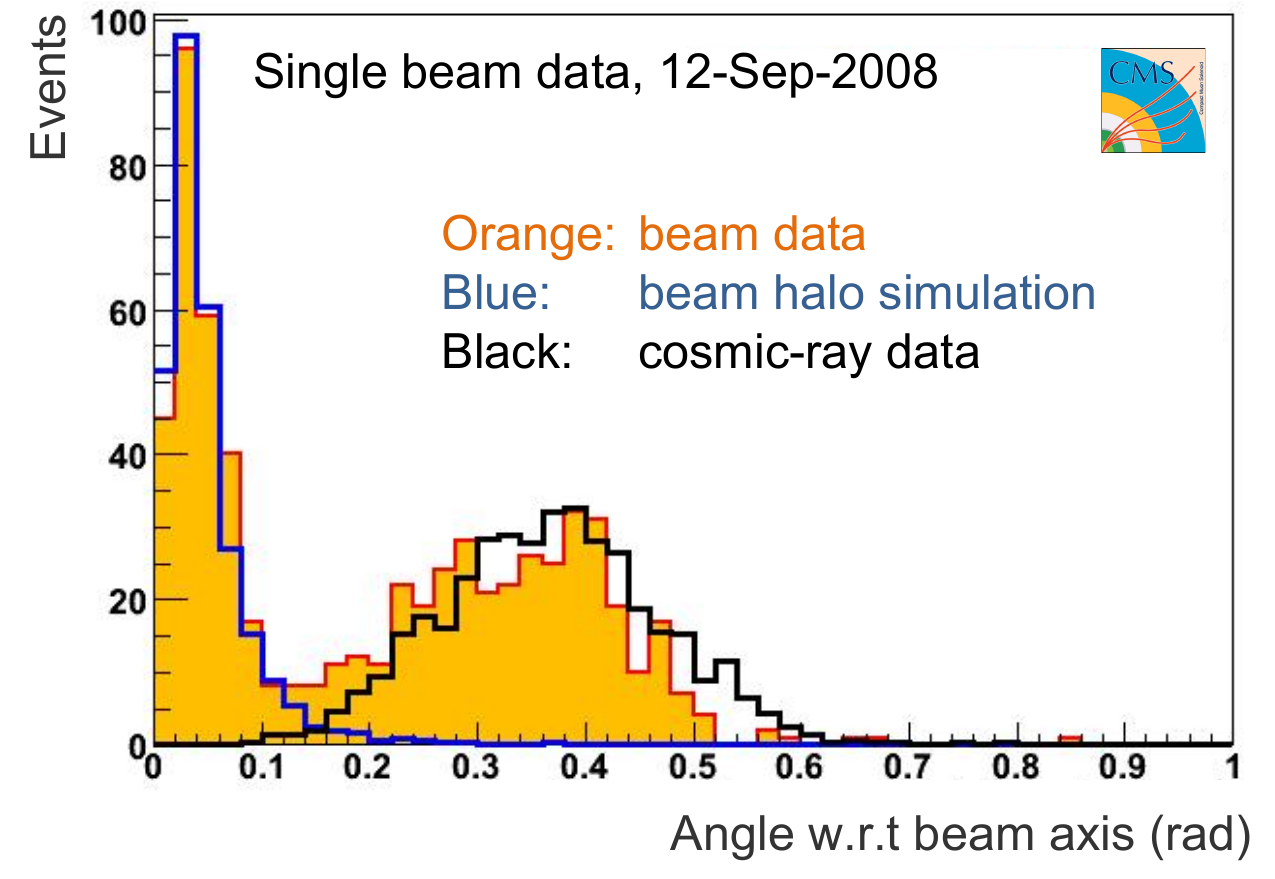}
  \end{center}
  \vspace{-0.4cm}
  \caption{Distributions of the track polar angle with respect to the beam axis
                obtained by CMS for single-beam data (orange shaded), beam-halo background 
                simulation (blue line), and cosmic ray data with no beam (black line).  }
  \label{fig:beamhalocms}
\end{figure}

Circulating single-beam bunches can also provide beam-related background 
particles that are measured by the experiments. At low beam intensities, there 
are two main sources of beam backgrounds referred to as `beam--gas interactions',
which are interactions of beam particles with residual gas in the beam pipe or 
with the beam pipe wall. Via the decay of pions such a process also produces
muons, which travel with the proton beam in what is called the 
`beam halo' (usually referred to as `beam-halo background', which is 
what seems to be the primary single-beam background seen so far in the detectors). 
Such beam related backgrounds originating from fixed-target collisions
are strongly boosted in the forward direction. 
Figure~\ref{fig:beamhalocms} shows the distributions of the track polar angle 
with respect to the beam axis for single-beam data, simulated beam-halo 
background, and cosmic ray events taken with no beam present in the LHC. Whereas the 
beam background peaks at small angles, cosmic ray tracks peak at larger values,
which are, however, much below the $\sim$1.5\,rad that would be expected, because
a forward trigger has been used to select these events. The orange shaded 
histogram shows the distribution of single-beam events accepted by the same 
trigger. One clearly distinguishes the beam-related background from the 
cosmic muon contamination. 

\begin{figure}[t]
  \begin{center}
	  \includegraphics[width=1\textwidth]{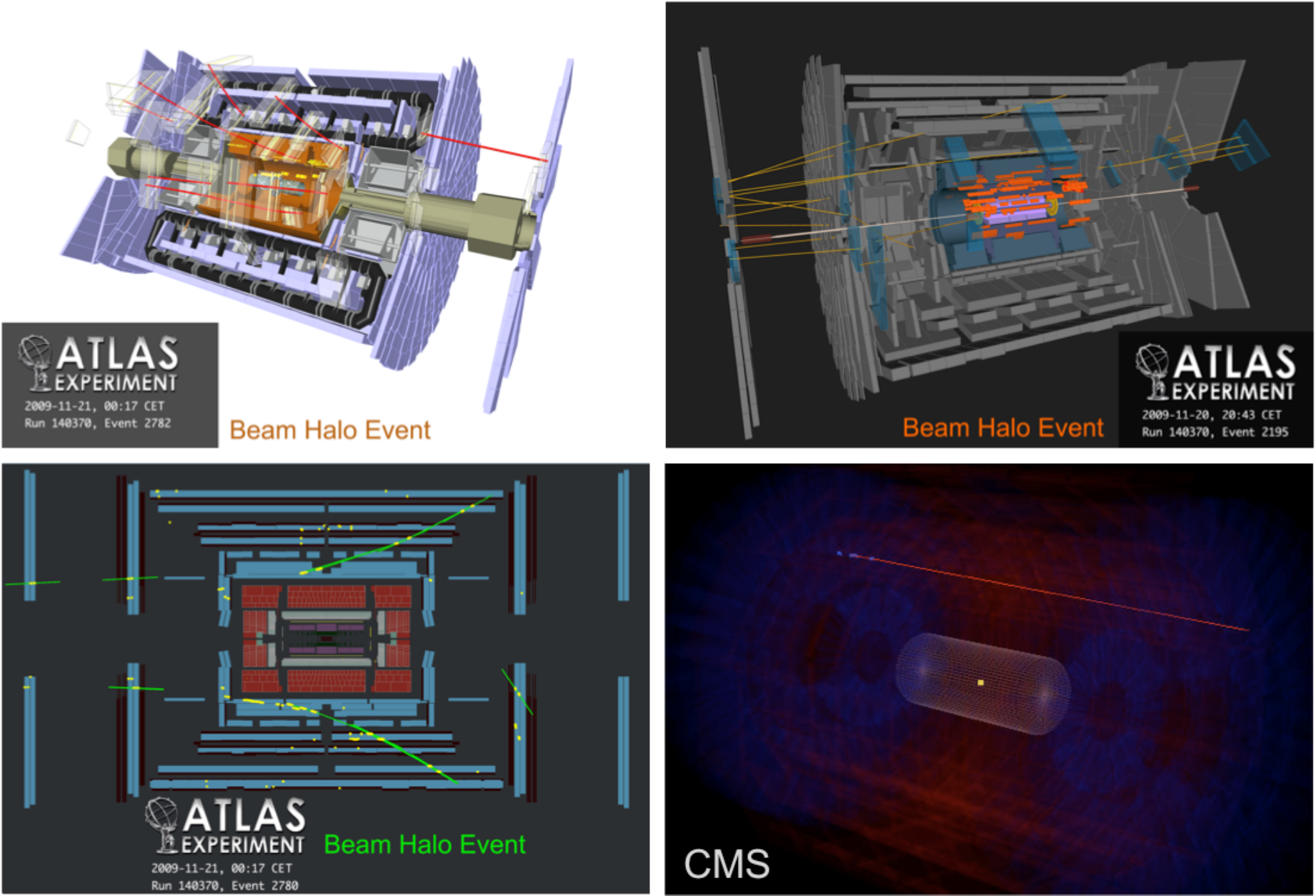}
  \end{center}
  \vspace{-0.4cm}
  \caption{Beam-related background events with halo muons taken by ATLAS and 
               CMS (lower right) in November 2009. }
  \label{fig:beamhaloeventdisplays}
\end{figure}

Event displays of beam background events with halo muons taken by ATLAS and CMS 
are shown in Fig.~\ref{fig:beamhaloeventdisplays}. In ATLAS the toroidal magnetic fields 
in the muon spectrometer bend the muon tracks longitudinally in the $z$ coordinate. 

\subsection{Radio-frequency bunch capture}

\begin{figure}[p]
  \begin{center}
	  \includegraphics[width=0.7\textwidth]{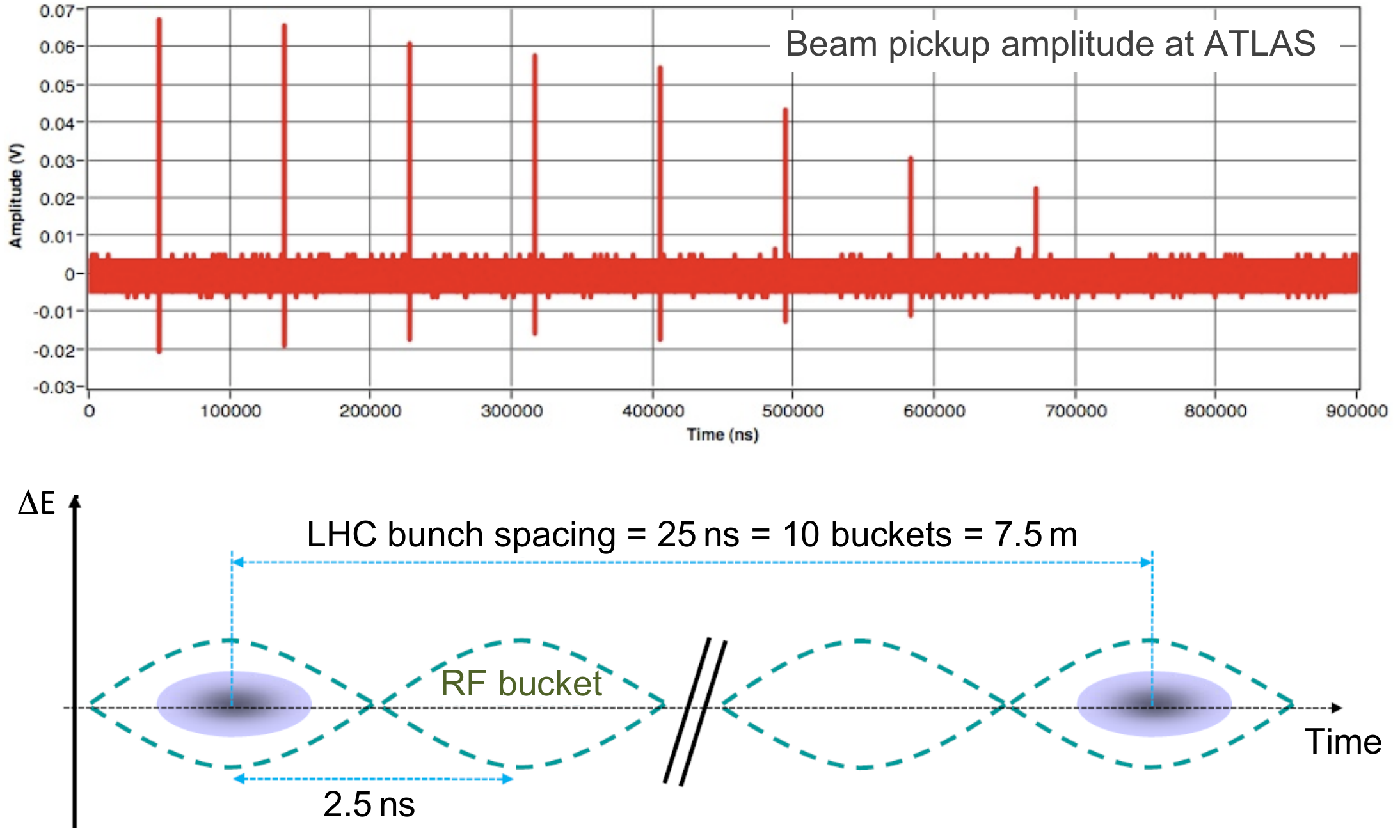}
  \end{center}
  \vspace{-0.4cm}
  \caption{\underline{Top}: decaying circulating beam signal in an ATLAS beam pickup
           detector due to beam debunching.
           \underline{Bottom}: bucket and bunch structure in the LHC. }
  \label{fig:beampickupsrfbucketsketch}
  \begin{center}
	  \includegraphics[width=1\textwidth]{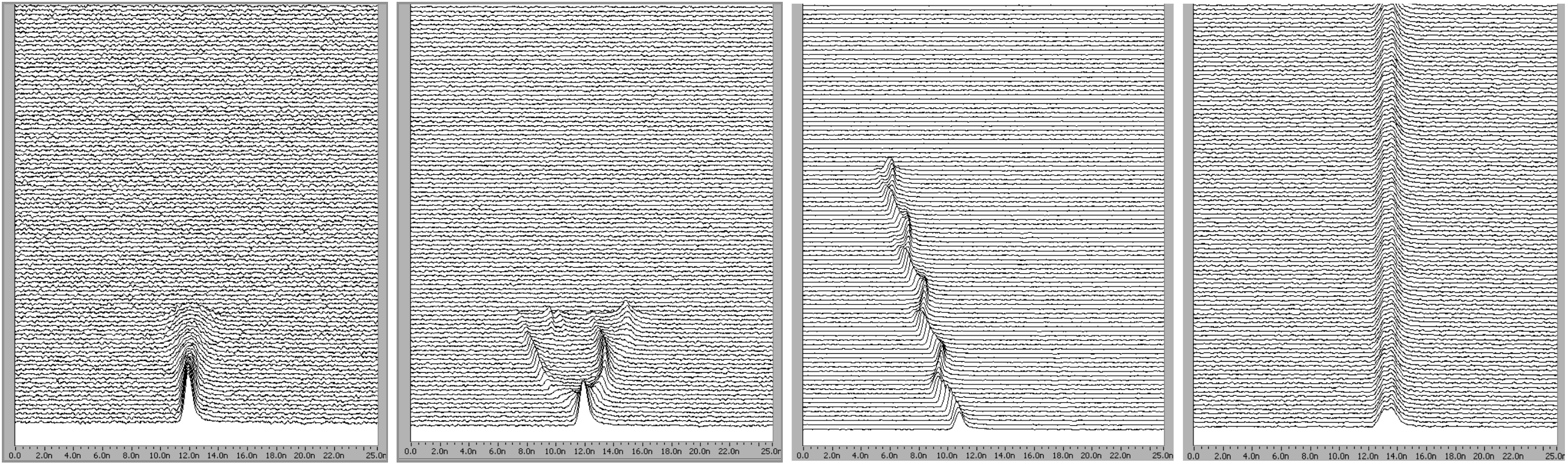}
  \end{center}
  \vspace{-0.4cm}
  \caption{Attempts and successful (rightmost plot) radio-frequency capture of 
           a bunch in the LHC. Each horizontal line represents 10 LHC turns. 
           See text for a discussion of the plots.}
  \label{fig:beamcapturelhc}
  \begin{center}
	  \includegraphics[width=0.65\textwidth]{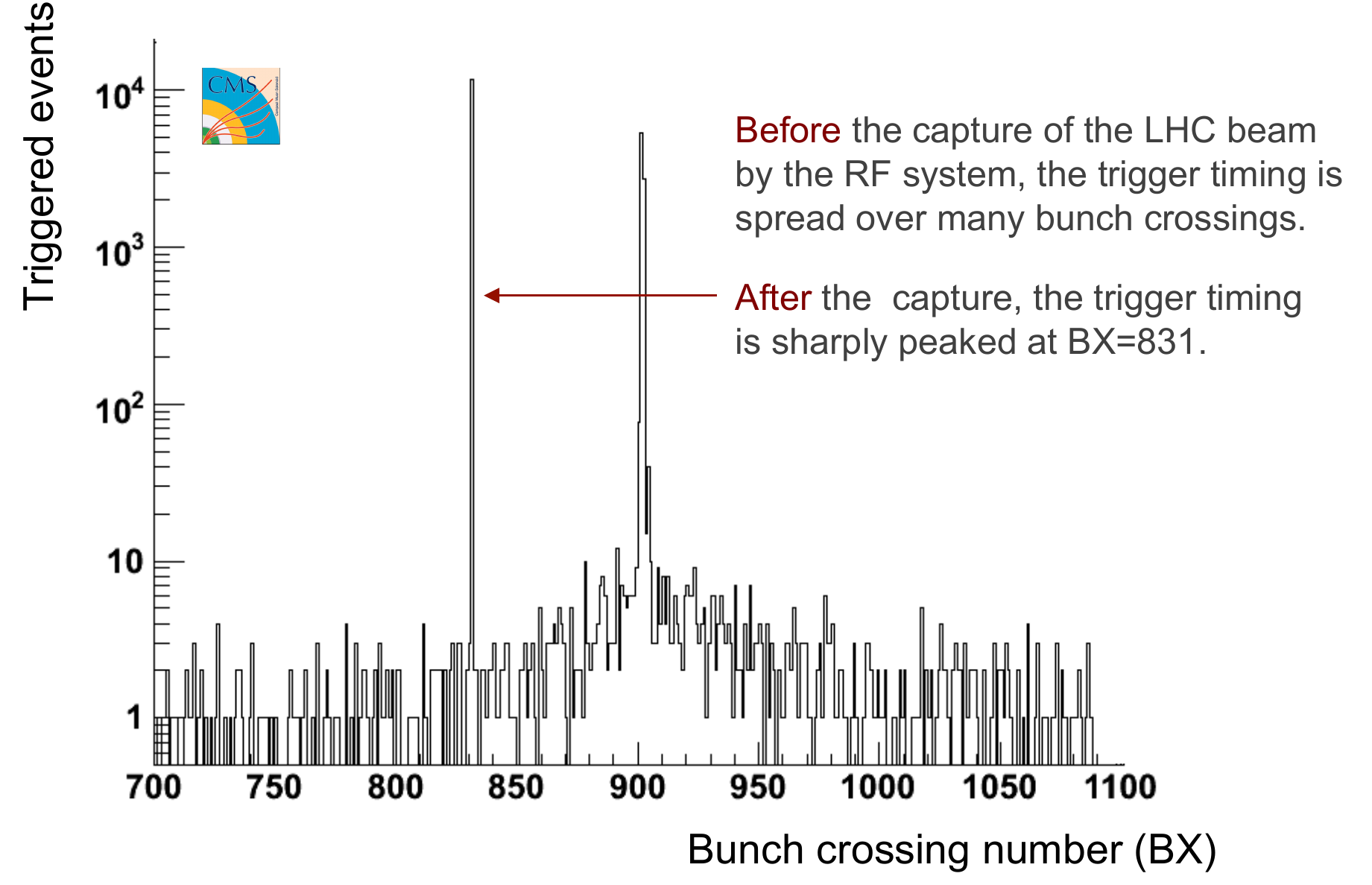}
  \end{center}
  \vspace{-0.4cm}
  \caption{LHC bunch decay and radio-frequency capture as measured by CMS. }
  \label{fig:bunchcapturebcidcms}
\end{figure}

After injection into the LHC, the protons in a bunch start to spread longitudinally
and transversely due to their mutual repulsion. Within milliseconds the bunch thus 
`debunches'.\footnote
{
   Debunching can also be useful. For example, controlled debunching and rebunching 
   can be used to split and multiply bunches in the injection chain of an accelerator. 
   This is, however, a delicate technique which is not used for bunch splitting
   in the LHC injector (PS). 
} 
Debunching can be directly observed by the experiments via a decaying beam pickup
signal during circulating beam. An example for this is displayed in the upper plot of 
Fig.~\ref{fig:beampickupsrfbucketsketch} showing the beam pickup signal amplitude 
in volts versus the time in nanoseconds as measured by ATLAS. The spikes represent the 
induced signal when a bunch passes nearby an ATLAS beam pickup 
detector. The time difference between adjacent spikes amounts to 
89\,$\mu$s, which corresponds to an LHC revolution period. The signal weakens while
the bunch disintegrates. The lower panel of Fig.~\ref{fig:beampickupsrfbucketsketch}, 
sketches the radio-frequency field bucket structure of the LHC. A bunch filled with 
protons is captured within a bucket of 2.5\,ns length (precisely: 2.495\,ns, \ie,
a radio frequency of 400.79\,MHz).\footnote
{
   The radio-frequency electrical field together with the relativistic contraction provide a stronger 
   longitudinal constraint on the bunch size than the bucket length. At 450\,\GeV 
   beam energy, the longitudinal RMS of the bunch is expected to be around 8\,cm,
   (the measured values were found to be significantly lower than that)
   decreasing to approximately 6\,cm at 7\,\TeV design energy. 
}
Only every tenth bucket is filled providing the design bunch period of 25\,ns. 

Figure~\ref{fig:beamcapturelhc} shows a series of attempts in September 2008 to capture a 
bunch in the LHC within a radio-frequency bucket. The horizontal lines
represent a measured beam pickup signal after 10 LHC turns. The leftmost plot 
shows the decaying bunch in absence of a radio-frequency (RF) field. The signal induction 
from the debunched beam 
becomes unmeasurable after 250 turns. The centre-left plot shows a first capture attempt, 
at a wrong injection phase, so that the bunch is split into two by the RF field, leading to 
a fast decay. For the centre-right plot the injection phase has been improved, but is still 
shifted with respect to the RF phase, leading to a moving proton package and a fast decay. 
Finally, the rightmost plot shows an accurate injection phase and a properly captured bunch. 
No decay of the signal due to limited lifetime can be noticed.

Since the experiments record events triggered by the beam pickup signals and, by 
running synchronously with the LHC clock, also store the bunch crossing number
that led to the trigger accept, it is possible to measure the beam debunching {\em and} 
its capture in an RF bucket. Such a measurement has been performed by CMS and the 
result is shown in Fig.~\ref{fig:bunchcapturebcidcms}. Before the RF capture the bunch 
crossing number of the triggered events is spread over many bunches. After successful 
RF capture all triggered events have the same bunch crossing number 831 as seen by the 
spike in the distribution at that point.

\newpage
\section{Early physics at the LHC --- Overview}

The majority of the LHC proton--proton physics programme can be grouped in the 
following grand themes.
\begin{enumerate}

\item {\bf Mass} --- search for the Higgs Boson.

\item {\bf Electroweak unification} --- precision measurements ($W$ and top masses) 
      and tests of the Standard Model.

\item {\bf Hierarchy in the TeV domain} --- 
      search for supersymmetry, extra dimensions, new symmetries in the \TeV domain, 
      and other exotic phenomena.

\item {\bf Flavour} --- {\em B} meson mixing, rare decays and \CP violation as tests of 
      the Standard Model.

\end{enumerate}
This programme is also reflected in the ATLAS and CMS physics organisation, separated
into so-called `physics objects groups' (CMS) or `combined performance groups' (ATLAS),
and `physics analysis groups'. The former groups provide the reconstruction of the 
objects that combine various detector systems and that are common input for physics 
analysis. They are subdivided in `e/gamma', `jets/missing transverse energy', 
`hadronic tau', `muons' and `flavour tagging' groups. The physics groups are 
organised in `Standard Model' containing QCD, electroweak and diffraction physics, 
`B physics', `Top', `Higgs', `SUSY', `Exotics', `Heavy ions', `Luminosity', 
and `Monte Carlo generators' subgroups. 

Since protons are made out of quark and gluon constituents (`partons'), collisions
of protons are complex scattering processes involving elastic, diffractive 
(single and double), inelastic non-diffractive and central diffractive interactions
(pomeron--pomeron scattering). The large majority of the proton--proton events are due 
to interactions at large distances. The inclusive sum of single and double diffractive, 
and non-diffractive processes are  called `minimum bias' events, in allusion to 
lowest transverse momentum events that can be selected by a trigger, and in contrast 
to `zero-bias events', which can only be obtained if all events or a random sample
of events are selected. The total minimum bias cross section at 14\,\TeV centre-of-mass
energy at the LHC is approximately 70\,mb. It dominates by orders of magnitude
the primary physics channels of interest. Minimum bias events are characterised by
tracks with small transverse momenta of $\langle p_T\rangle=0.5$\,\GeV
on average. 
\begin{center}
   \includegraphics[width=0.7\textwidth]{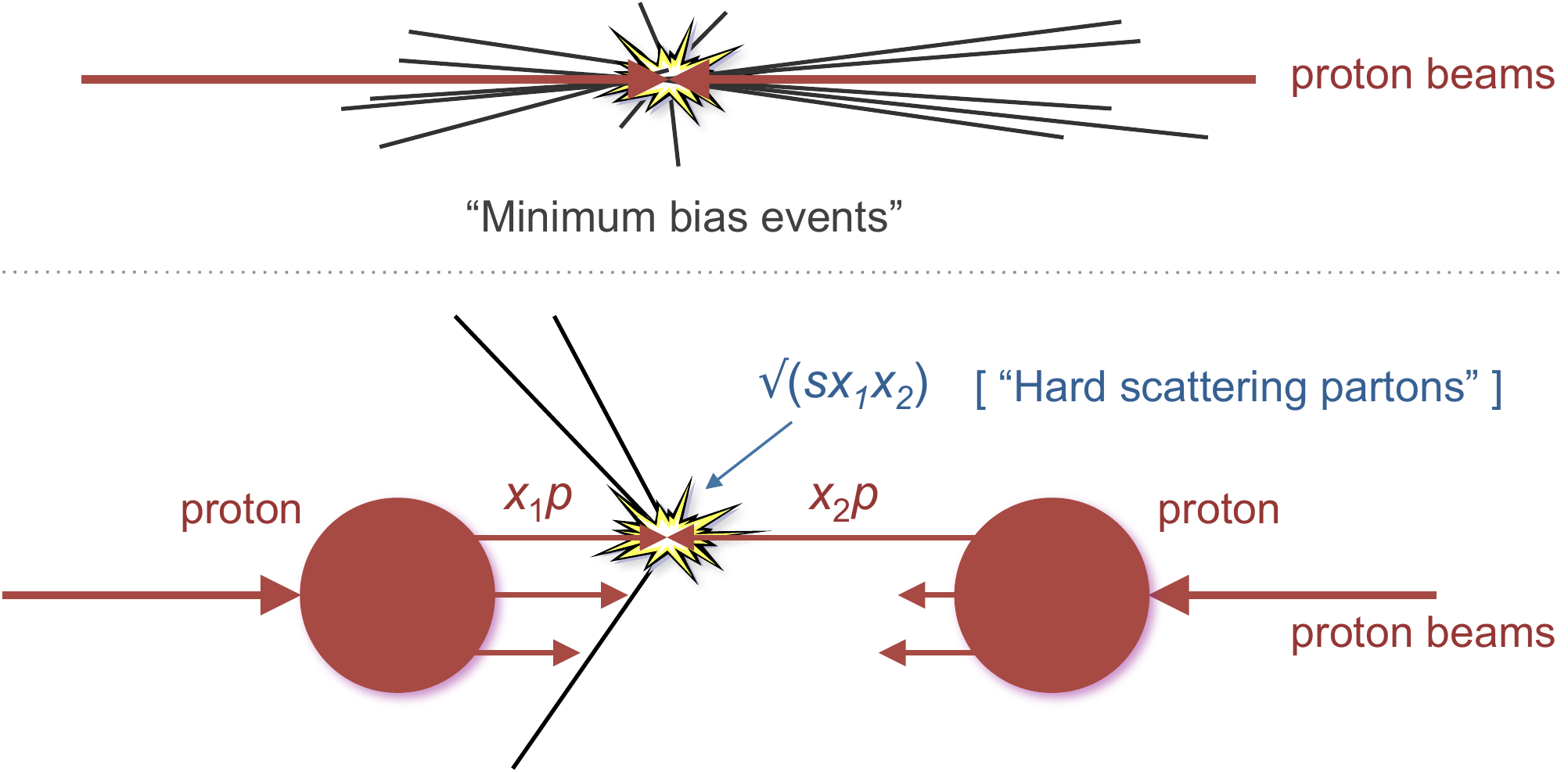}
\end{center}

The constituents of the protons participating in the interaction carry only a fraction 
of the proton's momentum. The fraction is governed by parton distribution functions that 
cannot be predicted from first principles and are taken from experiment. 
The complexity of describing proton--proton interactions includes, besides the 
hard scattering as described by parton-level perturbative QCD, 
the parton distribution functions of the proton, the underlying event (describing
the possibility of multiple parton interactions in the same proton--proton collision), 
initial- and final-state radiation, the definition of jets, and the minimum bias 
event properties.

\begin{wrapfigure}{R}{0.380\textwidth}
  \vspace{-15pt}
  \begin{center}
	  \includegraphics[width=0.380\textwidth]{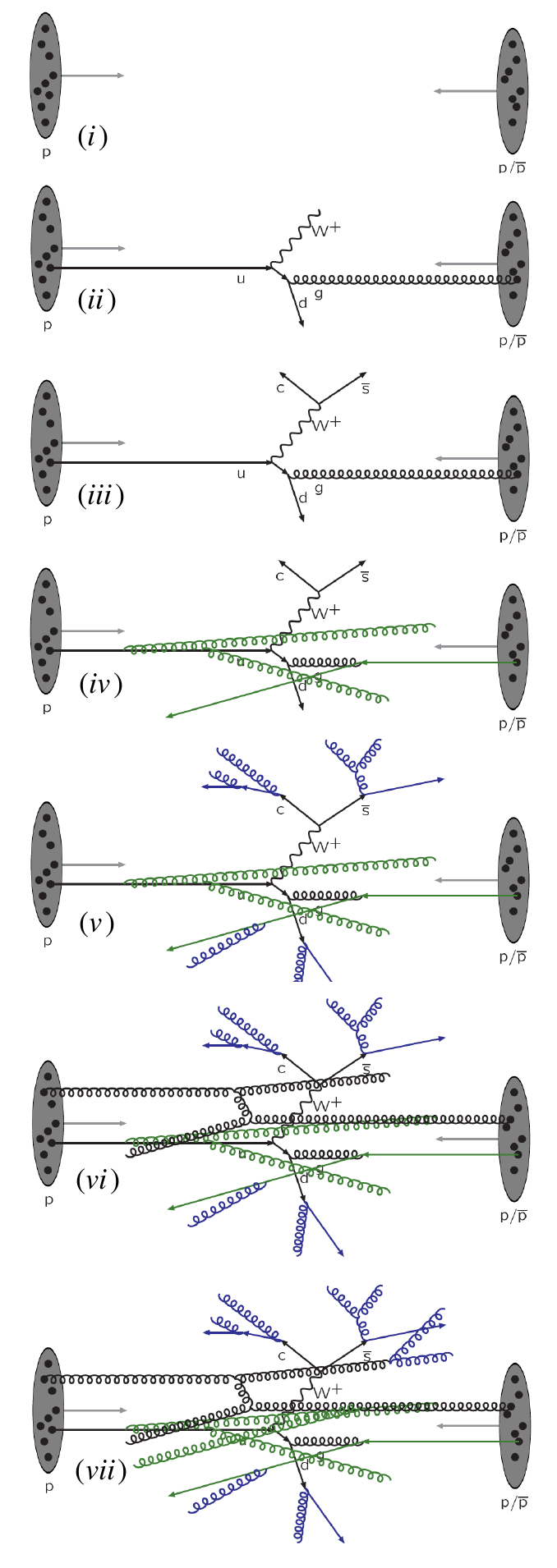}
  \end{center}
  \vspace{-15pt}
  \caption{Schematic Feynman graphs for proton--proton collisions corresponding to 
           (from top to bottom): 
           ($i$)  incoming proton beams: parton distributions;
           ($ii$) hard subprocess: described by matrix elements;
           ($iii$) resonance decays: correlated with hard subprocess;
           ($iv$) initial-state radiation: spacelike parton showers;
           ($v$) final-state radiation: timelike parton showers;
           ($vi$) multiple parton--parton interactions;
           ($vii$) multiple parton--parton interactions with its initial- and final-state radiation.
           Pictures and legend taken from Ref.~\cite{sjoestrand}.}
  \label{fig:protoncollisionfeynman}
  \vspace{-22pt}
\end{wrapfigure}
Figure~\ref{fig:protoncollisionfeynman} (taken from Ref.~\cite{sjoestrand}) illustrates 
the structure of a proton--proton collision event as it occurs in the LHC. Hard
subprocesses between partons need to be convolved with parton densities, the decays 
of the hard subprocesses, initial- and final-state radiation, and multiple parton
interactions (and their initial- and final-state radiation), as well as beam remnants 
and other outgoing partons (not shown) to arrive at a realistic description. 
All parton-level processes are connected through colour confinement, leading to a 
primary hadronisation, with many primary hadrons being unstable and further decaying. 

To reconstruct such an event in ATLAS or CMS it first needs to be triggered, \ie, 
the event must pass several trigger levels with increasing rejection power. Once
accepted, the event is written to disk and promptly reconstructed on large offline
computer farms comprising several thousand central processing units. The reconstruction
program reconstructs tracks of charged particles in the inner tracker and the muon 
systems, electromagnetic clusters in the electromagnetic calorimeter, hadronic clusters
and jets in the combined electromagnetic and hadronic calorimeters, missing transverse
energy in the calorimeters, and identifies particles and objects: muons, electrons, 
photons, taus, jets, and heavy quark flavour. All these steps in the reconstruction
chain involve tremendous challenges regarding efficiency, purity, accuracy and resolution
(calibration). The extensive commissioning work performed by the experiments will 
surely pay off when analysing the first collision data and comparing them with Monte 
Carlo simulations. With increasing statistics
data-driven analysis and calibration methods will take over and the experiments
will achieve the performance they have been designed for. 

After the reconstruction of the primary physics objects, the events are selected
according to topological criteria that characterise the physics channel of interest.
{\em Inclusive analyses} count events with leptons, photons, jets or missing transverse
energy. For example, a QCD analysis may select events with high-energetic (or many) 
jets. A combined QCD and electroweak analysis may select events with leptons or 
photons in the final state. A search for supersymmetry with $R$-parity conservation
will select events with large missing transverse energy, and may also require leptons
to reduce the contamination from Standard Model QCD events. 
{\em Exclusive analyses} kinematically combine reconstructed objects. For example,
an analysis using $W\to\mu\nu$ decays will identify a muon and compute the transverse
$W$ mass using the muon momentum and the transverse missing energy vector. To select
top--antitop events, where, for example, one top decays to $b e\nu$ and the other to 
$b qq$, one must identify the electron and two $b$-jets, and compute the top mass from 
the invariant mass of one of the $b$ jets and two hard light-quark jets, which originate 
from a $W$ decay. To identify Higgs decays into two photons one must identify two 
photons in the event and compute their invariant mass, which needs to accumulate at
the same value within the experimental errors to create a significant Higgs signal 
over backgrounds from random two-photon or misidentified photon-jet combinations. 
Similarly, to search for Higgs decays into two electrons and two muons, once must 
identify the corresponding leptons and compute their invariant mass. Intermediate
on-shell resonances with known mass can be used as additional kinematic constraints. Finally, 
to search for new high-mass resonances such as Kaluza-Klein graviton states decaying
into lepton pairs, as predicted in models with extra spatial dimensions, one must 
identify the leptons and compute their mass to obtain a signal over the dominant 
Drell-Yang di-lepton background. For many of these analyses it is beneficial to 
combine all the available object-level and event-level information using multivariate
statistical pattern recognition techniques.

\section{Physics commissioning}
\label{sec:physicscommissioning}

With emphasis at the beginning of the collision data taking, but also throughout
the whole lifetime of the experiments, physics commissioning such as the calibration 
and alignment of detector systems and physics objects, as well as the data-driven
(`{\em in-situ}') measurement of efficiencies, purities, calibration biases and 
resolutions, will represent a large part of the experimental work. We discuss in 
the following the {\em in-situ} calibration of the electromagnetic calorimeter, the 
determination of material in the inner tracking detector, and jet and missing
transverse energy calibration and reconstruction. 

\subsection{{\em In-situ} electromagnetic calorimeter calibration}

Among the primary measurements driving the performance requirements for the 
ATLAS and CMS electromagnetic calorimeters is the search for $H\to \gamma\gamma$.
Since this channel is important at low Higgs mass where the intrinsic width of
the Higgs is negligible,\footnote
{
   A Higgs of mass 120\,\GeV has an intrinsic width of 4\,\MeV, while at 
   200\,\GeV the Higgs has a width of 1.4\,\GeV due mainly to the opening 
   of the di-weak-boson channels. 
}
the measured width of the di-photon invariant mass, and hence the sensitivity 
for discovery, will be determined by the energy resolution of the 
electromagnetic calorimeter. We have already mentioned the importance of the 
constant term in the calorimeter energy resolution for Higgs searches in 
Footnote~\ref{ftn:higgsconstantterm} on page~\pageref{ftn:higgsconstantterm}. 
We can extend this by a back-on-the-envelope exercise. Let us consider a data 
sample for an integrated luminosity of 20\invfb containing
690 $H\to\gamma\gamma$ and $\sim$170\,000 background events with
di-photon invariant mass $110<m_{\gamma\gamma}<150$\,\GeV. With the 
nominal (design) ATLAS electromagnetic calorimeter resolution, assuming a 
constant term of 0.7\%, a fit to the di-photon mass would yield a signal 
significance of $2.9\sigma$. Worse constant terms of 1.0\% or even 2.0\% would
reduce this significance to $2.4\sigma$ and $1.8\sigma$, respectively.\footnote
{
   Note that this test assumes the simplest possible $H\to\gamma\gamma$ analysis 
   approach. A more sophisticated fit using more discriminating variables 
   and detector-specific `categories' boosts the fit performance significantly.
} 

\begin{figure}[t]
  \begin{center}
	  \includegraphics[width=1\textwidth]{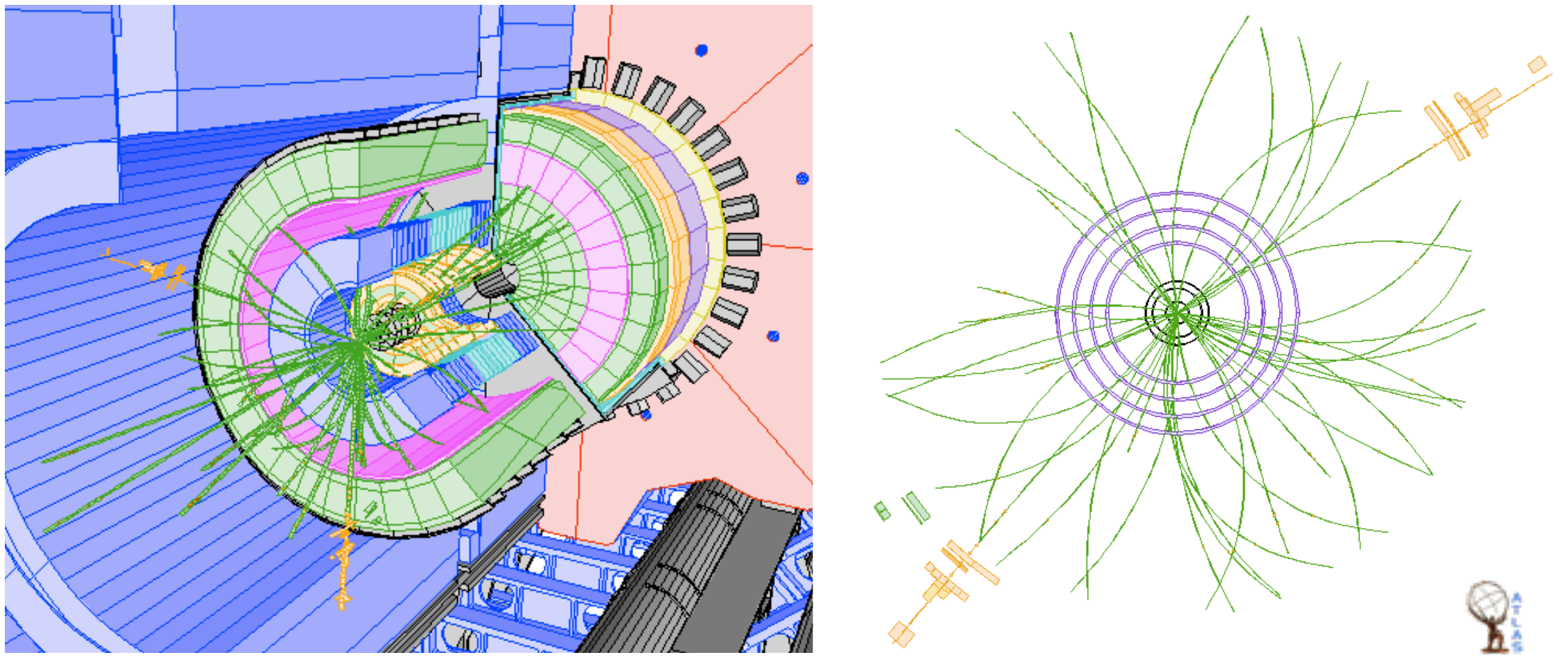}
  \end{center}
  \vspace{-0.3cm}
  \caption[.]{Event display of a simulated $Z\to ee$ event in ATLAS. The final-state 
              electrons have tracks in the inner tracker and large energy 
              depositions in the electromagnetic calorimeter. Their invariant 
              mass is consistent with that of a $Z$ boson. }
\label{fig:zeepersint}
\end{figure}
It is hence mandatory to keep the constant term, originating from non-uniformities 
in the calorimeter response due to inhomogeneities and non-linearities, as small
as possible by {\em intercalibrating} the calorimeter with physics events. 
Calorimeter intercalibration (which is {\em not} absolute scale calibration) can 
be performed with any physics events that provide a predicted or smooth energy 
deposition. 

The most favourable channel for {\em in-situ} intercalibration  
is $Z\to ee$. The $Z$ mass being precisely measured at LEP to 
$(91.1875 \pm 0.0021)$\,\GeV, the average reconstructed di-electron mass in the detector 
after calibration must reproduce it (per event, the detector resolution and 
the natural width of $2.5$\,\GeV will lead to a natural smearing). With 
sufficient statistics, the mass-constrained intercalibration can be done 
per geometrical detector units, which are suitably chosen regions in pseudorapidity 
and azimuth, typically $\Delta\eta$$\times$$\Delta\phi=0.2$$\times$$0.4$.
($Z\to ee$ decays also allow one to calibrate the absolute energy scale,
which is required to be known at the per mil level or less for most analyses, 
and should be at the 0.02\% level for the high-precision $W$ mass measurement.) 
For a given intercalibration region $i$, it is assumed that long-range non-uniformities, 
encoded in a parameter $\alpha_i$, have modified the measured electron energy as 
$E^{\rm reco}_i = E^{\rm true}_i\cdot (1+\alpha_i)$. Neglecting correlations between
the electrons and postulating that the opening angle between the two electrons 
is correctly measured on average, the effect on the di-electron invariant mass 
is $M^{\rm reco}_{ij}=M^{\rm true}_{ij}(1+(\alpha_i+\alpha_j)/2)$. The $\alpha_i$ 
can be extracted from a maximum-likelihood fit to $Z\to ee$ candidates, which 
must also incorporate a background component from events other than $Z\to ee$.

\begin{figure}[t]
  \begin{center}
	  \includegraphics[width=0.7\textwidth]{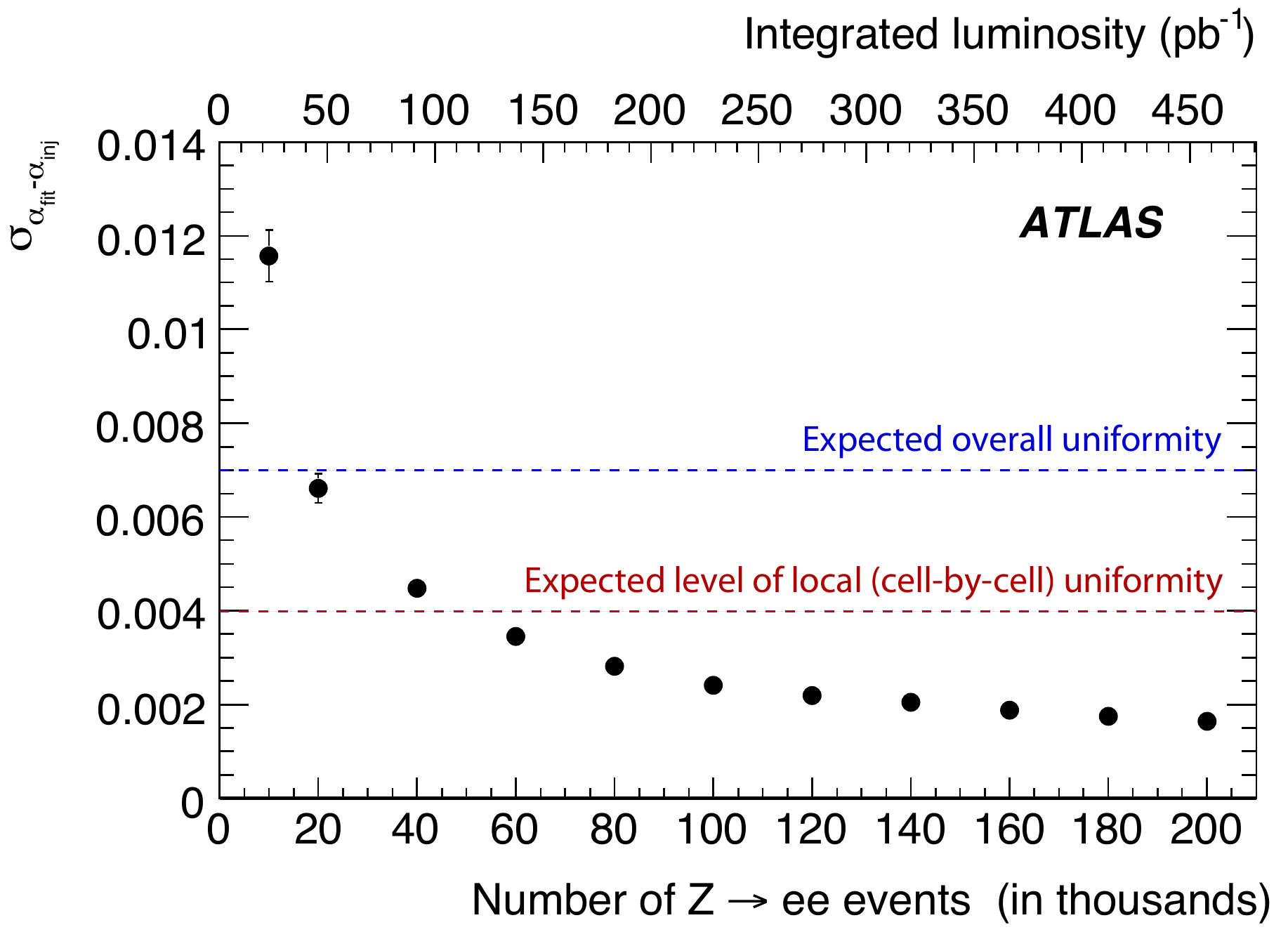}
  \end{center}
  \vspace{-0.3cm}
  \caption[.]{Statistical yield of the $Z\to ee$ electromagnetic 
              calorimeter intercalibration in ATLAS. Shown is the expected accuracy 
              achieved for the constant term versus the number of events used in 
              the intercalibration fit. The corresponding integrated luminosity
              is given on the upper abscissa. }
\label{fig:zeeemintercalib}
\end{figure}
Figure~\ref{fig:zeepersint} shows the event display of a simulated $Z\to ee$
event in ATLAS. The electrons leave large energy deposits in the electromagnetic
calorimeter and their invariant mass is consistent with that of a $Z$ boson. 
Approximately 10\,000 of these events (and approximately 10 times
more $W\to e\nu$) will be recorded in 10\invpb integrated luminosity (reconstruction 
efficiency not subtracted). Figure~\ref{fig:zeeemintercalib} depicts the 
expected statistical yield of the $Z\to ee$ electromagnetic calorimeter 
intercalibration in ATLAS. Shown is the expected accuracy achieved for the 
constant term versus the number of events used in the intercalibration fit.
Indicated by the dashed horizontal lines are the design value of 0.7\% for 
the constant term and the level of the local non-uniformity from cell-by-cell 
variations, estimated to be 0.4\%. Design calibration performance is expected 
to be reached with 20\invpb integrated luminosity.

\subsection{Inner detector material mapping}

\begin{wrapfigure}{R}{0.4\textwidth}
  \vspace{-24pt}
  \begin{center}
	  \includegraphics[width=0.4\textwidth]{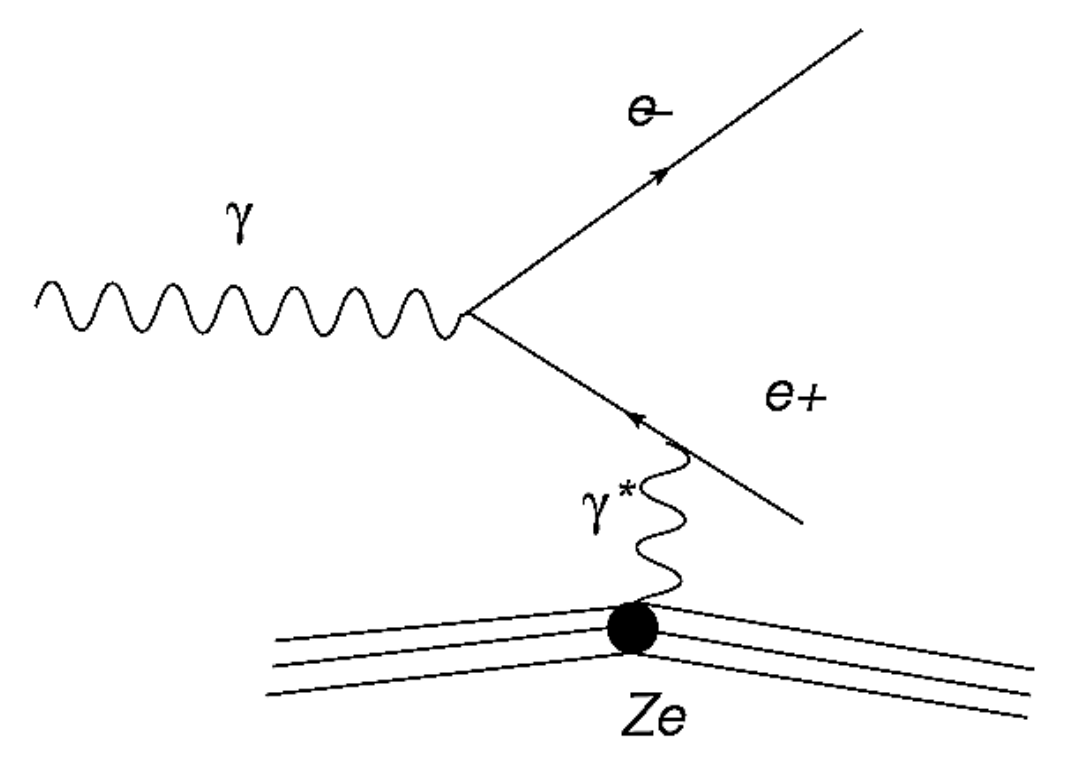}
  \end{center}
  \vspace{-15pt}
  \caption{Feynman diagram for the conversion of a photon to an electron--positron pair
           in presence of a nucleus.}
  \label{fig:photonconversion}
  \vspace{-6pt}
\end{wrapfigure}
\begin{figure}[t]
  \begin{center}
	  \includegraphics[width=0.9\textwidth]{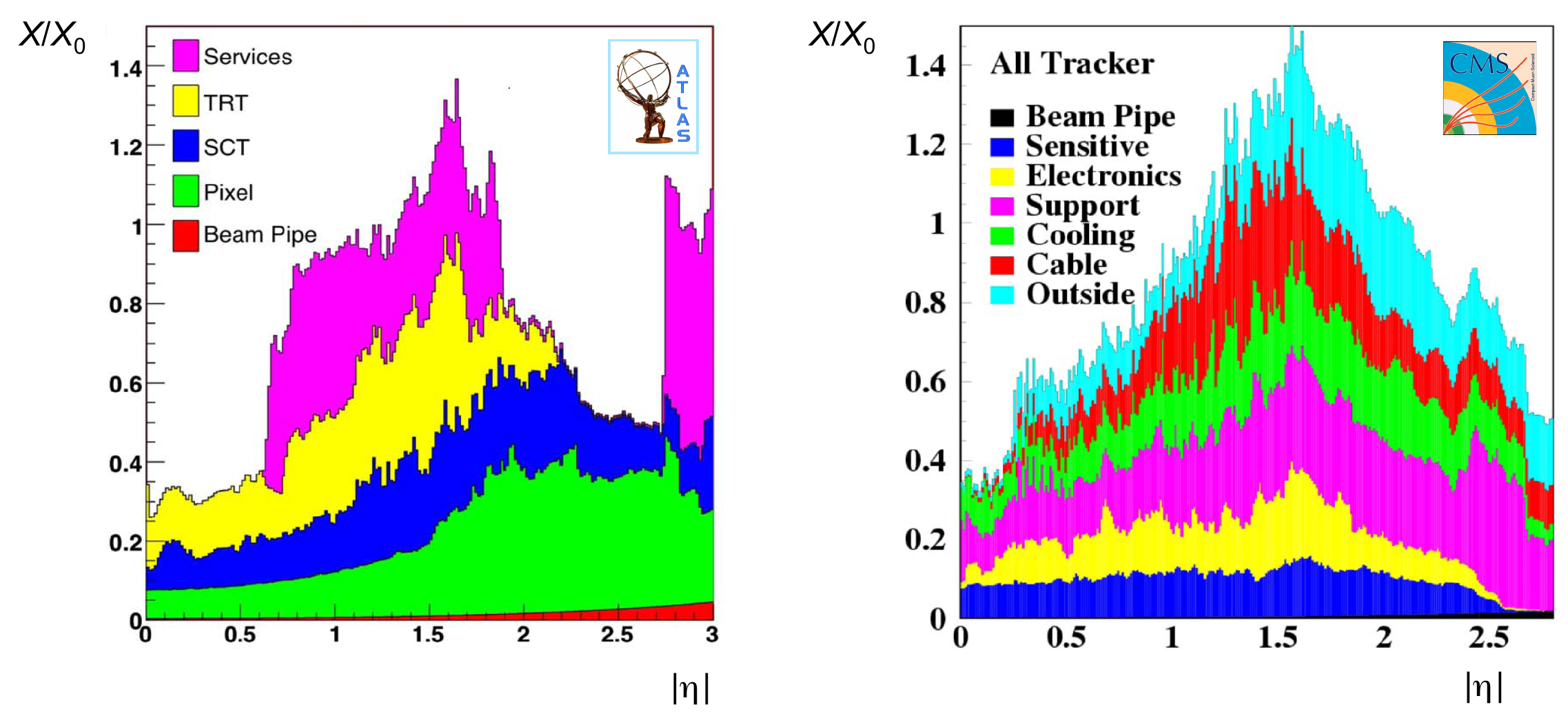}
  \end{center}
  \vspace{-0.5cm}
  \caption[.]{Material in the inner tracking system of ATLAS (left) and CMS (right)
              in terms of numbers of radiation lengths $X/X_0$. Solenoid and calorimeter
              cryostat add roughly $2X_0$ before the electromagnetic calorimeter
              presampler in ATLAS. }
\label{fig:innerdetectormaterialatlascms}
\end{figure}
The high-precision and redundant inner tracking systems of ATLAS and CMS come at 
the price of a significant amount of material the particles must traverse. 
Figure~\ref{fig:innerdetectormaterialatlascms} shows the material in the inner 
tracking system of ATLAS (left) and CMS (right) in terms of radiation lengths. It is 
remarkable that only a small part of it stems from active detector material, whereas the 
main contributions are due to services. The amount of radiation lengths in these services 
needed to be systematically reevaluated throughout the planning and construction phases
of both detectors. While the technical proposals in 1994 estimated about (in units
of $X_0$) 0.2 (0.6) at $\eta=0$ ($\eta=1.7$ corresponding to about $20^\circ$ polar
angle) for both ATLAS and CMS, it became 
0.2 (1.5 for ATLAS and 0.85 for CMS) at the time of the TDRs in 1997, to finally converge 
to 0.3 (1.3 for ATLAS and 1.5 for CMS)  at the time of the construction in 2006.
Note that in ATLAS objects need to traverse approximately an additional $2X_0$ before 
reaching the presampler (available for $|\eta|<1.8$), and roughly another $X_0$ 
before the electromagnetic calorimeter.

\begin{figure}[t]
  \begin{center}
	  \includegraphics[width=0.7\textwidth]{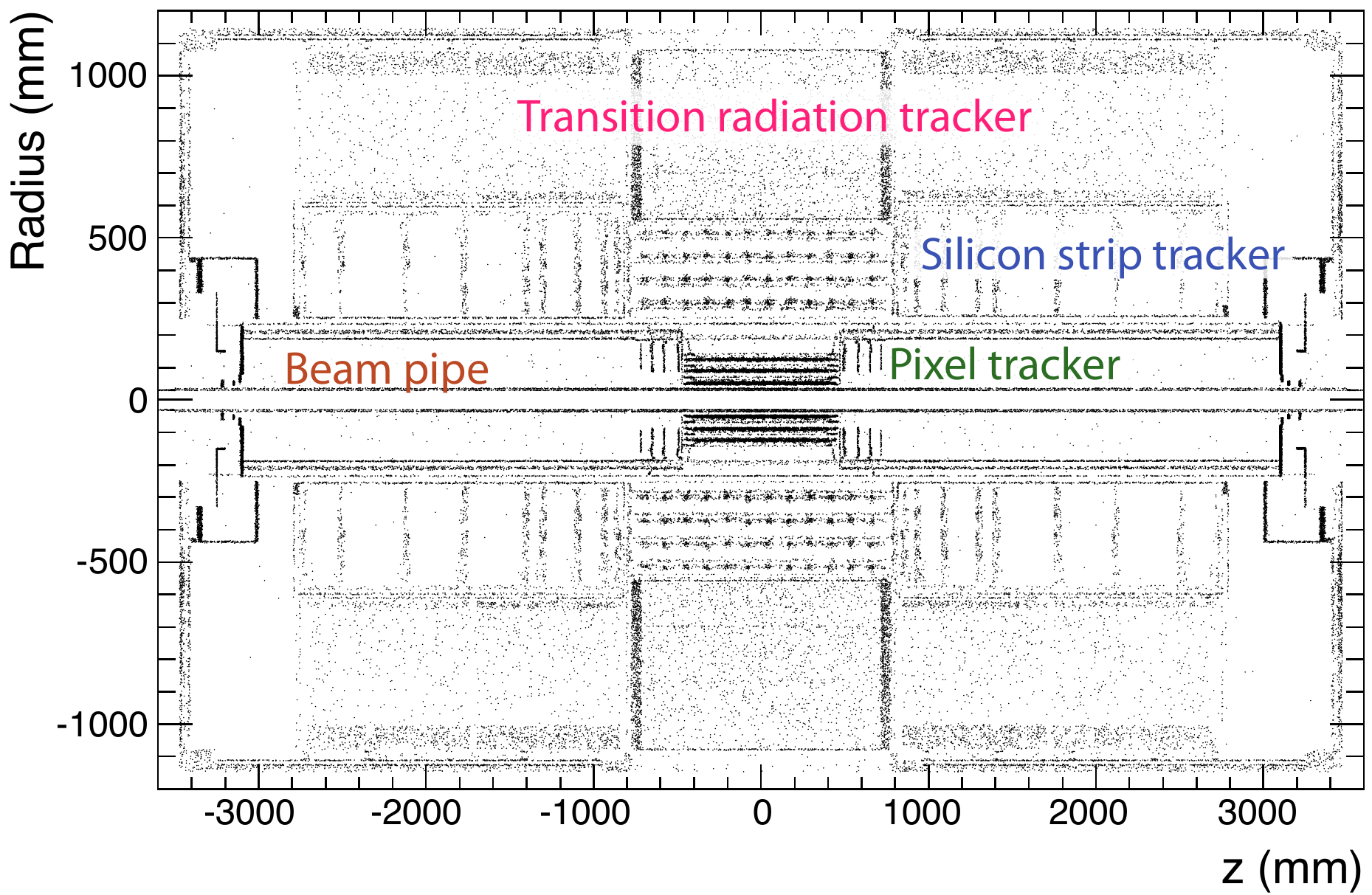}
  \end{center}
  \vspace{-0.5cm}
  \caption[.]{Mapping of photon to electron--positron 
              conversions as a function of $z$ and radius, 
              integrated over the azimuth angle, for the ATLAS inner tracking detector. 
              The mapping has been made from 500\,000 simulated minimum bias events 
              ($\sim$40 minutes of data taking at 200 Hz output rate), 
              using $\sim$90\,000 conversion electrons of transverse momentum 
              larger than 0.5\,\GeV, originating from photons from $\piz$ and $\eta$ 
              decays. Monte Carlo truth information is used for the conversion vertices.
              The plot shown does not represent the latest version of the ATLAS 
              detector description. In particular the beam condition monitor stations
              located at $z=\pm$1840\,mm are not yet included. }
\label{fig:conversionmaterialtrackeratlas}
\end{figure}

A good understanding and simulation of the inner detector material is crucial for 
precision measurements such as the $W$ mass, where the accurate calibration at 
the $Z$ mass needs to be transferred to the $W$ mass using Monte Carlo simulation. 
Many other physics analyses benefit from a precise material mapping. The best 
method to perform a radiography of the inner tracking detector is to use 
photon-to-electron--positron-pair conversion, which occurs only in the vicinity of a 
nucleus that recoils against the $\ee$ system and thus ensures momentum conservation
(\cf Fig.~\ref{fig:photonconversion}). The conversion needs to happen not too 
far from the interaction point so that sufficient tracking layers remain to 
reconstruct the electron and positron tracks and their common vertex position, 
which indicates matter. A photon-conversion-based radiography of the ATLAS inner 
tracking detector, obtained from Monte Carlo simulation, which implements a 
detailed modelling of the active and passive components, is shown in 
Fig.~\ref{fig:conversionmaterialtrackeratlas}. The photons originate from $\piz$ 
and $\eta$ decays, and Monte Carlo truth information has been used for the 
conversion vertices (the measured conversion map will look quite different). 

\subsection{Efficiency determination with the tag-and-probe method}
\label{sec:tagandprobemethod}

\begin{wrapfigure}{R}{0.3\textwidth}
  \vspace{-44pt}
  \begin{center}
	  \includegraphics[width=0.3\textwidth]{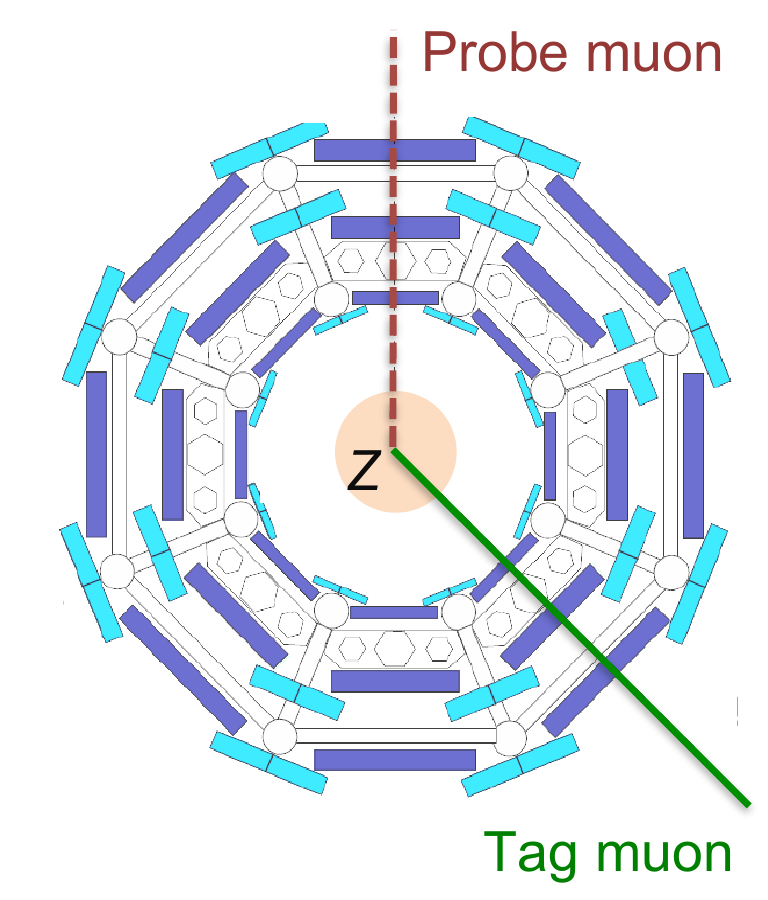}
  \end{center}
  \vspace{-15pt}
  \caption{Sketch illustrating the tag-and-probe method. }
  \label{fig:tagandprobemethodsketch}
  \vspace{-26pt}
\end{wrapfigure}
\begin{figure}[t]
  \begin{center}
	  \includegraphics[width=0.7\textwidth]{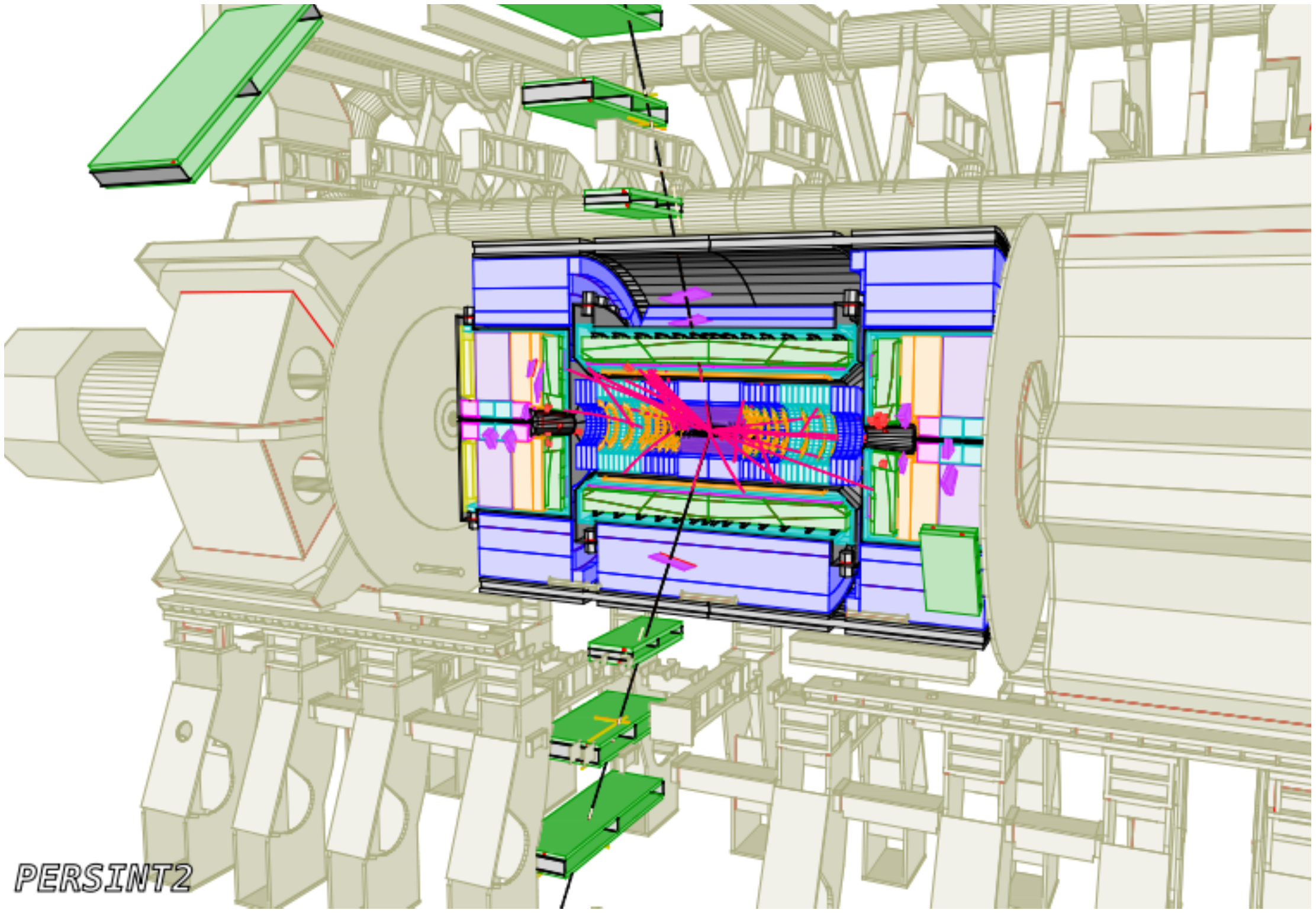}
  \end{center}
  \vspace{-0.3cm}
  \caption[.]{Event display of a simulated $Z\to\mu\mu$ event in ATLAS. The 
              final-state muons are measured in the muons spectrometer.}
\label{fig:zmumupersint}
\end{figure}

Decays of $Z$ bosons to leptons can also be exploited to measure trigger selection 
and offline reconstruction efficiencies from data. The primary method used for this
is denoted `tag-and-probe method', the principle of which is straightforward 
(see sketch in Fig.~\ref{fig:tagandprobemethodsketch}). 
Let us consider the example of determining the reconstruction efficiency of 
muons in the muon system using $Z\to\mu\mu$ candidate events.\footnote
{
   The expression `candidate' refers to the fact that for real data we do not 
   know whether a reconstructed $Z\to\mu\mu$ candidate is indeed the process we 
   believe it to be, or whether it is background from random combinations of 
   muons (`combinatorial background') or objects faking muons. Only a 
   statistical analysis allows us to separate signal from irreducible background. 
} 
The candidate event has been triggered by the `tag muon', which is a `golden' 
muon candidate with an isolated track from combined inner tracker and muon system 
reconstruction, and transverse momentum larger than 20\,\GeV. The probe muon
is another muon candidate, which is independent of the tag-muon selection.
To find the candidate we require a track reconstructed in the inner tracker and 
an invariant mass of tag and probe muons consistent with that of a $Z$ boson.
We now count how often the probe muon has been reconstructed in the muon spectrometer.
With sufficient statistics the efficiency of the probe muon reconstruction can be 
evaluated in bins of $p_T$, $\eta$ and $\phi$. Usually, the result has to be 
corrected for combinatorial background under the $Z$ peak. The most powerful 
approach combines background and efficiency determination in all regions
within a single unbinned maximum-likelihood fit. The tag-and-probe method is 
very flexible, and many versions of the same idea exist.
Figure~\ref{fig:zmumupersint} shows an event display of a simulated $Z\to \mu\mu$
event in ATLAS. The minimum ionising muon tracks traverse the calorimeters
and leave measured hits in the muon spectrometer. Approximately 10\,000 of these 
events (and approximately 10 times more $W\to \mu\nu$) will be recorded in 
10\invpb integrated luminosity (reconstruction efficiency not subtracted). 

\subsection{Jet calibration}

A precise knowledge of the absolute jet energy scale (JES) is needed by
many physics analyses. Typically a calibration of better than 1\% is required 
for the measurement of the top-quark mass, but also for supersymmetry signatures. 
Jets are complex phenomenological objects, and their reconstruction involves 
a large number of corrections and calibrations. Only a brief overview is 
given here.

The jet energy reconstruction and calibration can be divided in four steps: 
\begin{enumerate}

\item Calorimeter tower or cluster reconstruction.

\item Jet forming (cone, $k_t$, anti-$k_t$ or other `jet algorithms').

\item Jet calibration from calorimeter to the particle scale.

\item Jet calibration from particle to the parton scale.

\end{enumerate}
The discussion here concentrates on jet calibration, assuming jets have been 
formed by an algorithm with suitable experimental and theoretical properties for 
the physics measurement under study.

\begin{wrapfigure}{R}{0.3\textwidth}
  \vspace{-24pt}
  \begin{center}
	  \includegraphics[width=0.3\textwidth]{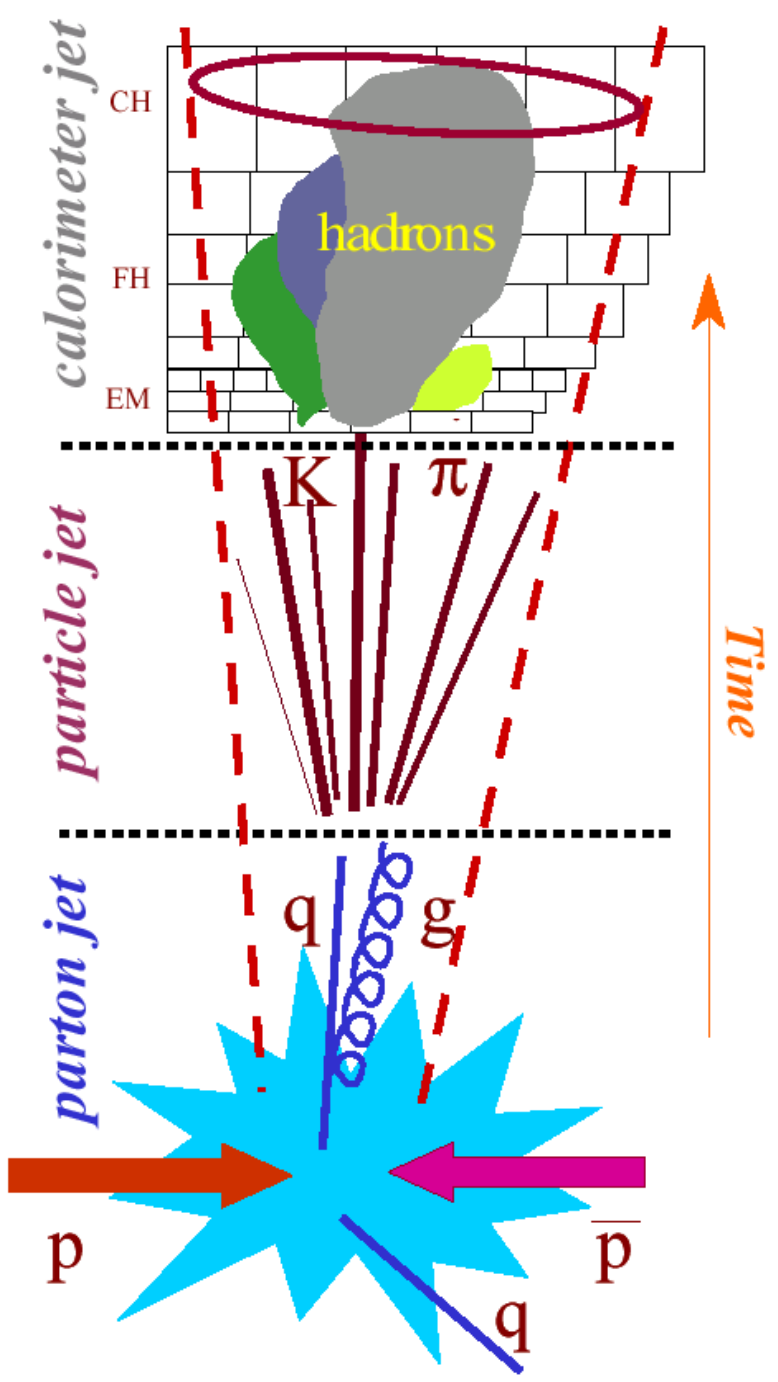}
  \end{center}
  \vspace{-15pt}
  \caption{Illustration of the various jet reconstruction levels from partons
           over hadrons to the calorimeter. }
  \label{fig:jescalibrationsketch}
  \vspace{-6pt}
\end{wrapfigure}
Several and conceptually quite different calibration approaches are considered
by the experiments. Monte Carlo based jet calibrations, transforming the 
electromagnetic energy scale to the hadronic scale, can be distinguished 
according to the level of detail with which the jet constituents are treated 
and separately corrected. The `global jet calibration' uses as input
clusters that have been properly calibrated at the electromagnetic scale, and which 
are matched in energy to the Monte Carlo truth particle jet for bins of $E_T$
and $\eta$. This calibration returns the jet energy at the hadronic scale
(\cf sketch in Fig.~\ref{fig:jescalibrationsketch}). On the contrary, the `local hadron 
calibration' calibrates clusters independently of the jet algorithm by making 
an assumption on their electromagnetic or non-electromagnetic nature. 
Jets are then formed out of calibrated clusters, and the jet energy is 
given at the hadronic scale. Finally, {\em in-situ} calibration methods are used
to match the hadron to the parton levels of the jet using known physics processes. 

\begin{figure}[t]
  \begin{center}
	  \includegraphics[width=1\textwidth]{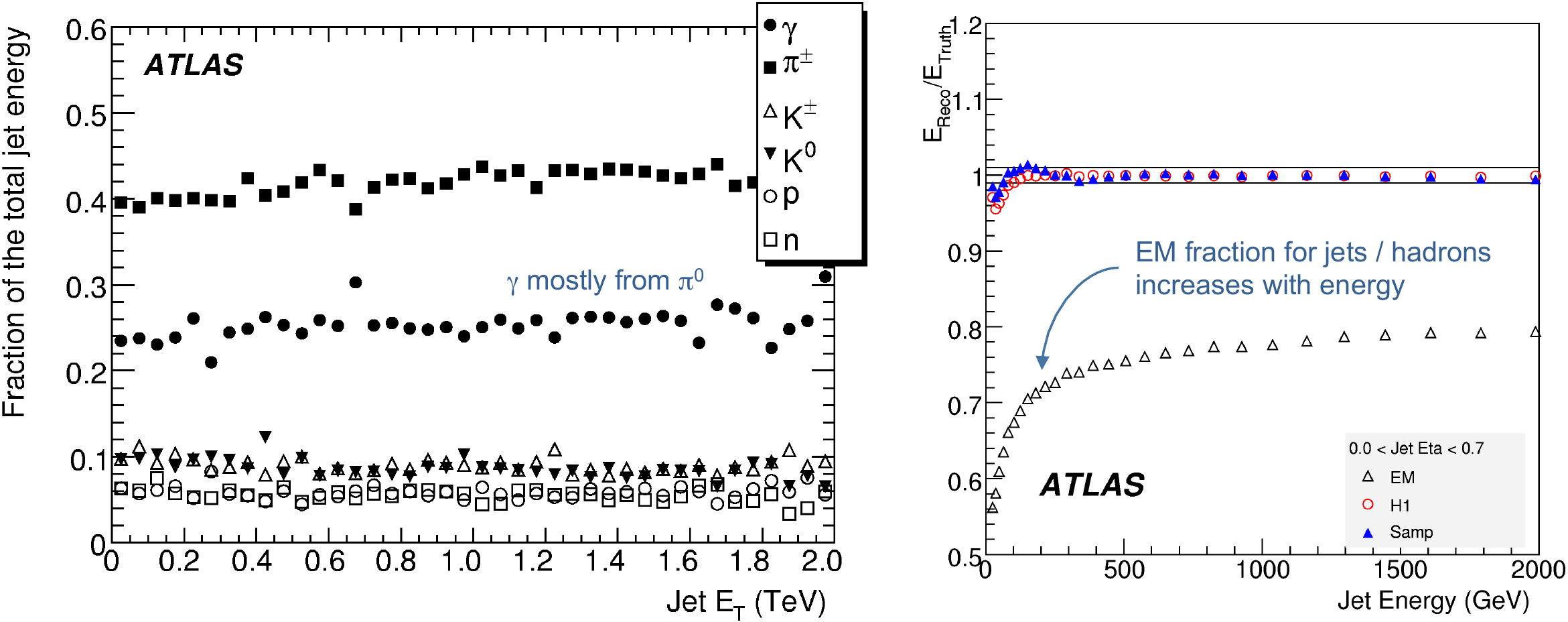}
  \end{center}
  \vspace{-0.3cm}
  \caption[.]{\underline{Left}: fractional energy carried by different particle 
              types as a function of the jet energy (ATLAS simulation). 
              \underline{Right}: jet energy linearity as a function of jet energy
               (ATLAS simulation). 
              Shown are jets reconstructed at the electromagnetic (EM) scale (open triangles), 
              and using global jet calibration algorithms (open circles and full triangles). 
              The jets have a large cone radius of 0.7.}
\label{fig:jetenergyplotsatlas}
\end{figure}
A large amount of contributions to the jet signal at the various jet levels must 
be considered in the calibration process. The parton level is governed by the 
physics process of interest. At the hadron level (particle jet), one must take into 
account the jet reconstruction algorithm efficiency, added tracks from in-time 
event pileup from minimum bias scattering interactions, 
added tracks from the underlying event, and lost soft tracks due to 
the magnetic field. At the calorimeter jet level one must account for longitudinal 
energy leakage, detector signal inefficiencies (\eg, dead channels, dead HV boards)
background from pileup events, electronic noise, the definition of the calorimeter 
signal (cluster algorithm, noise suppression, etc.), dead material losses (front
material, geometrical cracks in the active material, transition regions, etc.),
the detector response characteristics ($e/h\ne 1$), and the jet reconstruction 
algorithm efficiency. The left panel of Fig.~\ref{fig:jetenergyplotsatlas} 
shows the fractional 
energy that is carried by different particle types in a jet as a function of the 
jet energy. The largest contributors are charged pions, followed by photons 
originating mostly from \piz decays, so that the total pion component amounts to
roughly 70\% of the jet energy, with no significant jet energy dependence. 
The right plot shows the jet energy linearity and the electromagnetic fraction 
versus the jet energy. The 
electromagnetic fraction for jets or hadrons increases with the jet energy, 
asymptotically reaching 80\% for very hard jets. After calibration, the 
energy response is accurate above 300\,\GeV, whereas softer jets are
more difficult to calibrate due to the stronger impact of calorimeter noise
fluctuations and other effects.

\begin{wrapfigure}{R}{0.34\textwidth}
  \vspace{-24pt}
  \begin{center}
	  \includegraphics[width=0.34\textwidth]{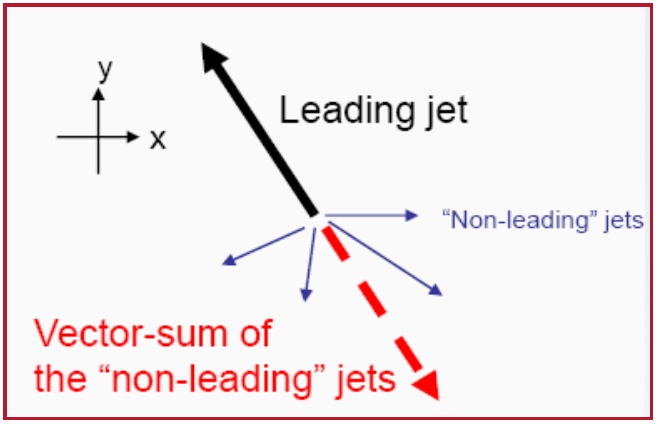}
  \end{center}
  \vspace{-15pt}
  \caption{Illustration of multi-jet energy calibration.}
  \label{fig:multijetcalibrationsketch}
  \vspace{-6pt}
\end{wrapfigure}
\begin{figure}[t]
  \begin{center}
	  \includegraphics[width=0.6\textwidth]{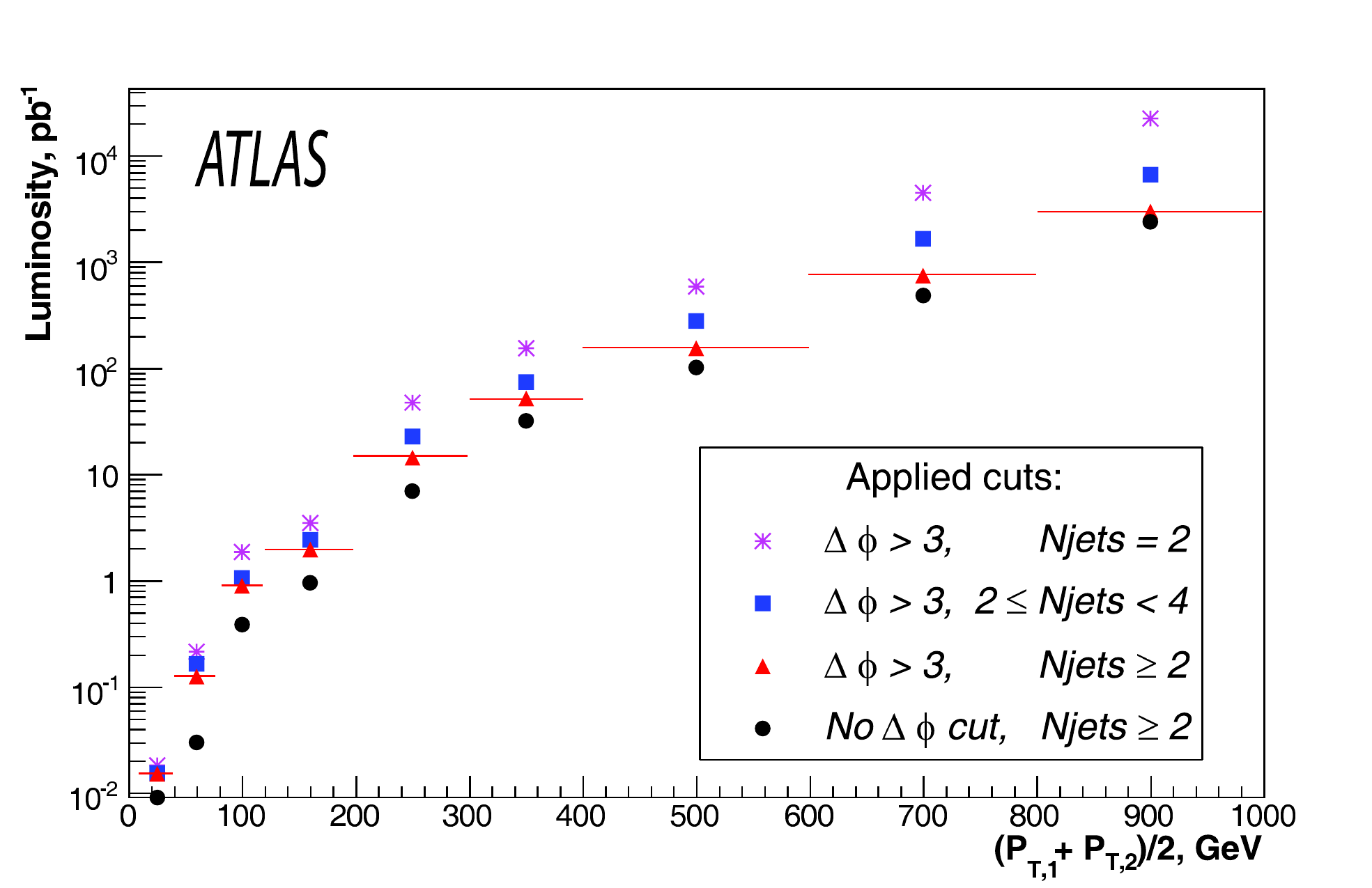}
  \end{center}
  \vspace{-0.3cm}
  \caption[.]{Integrated luminosity required to reach 0.5\% precision on the jet 
              energy scale with the multi-jet calibration method for various 
              $p_T$ ranges in the region $0.7 < \eta < 0.8$, and with different sets of 
              selection cuts (ATLAS simulation): all Pythia di-jet events (circles), 
              requiring $\Delta\phi > 3$\,rad between the two leading jets (triangles), 
              requiring in addition fewer than four reconstructed jets in an event (squares), 
              and requiring exactly two reconstructed jets (stars). }
\label{fig:multijetcalibrationresultsatlas}
\end{figure}
The ultimate goal of the jet reconstruction is to match the calibrated hadronic scale
to the initial parton momentum with the use of physics events, \ie, to perform {\em in-situ}
calibration. Several approaches exist. 
\begin{itemize}
\item Directly verify the hadronic energy scale with {\bf isolated prompt hadrons} from minimum 
      bias events, or hadrons from $\tau$ decays, by comparing the reconstructed 
      hadron energy with the momentum of the hadron track measured in the inner 
      tracker. Another possibility is to use track balancing in $\phi$ (energy 
      and momentum conservation of proton--proton collisions requires the event 
      to be transversely balanced, but not longitudinally) to intercalibrate the 
      hadronic scale with respect to different hadron energies. 
\end{itemize}
\vspace{-0.4cm}

\begin{itemize}
\item Use {\bf transversely balanced $\mathitbf{\gamma}$-jet or 
      $\mathitbf{Z}$($\mathitbf{\to\ell\ell}$)-jet events}. This 
      method assumes that electromagnetic objects have been properly calibrated
      beforehand. The jet energy calibration is performed with respect to the 
      average transverse momentum of photon (or $Z$) and jet. 
      Owing to the large cross section of 180\,nb for $\gamma$-jet 
      processes\footnote
      {
         The leading parton level processes contributing to the $\gamma$-jet 
         cross section are  $t$-channel quark--quark scattering 
         via fermion exchange into $g+\gamma$, and quark--gluon scattering via
         fermion exchange into $q+\gamma$, gluon--gluon scattering via a box 
         diagram into $g+\gamma$, and the $s$-channel quark--gluon-to-quark
         annihilation into $q+\gamma$.
      } 
      this method can be applied with initial data. The statistical yield 
      corresponding to an integrated luminosity of 
      10\invpb would allow a jet calibration of better than 1\% statistical 
      precision for $p_T<200$\,\GeV. However, the determination of systematic uncertainties
      is tricky, and requires careful studies. For example, initial- and final-state 
      radiation, underlying events and in-time event pileup, but also out-of-jet 
      particles have the potential to contribute to the $\gamma$-jet imbalance, and 
      these effects must be disentangled from miscalibration. Monte Carlo studies by
      ATLAS have shown that systematic imbalances of non-calibration origin
      contribute at the 10\% level for 20\,\GeV jets, whereas the effect is below 
      1\% for jets above 100\,\GeV.
\end{itemize}
\vspace{-0.4cm}

\begin{itemize}
\item Use {\bf QCD di-jet and multi-jet events} for $\Delta\eta$$\times$$\Delta\phi$ 
      intercalibration. Di-jet events cannot constrain the absolute jet energy scale, 
      but allow one to intercalibrate the calorimeter response. In case of more than two 
      jets in the event, the leading jet dominates the energy resolution of the event, 
      so that one may assume that the error in the vector sum of the `soft' jets is 
      negligible with respect to the hard jet, and hence `calibrate' $E_T$
      versus $\eta$ and $\phi$ (\cf sketch in Fig.~\ref{fig:multijetcalibrationsketch}). 
      This method benefits from huge statistics 
      (the di-jet cross section exceeds by a factor 100 to 5000 the $\gamma$-jet 
      cross section), but sizable systematic effects arise from soft jets, in 
      particular for the multi-jet approach, requiring detailed studies.
      Figure~\ref{fig:multijetcalibrationresultsatlas} shows the integrated luminosity 
      required to reach a precision in the jet energy scale of 0.5\% with the 
      multi-jet calibration method for various $p_T$ ranges in the region 
      $0.7 < \eta < 0.8$ and for different sets of selection cuts (see figure 
      caption). Requiring the jets to be back-to-back (\ie, applying a tight 
      $\Delta\phi$ cut) reduces systematic effects from initial- and final-state
      radiation and the underlying event. 
      The figure has been obtained by ATLAS with the use of simulated events.
\end{itemize}
\vspace{-0.4cm}

\begin{itemize}
\item Absolute jet energy scale calibration is possible by means of 
      {\bf $\mathitbf{W}$ decays into a pair of jets}, for clean $W$ from top decays. However, this 
      calibration applies to soft jets only (jet energies below 200\,\GeV). In 
      addition, the $W$ boson does not carry colour charge, which makes it differ
      from QCD jets. 
\end{itemize}

The LHC will explore energies that have never been reached before. Above 500 GeV, 
neither measurements nor test beam results are available for jet calibration. 
Multi-jet balancing should allow a few per cent jet energy scale accuracy in that 
range with 1\invfb integrated luminosity.

\subsection{Missing transverse energy reconstruction}

\begin{wrapfigure}{R}{0.45\textwidth}
  \vspace{-24pt}
  \begin{center}
	  \includegraphics[width=0.45\textwidth]{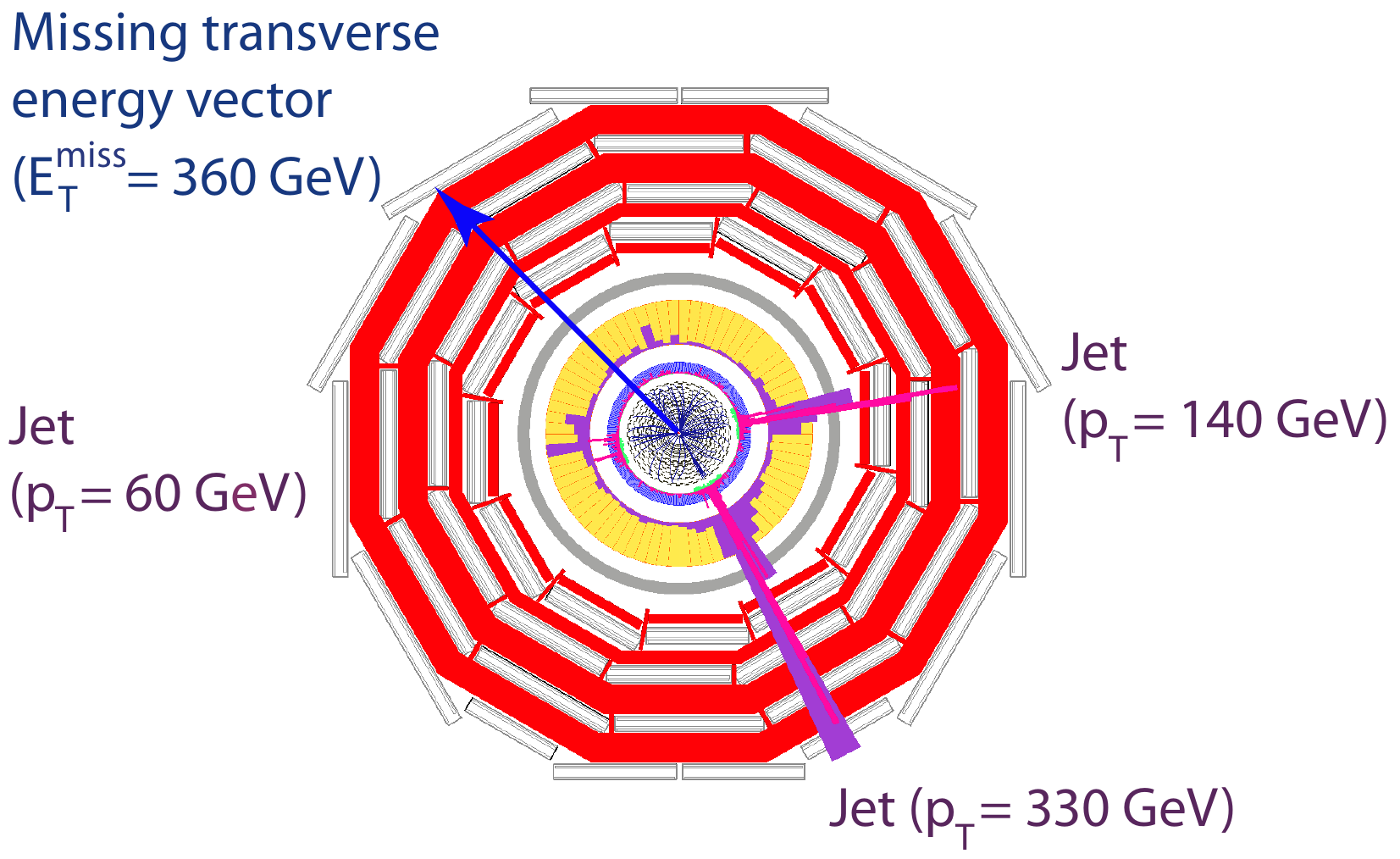}
  \end{center}
  \vspace{-15pt}
  \caption{Display of a simulated SUSY event in CMS. The arrow indicates the 
           missing transverse energy vector.}
  \label{fig:susyeventdisplaycms}
  \vspace{-6pt}
\end{wrapfigure}
A precise reconstruction of missing transverse energy (MET) in terms of energy scale, 
linearity, and resolution is essential for the ATLAS and CMS physics programme.
Large MET is predicted in many new physics scenarios, notably in supersymmetric
extensions of the Standard Model respecting $R$-parity, where a stable weakly 
interacting neutral particle is produced that --- just as neutrinos --- escapes
the detector without measurable interaction with the active material.
Figure~\ref{fig:susyeventdisplaycms} shows a simulated SUSY candidate 
event in CMS that exhibits significant MET of 360\,\GeV.
MET is also an ingredient of precision Standard Model measurements, such as 
semileptonic top reconstruction and the $W$ mass, and also of the search for 
$H\to \tau\tau$ decays, the cross section of which may or may not be 
enhanced by beyond Standard Model contributions. The MET measurement is particularly 
sensitive to systematic effects in the detector response and the reconstruction. 
Understanding MET in early data is therefore one of the primary physics commissioning 
challenges. 

The conceptually simplest way to reconstruct MET is to compute the transverse 
vector sum of all the electromagnetic and hadronic calorimeter cells and to correct 
for unaccounted contributions. In the case of ATLAS, one has
\beqn
   \Emiss_T    &=& \sqrt{\Emiss_x^2 + \Emiss_y^2}\,,\\
\label{eq:etmiss}
   \Emiss_{x,y} &=& \Emiss_{x,y}^{\rm Calo} + \Emiss_{x,y}^{\rm Cryo} + \Emiss_{x,y}^{\rm Muon}\,,
\eeqn
where the symbol \Emiss denotes missing energy. The calorimeter term 
\beqn
   \Emiss_{x,y}^{\rm Calo} &=& -\!\!\!\!\!\!\!\!\sum_{\rm EM~\&~Had~cells} \!\!\!\!\!\!\!\! E_{x,y}\,,
\eeqn
is calibrated at the hadronic energy scale. The electromagnetic scale would 
underestimate MET by roughly 30\% because the largest contributions to it 
stem from hadrons and jets.

The `cryostat' term in Eq.~(\ref{eq:etmiss})
corrects for energy loss (leakage) in the cryostats between 
the electromagnetic and hadronic calorimeters and becomes important for jets
with large transverse momentum (representing a 5\% contribution per jet with 
$p_T> 500$\,\GeV). It is given by
\beqn
   \Emiss_{x,y}^{\rm Cryo} &=& -\sum_{\rm Jets} w^{\rm Cryo}\cdot E_{x,y}^{\mbox{\footnotesize Jet-at-cryo}}\,,
\eeqn
where $w^{\rm Cryo}$ is a calibration weight determined empirically from Monte 
Carlo simulation, and $E_{x,y}^{\mbox{\footnotesize Jet-at-cryo}}$ is the average of the jet energies
deposited in the third layer of the electromagnetic calorimeter and in the 
first layer of the hadronic calorimeter.

Finally, the muon term sums over measured muon momenta within the muon spectrometer 
acceptance ($|\eta|<2.7$) 
\beqn
   \Emiss_{x,y}^{\rm Muon} &=& -\!\!\sum_{\rm Muons} \!\!E_{x,y}\,.
\eeqn

The MET reconstruction can be refined by associating reconstructed electrons, 
photons, muons, hadronically decaying $\tau$ leptons, $b$-jets and light jets to 
calorimeter cells, and replacing for these cells the global calibration by one 
that takes into account the nature of the identified objects.

\begin{wrapfigure}{R}{0.42\textwidth}
  \vspace{-24pt}
  \begin{center}
	  \includegraphics[width=0.42\textwidth]{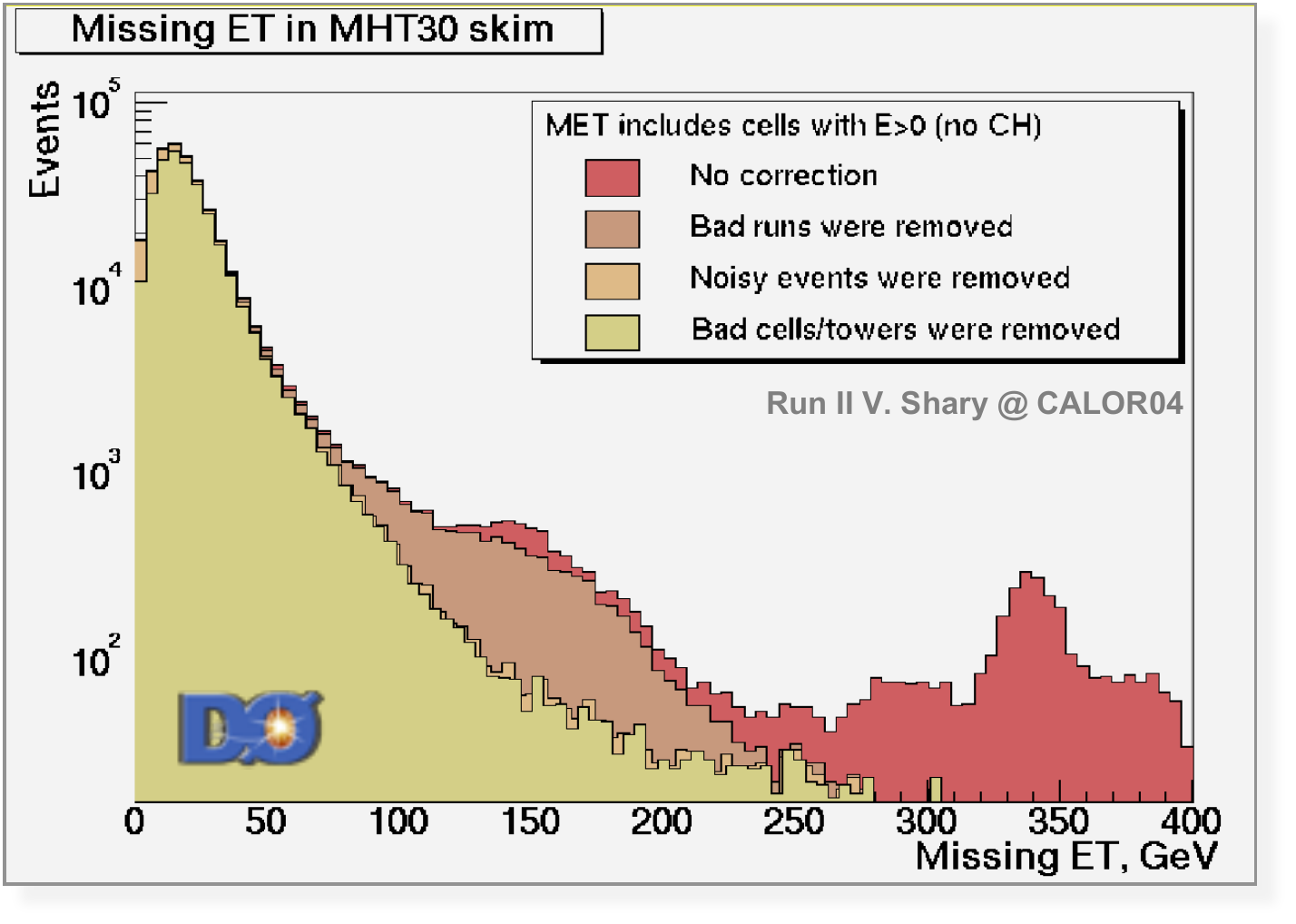}
  \end{center}
  \vspace{-15pt}
  \caption{Missing transverse energy distribution measured by the D0 experiment 
           at the Tevatron $\ppbar$ collider. Shown are the various 
           correction stages leading to the removal of fake MET tails that 
           could be misinterpreted as new physics.}
  \label{fig:metd0}
  \vspace{-6pt}
\end{wrapfigure}
\begin{figure}[t]
  \begin{center}
	  \includegraphics[width=0.8\textwidth]{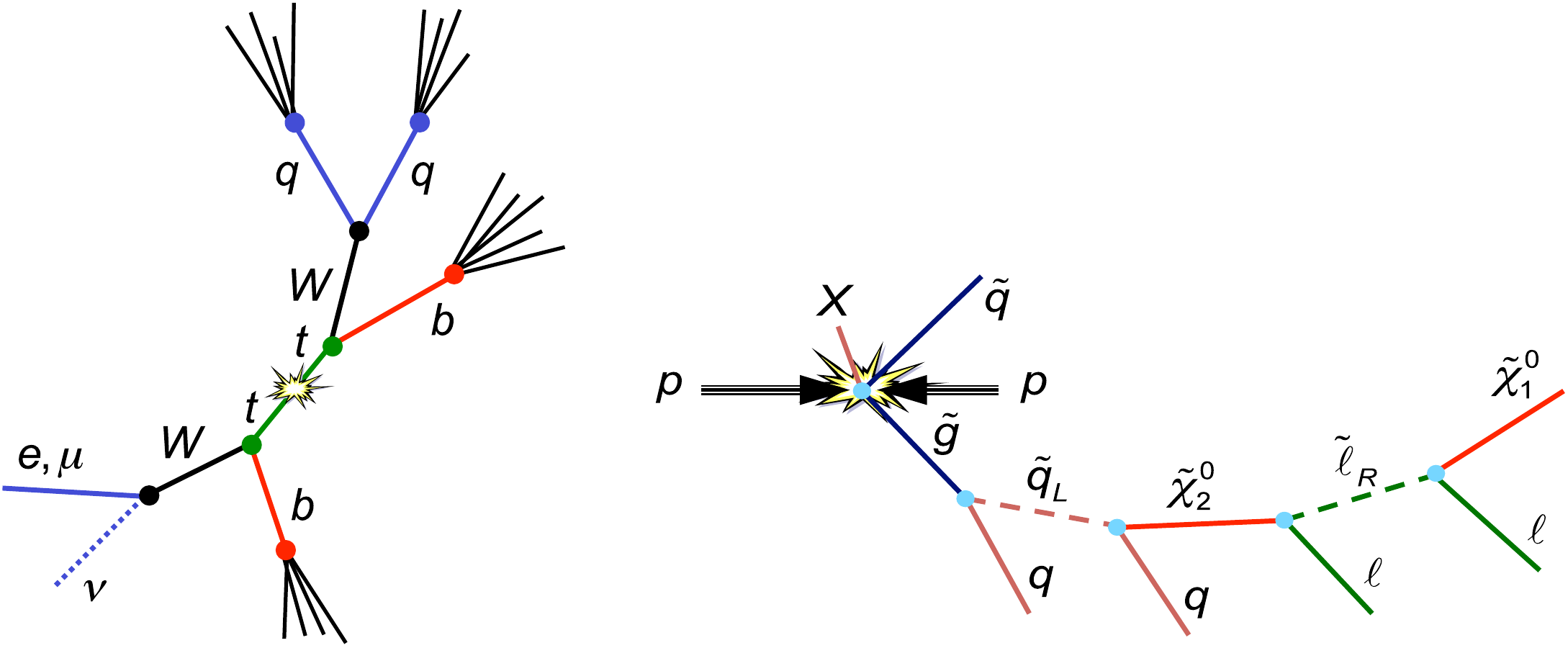}
  \end{center}
  \vspace{-0.3cm}
  \caption[.]{Schematic graphs of processes generating true MET. The left 
              graph corresponds to a $pp\to t\tbar (+ X)$ Standard Model 
              event (only the top part of the event is shown), where one 
              top-quark decays fully hadronically and the other semileptonically.
              The neutrino generates MET. The right graph depicts a typical
              decay cascade as obtained in $R$-parity conserving 
              supersymmetry. An initial gluino decays into a left-handed
              squark and a quark (giving a jet), the squark decays into 
              a heavy neutralino and a quark (giving another jet), the 
              heavy neutralino further decays into a slepton and a lepton,
              and the slepton finally decays into the lightest neutralino,
              which escapes detection, and a second lepton of opposite 
              charge with respect to the previous lepton. Note that the initial 
              supersymmetric particles are created in pairs, but only one 
              decay cascade is shown here.}
\label{fig:truemetgraphs}
\end{figure}
\begin{figure}[p]
  \begin{center}
	  \includegraphics[width=0.805\textwidth]{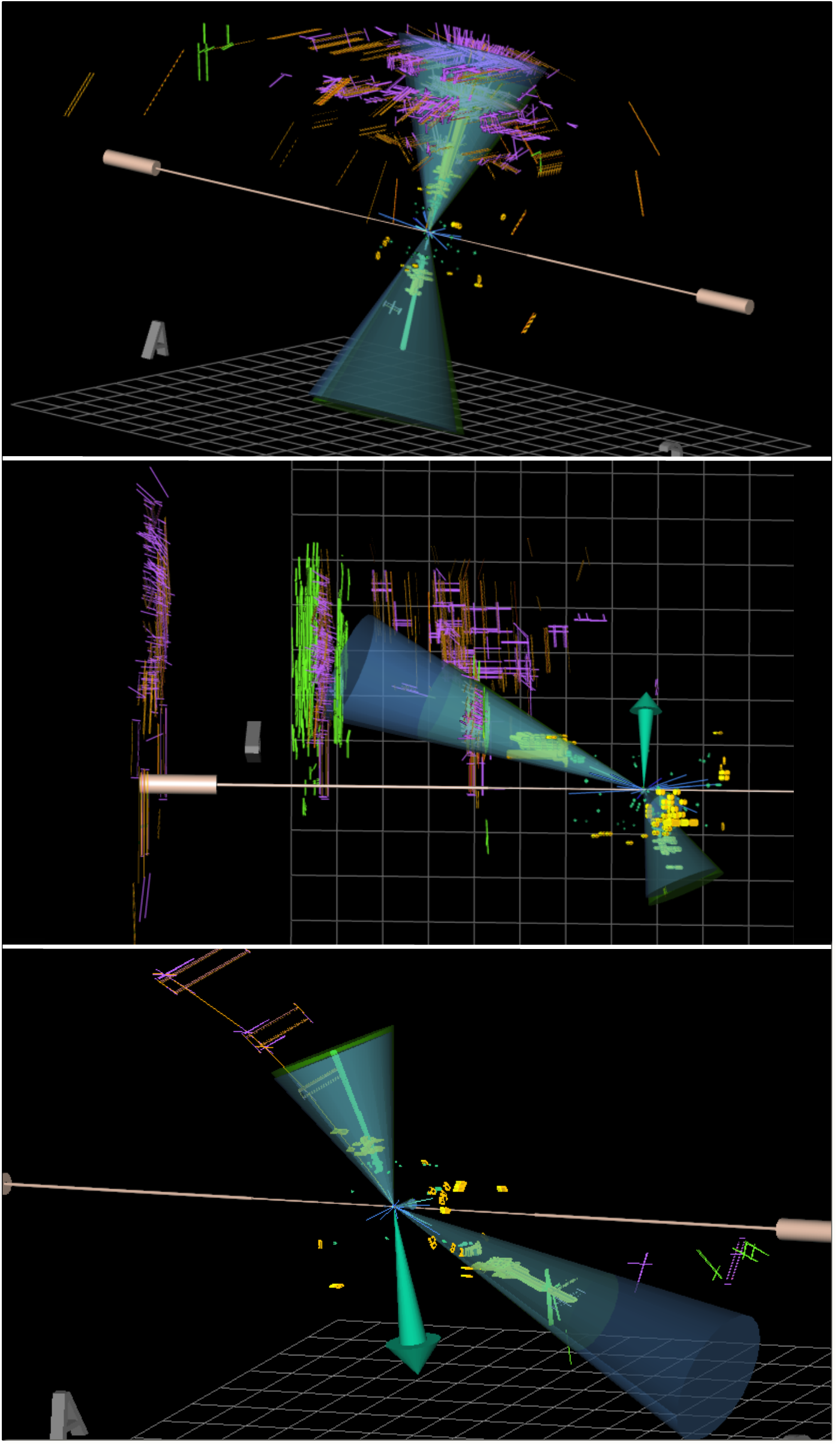}
  \end{center}
  \vspace{-0.3cm}
  \caption[.]{ATLAS event displays of simulated rare di-jet events creating large 
              amounts of fake MET (represented by the round arrow). 
              The upper two displays show hadron 
              punch-through from the calorimeter into the muon spectrometer. The 
              lower display shows large MET found in an event populating 
              the jet energy resolution tails. }
\label{fig:metdijeteventdisplayatlas}
\end{figure}
It is apparent from the above equations that all detector systems contribute to the 
MET measurement, which makes it vulnerable to hardware, reconstruction, and calibration 
problems. One distinguishes between `true' and `fake' MET. For example, 
weakly interacting neutral particles generate true MET (\cf Fig.~\ref{fig:truemetgraphs}). 
Even without systematic
effects, MET is created by the detector response resolution, giving rise to 
fake MET. Fake MET can also be introduced by detector problems or misreconstruction,
such as dead and noisy channels, particles falling out of the detector 
acceptance (\eg, muons for $|\eta| > 2.7$), unaccounted pile-up contributions to 
resolution effects, backgrounds from beams or cosmic rays, `punch-through' of 
hadron showers into the muon system faking a muon signal, and many more effects. 

The suppression and --- if not possible --- proper simulation 
of fake MET is crucial to increase the sensitivity to 
the true MET. This requires the best possible jet energy resolution and absolute 
scale, and a thorough classification through data quality bookkeeping and the
simulation of varying detector problems. Figure~\ref{fig:metd0} shows an extreme case of 
MET distortions due to detector noise and bad channel effects, provided for illustration 
purposes by the D0 experiment. The effects lead to large tails that --- if not 
properly corrected or simulated --- could be misinterpreted as a new physics
signal. Figure~\ref{fig:truemetgraphs} depicts schematically two processes
that generate true MET. The left one is a Standard Model $t\tbar$ event where 
one of the tops decays semileptonically, and the right one is a supersymmetric 
event with its typical decay cascades ending with two invisible lightest stable 
supersymmetric particles (only one of two decay cascades is shown). 
Figure~\ref{fig:metdijeteventdisplayatlas} shows event displays from 
simulated events in ATLAS that were selected for featuring pathologically 
large fake MET due to hadron punch-through (upper two displays) and 
jet mismeasurement (lower display). Both types of fake MET usually point towards
a jet, which allows such backgrounds to be reduced by eliminating events where the 
MET vector lies on a jet axis.
\begin{figure}[t]
  \begin{center}
	  \includegraphics[width=1\textwidth]{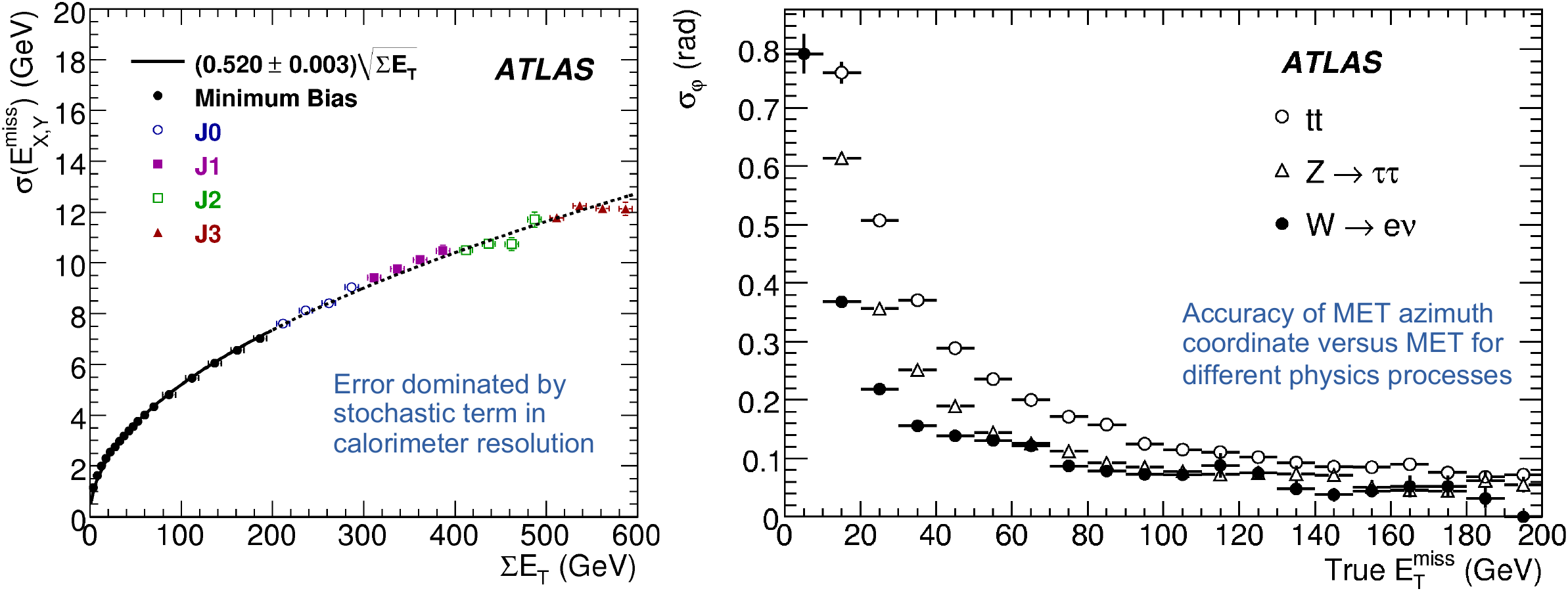}
  \end{center}
  \vspace{-0.4cm}
  \caption[.]{\underline{Left}: expected MET resolution for ATLAS versus the transverse
              energy sum for minimum bias events and various jet samples.
              \underline{Right}: accuracy of the MET azimuth angle versus 
              MET for $t\tbar$, $Z\to\tau\tau$ and $W\to e\nu$ events, obtained
              from Monte Carlo simulation in ATLAS. }
\label{fig:metreconstructionminbiastauatlas}
\end{figure}

Fake MET tails can be studied with early data using minimum bias events.
Ample statistics will be available thanks to the large minimum bias cross section,\footnote
{
  At 14\,\TeV centre-of-mass energy, a minimum bias event rate of 70\,kHz is expected
  to be produced at \Lumi30 peak luminosity, and owing to the logarithmic 
  $\sqrt{s}$ dependence similar orders of magnitude are expected at lower centre-of-mass
  energy. For example, at 900\,\GeV the minimum bias cross section is reduced by 
  a factor of only 1.8 with respect to 14\,\TeV.
}
allowing the experiments to select clean data samples. Because the expected true MET is negligible
($\sim$0.06\,\GeV) the measured MET will be dominated by fake effects from single hadron 
and jet energy resolution (82\%), and acceptance (18\%). ATLAS expects an MET average 
value of 4.3\,\GeV. The left panel in Fig.~\ref{fig:metreconstructionminbiastauatlas} 
shows the expected MET resolution for ATLAS versus the measured transverse energy sum
for minimum bias events and various jet samples providing increasing transverse 
energies. The resolution is dominated by the stochastic term in the jet energy
resolution, giving a square-root dependence on the transverse energy 
sum with the expected coefficient of approximately 50\% (see Section~\ref{sec:testbeams}).

True MET in early data can be measured in leptonic $W$ decays, which have good statistics,
but also in $Z\to\tau\tau$ events. In case of one $\tau$ decaying semileptonically 
(hadron(s) plus neutrino) and the other leptonically (electron or muon plus two 
neutrinos), one can reconstruct the $\tau$ mass by assuming that the $\tau$
decay products were emitted collinear with the $\tau$ flight direction in the lab
frame. This is a useful conjecture since the $\tau$ exhibits a strong boost. With this
one finds
\beq
   m_{\tau\tau}^2 \approx 2\cdot\left(E_h + E_{\nu(h)}\right)\left(E_\ell + E_{\nu(\ell)}\right)
                       \left(1-\cos\!\theta_{h\ell}\right)\,,
\eeq
where the neutrino energies are approximated by MET. Simulated $Z\to\tau\tau$ decays
in ATLAS showed that this method allows the $Z$ mass to be reconstructed with an average 
resolution of 12\,\GeV. The right panel of Fig.~\ref{fig:metreconstructionminbiastauatlas}
gives the expected accuracy of the MET azimuth angle versus the true MET for 
$t\tbar$, $Z\to\tau\tau$ and $W\to e\nu$ events simulated by ATLAS. With larger
true MET the signal-to-calorimeter-noise ratio increases and hence the quality of the 
MET reconstruction. Moreover, the more hadronic activity in the detector, the
worse the MET reconstruction.

\newpage
\section{Early physics with ATLAS and CMS}

Early physics measurements will be performed while the detectors are still being 
commissioned. Some of the commissioning tasks will
thus have to take priority to allow systematic uncertainties to be 
evaluated. An example is the determination of the absolute tracking efficiency,
which is an important ingredient of first QCD measurements such as the average 
number of produced tracks per pseudorapidity region, and which depends on 
basic detector properties such as the hit efficiency, the alignment of the 
inner tracking systems, and low-transverse-momentum track finding. 
Likewise, any physics measurement requires the determination
of at least the relative trigger efficiency, and in case of cross section 
measurements also the absolute trigger efficiency as well as the integrated 
luminosity. The latter quantity requires either an absolute
luminosity detector or, more importantly at the beginning of data taking, an 
LHC beam scan (`Van-der-Meer scan'\footnote
{
   The beam scan is used to measure the beam sizes and positions in a collider, 
   which, together with the known currents, can be used to compute the absolute 
   luminosity. The beams are scanned across each other at the collision point 
   and, using beam position measurements, the amount of motion is correlated 
   with detectors monitoring the relative luminosity of the collisions at 
   each scan point. This method has been successfully
   applied at the heavy-ion collider RHIC~\cite{rhicbeamscan}. 
}~\cite{vandermeerscan}).

The following paragraphs present a very brief and incomplete overview of initial
measurements that will be performed at ATLAS and CMS after the
collection of approximately 100\invpb integrated luminosity. Most of the 
prospective studies shown here are taken from ATLAS~\cite{atlascscbook}.
The CMS studies, documented in Ref.~\cite{cmsphysicstdr}, are very similar. 
All results shown are based on Monte Carlo simulation at 14\,\TeV centre-of-mass
energy. This is, however, not the energy at which the LHC will start. Because of problems 
with the magnet quench protection, the startup centre-of-mass energy in 2010 
will be 7\,\TeV, after a pilot run at LHC injection energy of 0.9\,\TeV in 
2009. The decision whether or not to raise the energy to 10\,\TeV in the course 
of the year 2010 will depend on the running experience. The design energy of 14\,\TeV
can only be reached after a shutdown of approximately one year, which may be 
scheduled in 2011 or 2012, when vulnerable parts of the quench protection system 
are exchanged. 

\subsection{Minimum bias studies}

Minimum bias events will dominate the first triggered data samples of all LHC
experiments. The total minimum bias cross section receives contributions from 
inelastic non-diffractive and diffractive collisions,\footnote
{
   Diffraction denotes the excitation of the proton(s) participating in the 
   inelastic scattering. One distinguishes single, double and central diffraction. 
   While single and double diffractive events produce activity in only one and both
   forward regions of the detector, respectively, central diffractive events, 
   (which are described by double pomeron exchange, and have small cross sections),
   give activity at small pseudorapidities. 
} 
where whether or not
single diffractive events are included is subject to the experiment's definition.
Experimentally, it is not possible to distinguish these classes of events on 
an event-by-event basis. 
Minimum bias triggers have usually large (medium, small) efficiencies for 
non-diffractive (double diffractive, single diffractive) events. If coincident
hits in both forward regions of the detector are required, the efficiency 
of single diffractive events becomes small. In-time coincidence is 
a useful requirement to eliminate beam related backgrounds (beam gas and 
beam halo events, \cf Section~\ref{sec:beambackgroundevents}). These 
backgrounds are, however, also eliminated when requiring the reconstructed 
tracks in the event to form a primary vertex. The minimum bias analysis
will most likely be the first paper published by ALICE, ATLAS and CMS.
Apart from the physics measurement, it will represent a first proof that 
the detectors (mainly the inner tracking systems) work and the data 
including the trigger and tracking efficiencies are understood.

\begin{figure}[t]
  \begin{center}
	  \includegraphics[width=1\textwidth]{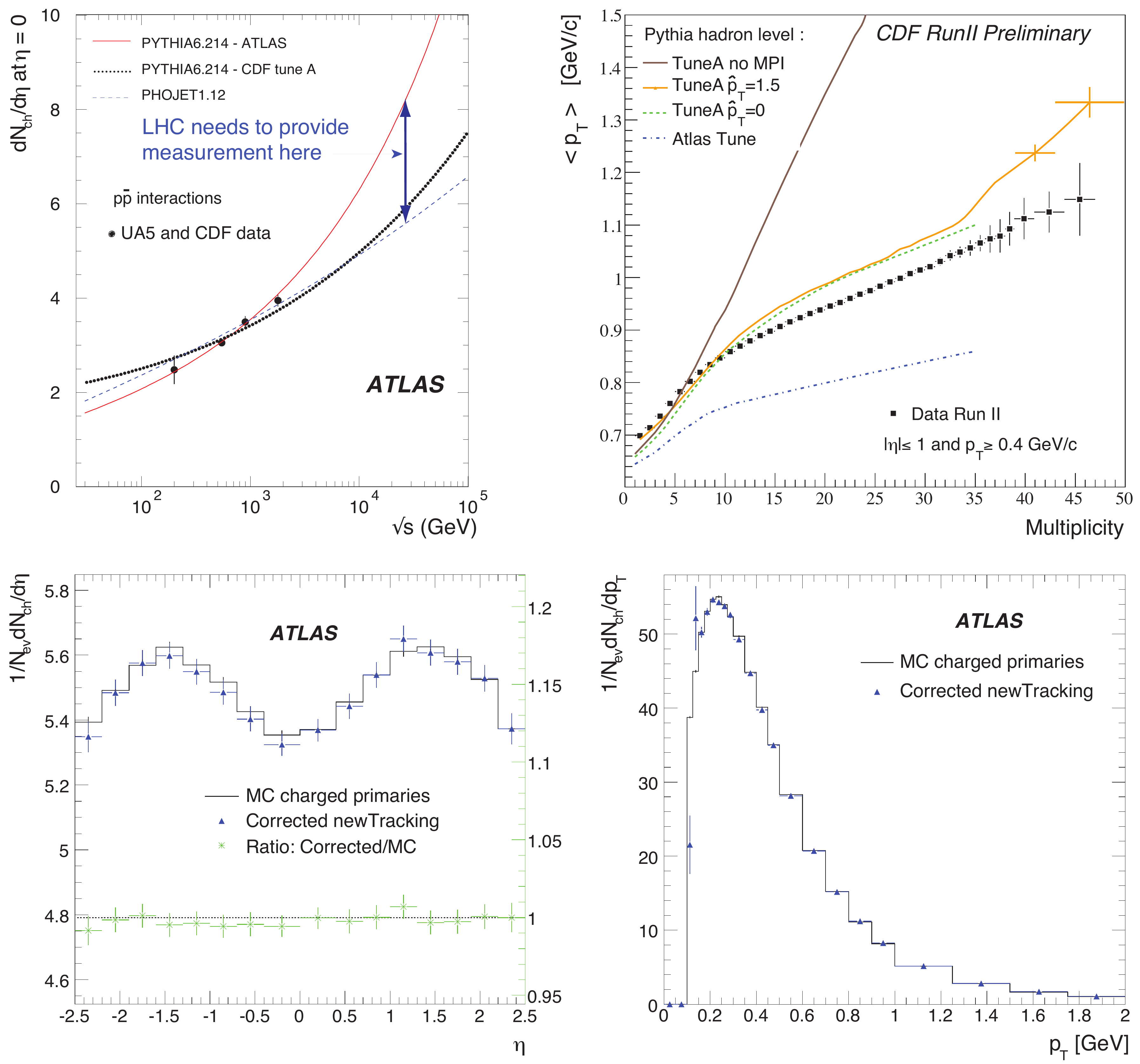}
  \end{center}
  \vspace{-0.5cm}
  \caption[.]{\underline{Top left}: central charged particle density for non-single 
              diffractive inelastic events in $\ppbar$ collisions 
              as a function of energy, extrapolated
              to large centre-of-mass energies. Shown are available measurements
              and Monte Carlo generator predictions. 
              \underline{Top right}: correlation between average track transverse 
              momentum and the charged particle multiplicity 
              for $\eta<1$ as measured by CDF~\cite{cdfminbias}, 
              and compared with various Pythia generator tunings. `No MPI' means
              that multiple parton interactions have been switched off in the generator. 
              \underline{Bottom plots}: particle density in non-diffractive minimum bias 
              events versus the pseudorapidity (left) and $p_T$ (right) in ATLAS
              with special low-momentum track reconstruction enabled. Systematic 
              errors on track reconstruction are not included in the right plot. }
\label{fig:physicsminbiasmctune}
\end{figure}
Multiparticle production is successfully described by phenomenological models 
with pomeron exchange, which dominates at high energies. These models relate 
the energy dependence of the total cross section to that of the multiplicity 
production using a small number of parameters, and are the basis for several 
Monte Carlo event generators describing soft hadron collisions. Minimum 
bias multiplicity measurements between 200\,\GeV and 
2\,\TeV centre-of-mass energies at the CERN ISR, CERN S${\rm p\overline p}$S, 
Fermilab's Tevatron, and BNL's RHIC colliders have been used to tune 
these generators for predictions of multiplicities at  LHC energies. 

The top left panel in Fig.~\ref{fig:physicsminbiasmctune} shows a comparison of 
model predictions for the central charged particle density in non-single-diffractive 
$\ppbar$ events for a wide range of centre-of-mass energies compared with measurements
that have been corrected for detector acceptance. Large extrapolation uncertainties 
exist that must be overcome by LHC measurements. Improved generator tunings at LHC
energies will directly feed into Monte Carlo predictions of many primary 
physics channels. A good minimum bias multiplicity description is 
also important because event pileup from minimum bias interactions is 
background to hard scattering processes at high luminosity. The top right panel 
in Fig.~\ref{fig:physicsminbiasmctune} is taken from CDF~\cite{cdfminbias}.
It shows the measured dependence of the average track transverse momentum 
on the charged particle multiplicity per event for $|\eta|<1$, compared with
various Pythia generator tunings. Without multiple parton interactions 
the average predicted $p_T$ multiplicities above 6 is grossly overestimated.
The bottom plots in Fig.~\ref{fig:physicsminbiasmctune} show the particle 
density versus pseudorapidity (left) and transverse momentum (right) in ATLAS for 
simulated minimum bias events. Special low-$p_T$ tracking reconstruction has been 
enabled for these plots, which allows one to lower the track measurement down 
to $p_T=150$\,\MeV (standard cut is 500\,\MeV), at the price of larger
systematic uncertainties (not included in the error bars). The statistics shown 
corresponds to 1 minute of data taking with \Lumi31 at 14\,\TeV.

\subsection{Di-jet studies}

Jet production has a roughly 1000 times lower cross section than non-diffractive
minimum bias scattering, but is still an abundant process for early physics 
measurements and performance studies. Apart from its importance for QCD studies
and Monte Carlo generator tuning at yet unexplored centre-of-mass energies, 
jet production can be used to probe the Standard Model. Inclusive di-jet 
production ($pp \to 2\;$jets$+X$) is the dominant LHC hard scattering process.
It is straightforward to observe and has a rich potential of new physics 
signatures. Restricting the leading jet (the jet with the largest $p_T$)
to the central detector region $|\eta|<1$
reduces the background from QCD $t$-channel processes, thus enhancing 
the sensitivity to new physics contributions to the $s$-channel at small
pseudorapidities. The main variables used for new physics searches are the 
transverse momentum of the leading jet and the di-jet invariant mass. 
Prospective studies from CMS show that the highest di-jet masses reached
with integrated luminosities of 100\invpb, 1\invfb, and 10\invfb are 
respectively 5, 6 and 7\,\TeV. The current limits from measurements by the 
Tevatron experiments will be almost immediately extended by ATLAS and CMS. 

\begin{figure}[t]
  \begin{center}
	  \includegraphics[width=1\textwidth]{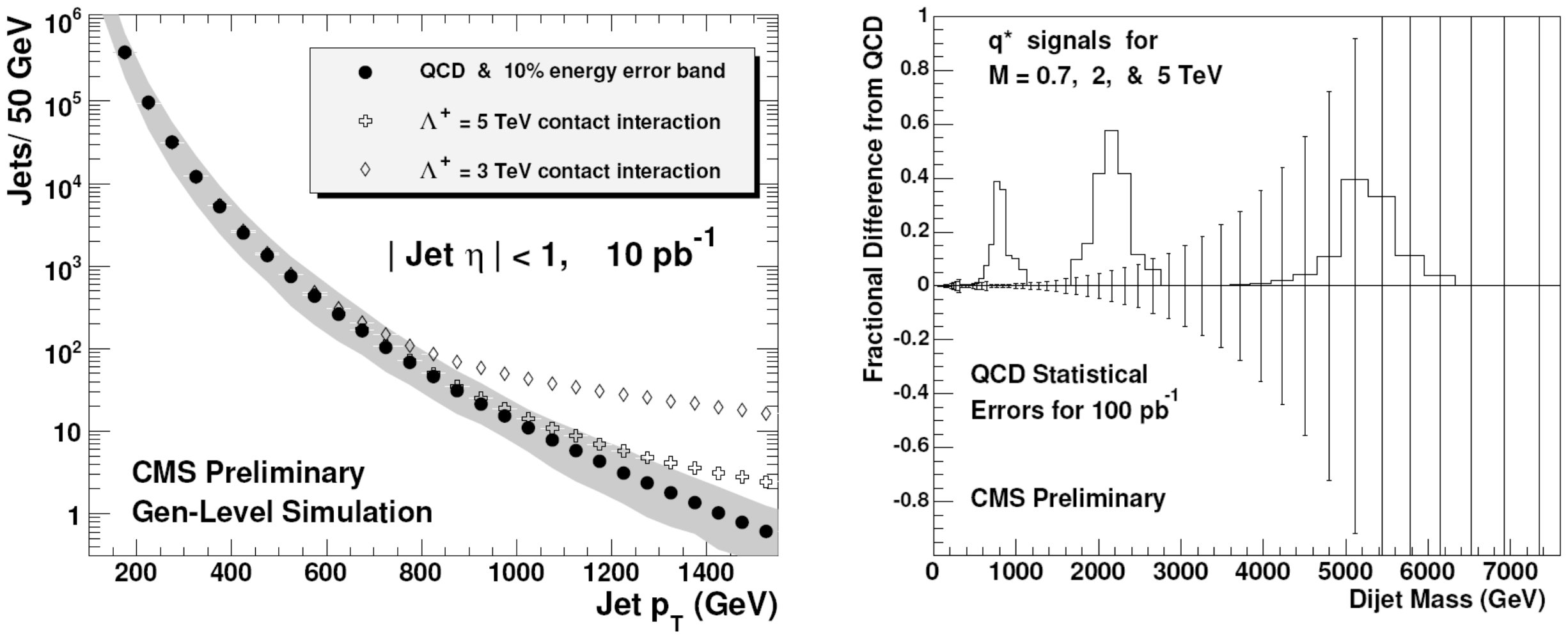}
  \end{center}
  \vspace{-0.5cm}
  \caption[.]{\underline{Left}: distribution of the transverse 
              momentum of the hardest central jet in simulated di-jet events 
              for 10\invpb integrated luminosity (CMS study). 
              The shaded band indicates the systematic uncertainties. The
              cross and diamond curves indicate the distortions in the 
              high-$p_T$ spectrum expected from contact interactions at the
              scales  $\Lambda^+=3$\,\TeV and 5\,\TeV, respectively. 
              \underline{Right}: fractional difference from the QCD
              expectation of the di-jet invariant mass (CMS). Also shown 
              are the contributions to the difference from heavy excited
              quarks decaying into jet pairs.  }
\label{fig:dijetnewphysicscms}
\end{figure}
The left panel in Fig.~\ref{fig:dijetnewphysicscms} shows the distribution 
of the leading central-rapidity jet $p_T$ in simulated di-jet events as 
expected by CMS for an integrated luminosity of 10\invpb.
The shaded band indicates the estimated systematic uncertainties. 
Also shown are the distortions in the spectrum expected from contact 
interactions\footnote
{
   New physics models with fermion substructure (`compositness') at high scale 
   lead to excitations of these fermions which modify scattering cross sections. 
   The interaction can be parametrised by an effective four-fermion 
   contact term
   \beq
      {\cal L}_{\rm eff} = \frac{4\pi^2}{\Lambda^2}
                        \!\!\sum_{i,k=L,R}\!\!\!\!\alpha^{ik}
                        \left(\qbar_i\gamma^\mu q_i^\prime\right)
                        \left(\fbar_k\gamma^\mu f_k^\prime\right)\,,
   \eeq
   where $\Lambda$ is the mass scale of the new interaction. Experimental 
   limits exclude excited fermions up to a few \TeV.
}  
at the characteristic scales $\Lambda^+=3$\,\TeV and 5\,\TeV, respectively.
A quantitative sensitivity study shows that contact interactions  
up to $\Lambda=3$\,\TeV can be discovered with the first 10\invpb.
However, the analysis requires excellent understanding of the jet 
resolution in the tails and the jet energy scale. Systematic errors 
dominate over the statistical ones and over uncertainties from the 
parton density functions.

The right plot in Fig.~\ref{fig:dijetnewphysicscms} shows a Monte Carlo
study by CMS of a search for strongly produced heavy excited quarks 
decaying into a quark pair. The most sensitive observable here is the 
di-jet invariant mass. Shown in the plot is the fractional difference 
between measurement (here: simulation) and Standard Model expectation for 
100\invpb integrated luminosity. Shown by the resonances are the 
contributions from excited quarks to that difference, which can be clearly
separated below 3\,\TeV. Other variables can also be looked at. For example,
the ratio of di-jet abundances between different regions of pseudorapidity
versus the di-jet invariant mass benefits from reduced systematic 
uncertainties compared with absolute cross section measurements. Also 
angular distributions exhibit sensitivity to new physics.

\subsection{Quarkonia production}

\begin{wrapfigure}{R}{0.3\textwidth}
  \vspace{-24pt}
  \begin{center}
	  \includegraphics[width=0.3\textwidth]{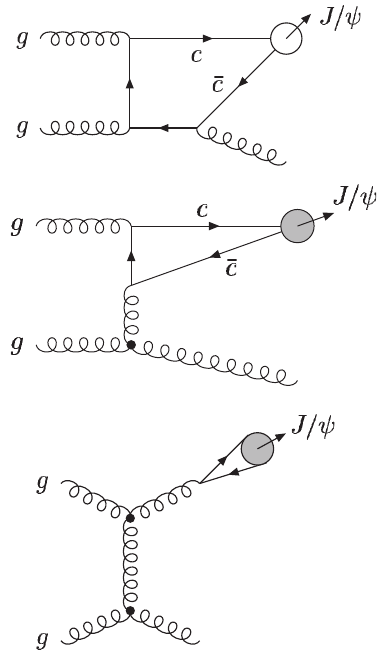}
  \end{center}
  \vspace{-10pt}
  \caption{Examples of Feynman diagrams for the singlet and octet
           production of a $J/\psi$ resonance (see text). }
  \label{fig:jpsiproductionfeynmangraphs}
  \vspace{-6pt}
\end{wrapfigure}
Quarkonia ($q\qbar$ resonances such as $J/\psi$, $\psi^\prime$, $\Upsilon$, 
$\Upsilon^\prime$, etc.) are abundantly produced at the LHC (see the Feynman 
graphs in Fig.~\ref{fig:jpsiproductionfeynmangraphs}) and excellent 
sources for early physics commissioning, 
but also for early physics measurements, \eg, prompt versus non-prompt 
production distinguished via different lifetimes, ratios of cross 
sections, polarisation, etc. Examples of Feynman diagrams for the singlet and octet
production of a $J/\psi$ resonance are drawn in 
Fig.~\ref{fig:jpsiproductionfeynmangraphs}. The upper diagram
describes the leading colour-singlet process, which has a small
cross section. The middle diagram, which dominates at low $p_T$,
can be produced through both singlet and octet $c\cbar$ states
with various quantum numbers.
At high $p_T$, the gluon fragmentation subprocess shown in the lower 
plot becomes increasingly important.

Quarkonia in ATLAS and CMS are mainly studied 
through their decays into muon and also electron pairs. Since they are 
narrow resonances they can be used as commissioning tools for the alignment 
and calibration of the trigger, tracking, and muon systems. Efficiency 
studies can employ the `tag-and-probe method' (see 
Section~\ref{sec:tagandprobemethod}). Owing to the low mass of the resonances,
trigger considerations are crucial to estimate the available cross section
for analysis. Using a di-muon trigger with 4\,\GeV thresholds for each muon,
the overall rate of events from all quarkonium states is likely to 
remain below the rate of 1\,Hz at a luminosity of \Lumi31. (The trigger rates
may be dominated by background processes.)

\begin{figure}[t]
  \begin{center}
	  \includegraphics[width=1\textwidth]{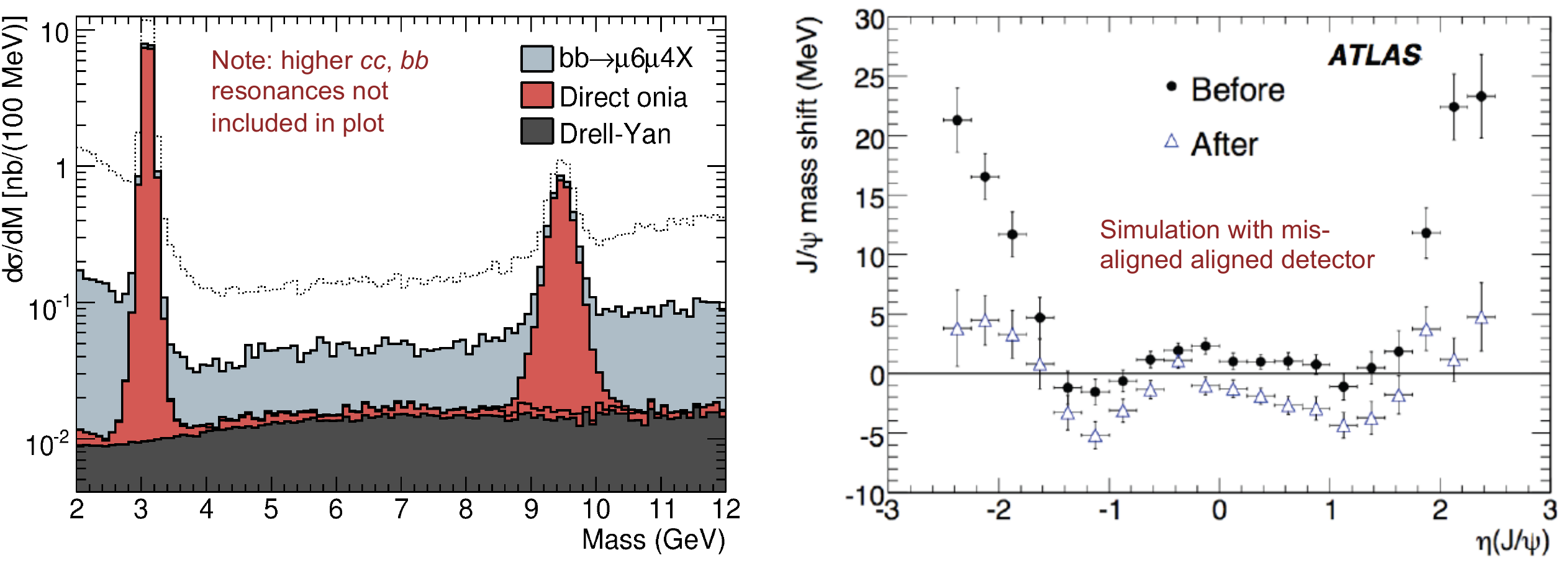}
  \end{center}
  \vspace{-0.5cm}
  \caption[.]{\underline{Left}: cumulative differential cross section 
              versus the invariant mass of muon pairs from various quarkonia 
              signal and combinatorial background sources (ATLAS study). 
              A primary vertex and pseudo-proper time requirement of 0.2\,ps 
              has been applied. The dotted line shows the cumulative 
              distribution without these cuts. The quarkonia simulation used
              for the plot does not include higher radial excitations.
              \underline{Right}: invariant mass of di-muons from $J/\psi$ decays 
              versus the $J/\psi$ pseudorapidity for simulated ATLAS data where a 
              severe misalignment of the inner tracking system has been introduced. 
              Shown are results before and after alignment. }
\label{fig:jpsiupsilonmassandalignmentatlas}
\end{figure}
The left-hand plot in Fig.~\ref{fig:jpsiupsilonmassandalignmentatlas} shows
the cumulative differential cross section of the invariant di-muon invariant 
mass for $J/\psi$ and $\Upsilon(1S)$ signal events and various combinatorial 
backgrounds from an ATLAS Monte Carlo study. 
The plot includes trigger requirements of at least one muon with 
6\,\GeV and another one with 4\,\GeV transverse momentum, and that these muons 
must originate from a common primary vertex. In addition a lifetime requirement
has been applied. (The dotted line shows the cumulative distributions without 
these latter two requirements). Backgrounds from Drell--Yan processes and leptonic 
heavy-quark decays are of similar size. 

The right panel of  Fig.~\ref{fig:jpsiupsilonmassandalignmentatlas} shows a
simulated commissioning result from ATLAS. Events of the type
$pp\to J/\psi(\to\mu\mu)+X$ with (somewhat unrealistically) severe misalignment 
in the inner tracker have been simulated and run through the alignment procedure 
based on hits-on-track residual minimisation. As discussed in 
Section~\ref{sec:innertrackeralignment}, this method suffers from so-called
weak modes, which denote misalignments that leave the global $\chi^2$
function, used to minimise the hit residuals, invariant. As seen in the plot, 
the reconstructed invariant di-muon mass versus the pseudorapidity of the 
di-muon system exhibits a strong non-uniformity before the alignment, but 
remaining effects caused by weak modes after the alignment. 

\subsection{$W$ and $Z$ boson production}

Inclusive production of $W$ and $Z$ bosons ($pp\to W(Z)+X$) has large cross
sections so that interesting data-driven cross section measurements can 
be performed with early data (10--50\invpb). The weak bosons are also important
ingredients for commissioning studies: $Z$ bosons are most important for various 
{\em in-situ} calibrations (\cf Section~\ref{sec:physicscommissioning}), and $Z+\,$jets 
and $W+\,$jets are sensitive probes of higher order QCD calculations. Inclusive
weak boson production is also precisely predicted by theory so that a 
cross section measurement in particular of the more abundant $W$ production
can be used to infer the absolute integrated luminosity recorded.

\begin{figure}[t]
  \begin{center}
	  \includegraphics[width=1\textwidth]{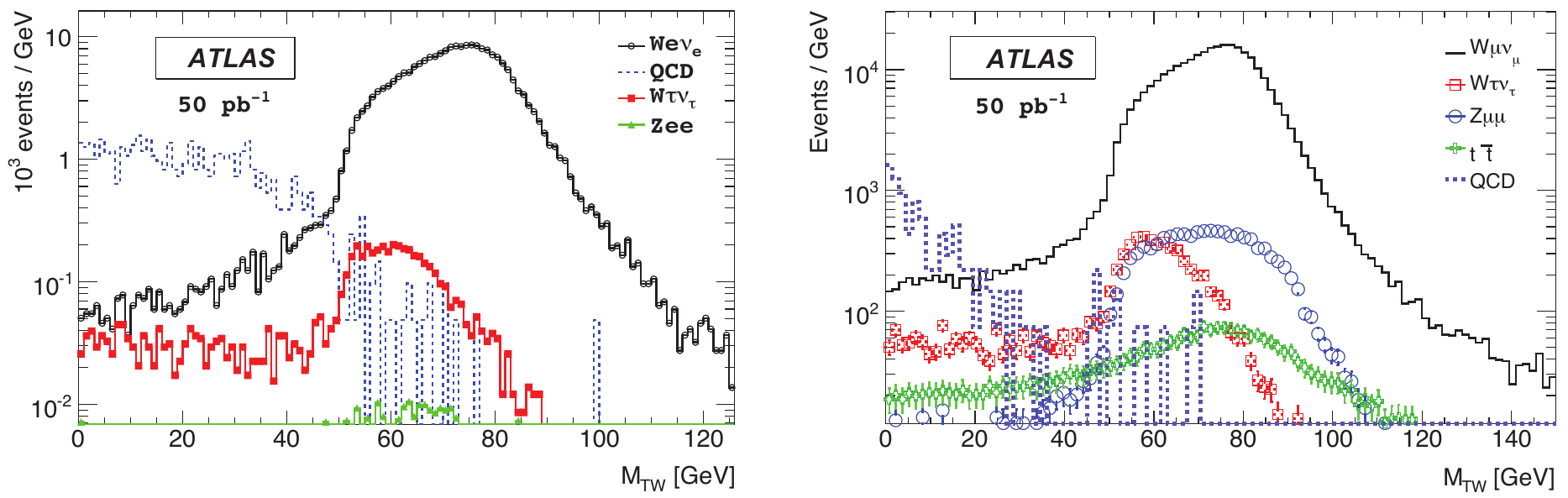}
  \end{center}
  \vspace{-0.5cm}
  \caption[.]{$W$-boson transverse mass distribution for $W\to e\nu$ (left) 
              and $W\to\mu\nu$ (right) and backgrounds after full selection
              except for the $M_T$ cut, for simulated data corresponding to 
              an integrated luminosity of 50\invpb.}
\label{fig:wandzbosonproduction}
\end{figure}

Figure~\ref{fig:wandzbosonproduction} shows the distribution of the 
$W$ transverse mass for $W\to e\nu$ (left) and $W\to\mu\nu$ (right) decays
together with their dominant backgrounds for simulated data corresponding to 
50\invpb integrated luminosity (ATLAS study). The transverse mass is defined
by
\beq
\label{eq:transversemass}
   m_T = \sqrt{E_T^\ell \Emiss_T\left(1-\cos\!\Delta\phi\right)}\,,
\eeq
where $\Delta\phi$ is the angle between the transverse lepton and missing 
energy vectors, and $E_T^\ell$ is the transverse energy of the lepton.
The transverse $W$ mass is also used as an ingredient for the precision 
measurement of the $W$ mass, which, however, requires much larger data samples
for a competitive measurement, because of the required mass calibration with 
respect to the accurately known $Z$ boson, which has a ten times smaller 
leptonic cross section. 

\begin{wrapfigure}{R}{0.4\textwidth}
  \vspace{-14pt}
  \begin{center}
	  \includegraphics[width=0.4\textwidth]{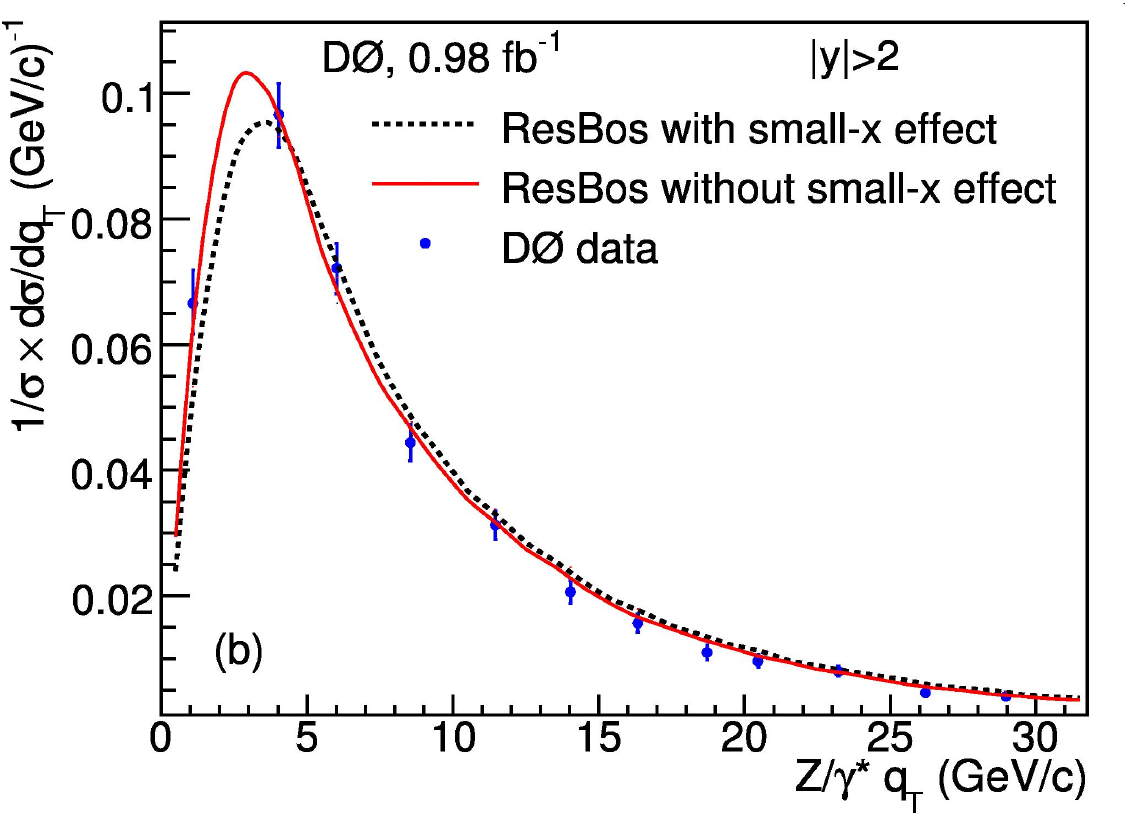}
  \end{center}
  \vspace{-15pt}
  \caption{$Z$-boson transverse momentum distribution,
           measured by the D0 experiment at the Tevatron, compared
           with Monte Carlo generator models.  }
  \label{fig:zptd0}
  \vspace{-6pt}
\end{wrapfigure}
Figure~\ref{fig:zptd0} shows the $Z$-boson transverse momentum distribution 
as measured by D0 at the Tevatron for a centre-of-mass energy of 1.96\,\TeV.
It is compared with Monte Carlo generator models including next-to-leading 
order QCD calculations. Good control of the transverse momentum of weak bosons
is important for many physics studies. Specifically in multivariate Higgs 
searches, the Higgs transverse momentum can be used as a discriminating 
variable since Higgs production is expected to have a harder spectrum 
than QCD backgrounds. 

\subsection{Top-quark production}

The roughly 100 times larger $\ttbar$ production cross section of $\sim$830\,pb 
at the LHC (at 14\,\TeV centre-of-mass energy) compared with $\sim$7.5\,pb at 
the Tevatron, makes it possible to observe top quarks in early data. Also the 
electroweak single-top production cross section of $\sim$300\,pb is similarly 
enlarged. Apart from having important physics potential, top quarks represent 
excellent objects for data-driven 
commissioning and calibration analyses, notably $b$-tagging and jet energy 
scale fits. The leading processes contributing to $\ttbar$ production
are gluon--gluon scattering ($s$ and $t$-channels) and quark--antiquark annihilation 
($s$-channel). Single-top production is dominated by $W$--gluon fusion ($t$-channel),
$W$ exchange between $b$ quarks ($t$-channel), associated production of top
and $W$, and quark--antiquark annihilation ($s$-channel, smaller cross section).

\begin{figure}[t]
  \begin{center}
	  \includegraphics[width=1\textwidth]{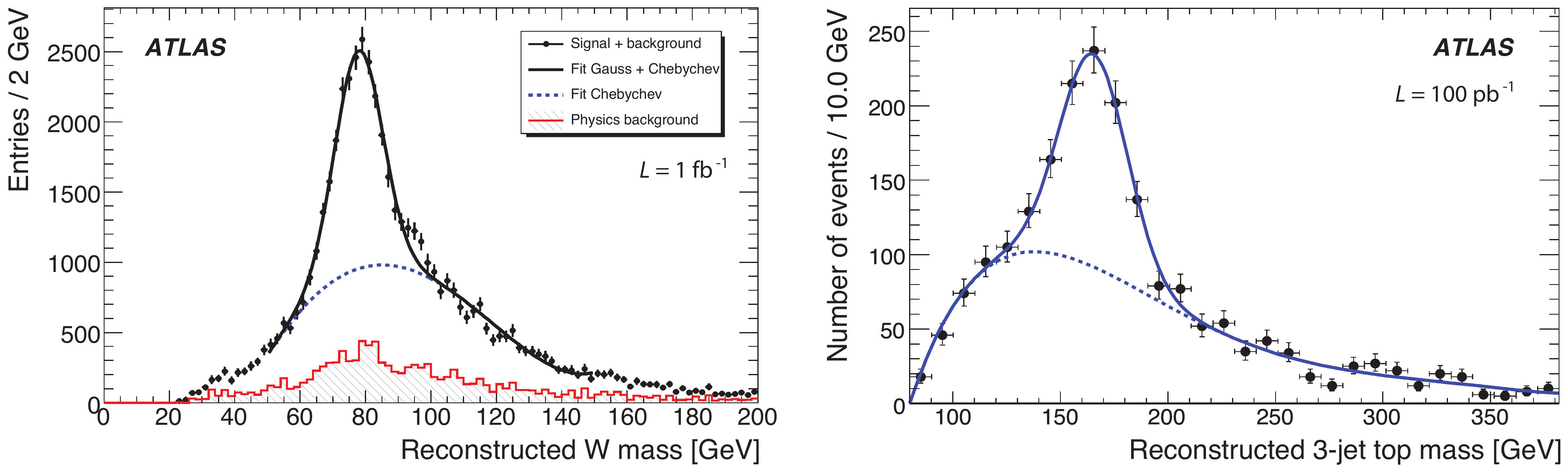}
  \end{center}
  \vspace{-0.4cm}
  \caption[.]{\underline{Left}: distribution of the invariant di-jet mass 
              of light-flavoured jets
              as an estimate of the hadronically decaying $W$-boson mass
              in a simulated data sample corresponding to 1\invfb integrated
              luminosity for signal $\ttbar$ and background processes (ATLAS study).
              \underline{Right}: reconstructed hadronic top mass from the 
              combination of three jets in simulated data corresponding to 
              100\invpb integrated luminosity signal $\ttbar$ and 
              background events after full selection. }
\label{fig:topmassatlas}
  \begin{center}
	  \includegraphics[width=1\textwidth]{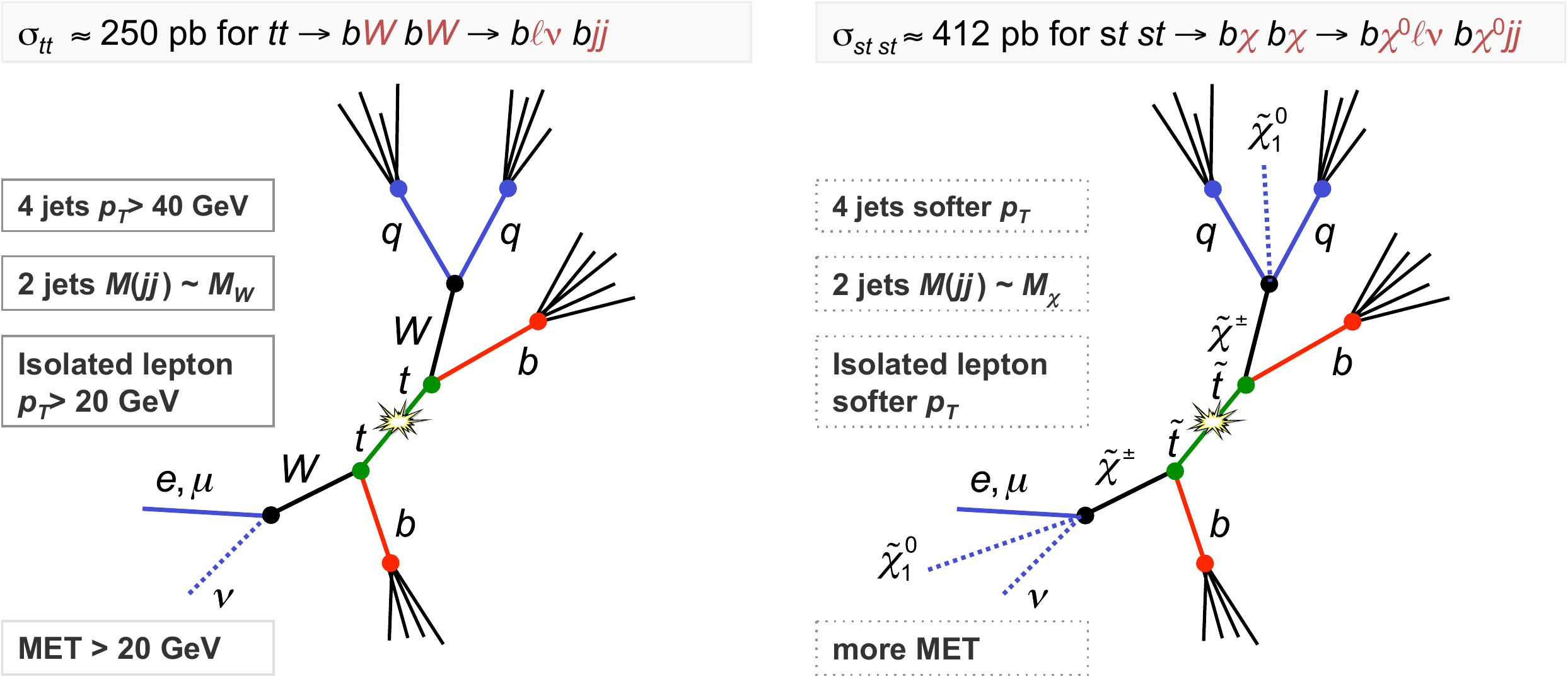}
  \end{center}
  \vspace{-0.3cm}
  \caption[.]{\underline{Left}: 
              schematic graph of a $pp\to \ttbar (+ X)$ process where one 
              top decays fully hadronically and the other semileptonically.
              The neutrino generates missing transverse energy.
              \underline{Right}: 
              schematic graph of a supersymmetric stop--antistop production.
              The $W^\pm$ propagators in top--antitop production are replaced
              by charginos that decay into three bodies of which one 
              (the neutralino) is a stable weak interacting neutral particle. }
\label{fig:topandstopdiagrams}
\end{figure}
Data corresponding to an integrated luminosity of 100\invpb should allow the
experiments to measure the \ttbar production cross section, with events where
both $W$ bosons decay leptonically, to an accuracy of 3\% statistical 
and 5\% systematic error (dominated by the uncertainty in the integrated 
luminosity value). The measurement provides an important probe of the 
validity of the Standard Model at unexplored centre-of-mass energy. 
Figure~\ref{fig:topmassatlas} shows on the right panel the reconstructed 
hadronic top mass from a combination of three jets as found 
in a simulated signal and background sample corresponding to 
100\invpb integrated luminosity after full event selection. The 
left panel shows the corresponding di-jet mass formed by light-flavour jets, 
representing the $W$ signal and combinatorial background. This plot can be used 
to determine and adjust the jet energy scale. A kinematic fit to the 
true $W$ mass can be used to improve the accuracy of the three-jet 
top-mass reconstruction.

Single-top production is of particular interest due to its sensitivity to 
charged new physics fields, such as a charged Higgs replacing the $W$ in 
the weak propagator as occurs in two-Higgs-doublet models.
Single-top production has been observed by the Tevatron in 2009 with
the use of advanced multivariate analysis techniques~\cite{tevsingletop}. 
The measured cross section of $(2.3^{\,+0.6}_{\,-0.5})$\,pb (CDF), 
$(3.94 \pm 0.88)$\,pb (D0), is in agreement with the Standard Model 
expectation. 

Figure~\ref{fig:topandstopdiagrams} shows on the left diagram a schematic
drawing of a Standard Model top--antitop event, where one top 
decays fully hadronically and the other semileptonically. The right diagram 
shows the production and decay of a light supersymmetric ({\em R}-parity conserving) 
stop--antistop pair, which follows a similar decay cascade with, however, 
an additional weak interacting neutral particle in the final state that
escapes the detector. As a consequence, the four particle jets and the isolated
lepton are softer than in the $\ttbar$ case, the two light jets originating 
from the heavy neutralino decay form the invariant mass of a neutralino, instead 
of that of a $W$, and significantly more missing transverse energy 
is produced in the supersymmetric event. The experimental separation of 
the $\ttbar$ and $\ststbar$ processes is difficult and requires more statistics
than available in early data taking. The analysis requires $b$-flavour tagging
to be commissioned and proceeds by plotting the minimum three-particles 
invariant mass that can be formed of a $b$-jet and the two light-flavoured jets.
Subtracting from it the expected $\ttbar$ Standard Model contribution 
a $\ststbar$ contamination would show up by a peak below the top (and below
the stop mass, due to the escaped neutralino). A study performed by ATLAS
shows that with 1.8\invfb and a stop mass of 137\,\GeV, for which the
$\ststbar b\chi b\chi\to (b\chi^0\ell\nu)(b\chi^0qq)$ cross section amounts 
to 412\,pb (depending also on other model parameters), exceeding by a factor
1.6 the corresponding $\ttbar\to bWbW\to (b\ell\nu)(bqq)$ cross section,
a clear signal can be derived. 

\subsection{Standard Model Higgs boson search}

The observation of a Standard Model Higgs boson is inverse femtobarn rather than picobarn 
physics, and hence not of primary importance for early physics. However, new physics
may enhance Higgs-like signals and the experiments must be prepared for surprises. 
It is also important to begin early with the understanding and improvement of electron
muon, tau and photon selection efficiencies and purities, and the study of $b$-jet
and forward-jets tagging, and a thorough categorisation of the relevant
Higgs backgrounds to tune the multivariate analyses that will be used to extract a signal. 

\begin{wrapfigure}{R}{0.35\textwidth}
  \vspace{-24pt}
  \begin{center}
	  \includegraphics[width=0.35\textwidth]{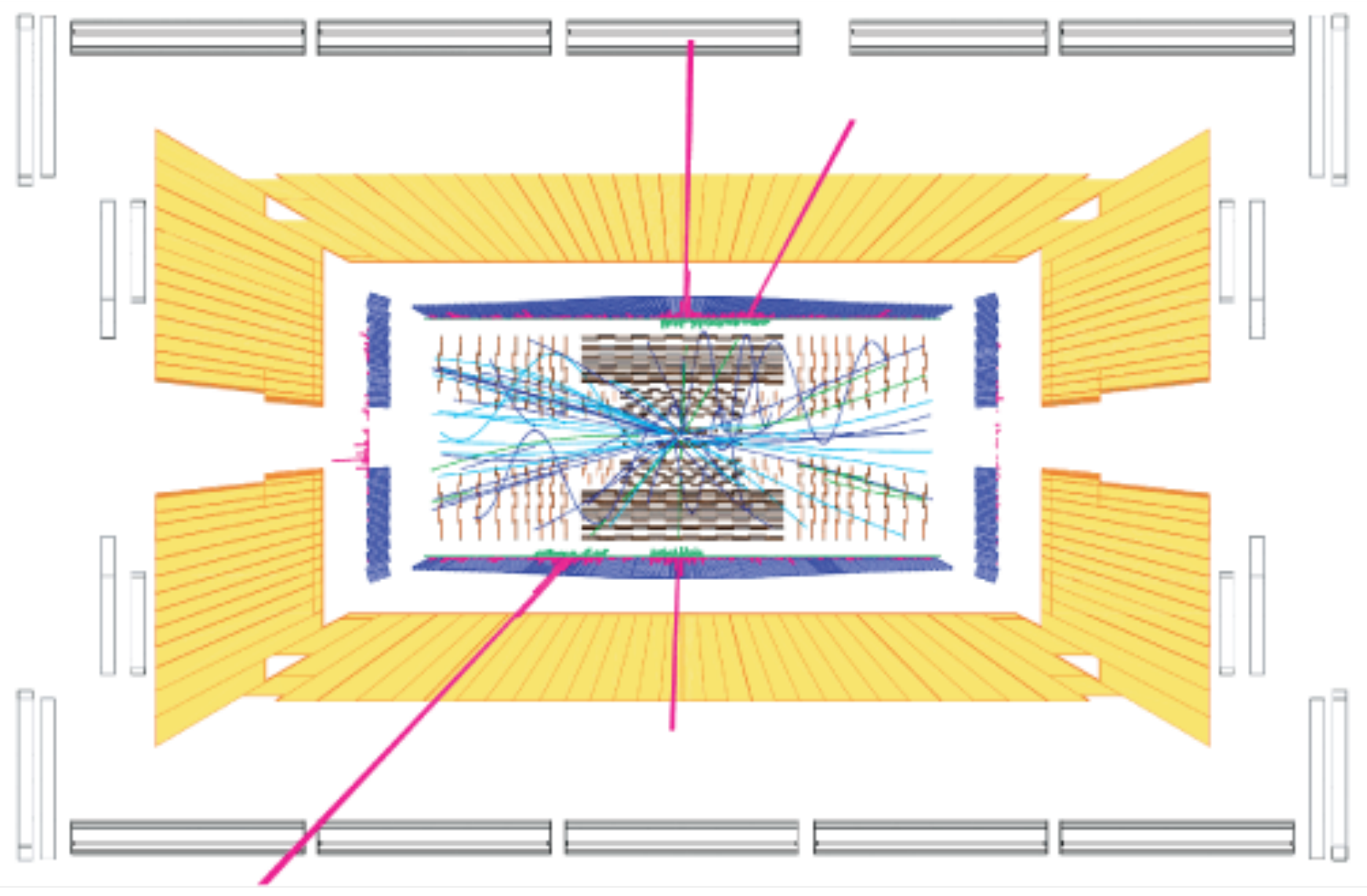}
  \end{center}
  \vspace{-10pt}
  \caption{A simulated $H\to ZZ^\star\to 4e$ event with $m_H=150$\,\GeV in CMS. }
  \label{fig:hto4eeventdisplaycms}
  \vspace{-6pt}
\end{wrapfigure}
\begin{figure}[t]
  \begin{center}
	  \includegraphics[width=1\textwidth]{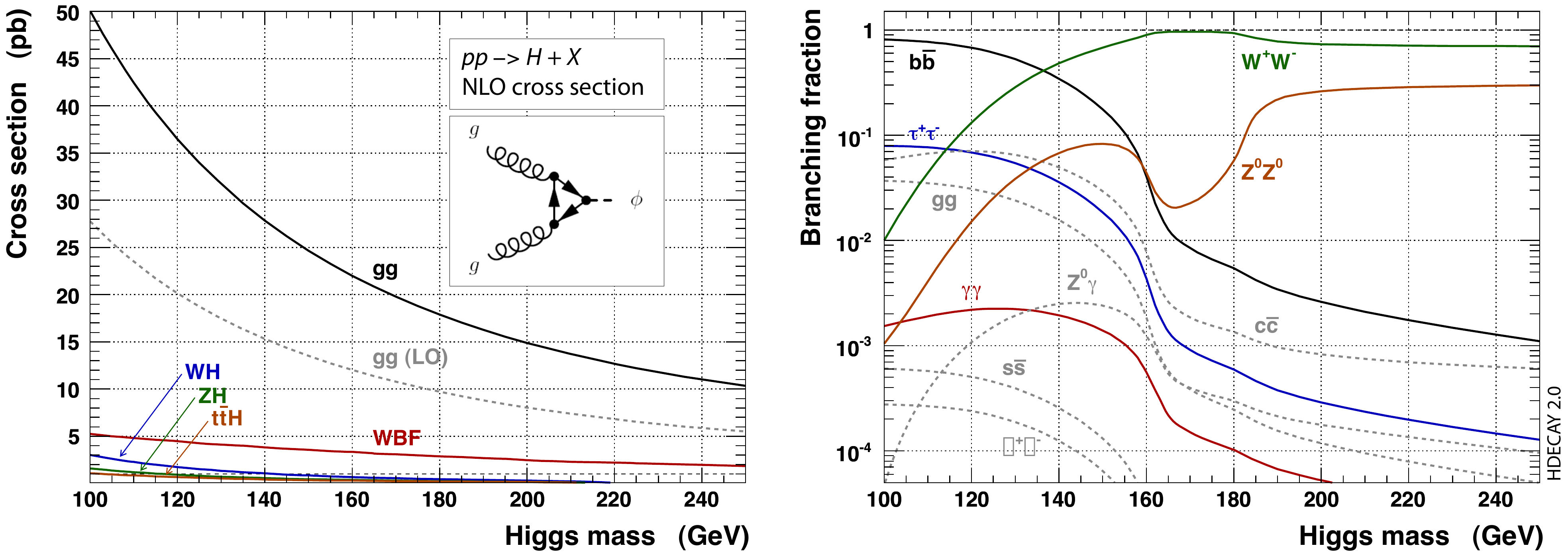}
  \end{center}
  \vspace{-0.3cm}
  \caption[.]{Expected inclusive Standard Model Higgs boson production cross section 
              for the various production modes (left) and Higgs branching fractions (right) 
              versus the Higgs mass at 14\,\TeV centre-of-mass $pp$ collisions.   }
\label{fig:higgsxsecbr}
\end{figure}
Figure~\ref{fig:higgsxsecbr} shows the dependence of the inclusive Standard Model 
Higgs boson production cross section and branching fractions on the Higgs mass.  
The dominant production mode is the fusion of two gluons into a scalar Higgs via
triangular top loop. Next-to-leading order (NLO) corrections give a sizable $K$-factor
(the factor with which the leading order result needs to be multiplied to include
higher orders) in this process. The second most important process is weak boson 
fusion that is accompanied by two forward jets. Since it is a weak process,
next-to-leading order corrections are less important. Following are the associated 
Higgs production with a $W$ or a $Z$ boson, or with a $\ttbar$ pair. The strong
rise in the branching fractions to the heavy weak boson pairs is due to the kinematic 
opening of these channels, which are favoured because the Higgs couples to the 
masses of the particles (if the Higgs boson were heavy enough to be able to 
decay into a top--antitop pair (not shown in the plot), it would reach a 
branching fraction of up to 20\% at around $m_H\sim500$\,\GeV).

\begin{figure}[t]
  \begin{center}
	  \includegraphics[width=1\textwidth]{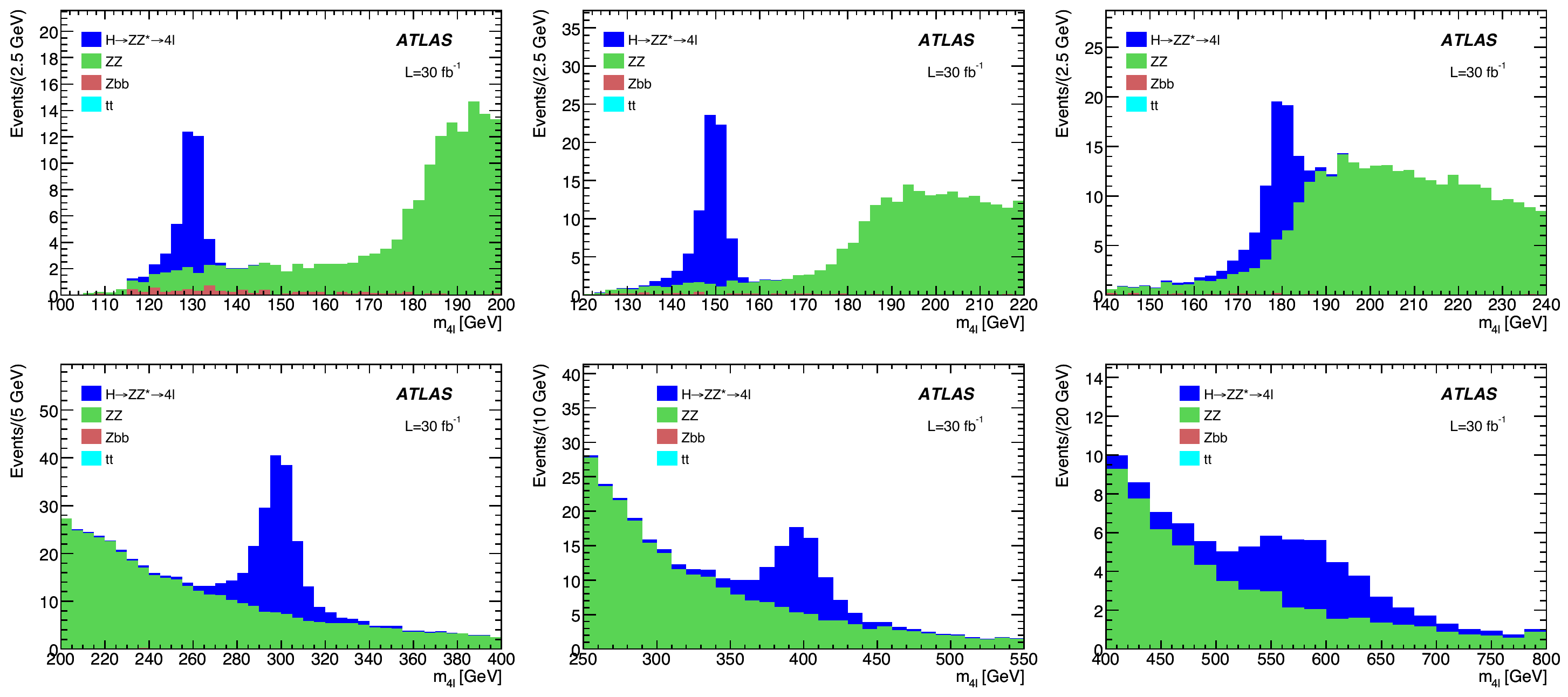}
  \end{center}
  \vspace{-0.4cm}
  \caption[.]{Reconstruction of the four-lepton invariant mass for simulated
              $H\to ZZ^{(\star)}\to4\ell$ ($\ell=e,\mu$) signal and background
              events corresponding to an integrated luminosity of 30\invfb (ATLAS study). 
              From upper left to lower right are shown analyses 
              for the true Higgs masses 130, 150, 180, 300, 400, and 600\,\GeV,              
              respectively. 
  }
\label{fig:higgstozto4lepallmasses}
\end{figure}

Figure~\ref{fig:higgstozto4lepallmasses} illustrates the results of a simulated 
search for $H\to ZZ^{(\star)}\to4\ell$ in ATLAS with 30\invfb for different 
true Higgs masses. Because the Higgs partial width into two vector bosons
increases with $m_H^3$, but only linearly for a Higgs decaying into two fermions,\footnote
{
   The leading order width of the Higgs boson decay into a fermion--antifermion
   pair is given by
   \beq
      \Gamma^{({\rm LO})}(H\to f\fbar) = \frac{G_F N_C}{4\sqrt{2}\pi} m_H m_f^2 \beta_f^3\,,
   \eeq
   where $G_F=1.16637\cdot 10^{-5}$\,\GeV$^{-2}$ is the Fermi constant, 
   $\beta_f=\sqrt{1-4 m_f^2/m_H^2}$ is the fermion velocity in the Higgs rest
   system, and $N_C=3(1)$ is the number of colours for quarks (leptons). 
   Large next-to-leading order corrections can occur in the case of quarks.
   The leading order width of the decay into two on-shell weakly interacting 
   vector bosons reads 
   \beq
      \Gamma(H\to VV ) = \frac{G_F m_H^3}{16\sqrt{2}\pi}\cdot\delta_V\cdot A(x)\,,
   \eeq
   where $\delta_V=2(1)$ for $V=W(Z)$, and $A(x)=\sqrt{1-4x}\cdot(1-4x+12x^2)$ with 
   $x=m_V^2/m_H^2$. For masses much larger than $2m_Z$ the width $\Gamma(H\to WW)$ 
   is twice as large as $\Gamma(H\to ZZ)$. Very roughly one finds
   $\Gamma(H\to WW+ZZ )\approx0.5\,{\rm \TeV}\cdot(m_H/1\,{\rm \TeV})^3$, so that
   for a Higgs mass of 1\,\TeV the Higgs width becomes of the same order of magnitude. 
}
the total Higgs width grows fast beyond the $H\to WW$ and $H\to ZZ$ openings. 
For example, while the Standard Model Higgs width is only 3.6\,\MeV at $m_H=120$\,\GeV
and 76\,\MeV at 160\,\GeV, it grows to 1.4\,\GeV at 200\,\GeV and 8.5\,\GeV at 300\,\GeV.
In the region favoured by the electroweak fit (see below) the Higgs intrinsic
width is much smaller than the experimental resolution and hence negligible.

\begin{figure}[t]
  \begin{center}
	  \includegraphics[width=0.49\textwidth]{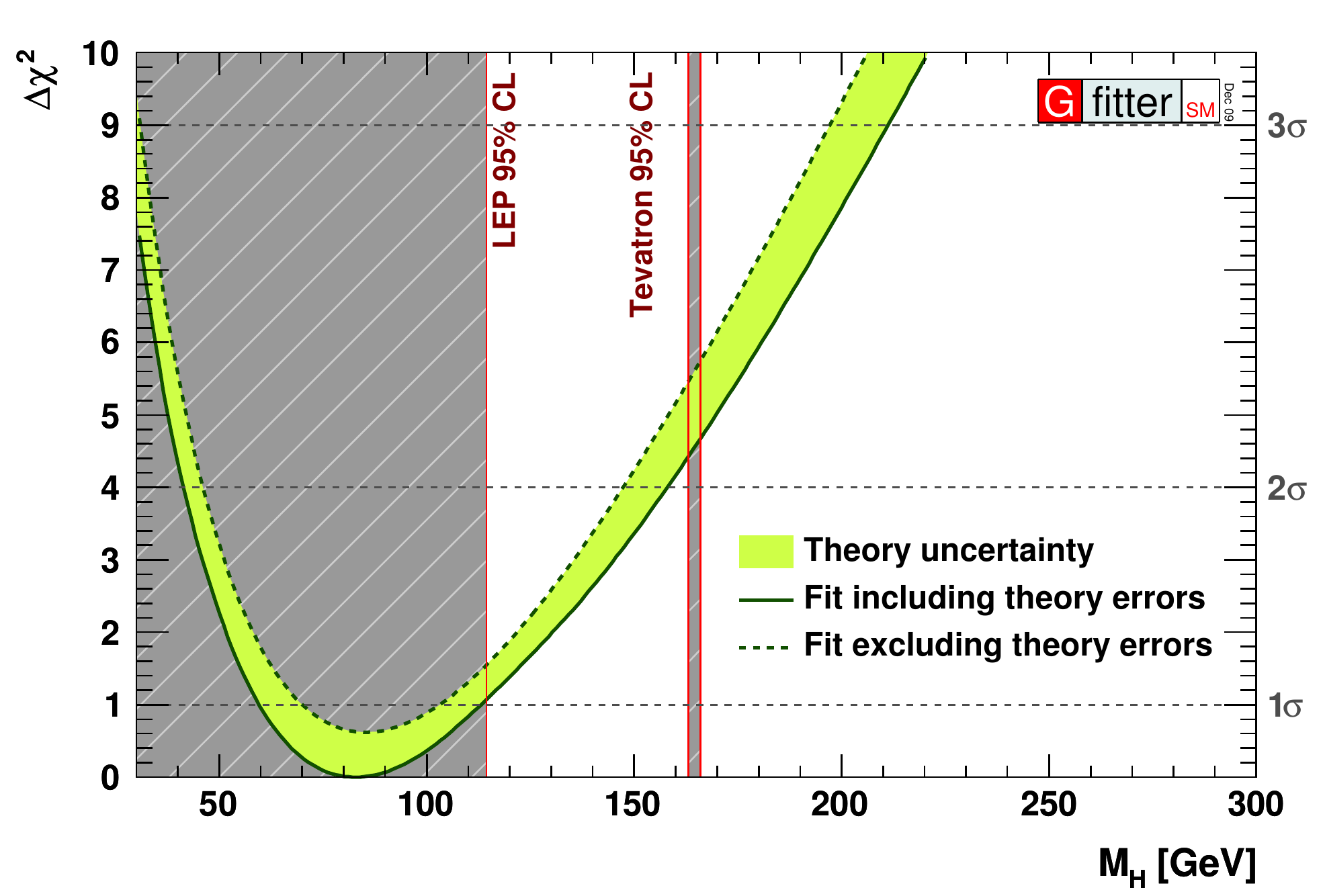}
	  \includegraphics[width=0.49\textwidth]{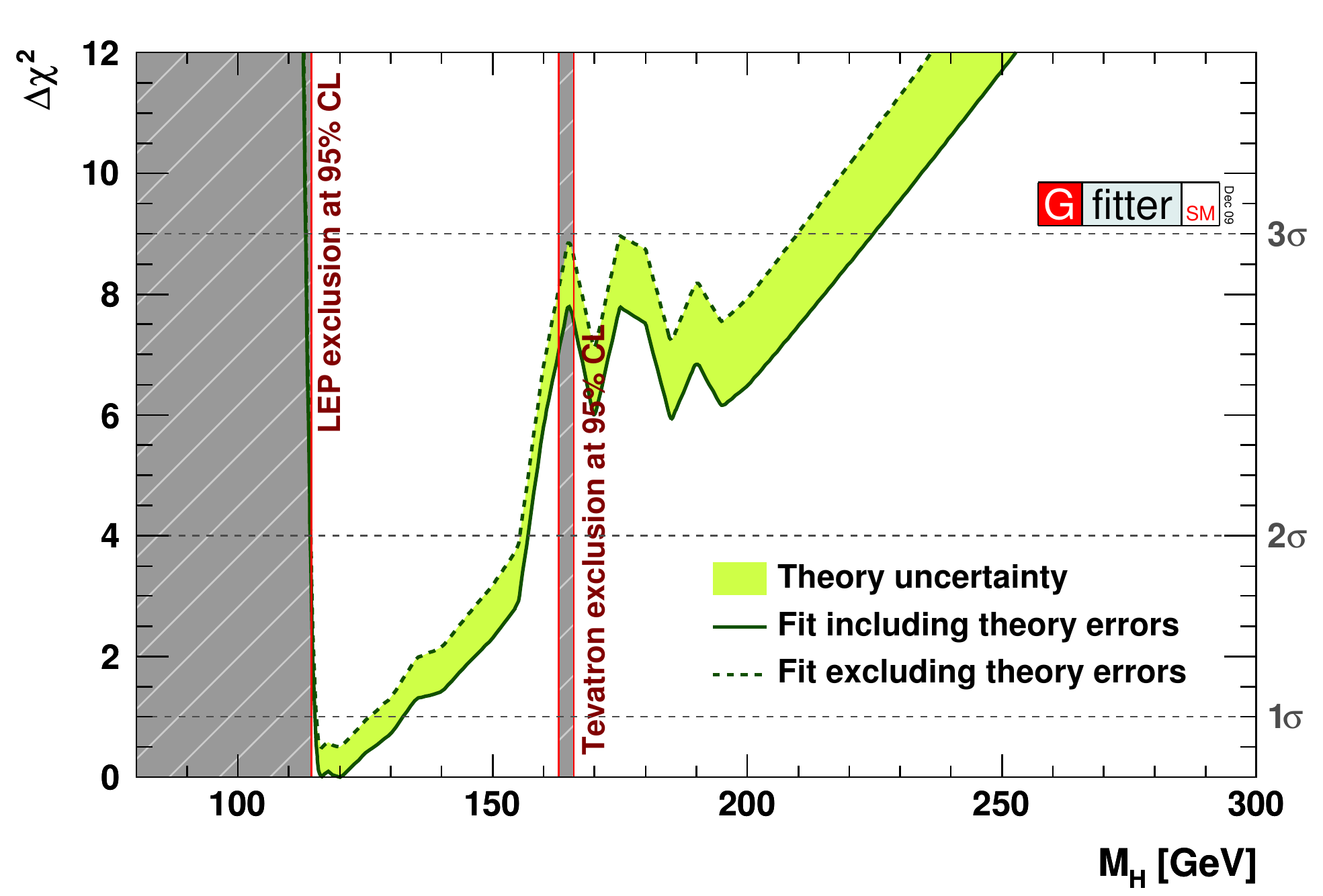}
  \end{center}
  \vspace{-0.4cm}
  \caption[.]{Curves of $\Delta\chi^2$ obtained from the global fit to electroweak 
              precision data. The right plot includes in addition 
              the results from the direct Higgs boson searches at LEP and Tevatron. 
              The plots are taken from Ref.~\cite{gfitter}.}
\label{fig:higgsewfit}
\end{figure}

Electroweak precision observables, measured by experiments at the LEP (CERN), SLC (SLAC) 
and Tevatron (FNAL) accelerators, can be used in a global Standard
Model fit to derive a constraint on the Higgs mass. The resulting $\Delta\chi^2$
curves versus the Higgs boson mass, without (left) and with (right) 
results from direct Higgs boson searches at LEP and the Tevatron included in the fit, 
are given in Fig.~\ref{fig:higgsewfit}. The result including all the available 
information yields the allowed range $114<m_H<157$\,\GeV at 95\% confidence level. 
Although this represents an important indication, experimentalists cannot afford
to disregard the high-mass region. The analyses must cover all Higgs masses 
that are not yet excluded by direct searches. 

\begin{figure}[t]
  \begin{center}
	  \includegraphics[width=1\textwidth]{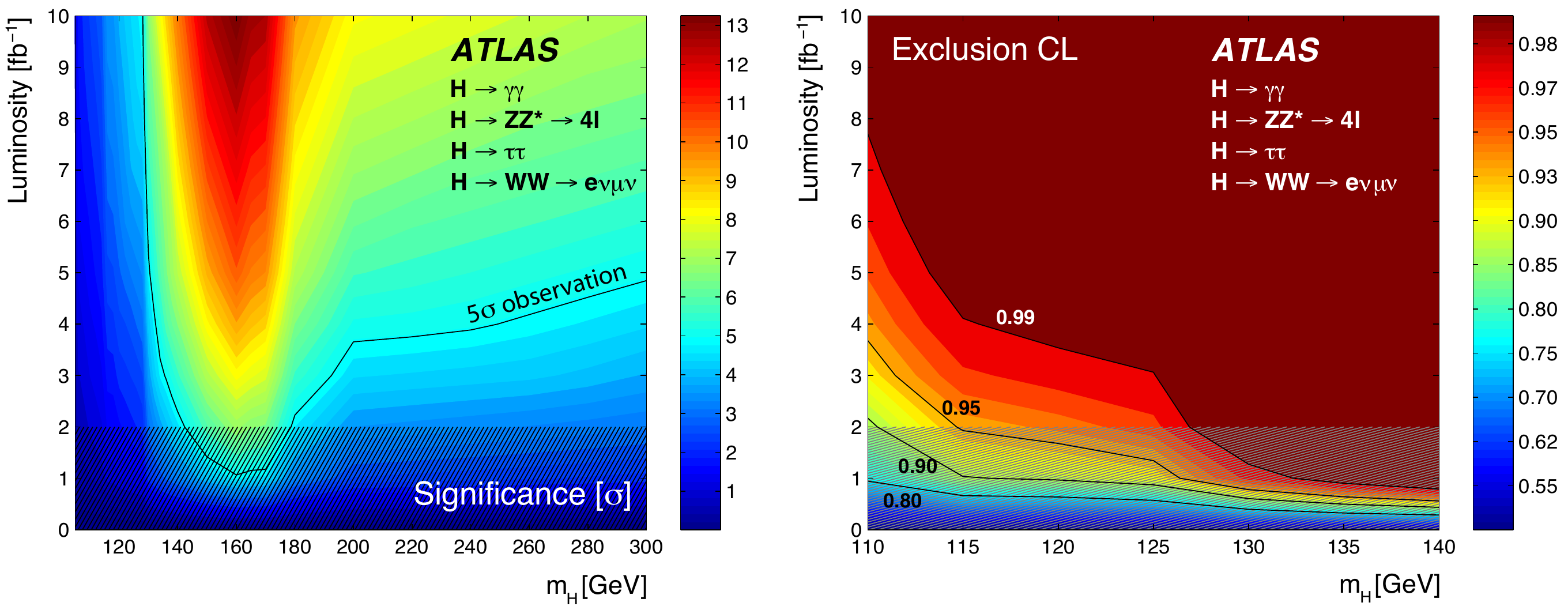}
  \end{center}
  \vspace{-0.4cm}
  \caption[.]{Standard Model Higgs boson discovery (left) and exclusion potential (right)
              for given integrated luminosity versus the Higgs mass (ATLAS study). 
              The shaded bands at low integrated luminosity indicate that the 
              results are less accurate (but are expected to be pessimistic).}
\label{fig:higgsdiscoverpotentialatlas}
\end{figure}

From Fig.~\ref{fig:higgsewfit} it becomes clear that very different experimental 
search strategies need to be pursued depending on the Higgs mass hypothesis. The 
golden discovery modes are $H\to\gamma\gamma$ for masses below $\sim$150\,\GeV (grand 
maximum), which is a very rare channel (branching fraction of about 0.2\%)  with a 
clean signature, $H\to WW^{(\star)}\to\ell\nu\ell\nu$ for high masses, which is an 
abundant but not a clean mode, and $H\to ZZ^{(\star)}\to 2\ell2\ell^\prime$
which has a sizable branching fraction above $m_H\simeq 130$\,\GeV, and which 
is clean at relatively low mass. We have no space here to discuss all these measurements. 
Early searches will concentrate on the high-cross-section modes leading to a successive 
exclusion (or discovery) of smaller and smaller Higgs masses. ATLAS and CMS have performed
studies to evaluate the discovery reach of the various Higgs search analyses as a 
function of the Higgs mass. Figure~\ref{fig:higgsdiscoverpotentialatlas}
shows an ATLAS study for the Higgs boson discovery (left panel) and exclusion potential 
(right panel) for given integrated luminosity versus the Higgs mass. At 1\invfb
a Higgs of mass between 150\,\GeV and 170\,\GeV could be observed with five standard 
deviations significance, and Higgs masses above $\sim$127\,\GeV can be excluded
to at least 95\% confidencel level.

\subsection{Search for phenomena beyond the Standard Model}

The primary motivation for the LHC construction is --- beyond the discovery of the 
Higgs boson --- the search for signatures from unknown physics at the high-energy 
frontier, which it is hoped will
provide answers to at least part of the current unknowns and problems outlined 
in Section~\ref{sec:motivation}. There is a wealth of models introducing new 
physics, which is also driven by the relatively few constraints that the high-energy 
sector must comply with. At any order of magnitude beyond the \TeV scale may lurk 
new symmetries, the breaking of which creates partners of the known Standard 
Model fields, but which also may lead to a profusion of new particles at 
ever higher mass scales. Alternatively, in case we live in an apparently 
severely fine-tuned world, no new physics exists at least in the quark sector 
up to the reduced Planck scale, leaving a desert of 16 orders of magnitude 
all described by the Standard Model interactions. This latter picture must
probably be regarded as disfavoured, not only by the fine-tuning 
argument, but since it also contradicts our experience: up to now, each ascent 
of an order of magnitude in energy has afforded new phenomena in particle 
physics.

\subsection*{Di-lepton resonances at high mass}

\begin{wrapfigure}{R}{0.32\textwidth}
  \vspace{-24pt}
  \begin{center}
	  \includegraphics[width=0.32\textwidth]{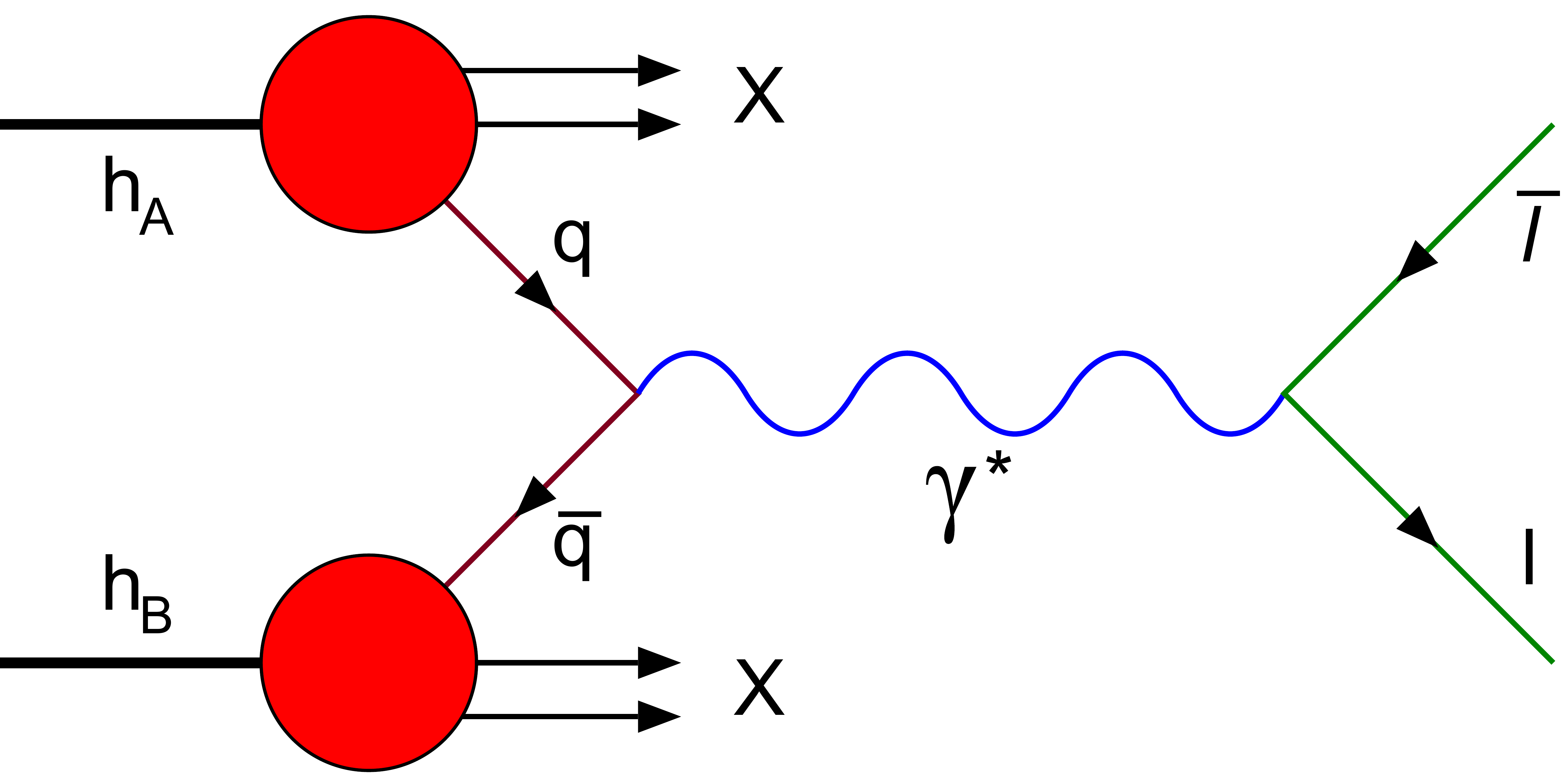}
  \end{center}
  \vspace{-10pt}
  \caption{Feynman graph of a Drell--Yan process (quark--antiquark 
           annihilation to a virtual photon or $Z$ boson) producing 
           a final-state lepton pair. }
  \label{fig:drellyanprocess}
  \vspace{-6pt}
\end{wrapfigure}
\begin{figure}[t]
  \begin{center}
	  \includegraphics[width=1\textwidth]{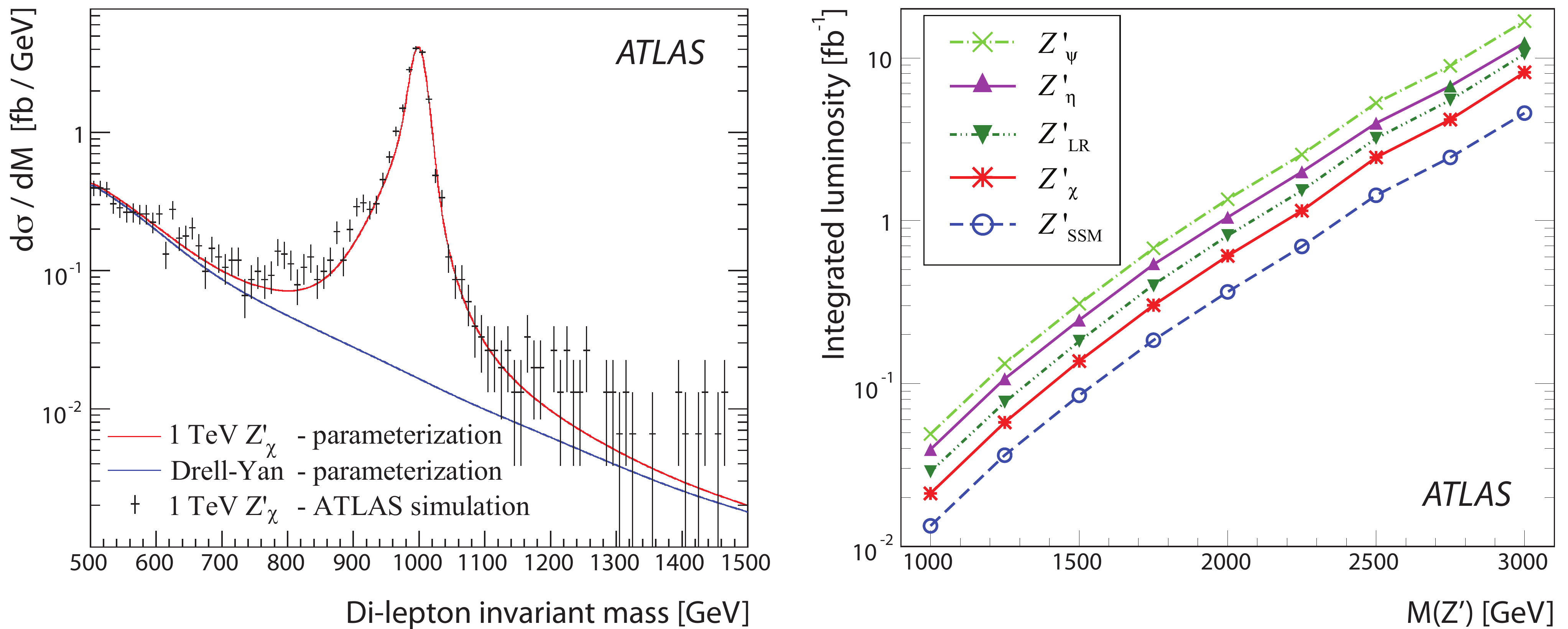}
  \end{center}
  \vspace{-0.3cm}
  \caption[.]{\underline{Left}: 
              distribution of the di-electron mass for fully simulated ATLAS 
              data (dots) in presence of a 1\,\TeV $Z^\prime_\chi$ (solid line) 
              and Drell--Yan background (dashed line). The statistics used 
              correspond to 21\invfb.
              \underline{Right}: 
              Required luminosity versus the $Z^\prime$ mass for a $5\sigma$ observation 
              according to various $Z^\prime$ models (ATLAS study).}
\label{fig:zprimestudiesatlas}
\end{figure}
Popular early searches for new physics involve di-lepton invariant 
mass spectra, which may exhibit peaks originating from generic $Z^\prime$
resonances present in many beyond the Standard Model scenarios, such 
as grand unified theories, little Higgs models, Technicolour, and 
models featuring extra spatial dimensions. The widths of the new resonances 
may be narrow (such as for Randall--Sundrum gravitons), or broad enough so that 
they may be resolved in the detector (for example heavy resonances in 
grand unified theories and little Higgs models, as well as in models with 
small extra dimensions where the gauge fields are allowed to propagate into 
the extra-dimensional bulk). 
The most rigorous direct limits on the existence of heavy neutral particles 
decaying into di-leptons come from direct searches at the Tevatron, excluding
mass scales until approximately 1\,\TeV (model dependent).

\begin{wrapfigure}{R}{0.32\textwidth}
  \vspace{-24pt}
  \begin{center}
	  \includegraphics[width=0.32\textwidth]{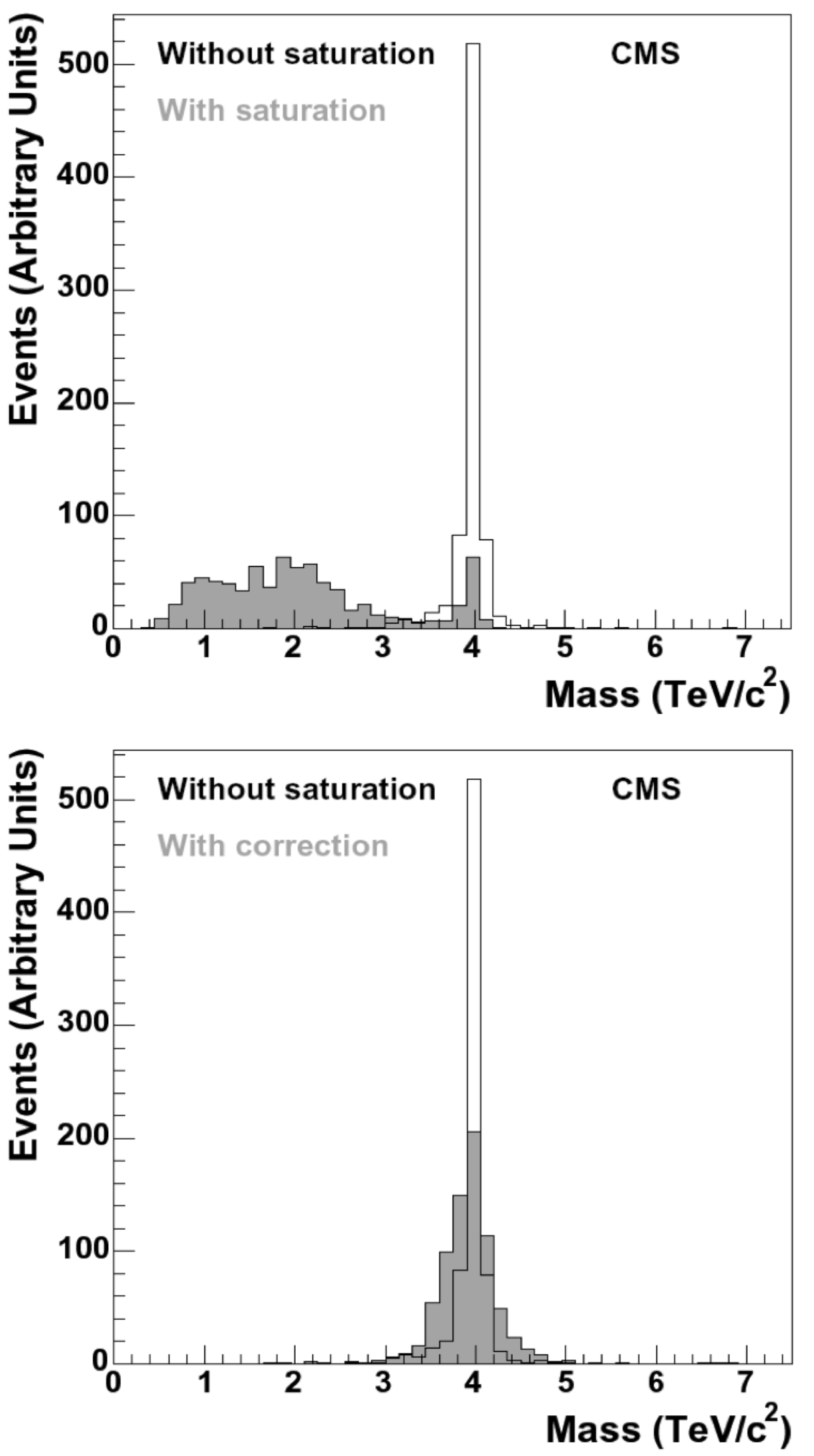}
  \end{center}
  \vspace{-10pt}
  \caption{High energy saturation effect (upper plot) and its correction (lower) 
           in the CMS electromagnetic calorimeter for 4\,\TeV Randall--Sundrum 
           gravitons decaying to \ee.}
  \label{fig:ecalsaturationcms}
  \vspace{+55pt}
\end{wrapfigure}
Contrary to searches with missing transverse energy, which usually do 
not exhibit clear-cut kinematic signatures, the observation of a di-electron
mass peak over (mostly) irreducible Drell--Yan background (Fig.~\ref{fig:drellyanprocess})
does not require the design calorimeter performance (this is somewhat different 
for heavy di-muon resonances, where the alignment of the muon system must be 
well understood to reach good resolution and charge measurement). In case 
of the search for very high-mass resonances (not early physics), 
electromagnetic calorimeter saturation must be corrected 
(Fig.~\ref{fig:ecalsaturationcms} for CMS). It is 
also not required to predict the background shapes with Monte Carlo 
simulation. It can be determined from data by means of a parametrised 
maximum-likelihood fit with parameters determined simultaneously with 
the signal abundance by the fit.

The left panel in Fig.~\ref{fig:zprimestudiesatlas} shows a $Z^\prime_\chi\to ee$
peak in ATLAS for a simulated $Z^\prime$ with mass 1\,\TeV, over Drell--Yan
background. The right panel gives the luminosity that is required  for a 
$5\sigma$ observation according to various $Z^\prime$ models, as a function 
of the $Z^\prime$ mass. With 100\invpb of data, and 14\,\TeV centre-of-mass
energy, $Z^\prime$ (and also $W^\prime$) resonances until a mass of roughly 1\,\TeV
could be discovered. The ultimate goal for ATLAS and CMS reaches about 
7\,\TeV (SLHC prospective).

\subsubsection*{Statistical considerations}

\begin{wrapfigure}{L}{0.32\textwidth}
  \vspace{-24pt}
  \begin{center}
	  \includegraphics[width=0.32\textwidth]{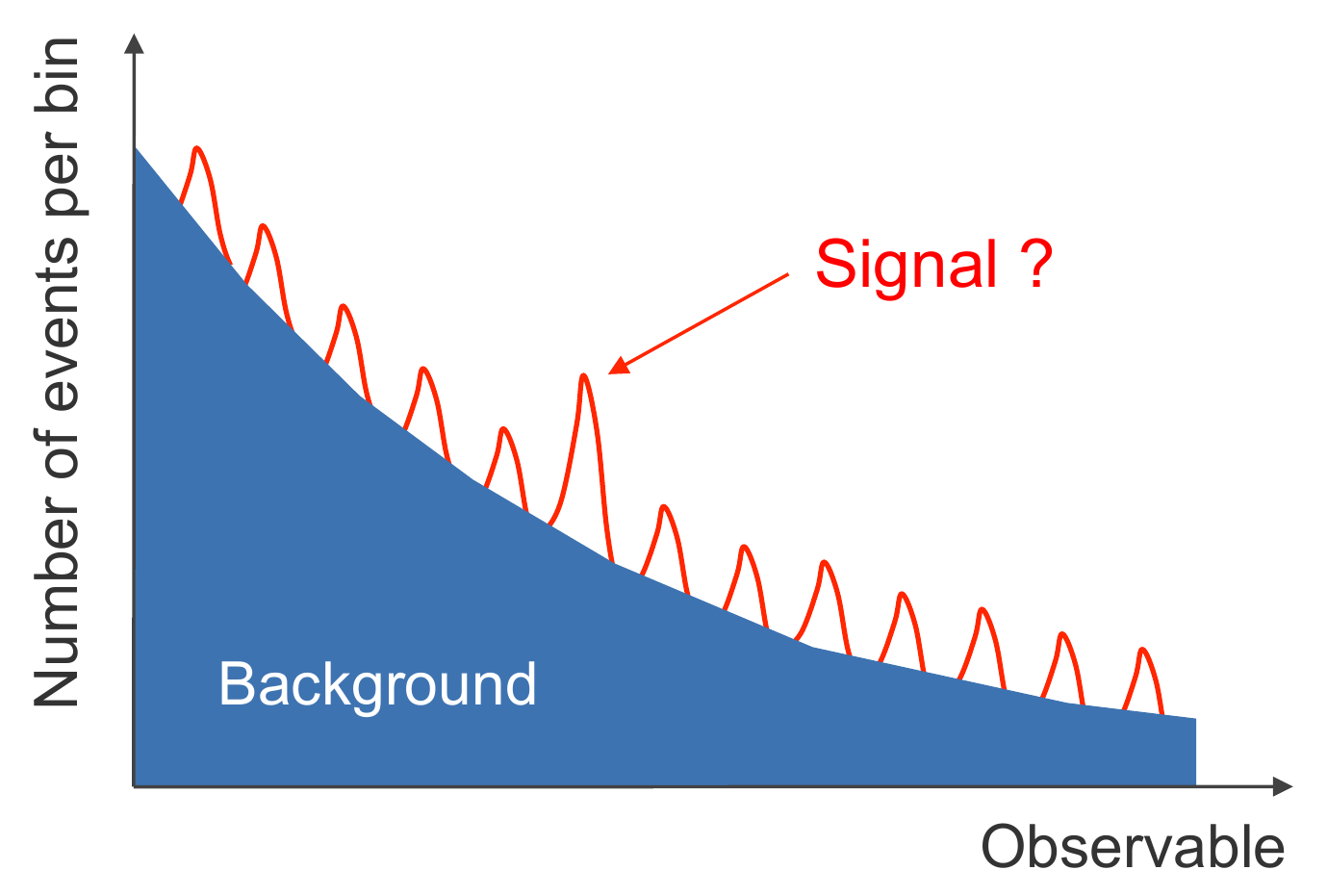}
  \end{center}
  \vspace{-10pt}
  \caption{The p-value quantifying the statistical significance of an observation 
           must be corrected for the statistical trials factor. }
  \label{fig:statisticaltrialsfactor}
  \vspace{-6pt}
\end{wrapfigure}
The search for di-lepton resonances in a mass range that is large compared 
with the experimental resolution, and without using prior knowledge about
which mass the resonance should have, introduces a statistical 
`look-elsewhere effect'. The probability of finding a 6 when playing a dice
is 1/6. The probability of finding at least one 6 when playing 2 dice is 
$2/6\cdot(1-5/6)$. In case of a small single-occurrence probability $p$ and 
at least one occurrence required, the binomial probability for an occurrence
with $n$ trials can be approximated by $n\cdot p$. What counts in the case
of the di-lepton invariant mass is the size of the search range in terms
of the mass resolution (assuming a negligible intrinsic width of the 
resonance that is searched for). The number of trials is thus roughly
the number of times the resonance `fits' in the given mass range. 
Assuming the `discovery' of a mass peak with a single p-value\footnote
{
   Terminology: the {\em significance level} of a statistical hypothesis test is 
   the fixed probability of wrongly rejecting the null hypothesis, if it is true. 
   It is the probability for a {\em Type-I error} to occur. The {\em p-value} is 
   compared with the significance level and, if it is smaller, the result is 
   significant. It is hence the significance of a single trial.
}
of $3.0\sigma$ ($5.0\sigma$) somewhere in the allowed mass range, and assuming 13 
independent trials fit into the mass range, the p-value must be corrected 
by the corresponding {\em trials factor}. In this case, we find a corrected p-value 
of $2.1\sigma$ ($4.5\sigma$). Because of the non-linearity in the relation between
probability and number of standard deviations, the effect of the correction
appears larger at smaller significance of the observation. 

The above exercise is a very rough approximation. In practice the evaluation
of the trials factor is complex, and the conceptually simplest way to take 
it into account is via toy Monte Carlo simulation. A natural way to proceed is
to perform an unbinned maximum-likelihood fit by describing the background by 
a simple parametrised function, with parameters determined by the fit, and the 
signal by a Gaussian or crystal-ball shaped function with predetermined width 
(obtained from Monte Carlo simulation, but taking into account the mass dependence 
of the calorimeter or tracking resolution) of which only the mean mass parameter 
is free to vary in the fit. Also determined by the fit are the signal and 
background abundances. The fit will converge towards `some' signal yield
at `some' mass value. To obtain a relative likelihood estimator, the fit 
is repeated by fixing the signal yield to zero, and the difference between
the log-likelihood estimators of the two fits is computed (the fit with 
free signal yield and mean mass always has a larger log-likelihood value,
so that the difference is positive). The p-value of the observed log-likelihood
difference is obtained by repeating the same exercise many times with a 
background-only Monte Carlo model faithfully describing the data. This 
Monte Carlo model is obtained by using the results from the background 
parametrisation obtained by the fit to data. The p-value is given by 
the ratio of the number of cases the log-likelihood difference in the 
Monte Carlo is found to be larger than the one in the data, divided by 
the total number of trials. 

\subsection*{Supersymmetry}

In spite of the many creative and interesting new physics models that appeared in 
recent years, supersymmetry remains the most popular Standard Model extension. 
It features an elegant solution of the hierarchy problem by 
cancelling the diverging weak boson radiative corrections to all orders
(where, however, a logarithmic divergence remains due to supersymmetry breaking),
a dark matter candidate, natural elementary scalar particles, the 
democratisation of the fermionic and bosonic degrees of freedom, and grand 
unification of the electroweak and strong forces. The Minimal supersymmetric
Standard Model introduces a conserved supersymmetry-parity, denoted {\em R-parity}, 
which is even for all Standard Model particles (including a Higgs doublet), and 
odd for all supersymmetric partners of these.\footnote
{
  {\em R}-parity, defined by $R = (-1)^{2S+3B+L}$, has been originally introduced 
  to avoid the proton decay $p \to e^+\piz$, which is possible in supersymmetry. 
} 
A consequence of {\em R}-parity conservation is that the lightest supersymmetric
particle (LSP) is stable. Since we have not observed any strongly or 
electromagnetically interacting particles in the universe that are not 
included in the Standard Model, and because we need a cold dark matter candidate,
it is assumed that the LSP is weakly interacting only (as are neutrinos). The
primary LSP candidate is the lightest neutralino, a linear combination 
of gauginos. In much of the supersymmetry parameter space the neutralino is 
a mixture of photino and zino, but could also be a gravitino. {\em R}-parity 
conservation also implies that supersymmetric particles can only be produced
in pairs. Hence, to produce supersymmetry in a hadronic interaction the 
centre-of-mass energy of the colliding partons must be twice the characteristic 
supersymmetric mass scale.

\begin{figure}[t]
  \begin{center}
	  \includegraphics[width=1\textwidth]{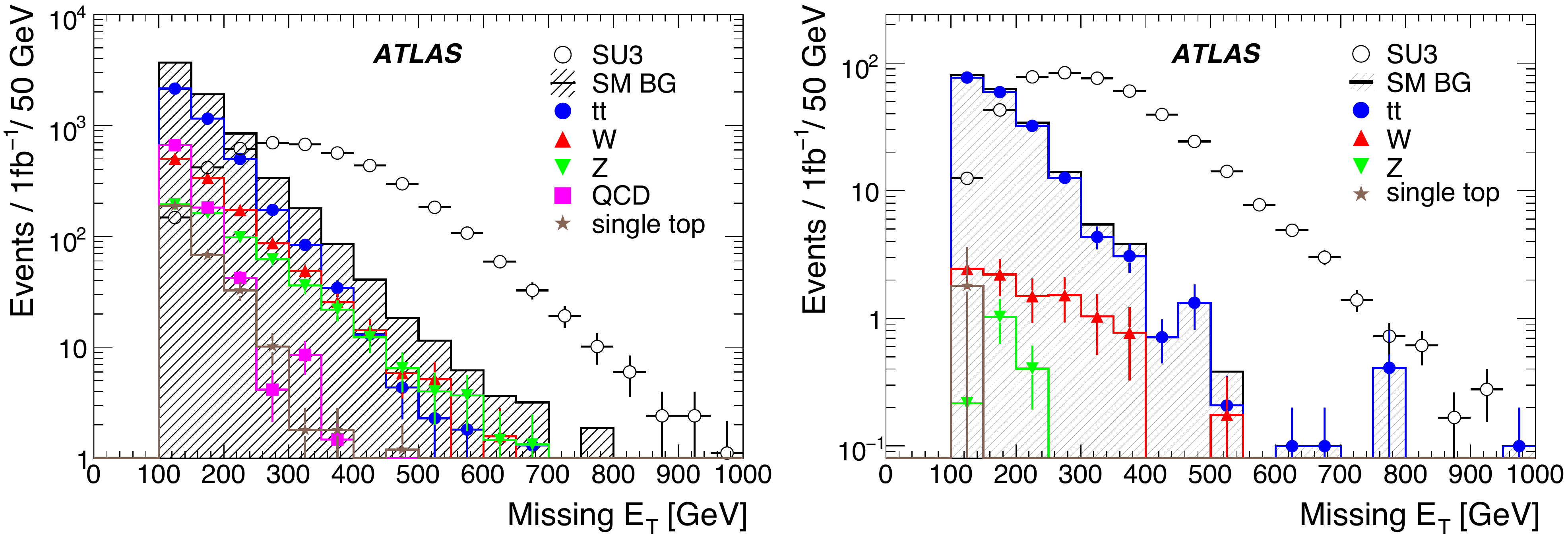}
  \end{center}
  \vspace{-0.4cm}
  \caption[.]{Simulated distributions of missing transverse energy in ATLAS for 
              analyses without (left) and with (right) requiring a reconstructed
              lepton (electron or muon) in the detector. Shown are the contributions
              from Standard Model processes and for {\em R}-parity conserving 
              supersymmetry using a minimal supergravity model (open circles) 
              with the parameters $m_0=100$\,\GeV and $m_{1/2}=300$\,\GeV. The 
              number of events corresponds to 10\invfb integrated luminosity. }
\label{fig:susymissingetatlas}
\end{figure}
A typical decay cascade of a supersymmetric squark or gluino is depicted
in the right-hand plot of Fig.~\ref{fig:topandstopdiagrams} on 
page~\pageref{fig:topandstopdiagrams}. From the diagram one notices that 
supersymmetric events produce many high-$p_T$ jets, sometimes leptons, and 
always missing transverse energy due to the escaping LSP (unless it escapes along
the beam pipe). Since 
squark and gluinos are produced by strong interactions with ${\cal O}$(picobarn)
cross sections if their masses are well below a \TeV, and because supersymmetric
events have a clear experimental signature, supersymmetry could be detected quite 
early. An integrated luminosity of 100\invpb is expected to be sufficient for 
a discovery of relatively low-mass supersymmetry, provided that the 
Standard Model backgrounds can be well controlled. 

\begin{wrapfigure}{R}{0.37\textwidth}
  \vspace{-24pt}
  \begin{center}
	  \includegraphics[width=0.37\textwidth]{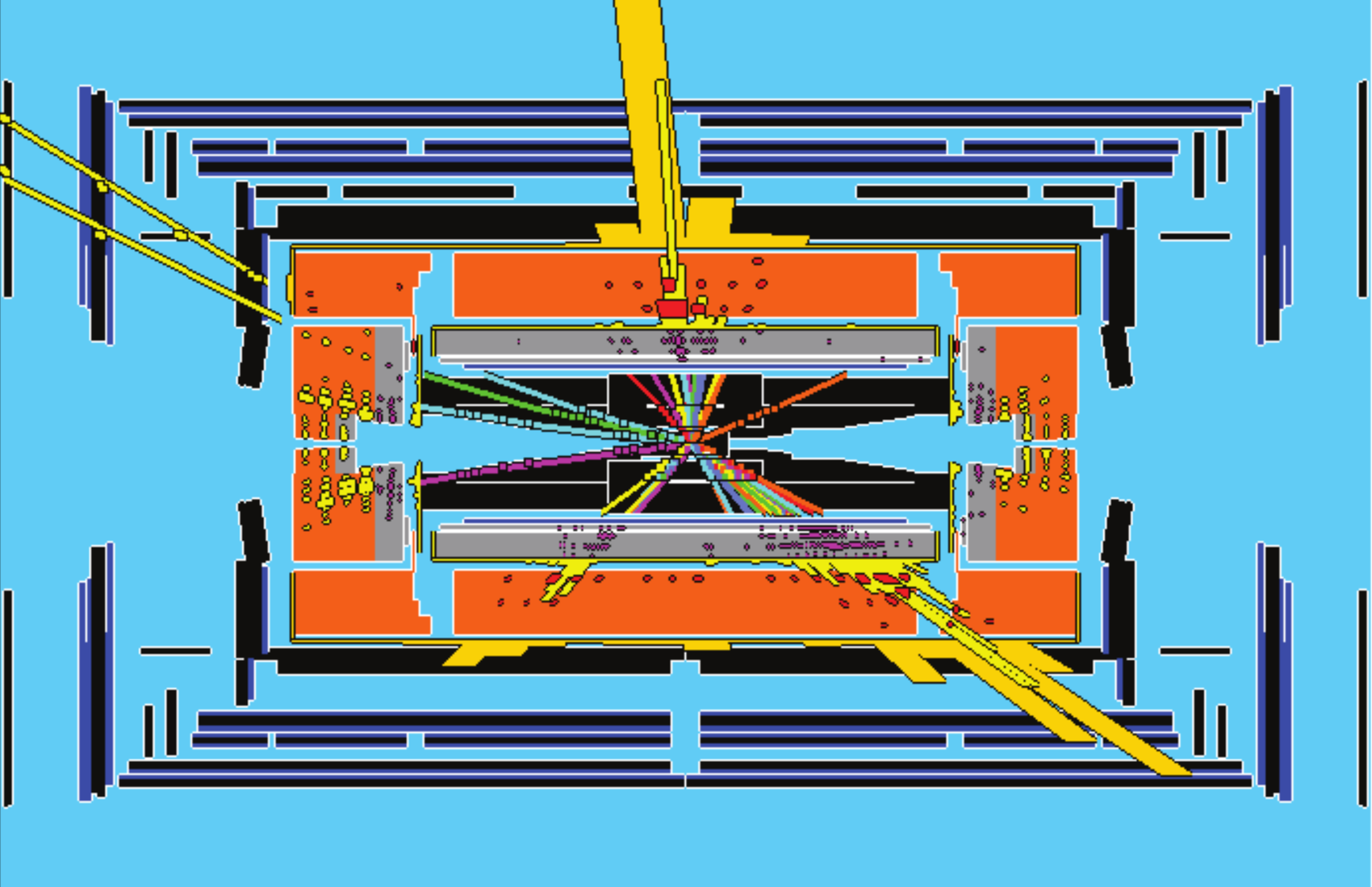}
  \end{center}
  \vspace{-10pt}
  \caption{Simulated supersymmetric event in ATLAS with six particle jets and 
           two muons with opposite charges in the final state, 
           and with large missing transverse energy. }
  \label{fig:susyeventdisplayatlas}
  \vspace{-6pt}
\end{wrapfigure}
Figure~\ref{fig:susymissingetatlas} shows distributions of missing transverse 
energy in ATLAS for simulated supersymmetry signal and Standard Model background
events, and for analyses with (right panel) and without (left panel) requiring 
a reconstructed electron or muon. A clear signal excess is perceptible
in both analyses, but the main Standard Model backgrounds differ significantly
between the two. Whereas without 
lepton requirement, $\ttbar$, and $W$ and $Z$ plus jets backgrounds are of 
similar size in the large-$\Emiss_T$ tails, and background from jets (QCD)
is also present, the background in the one-lepton analysis is entirely 
dominated by $\ttbar$, with some small contributions from $W$ and jets, but 
no QCD jets background. This makes the one-lepton analysis particular interesting
for the initial running period, when the understanding of the inclusive QCD 
background is still immature. 

Other discriminating
variables used in supersymmetry searches are the `effective mass', which is 
the scalar sum of the transverse momenta of all jets and leptons (other variations
of this variable also include $\Emiss_T$, or do not include the lepton momentum),
and the transverse mass (see Eq.~(\ref{eq:transversemass}) on page~\pageref{eq:transversemass})
which is particularly useful to reduce background from events with a $W$. 
Figure~\ref{fig:susyeventdisplayatlas} shows and event display of a typical 
supersymmetric event with jets, muons and large $\Emiss_T$ in ATLAS. 

\begin{figure}[t]
  \begin{center}
	  \includegraphics[width=0.7\textwidth]{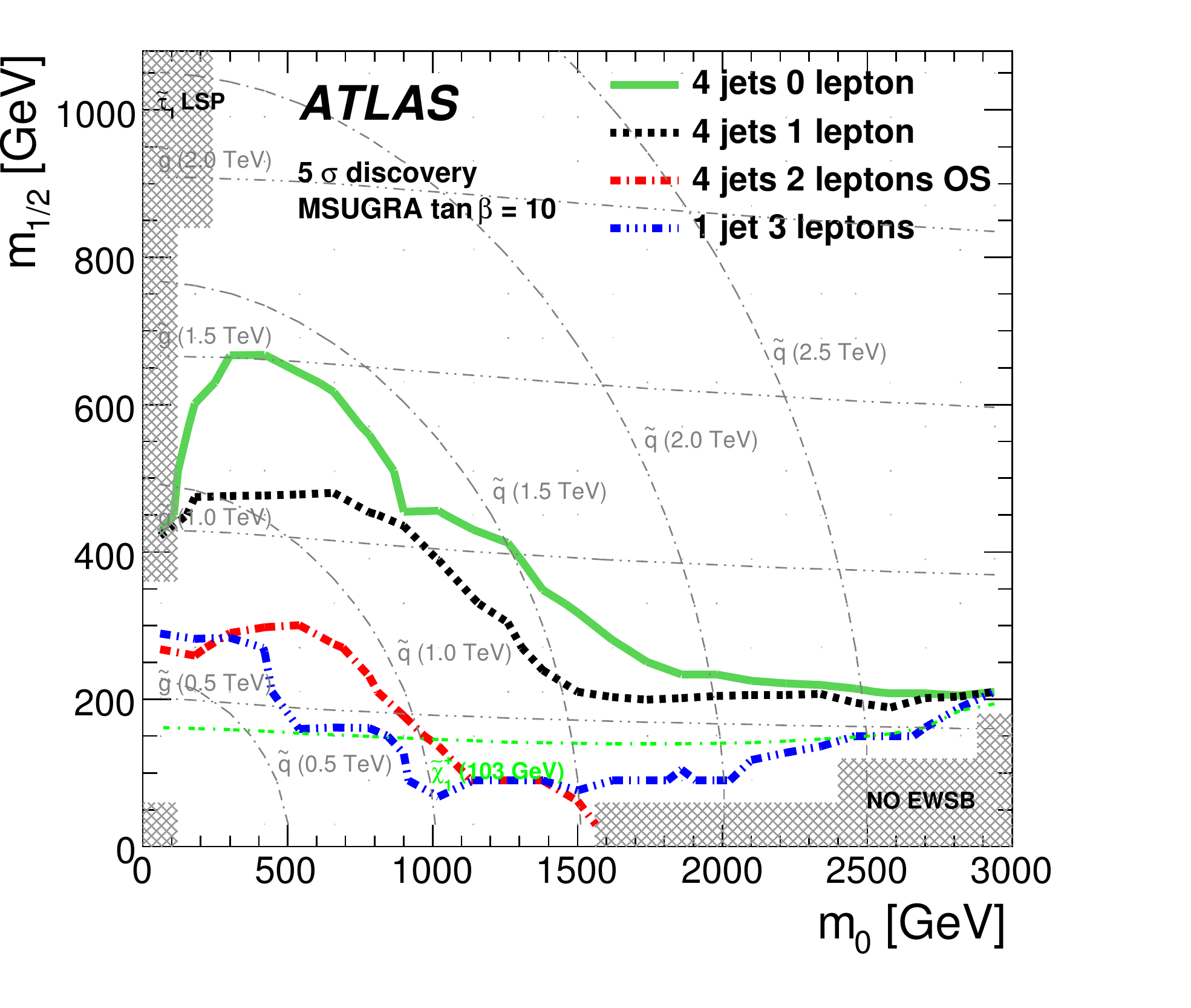}
  \end{center}
  \vspace{-0.8cm}
  \caption[.]{Expected contours for $5\sigma$ minimal supergravity discovery versus 
              the GUT mass parameters $m_{1/2}$ and $m_0$, for analyses 
              with various lepton requirements and for an integrated luminosity of 1\invfb.
              The grey dashed contour lines indicate the corresponding squark and 
              gluino masses. The zero-lepton analysis has the best discovery reach.}
\label{fig:susydiscoverypotentialatlas}
\end{figure}
Figure~\ref{fig:susydiscoverypotentialatlas} shows the expected discovery potential
for the minimal supergravity model as a function of the GUT mass parameters 
$m_0$ and $m_{1/2}$ (ATLAS study). The zero-lepton analysis has the best discovery reach.
However, taking into account the experimental difficulties of this mode, the 
one-lepton mode may reveal competitive. Squarks and gluinos with masses up to 
0.75, 1.35, 1.8\,\TeV can be discovered with integrated luminosities of 0.1, 
1 and 1\invfb, respectively, using the four-jets, zero-lepton analysis.

We should note that supersymmetry could also break {\em R}-parity. The signature 
could be taus originating from $\chi^0_1\to\tilde{\tau}\tau$ decays. Moreover, signals 
due to other phenomena could appear like supersymmetry so that a (challenging) 
neutralino spin analysis needs to be performed to reveal their fermionic nature.
Experimentalists should proceed the search for supersymmetry as model-independently 
as possible, and watch out for anomalies, \eg, the occurrence of photons, taus or 
strange tops

\subsection*{Strong gravity}

\begin{wrapfigure}{R}{0.37\textwidth}
  \vspace{-24pt}
  \begin{center}
	  \includegraphics[width=0.37\textwidth]{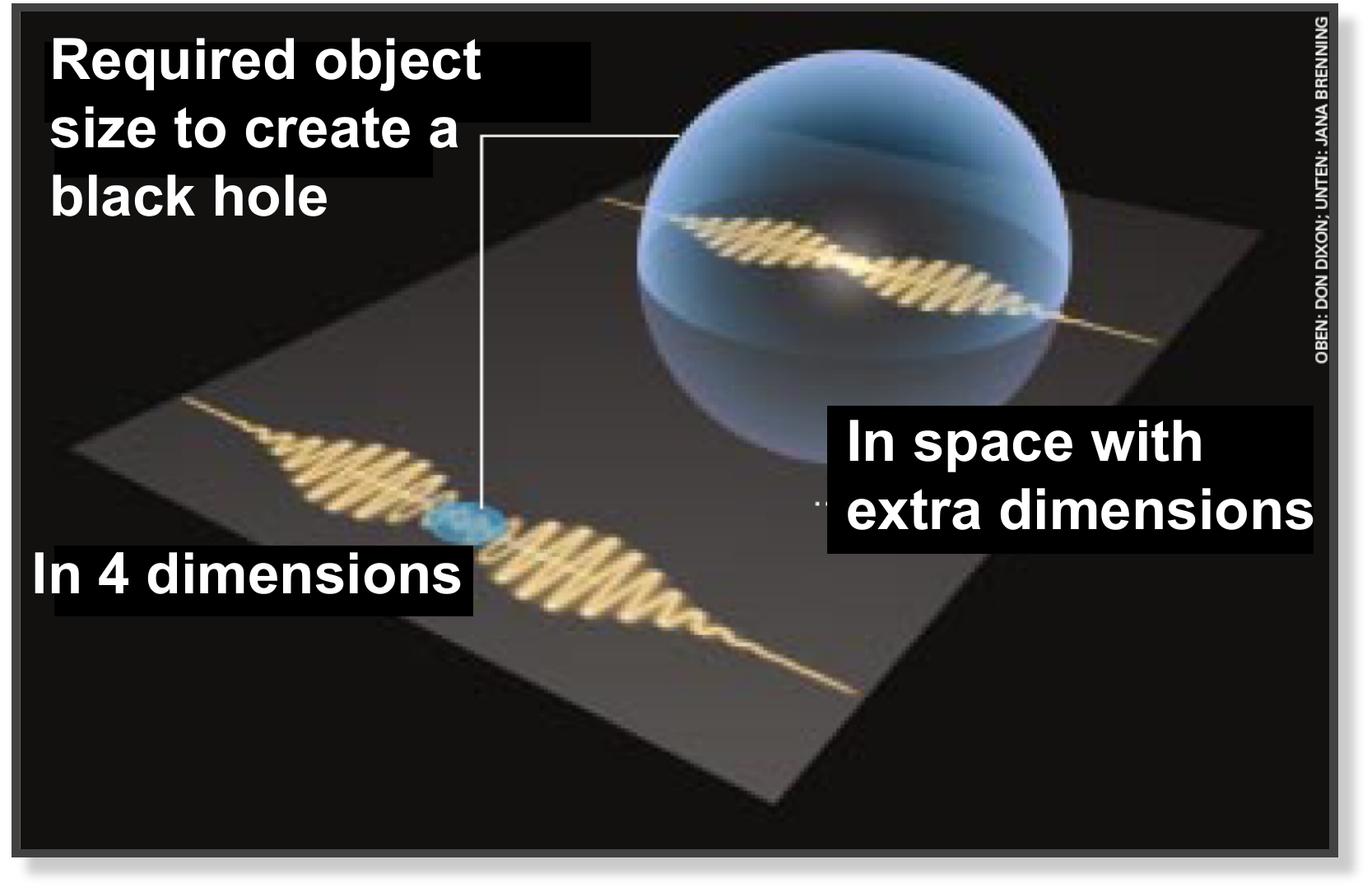}
  \end{center}
  \vspace{-10pt}
  \caption{Schwarzschild radius in 4 and $4+d$ dimensions. }
  \label{fig:blackholesketch}
  \vspace{+25pt}
  \vspace{-24pt}
  \begin{center}
	  \includegraphics[width=0.37\textwidth]{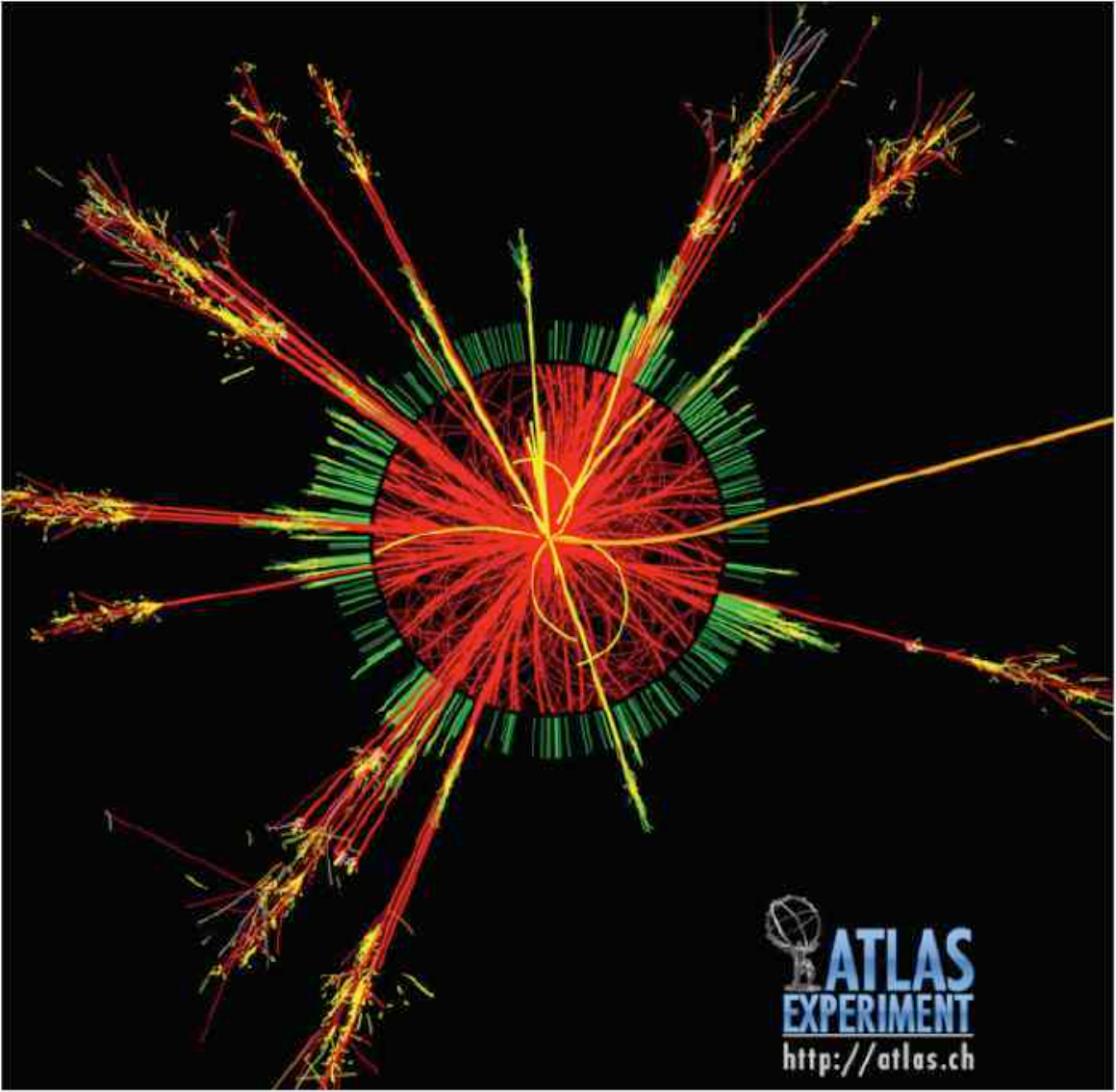}
  \end{center}
  \vspace{-10pt}
  \caption{Simulation of a black hole decay in ATLAS.}
  \label{fig:bheventdisplayatlas}
  \vspace{-6pt}
\end{wrapfigure}
Finally, if we are allowed to enter trans-Planck scales, that is, gravity in compact extra 
spatial dimensions is strong enough to reduce the Planck scale to energies reached
by the LHC, hard-scattering proton-proton collisions may produce microscopic black holes. 
An object becomes a black hole if it is smaller than the Schwarzschild radius $r =  2G M / c^2$.
In $4+d$ spatial dimensions the Schwarzschild radius\footnote
{
   The Schwarzschild radius is the radius below which the gravitational attraction between 
   the particles of a body is so strong that the body undergoes gravitational collapse. 
   For a typical star such as the Sun, the Schwarzschild radius is about 3\,km.
} 
becomes  $r = 2G^{(4+d)} M_D / c^2$,
where $G^{(4+d)}$ is a gravitational constant in the full-dimensional space.
The four-dimensional constant $G$ is thus only a reflection of the real gravitational 
constant $G^{4+d)}$, reduced (`diluted')  by the extra dimensions. The Planck scale 
is no longer fundamental. If $M_D \approx M_{\rm Planck}^{(4+d)} \approx 1$\,\TeV, a black 
hole can be produced by the LHC if the momentum transfer of the hard scattering 
reaction exceeds $M_D$. The cross section of the black hole production 
is $\sigma_{\rm BH} \approx \pi r^2$. With $M_D \sim 2$--3\,\TeV one finds 
$\sigma_{\rm BH} \sim {\cal O}$(pb) allowing a fast discovery for $M_{\rm BH} < 4$\,\TeV,
and $ d = 2$--6. 

The black hole undergoes a fast ($\tau_{\rm BH}\sim 10^{-27}$\,s) thermal decay via 
Hawking radiation of temperature $T_H \sim M_D \cdot (M_D/M_{\rm BH})^{1/(d+1)}$ 
(a microscopic black hole is not black at all!). The life cycle of a 10\,\TeV 
black hole could be sketched as follows: 
$(i)$ $\Delta t= 0$, $M_{\rm BH}(\Delta t)= 10$\,\TeV:  {\em creation}\, ---
the micro black hole is created in a $pp$ collision: it is asymmetric, may 
vibrate and rotate, and may be electrically charged;
$(ii)$ $\Delta t= 0$\,--\,$1\cdot10^{-27}$\,s, $M_{\rm BH}(\Delta t) 10\,--\,8$\,\TeV:  {\em `baldness phase'} ---
emission of gravitational and electromagnetic waves, and charged particles,
the black hole is solely characterised by mass and angular momentum;
$(iii)$ $\Delta t= 1$\,--\,$3\cdot10^{-27}$\,s, $M_{\rm BH}(\Delta t)= 8$\,--\,6\,\TeV:  {\em slowing down}\, ---
the black hole radiates by reducing its angular momentum, its form becomes spherical\,;
$(iv)$ $\Delta t= 3$\,--\,$20\cdot10^{-27}$\,s, $M_{\rm BH}(\Delta t)= 6$\,--\,2\,\TeV:  {\em Schwarzschild phase}\, ---
after loosing its angular momentum, the micro black hole evaporates its mass via 
Hawking radiation;
$(v)$ $\Delta t= 20$\,--\,$22\cdot10^{-27}$\,s, $M_{\rm BH}(\Delta t)= 2$\,--\,0\,\TeV:  {\em Planck phase}\, --- 
the black hole shrinks down to the Planck mass ($M_D$) and fully decays into 
all particles with probabilities according to their degrees of freedom
(\cf Fig.~\ref{fig:bheventdisplayatlas}). The spectacular decay signature 
cannot be missed by the experiments.

\newpage
\section{Conclusions and outlook}
\label{sec:outlook}

\begin{wrapfigure}{R}{0.37\textwidth}
  \vspace{-24pt}
  \begin{center}
	  \includegraphics[width=0.37\textwidth]{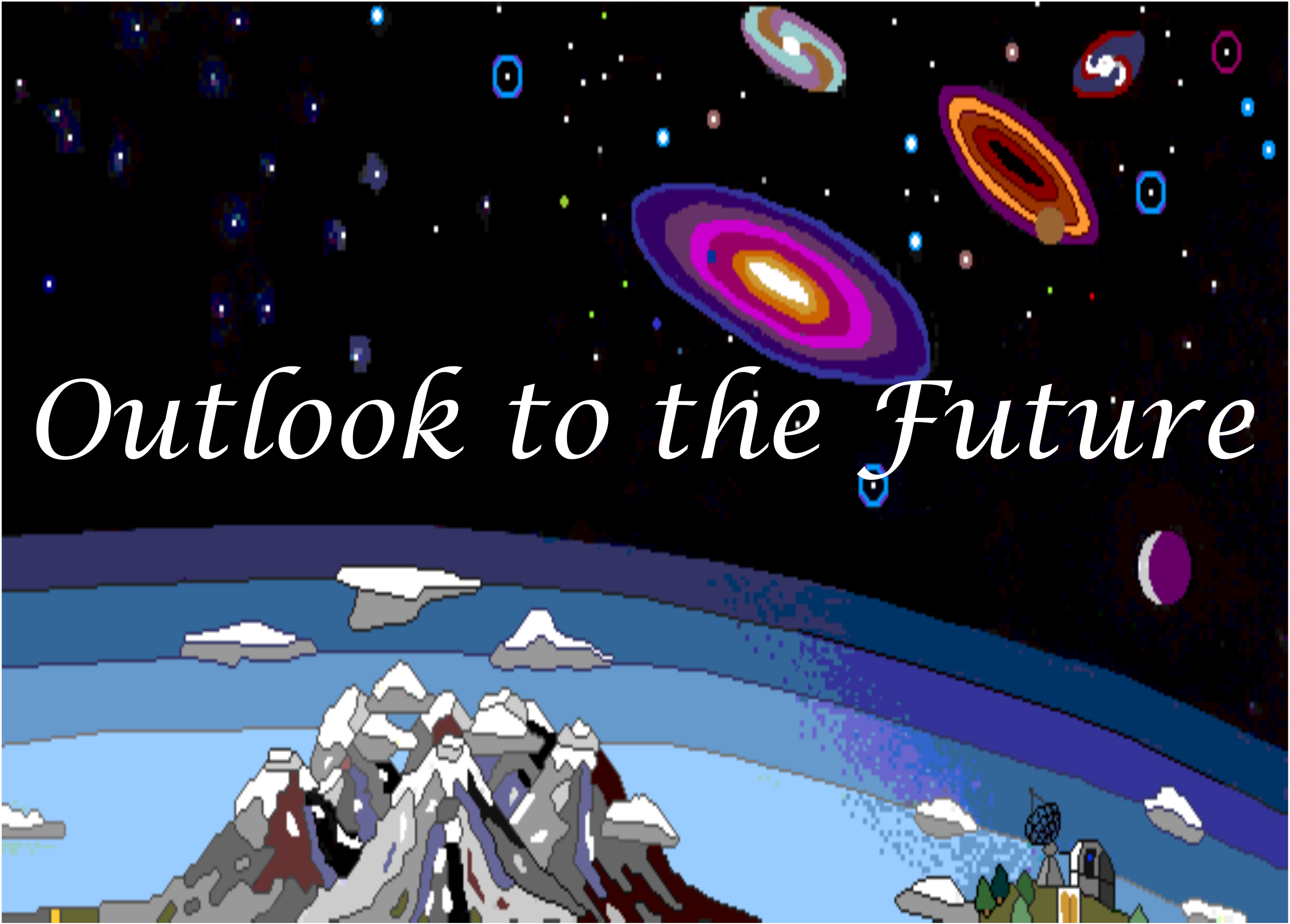}
  \end{center}
  \vspace{-10pt}
  \vspace{-6pt}
\end{wrapfigure}
Commissioning such tremendously complex apparati as the LHC high-$p_T$ experiments
ATLAS and CMS is a continuous challenge. It starts far earlier than with the installation
of the experiments in their underground caverns. To some extent it already 
begins with the design phase, when prototypes are drawn, simulated and eventually 
built for the purpose of testing and optimisation. Commissioning continues
in dedicated test beams where parts or even complete slices of the detectors, 
modelling as accurately as possible the final geometry, are assembled. While installing
the detectors at their final locations commissioning campaigns with cosmic ray events
are undertaken. Hundreds of million cosmic rays have been recorded by both experiments 
in roughly three years of data taking with more and more complete detectors. 
Finally, with the start of the LHC commissioning, single beams with 
900\,\GeV injection energy are sent through both LHC beam pipes, 
circulating or as beam-on-collimator `splash' dumps, radio-frequency
captured or not. Later two beams are injected, again at injection energy, radio-frequency
captured and brought to collision. These collisions produce for the first time
so-called minimum bias events, producing roughly 20 tracks in the inner tracking 
systems, some photons from \piz and $\eta$ decays, and electrons from photon 
conversion, as well as rare jet events and muons from pion and kaon decays. The
beams will not be squeezed at this initial stage so that due to the large beam 
spot, the small number of bunches in the machine, and the low bunch intensity 
the peak luminosity will not exceed \Lumi27. However, once the LHC 
decides to ramp the energy, the relativistic contraction of the beam will lead to
an increase in the luminosity, and the experiments will see jet rates, as well as electrons 
and muons mainly from heavy quark decays and quarkonia go up. Moreover, beam 
squeezing (\ie, the reduction of the beam envelope by the magnet optics)
and a crossing angle between the colliding beams 
will further allow to increase the luminosity of the LHC at higher energy. 

With the data taken during these commissioning phases, the experiments have 
gained experience and obtained a good initial understanding of the detector response, 
and improved the quality of the data by calibrating and aligning the detector 
subsystems, which will pay off when analysing the first collision data for 
physics and detector performance analysis. With the arrival 
of physics data it is very important to continue improving the detector understanding,
and the faithfulness of its description by the detector response simulation. It is the key 
to a longterm success of the experiments, and to physics results with the 
smallest possible bias and systematic errors. It is also important that the experiments 
optimise the fraction of useful data taken, by steadily improving the data taking
efficiency of all detector systems, and aiming at the best achievable data quality. 

\begin{figure}[t]
  \begin{center}
	  \includegraphics[width=0.8\textwidth]{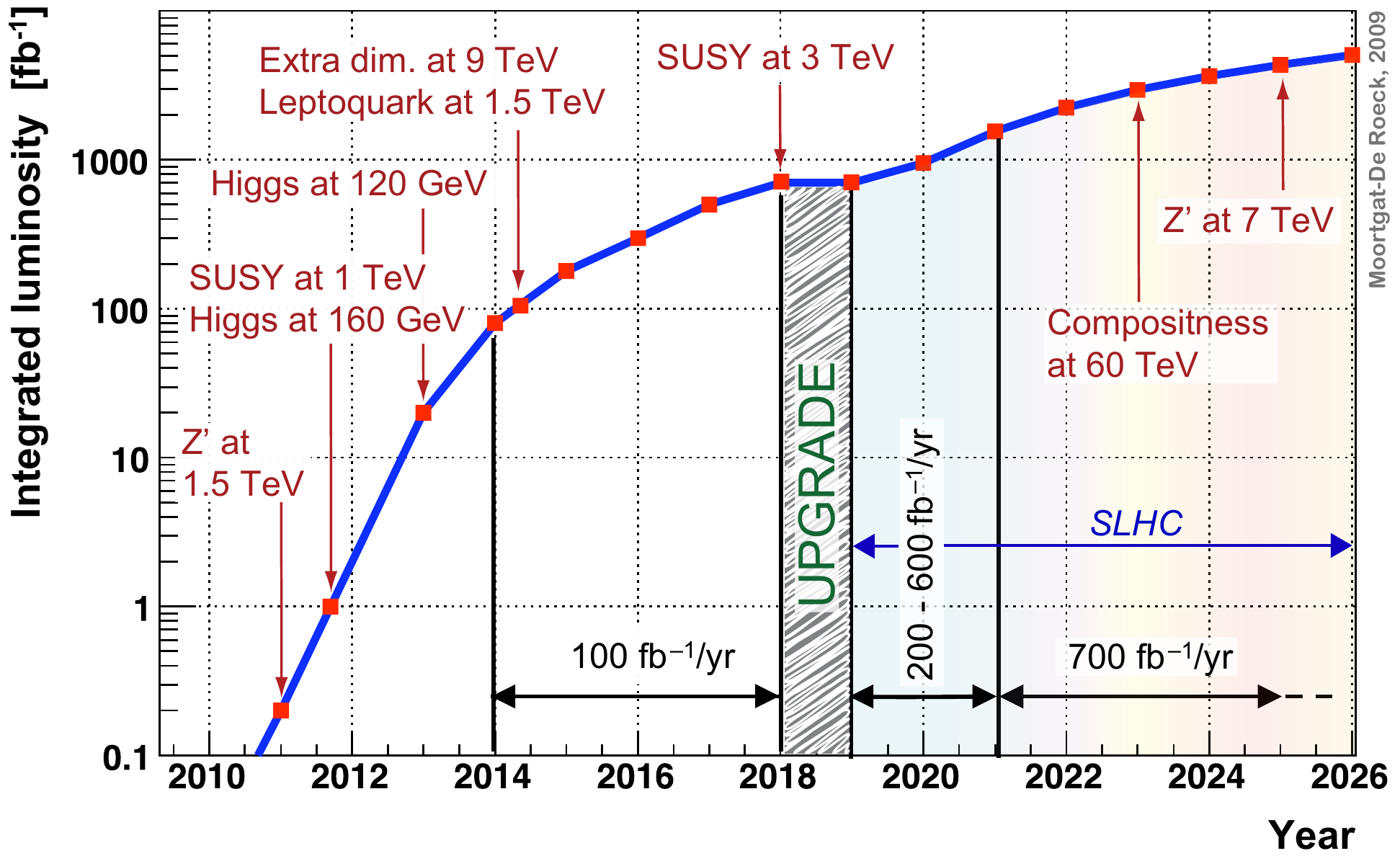}
  \end{center}
  \vspace{-0.5cm}
  \caption[.]{The LHC programme in a 
              nutshell~\cite{outlookfigure} (see text for discussion). }
\label{fig:lhcoutlook}
\end{figure}

Figure~\ref{fig:lhcoutlook} gives an exploratory view of the expected LHC performance
versus year of (design) operation at 14\,\TeV centre-of-mass energy~\cite{outlookfigure}, 
and the corresponding sensitivity for discovery of various phenomena by the ATLAS and 
CMS experiments. After accumulating 1\invfb integrated luminosity, 
minimal supersymmetry with up 1\,\TeV characteristic mass scale could be discovered. 
The Standard Model Higgs boson is expected to be observed at any mass with 30\invfb. 
With the ultimate integrated luminosity of possibly 500\invfb around the year 2018, 
the discovery reach for many 
new physics models can be pushed deep into the \TeV scale, and properties of earlier 
discoveries may be studied. Among these are the coupling strengths of the Higgs boson 
in various production and decay channels. If the Higgs is observed to decay into 
either $\gamma\gamma$ or $ZZ^{(\star)}$, one will know that it cannot have spin 1.
Observations of angular distributions and correlations in $ZZ^{(\star)}$ decays 
will enable the spin and \CP properties of the Higgs to be determined.
It should also be possible to constrain masses of supersymmetric particles, 
possibly even the spin of a heavy neutralino. A spin analysis of heavy resonances
decaying to di-leptons could be performed in case of a discovery. 
After 4 years at the highest peak luminosity with approximately 
100\invfb of data recorded each year, the increase in sensitivity becomes asymptotic
(recall the $1/\sqrt{\cal L}$ scaling of statistical errors),
which is the opportunity to undertake an upgrade of machine and detectors to the 
Super-LHC (SLHC). The SLHC programme proposes to increase the LHC peak luminosity to 
$1.5 \cdot$\Lumi35, \ie, 10 times the nominal LHC peak luminosity~\cite{slhc}. At 
nominal bunch pattern, it will compel the experiments to cope with 250 pile-up minimum 
bias interactions occurring in time with the hard-scattering event. This requires 
many changes to the detectors: $(i)$ reduce background rates by changing the beam pipe and 
improving the shielding, $(ii)$ improve the radiation and occupancy tolerance of the 
detectors and electronics, in many cases by replacing entire subsystems, and $(iii)$ 
increase the bandwidth of front-end and readout electronics to minimise pile-up 
and handle a 10 times increase in the event rate. A successful SLHC upgrade would 
allow the experiments to extend their discovery reach for supersymmetry and 
$Z^\prime$ bosons to 4\,\TeV and 7\,\TeV, respectively. 

The current situation however is that, due to limitations in the quench protection
system, the LHC will begin in 2010 with 7\,\TeV centre-of-mass energy, which later in 
the year may or may not be increased to a maximum of 10\,\TeV. ATLAS and CMS have performed
indicative studies to evaluate the impact of the reduced energy on their physics 
programme. It is expected that up to half an inverse femtobarn of data will be 
delivered in 2010. (If the run is continued through 2011 a total of one inverse femtobarn
of data could be delivered.)
Between 14\,\TeV and 10\,\TeV the number of selected $Z\to ee$ 
events will decrease from roughly 5,000 per 10\invpb integrated luminosity 
to 3,600 (linear relationship). The number of produced \ttbar 
events will drop by roughly a factor of 2, so that the sample size will attain
the one from the Tevatron after approximately 100\invpb at 10\,\TeV. The exclusion
of a Higgs boson requires about twice more integrated luminosity at 10\,\TeV
than at 14\,\TeV. A $5\sigma$ discovery of a Higgs with mass of 160\,\GeV (which is
unlikely) would require roughly 1\invfb of recorded physics data. To challenge
the Tevatron Higgs searches, a sample of about 200\invpb at 10\,\TeV is needed. The
sensitivity of the search for a heavy $Z^\prime$ is reduced by a factor of roughly 
3 at 10\,\TeV. A $5\sigma$ observation of a 1\,\TeV (Tevatron limit) weighing 
$Z^\prime$ would require roughly 100\invpb of 10\,\TeV collision data. To achieve
equivalent discovery reach for supersymmetry, a factor of 2 more integrated luminosity 
is required at 10\,\TeV centre-of-mass energy. Nevertheless, the current Tevatron 
limits can be improved with as little as 20\,\invpb of 10\,\TeV data. What would
be the impact of a 7\,\TeV centre-of-mass energy compared to 10\,\TeV? The number
of $Z\to ee$ will drop by a another factor of 1.4. The \ttbar rate will further drop 
by approximately a factor of 2. The required luminosity for equal search sensitivity 
for a $Z^\prime$ will increase by a factor of 3, similarly for supersymmetry
searches, and a factor of 2--3 for Higgs searches.




\end{document}